%% file: thesis.tex
\newif\ifDissLarge
\newif\ifDissDim
\newif\ifDissHead

\DissLargetrue  
\DissDimtrue     
\DissHeadtrue    

\ifDissLarge
  \documentclass[12pt,twoside,a4paper]{report}
  \IfFileExists{a4.cls}{\usepackage{a4}}{\relax}
\else
  \documentclass[11pt,twoside,a4paper]{report}
  \IfFileExists{a4.cls}{\usepackage{a4}}{\relax}
\fi

\newcommand{\MaquaHeadings}
{
  \ifDissHead  
    \pagestyle{fancyplain}
    \lhead[\fancyplain{}{\bfseries\thepage}]
          {\fancyplain{}{\bfseries\sffamily\let\uppercase\relax\leftmark}}
    \rhead[\fancyplain{}{\bfseries\sffamily\let\uppercase\relax\rightmark}]
          {\fancyplain{}{\bfseries\thepage}}
    \cfoot{\fancyplain{\bfseries\thepage}{}}
  \else
    \pagestyle{headings}
  \fi
}

\ifDissDim
   \IfFileExists{typearea.sty}{\relax}{\DissDimfalse}
\fi
\ifDissDim
  \usepackage[a4paper,headinclude,DIV14,BCOR5mm]{typearea}
\else
\fi

\ifDissHead
   \IfFileExists{fancyheadings.sty}{\relax}{\DissHeadfalse}
\fi
\ifDissHead
  \usepackage{fancyheadings}
\fi
\MaquaHeadings

\usepackage{latexsym}
\usepackage[dvips]{graphicx}
\usepackage{bbm}                      
\usepackage{ifthen}                   
\usepackage{exscale}                  
\usepackage[intlimits]{amsmath}       
\usepackage{amsfonts}
\usepackage{amssymb,amscd}
\usepackage[dvips]{epsfig}                   
\usepackage{array}
\usepackage{wrapfig}
\usepackage{graphics}

\newcommand*{\pd}{\partial}
\newcommand*{\pdm}{\pd_{\mu}}
\newcommand*{\pdn}{\pd_{\nu}}

\newcommand*{\be}{\begin{equation}}
\newcommand*{\ee}{\end{equation}}

\newcommand*{\bea}{\begin{eqnarray}}
\newcommand*{\eea}{\end{eqnarray}}

\newcommand*{\pref}[1]{(\ref{#1})}

\newcommand*{\prefr}[2]{(\ref{#1}-\ref{#2})}

\newcommand*{\brst}{\mathrm{BRST}}
\newcommand*{\tl}{\mathrm{tl}}

\newcommand*{\indexsep}{,}

\newcommand*{\piclinecol}{red }

\begin{document}
\include{frontmatter}
\include{abstract}
\vspace*{0.1cm}\newpage
\include{eprintnote}
\pagenumbering{roman}
\tableofcontents
\newpage
\pagenumbering{arabic}
\include{intro}
\include{theory}
\include{temperature}
\include{3dlimit}
\include{ft}
\include{derived}
\include{summary}
\include{ack}
\appendix
\include{conventions}
\include{dse}
\include{kernels}
\include{ir}
\include{uv}
\include{num}
\include{bib}
\end{document}

%% file: frontmatter.tex
\begin{titlepage}
\begin{center}
\vspace*{1cm}
{\bf \LARGE The High-Temperature Phase\\ of\\\vspace*{0.3cm} Yang-Mills Theory in Landau Gauge }\\
\vspace*{3cm}
Doctoral Thesis\\
\vspace*{2cm}
Axel Maas\\
\vspace*{2cm}
Institute of Nuclear Physics\\
Darmstadt University of Technology\\
Schlo{\ss}gartenstra{\ss}e 9\\
D-64289 Darmstadt\\
\vspace*{2cm}
2004\\

\end{center}
\end{titlepage}
\newpage
\vspace*{0.1cm}

%% file: abstract.tex
\thispagestyle{empty}

\vspace{0cm}

\begin{center}
\begin{large}
{\bf The high-temperature phase of\\ Yang-Mills theory in Landau gauge}
\end{large}
\vspace{0.5cm}
\end{center}

\vspace{0.3cm}

\centerline{\large Abstract}
\normalsize
\noindent The finite and high temperature equilibrium properties of Yang-Mills theory in Landau gauge are studied. Special attention is paid to the fate of confinement and the infrared properties at high temperatures. The method implemented are the equations of motion, the Dyson-Schwinger equations. A specific approximation scheme is introduced, which was previously applied successfully to the vacuum.

In a first step, the infinite temperature limit is taken. The theory reduces to a 3-dimensional Yang-Mills theory coupled to a massive adjoint Higgs field. The equations for the propagators of the Higgs, the gluon, and the Faddeev-Popov ghost are obtained. They are solved in the infrared analytically and at least in the Yang-Mills sector confinement is found. Therefore the high-temperature phase of a 4-dimensional Yang-Mills theory is non-trivial and strongly interacting. Solutions for all propagators are obtained numerically at all momenta. Thereby also the propagators of a pure 3-dimensional Yang-Mills theory are determined. Systematic studies find only quantitative effects of the errors, which are induced by the approximations. Good agreement to lattice calculations is found.

Finite temperatures down to the regime of the phase transition are investigated. It is found that the infrared properties are only quantitatively affected, and confinement of gluons transverse to the heat bath is established. The hard modes are nearly inert even at temperatures of the order of the phase transition temperature. Therefore the infinite temperature limit is a good approximation already at temperatures a few times the critical temperature, in agreement with lattice calculations. 

Finally quantities derived from the propagators are studied. The Schwinger functions are calculated. It is found that also the gluons longitudinal with respect to the heat bath are strongly influenced by higher order or even genuine non-perturbative effects, even in the infinite temperature limit. The analytic structure of the gluon propagator is investigated, and it is found that at least gluons transverse to the heat bath comply with the Kugo-Ojima and Zwanziger-Gribov confinement scenarios. Investigating the thermodynamic potential, an approximate Stefan-Boltzmann-like behavior is found. The thermodynamic potential, but not necessarily the pressure, is dominated by the hard modes.

By comparison with calculations below the phase transition and lattice calculations it is conjectured that Yang-Mills theory likely undergoes a first order phase transition, which changes a strongly interacting system into another. The phases differ mainly by the properties of the chromoelectric sector.

\vspace{0.3cm}

\thispagestyle{empty}

%% file: eprintnote.tex
\thispagestyle{empty}

\vspace{0.3cm}

\centerline{\large Note added to the e-print version}

This thesis has been submitted to the faculty of physics of the Darmstadt University of Technology on the $19^{th}$ of October 2004. It has been defended successfully on the $1^{st}$ of December 2004. The supervisor was Prof.\ Jochen Wambach. This e-print version has some formal changes compared to the accepted version, e.g.\ the german abstract has been removed. The accepted offical version can be obtained from the Universit\"ats- und Landesbibliothek of the Darmstadt University of Technology online. The URL is ``http://elib.tu-darmstadt.de/diss/000504/''.

Most parts of the chapters \ref{c3d} and \ref{cderived} have been published in \cite{Maas:2004se}. The unsettled problem discussed in subsection \ref{scttrunc} has been resolved since the submission of this thesis. The method implemented here has been demonstrated to yield the correct limit for the number of Matsubara frequencies $N$ going to infinity, $N\to\infty$. This will be discussed in some more detail in an upcoming publication by A.~Maas, J.~Wambach, and R.~Alkofer, as will be most of chapter \ref{cft}.


%% file: intro.tex
\chapter{Introduction}

\section{Strong Interactions}

Nature is described by only three fundamental forces according to the current knowledge. Each of those covers an own realm of physics and each one comes with its own problems. These forces are the electroweak and strong interactions, building together the standard model of particle physics, and gravitation \cite{Bohm:yx,Gravitation}.

Gravitation describes the behavior of macroscopic objects to the largest distances known and it determines the gross properties of the universe at the present time. Its formulation in the general theory of relativity \cite{Gravitation} has been supported by overwhelming experimental evidence. It is the weakest of the forces and quite well understood in the classical regime. On the other hand, there is still no successful implementation of a quantum theory of the tensorial gravitation field. Also several observations indicate that the current state of the universe is not only determined by the matter fields of the standard model. The nature and interactions of approximately 95\% of the universe are not known today.

Electromagnetism, electroweak symmetry breaking, weak processes like the $\beta$-decay, neutrino-oscillations and various other effects are due to electroweak interactions as formulated by the Glashow-Salam-Weinberg theory \cite{Bohm:yx}. It is rather well understood and its treatment using perturbative methods has been tested with high experimental precision. Although being rather tractable in general it is demanding in detail. Fundamental questions are still posed by the absence of the Higgs in experiments up to now and the origin of the large number of parameters.

The interaction binding together nucleons to build nuclei, thus forming the core of atoms, is known as the strong interaction. It is the strongest of all known forces. It describes also the way in which nucleons and all other hadrons are made out of their constituents, the quarks and gluons. The theory of strong interactions in its quantized form is termed quantum chromodynamics (QCD) \cite{Bohm:yx}, as quarks and gluons carry a charge termed color. In contrast to the electroweak theory, which includes electromagnetism, and gravitation, the strong force does not appear on a macroscopic level beyond its bound state spectrum.

QCD offers a rich set of phenomena, which are not yet really understood. Primarily, a complete first-principle calculation of the bound-state spectrum of QCD is still lacking. Such a calculation must be able to explain why some bound-states like mesons and baryons appear in large numbers but why others like glue-balls and hybrids have not yet been convincingly found or appear only in rather small numbers like penta-quarks and meson-molecules possibly observed recently \cite{Hicks:2004vd}.

At a more fundamental level, three properties of QCD are most striking. Two are connected to the fermionic content of the theory: The breaking of chiral symmetry and the axial anomaly. The first phenomenon gives rise to the proton mass of nearly 1 GeV, while the quarks making it up have masses only of the order of MeV. This can be understood as due to the spontaneous breaking of the approximate chiral symmetry of the lightest quarks and is also well known in other models. The axial anomaly is connected to a true quantum anomaly, and e.g. gives rise to the anomalous large mass of the $\eta'$-meson.

The third property is confinement, the absence of the colored degrees of freedom from the physical spectrum. It is this property which significantly shapes the low-energy reality of daily life while simultaneously being least understood of all the genuine non-perturbative effects of QCD. At the same time it is one of the properties which has been measured with the highest precision available: Free quarks have a unique experimental signature due to their fractional electric charge. The absence of such objects in nature has been established at a precision of the order of $1:10^{30}$ \cite{Hagiwara:2002fs}.

The true non-perturbative nature of QCD is the reason for the complications in the study of these phenomena. Perturbation theory describes strong effects reasonably well only at energies larger than a few GeV, the precise scale depending on the process. At smaller energies, which ultimately govern hadron and nuclear physics, the interaction is so strong that perturbation theory is not applicable. Therefore different approaches are needed.

Extensive results are available from model calculations. However, only few attempts of first-principle approaches have been employed. Out of these, lattice gauge theory \cite{Montvay:1994cy} is by far the most successful and widespread, and several achievements have been made using it. Lattice calculations are a numerical tool which discretizes space-time and thus is able to calculate the partition function directly, albeit numerically expensively. Especially fermionic contributions pose significant problems. Furthermore lattice calculations are restricted at best to two orders of magnitude in momenta and rather small volumes due to the limitation by computing time.

As many interesting phenomena of QCD are likely to be generated in the far infrared and include singularities, a continuum formulation is desirable. It is provided e.g.\/ by the equations of motions in form of the Dyson-Schwinger equations. It is this method which is implemented here and it will be discussed in more detail in chapter \ref{cdse}. Compared to lattice calculations, this approach suffers from a larger number of necessary assumptions to make it technically treatable. At the same time it allows the direct investigation of the continuum manifestation of non-perturbative effects. Both these methods and several other approaches are complementary and necessary to untangle the complex structure of low-energy QCD.

\section{Thermodynamics of QCD}\label{stdqcd}

The phase structure of QCD is a long-standing problem \cite{Cabibbo:1975ig}. Naively, deconfinement would be expected in a sufficiently hot medium in a simple picture like the bag-model \cite{Wong:1995jf}. The properties of such a different phase of QCD would grant additional insight into the mechanisms of strong interactions. Such a medium is also expected to have existed at the beginning of the universe, and its specific properties are of relevance for cosmology. Therefore great experimental efforts have been undertaken in the last decades to generate it. Temperatures of around 100-200 MeV have been reached in heavy-ion collisions, but with large experimental uncertainties on the highest temperature achieved, especially as the thermalization process is not yet clear \cite{Andronic:2004tx}.

This has also led to great theoretical efforts to solve the many-body problem of QCD. However, up to now the results generate more questions than answers. This is especially true for the non-equilibrium and dynamical part of the experiments, but even for the equilibrated medium it is still the case.

\begin{figure}[ht]
\epsfig{file=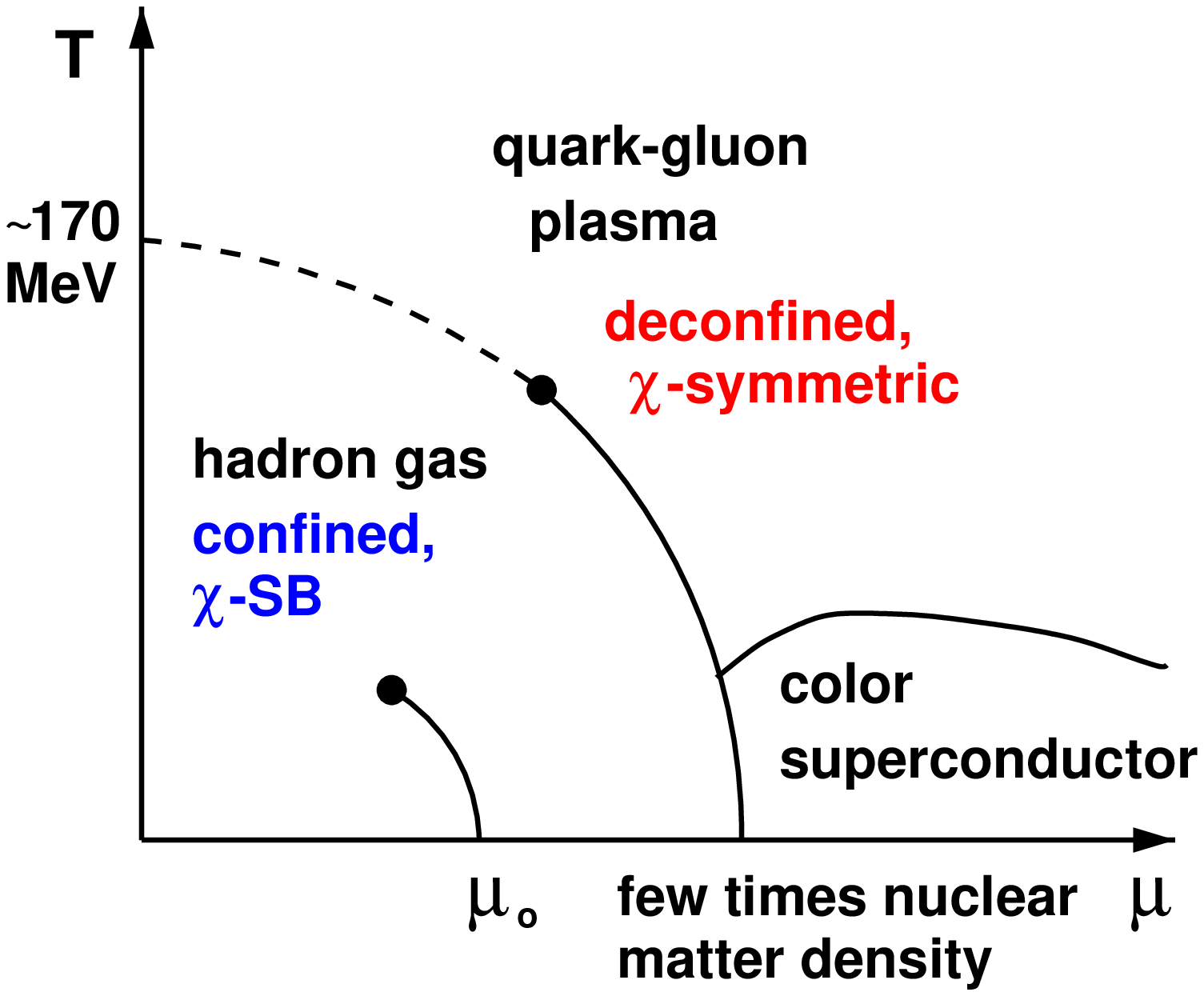,width=0.5\linewidth}\epsfig{file=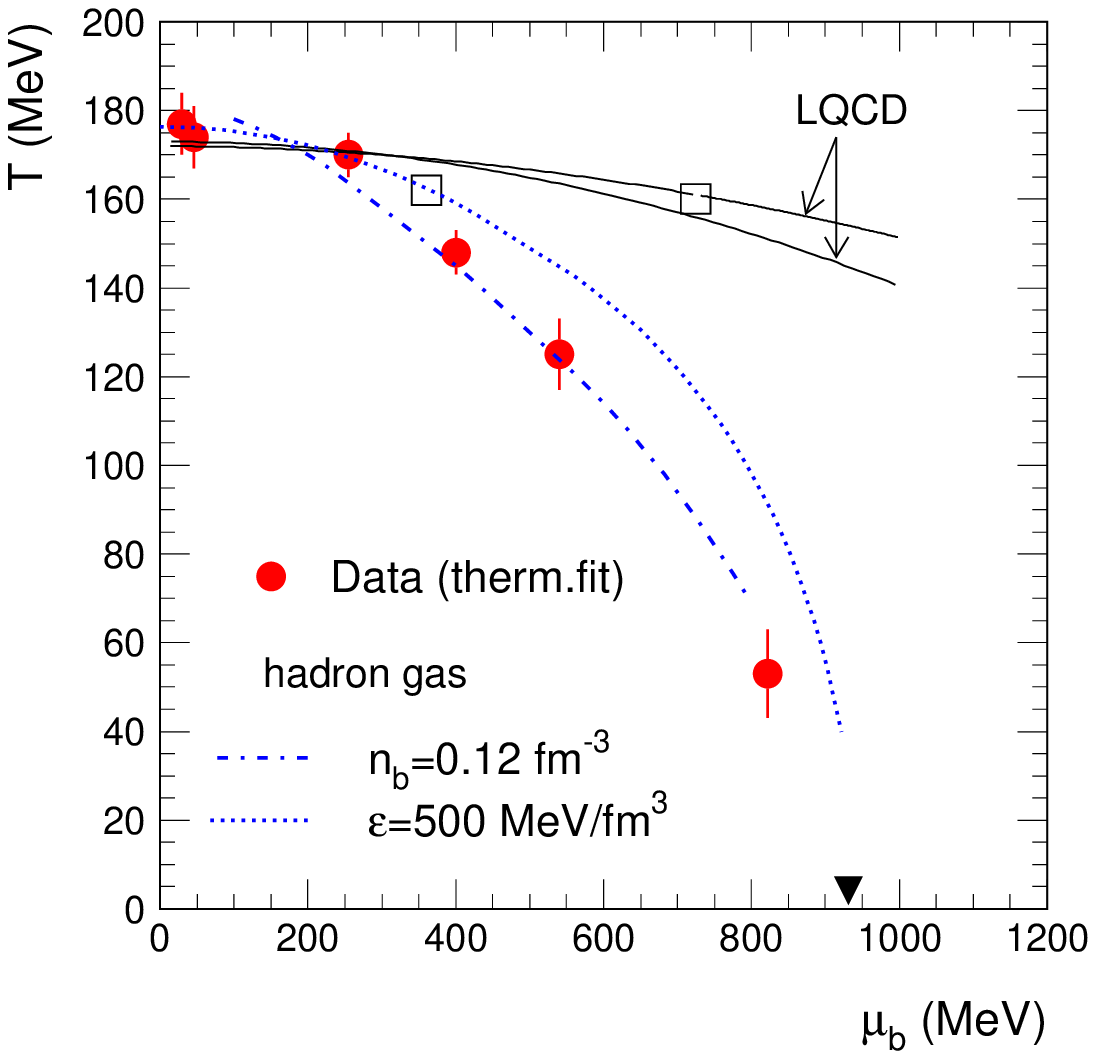,width=0.5\linewidth}
\caption{The left figure from \cite{Karsch:2003jg} sketches the current idea of the phase diagram of QCD in the temperature ($T$)-baryon chemical potential ($\mu$)-plane. It is not yet settled whether the transition at $\mu=0$ is a true phase transition or merely a cross-over. $\mu_0$ corresponds to normal nuclear matter density, $\chi$ denotes chiral. The right panel shows a more quantitative picture with $\mu_b\equiv\mu$. This figure from \cite{Andronic:2004tx} contains the available experimental data. The full triangle corresponds to normal nuclear matter and the open squares indicate the position of a possible tri-critical point from lattice calculations \cite{Fodor:2001pe}. The circles are from fits of particle ratios with a thermal model \cite{Andronic:2004tx,Braun-Munzinger:2001mh}. They have been obtained by the experiments at RHIC at high and low energies, SPS at high and low energies, AGS and SIS in order of increasing $\mu$. The dashed and dotted curves show freeze-out trajectories for a hadron gas at constant density or energy density, respectively \cite{Andronic:2004tx}.}\label{figintro}
\end{figure}

The general aim of  experiment and theory is to map out the phase diagram of QCD in the temperature-(baryon) chemical potential plane. Roughly sketched it is shown in figure \ref{figintro}, together with the currently available experimental data. The hadronic phase with its gas-liquid phase-transition occupies the small temperature and density/chemical potential region. A rich set of solid-state-like phases are expected at large chemical potentials, such as color-superconductors, see e.g.\/ \cite{Rajagopal:2000wf}. At high temperatures, a so-called quark-gluon-plasma is expected. It is not yet settled whether it is reached by a genuine phase transition or a cross-over in the real world. A pure gluon system like the one to be studied here exhibits a first order phase transition \cite{Karsch:2003jg}.

It is expected that chiral symmetry is restored in the high-temperature phase. This is supported by lattice calculations, see e.g.\/ \cite{Karsch:2003jg}. Concerning the other properties of this phase, even the nature of the effective degrees of freedoms is currently under debate. The initial picture made was a perturbative gas of free quarks and gluons, leading to the name deconfinement. As colored currents are gauge dependent, this view can only be true in a figurative sense. However, the idea of almost local colorless objects with otherwise the quantum numbers of quarks and gluons as weakly interacting quasi-particles remains.

The main aim of this work is an analysis of the properties of gluons in the high-temperature phase as well as the validity of this simple picture. Still addressing merely equilibrium dynamics, the fate of the confining properties at temperatures above the phase transition are investigated. In addition, there is recent evidence that at temperatures just above the phase transition the matter is in a non-trivial, strongly interacting phase. Whether this extends to all temperatures is also a main subject of this work.

The basic field theoretical concepts used will be compiled in chapter \ref{ctheory}. The Dyson-Schwinger equations used to study this theory including the introduction of thermodynamic features will be laid out in chapter \ref{cdse}. A first step will be concerned with investigations regarding the infinite temperature limit in chapter \ref{c3d}. This will be followed by introducing a high temperature expansion in chapter \ref{cft}. The thermodynamic potential, analytic properties and other aspects of the solutions will be investigated in chapter \ref{cderived}. The results will be interpreted and summed up in chapter \ref{csummary}. Appendix \ref{aconventions} contains conventions and commonly used symbols. Five further appendices contain technical details which would have broken the line of argument inside the main text.


%% file: theory.tex
\chapter{Aspects of QCD as a Gauge Theory}\label{ctheory}

\section{Formulation}

QCD describes the interaction of quarks, massive fermions, through gauge bosons, the gluons. The gauge group for the physical QCD is SU$(3)$. In this work, the scope is enlarged to an arbitrary semi-simple, compact Lie-group. However, only some of the assumptions made later have been tested by lattice calculations and just with respect to SU$(N_c)$ gauge groups at small $N_c$. Thus although the results will turn out to be independent of the gauge group, the assumptions made are potentially not.

The classical Lagrangian of such a general gauge theory is given by \cite{Bohm:yx}
\bea
{\cal L}_m&=&-\frac{1}{4}F_{\mu\nu}^aF^{\mu\nu\indexsep a}+\sum_f\left(\bar\psi^\alpha_f\gamma^\mu iD_\mu\psi_f^\alpha-m_f\bar\psi_f^\alpha\psi_f^\alpha\right)\label{qcdlagrangianm}\\
F^a_{\mu\nu}&=&\pdm A_\nu^a-\pdn A_\mu^a+g_df^{abc}A_\mu^bA_\nu^c\nonumber\\
D_\mu&=&\pdm-ig_dt^aA_\mu^a.\nonumber
\eea
\noindent $A_\mu^a$ denotes the gluon fields with color $a$ and Lorentz index $\mu$. $\bar\psi^\alpha_f$ and $\psi^\alpha_f$ are the anti-quark fields and quark fields, respectively, of flavor $f=1,...,N_f$ and color $\alpha$. For a SU$(N_c)$ gauge group $a=1,...,N_c^2-1$ and $\alpha=1,...,N_c$. $F_{\mu\nu}^a$ is the field strength tensor and $D_\mu$ the fundamental covariant derivative, $t^a$ are the generators of the gauge group and $f^{abc}$ its structure constants. The $m_f$ are the masses corresponding to the quark flavor $f$, and $g_d$ is the gauge coupling where $d$ is the space-time dimension. $\gamma^\mu$ denotes the Dirac $\gamma$-matrices \cite{Bohm:yx}. 

The introduction of equilibrium thermodynamics in chapter \ref{cdse} includes a Wick rotation \cite{Kapusta:tk} and hence the Euclidean version of \pref{qcdlagrangianm} will be more important throughout. It is given by \cite{Alkofer:2000wg}
\bea
{\cal L}&=&\frac{1}{4}F_{\mu\nu}^aF^{\mu\nu\indexsep a}-\sum_f\left(\bar\psi^\alpha_f\gamma^\mu D_\mu\psi_f^\alpha-m_f\bar\psi_f^\alpha\psi_f^\alpha\right)\label{lym}\\
F^a_{\mu\nu}&=&\pdm A_\nu^a-\pdn A_\mu^a-g_df^{abc}A_\mu^bA_\nu^c\nonumber\\
D_\mu&=&\pdm+ig_dt^aA_\mu^a.\nonumber
\eea
\noindent All further expressions will be in Euclidean space-time, if not otherwise noted. In the remainder of this work only the gauge-subsector of \pref{lym} will be investigated and the matter content discarded. This reduced theory is known as Yang-Mills theory \cite{Bohm:yx,Yang:ek}. In section \ref{ssfermions} it will be discussed to which extent this is justified. In section \ref{ssumprop} a short assessment of the influence of quarks on the results presented will be given.

In chapter \ref{c3d}, the Yang-Mills sector will be coupled to an adjoint scalar field. For the sake of completeness, its Lagrangian is given here in Euclidean space as
\bea
{\cal L}&=&\frac{1}{4}F_{\mu\nu}^aF^{\mu\nu\indexsep a}+\frac{1}{2}(D_\mu^{ab}\phi^b D_\mu^{ac}\phi^c+m_h^2\phi^a\phi^a)+\frac{h}{4}\phi^a\phi^a\phi^b\phi^b\label{l3d}\\
D_\mu^{ab}&=&\delta^{ab}\pdm+g_df^{abc}A_\mu^c.\nonumber
\eea
\noindent $\phi^a$ is the scalar field of color a, $m_h$ its mass and $h$ its self-coupling. For an SU$(N_c)$ gauge group $a=1,...,N_c^2-1$. $D_\mu^{ab}$ is the adjoint covariant derivative. A three-Higgs coupling is not present due to the antisymmetry of the coupling constants and the color symmetry at tree-level.

The next step is to quantize the Lagrangians \pref{lym} and \pref{l3d}.

\subsection{Quantization, Gauge Fixing and the Gribov Problem}\label{squant}

Quantization can be performed using either canonical or path integral methods. The latter will be used here, because they give a somewhat more sophisticated access \cite{Rivers:hi}. In this case, the generating functional is given by
\be
Z[j_\mu^a]=\int{\cal D}A_\mu^a\exp\left(\int d^dx\left(-{\cal L}+A_\mu^a(x)j^{\mu\indexsep a}(x)\right)\right)=:e^{W[j_\mu^a]},\label{part}
\ee
\noindent which depends on the classical sources $j_\mu^a$ for the gauge field; the 'free energy' $W$ is the generating functional of connected correlation functions. The integration is over the set of all configurations of the gluon field $A_\mu^a$. This set is termed gauge space. The classical sources have to be set to 0 at the end of all calculations.

The prescription \pref{part} has the problem that not all possible fields $A_\mu^a$ contribute independently to the path integral, since Yang-Mills theory is a gauge theory. ${\cal L}$ is invariant under local gauge transformations
\bea
A_\mu^a&\to& A_\mu^a+D_\mu^{ab}\delta\theta^b\label{gtrans1}\\
\phi^a&\to&\phi^a+g_df^{abc}\phi^c\delta\theta^b,\label{gtrans2}
\eea
\noindent where the second transformation is only relevant for \pref{l3d}. $\delta\theta^a$ are independent functions. Field configurations, which are equivalent up to a gauge transformation \pref{gtrans1} or \pref{gtrans2} are said to lie on the same gauge orbit. This gauge freedom leads to over-counting in the path integral \pref{part}. It is hence necessary to take the gauge freedom into account when quantizing.

The most powerful, albeit technically complicated method to circumvent these problems is stochastic quantization \cite{Damgaard:1987rr}. It is in generally well suited to calculate gauge invariant quantities, including all observables. Furthermore it is possible to obtain gauge dependent objects like gluon propagators corresponding to a conventional gauge choice \cite{Rivers:hi,Zwanziger:1981kg}. For Landau gauge, this process yields the same equations describing the gluons \cite{Zwanziger:2003cf} as the approach followed here within the approximation scheme employed.

The conventional approach followed here is to fix the gauge prior to quantization \cite{Rivers:hi}. The aim is to include only one configuration on each gauge orbit. This is performed by introducing an appropriate $\delta$-function in the path integral \pref{part}. The argument of this function is a local functional of the fields of the form $C^a[A_\mu^b,x]$, and its equality to 0 defines the gauge. The remaining steps are a standard procedure and will not be detailed further \cite{Bohm:yx}.

There is still one point to be mentioned. The gauge fixing usually used is an algebraic or differential condition, such as 
\be
\pdm A_\mu^a=0\label{lgauge}
\ee
\noindent for Landau gauge or
\be
\vec\nabla \vec A^a=0\label{cgauge}
\ee
\noindent for Coulomb gauge. Evidently, \pref{lgauge} and \pref{cgauge} are not complete gauge-fixings. Any harmonic gauge transformation in \pref{lgauge} and any transformation depending only on time in \pref{cgauge} are still allowed. It is possible to fix this residual classical gauge freedom. However, the conditions \pref{lgauge} and \pref{cgauge} are still not unique. In non-abelian gauge theories there is more than one gauge-equivalent solution to them, i.e. more than one configuration on each gauge orbit satisfies the gauge condition. E.g.\/ in the case of Coulomb gauge \pref{cgauge} a gauge-equivalent solution to the vacuum $A_\mu^a=0$ is a hedgehog configuration \cite{Gribov:1977wm}. The consequence is over-counting in \pref{part}. This is known as the Gribov problem \cite{Gribov:1977wm} and it can be shown to extend to a large class of local gauges \cite{Singer:dk}. As the additional copies in general involve large gauge field fluctuations, this problem does not appear in perturbation theory.

It is not yet known whether there exists any local gauge condition which resolves this problem\footnote{There are arguments that Landau gauge is still well-defined, if a summation is made over all signed Gribov copies \cite{Hirschfeld:yq}. It is however probable that the truncations introduced in section \ref{strunc} will interfere with these cancellations. Thus other techniques are necessary here.}. With regard to the problems induced by non-local gauge conditions, it is worthwhile to investigate if there is a way to circumvent this problem. Indeed it has been shown that the zeros of the Faddeev-Popov determinant
\be
M=\det(-\pd_\mu D_\mu^{ab})\label{fpdet}
\ee
\noindent define non-intersecting convex and compact regions of gauge-space. Each region is intersected at least once by each gauge orbit. The one enclosing the origin and thus perturbation theory is called the first Gribov horizon \cite{Gribov:1977wm,Singer:dk}. Within it, \pref{fpdet} is positive. It is possible in the approach followed here to ensure $M>0$ and thus to be inside the first Gribov horizon. This will be discussed in section \ref{strunc}. However, this condition is not sufficient \cite{vanBaal:1997gu}, and gauge copies are still present. The copy-free region contained inside the first Gribov horizon, called the fundamental region, can be defined using a non-local minimalization condition \cite{Zwanziger:1993dh}. This in turn implies that there is again no local condition like the Gribov horizon to eliminate the copies, and a full solution is still missing.

Nevertheless, it has been argued that a bounded volume in an infinite dimensional space, like gauge space, is dominated by its boundary. Therefore only this region contributes to objects constructed from a finite number of operators\footnote{So the following argument does not necessarily apply to the Polyakov loop or other exponentials of operators.} \cite{Zwanziger:2003cf}. Using this entropy argument, the common boundary of the fundamental region and the first Gribov horizon is the only region contributing and the Gribov horizon condition is sufficient. For the remaining part of this work, this argument will be accepted as an assumption and hence the Gribov horizon condition will be used to restrict the space of possible solutions.

The impact of Gribov copies can be studied by lattice calculations. A relatively weak effect on the gluon propagator and a stronger quantitative effect on the ghost propagator, to be discussed shortly, is found \cite{Cucchieri:1997ns}. It is therefore likely that even if the assumption is incorrect, the results found here are still qualitatively reliable.

\subsection{Landau Gauge}

Most calculations based on perturbation theory restrict the gauge condition as weakly as possible and use the requirement of gauge independence to check the results for errors. In a well-defined scheme such as perturbation theory, this is clearly advantageous. In the case of non-perturbative calculations no such rigorous scheme exists up to now. Hence, the requirement of approximate gauge invariance will actually be used to investigate the quality of the solution. To this end, Landau gauge \pref{lgauge} turns out to be well suited for reasons to be described shortly.

Landau gauge belongs to the class of covariant gauges. Using the standard prescription for gauge fixing \cite{Bohm:yx}, the generating functional \pref{part} becomes
\be
Z[j_\mu^a]=\int{\cal D}A_\mu^a M\exp\left(-\int d^dx\left({\cal L}+A_\mu^a(x)j^{\mu\indexsep a}(x)-\frac{1}{2\xi_g}\pdm A_\mu^a(x)\pd^\mu A_\mu^a(x)\right)\right),\label{lpart}
\ee
\noindent where $\xi_g$ is an arbitrary constant, the gauge constant. Landau gauge is obtained by the limit $\xi_g\to 0$, in which case the last term in \pref{lpart} is dismissed. The Faddeev-Popov determinant \pref{fpdet} emerges from an intermediate change of variables as a Jacobian. Faddeev and Popov showed \cite{Faddeev:fc} that it can be rewritten as a path integral over scalar Grassmann fields, leading in Landau gauge to
\be
Z[j_\mu^a]=\int{\cal D}A{\cal D}c{\cal D}\bar c\exp\left(-\int d^dx\left({\cal L}+\bar c^a(x)\pdm D^{\mu\indexsep ab}c^b(x)-A_\mu^a(x)j^{\mu\indexsep a}(x)\right)\right),\label{fppart}
\ee
\noindent where $c^a$ and $\bar c^a$ are the (Faddeev-Popov-) ghost and anti-ghost fields. As they are anti-commuting scalars, they may not appear in final states, since they would violate the CPT-theorem. This can be guaranteed in both, perturbative calculations and also in the non-perturbative calculations in this work, as will be detailed in subsection \ref{ssbrst} and section \ref{strunc}. Consequently, no external sources are associated with the ghosts at this level. They will be introduced in chapter \ref{cdse}, when off-shell ghosts will be investigated in more detail. Similar to time-like gluons, ghosts contribute indefinite norm states to the Hilbert-space already in perturbation theory. This is directly visible from the fact that ghosts act as `negative degrees of freedom' in scattering processes \cite{Peskin:ev}.

The ghost fields have a new global symmetry, the ghost number symmetry. Rescaling the ghost fields by a scale transformation $\exp(s)$ and its anti-field by $\exp(-s)$, leaving all other fields unchanged, is a symmetry of the Lagrangian\footnote{It is a left-over from the original local symmetry, which was broken by gauge fixing.}. It gives rise to the conserved ghost number $Q_{G}$, in analogy to the fermion number. As the ghosts are the only fields carrying them, it is necessary that all observable final states must have ghost number 0.

Note that the hermiticity assignment of the ghosts in \pref{fppart} is different from the common choice, which is reflected in the different sign for the ghost term in \pref{fppart}. This would lead to problems in general covariant gauges, but in ghost-anti-ghost symmetric gauges like Landau gauge this is permissible and for technical reasons advantageous \cite{Alkofer:2000wg,Alkofer:2003jr}. Of these gauges, Landau gauge turns out to be favorable as it is less singular than other gauges \cite{Alkofer:2003jr}. This manifests itself in the non-renormalization of the ghost-gluon vertex, which will be discussed below. A further advantage of Landau gauge is that as long as multiplicative renormalizability holds\footnote{Note that multiplicative renormalizability of Yang-Mills theory with and without matter fields is only proven perturbatively order by order. It is unknown if it also holds non-perturbatively. In the results presented here it does hold.}, it is a fixed point of the gauge parameter, since the latter is exactly 0.

\subsection{BRST Symmetry}\label{ssbrst}

With the introduction of ghosts in \pref{fppart} a further global symmetry arises, which is also a residual of the local gauge symmetry. It is named BRST-symmetry after its discoverers Becchi, Rouet, Stora, and Tyutin \cite{Becchi:1975nq}. Consequently it is possible to define the BRST charge $Q_{\brst}$, which has ghost number 1.

The corresponding symmetry transformations are
\bea
\delta_{\brst}A_\mu^a(x)&=&\delta\lambda D_\mu^{ab}c^b(x)\label{brstrans1}\\
\delta_{\brst}c^a(x)&=&-\delta\lambda\frac{1}{2}g_df^{abc}c^b(x)c^c(x)\\
\delta_{\brst}\bar c^a(x)&=&\delta\lambda\frac{1}{\xi_g}\pdm A_\mu^a(x)\\
\delta_{\brst}\phi^a(x)&=&\delta\lambda g_df^{abc}\phi^c(x) c^b(x)\label{brstrans4},
\eea
\noindent where for the Landau gauge the appropriate limit has to be taken. The transformation rule for the adjoint field $\phi^a$ was added for completeness. $\delta\lambda$ is an infinitesimal constant Grassmann parameter. This defines the BRST-operator $s$ as
\be
\delta_{\brst}F=\delta\lambda sF,\nonumber
\ee
\noindent by its action on any field $F$. As it is defined as the left-derivative of the transformed field with respect to $\delta\lambda$, it directly obeys the generalized Leibniz rule. Since it is Grassmann in nature, as can be seen from the fact that it changes the number of Grassmann fields in the transformations \prefr{brstrans1}{brstrans4} by 1, this rule reads
\be
s(FG)=(sF)G\mp F(sG).\label{brstproductrule}
\ee
\noindent The sign depends on whether $F$ is Grassmann or not. It further carries ghost number 0, as the transformation parameter $\delta\lambda$ has to carry ghost number -1 since the BRST charge carries ghost number 1. As \pref{brstrans1} is a global Grassmann valued gauge transformation, the gauge part of \pref{lym} is invariant on its own. The combination of the gauge fixing part and the ghost contribution is also invariant, where in Landau gauge the limit $\xi_g\to 0$ has to be taken. It is possible to linearize the transformation rules \prefr{brstrans1}{brstrans4} by introducing an auxiliary field, the Nakanishi-Lautrup field $B^a$ \cite{Peskin:ev}. Without going into details, this establishes manifestly the nil-potency of the BRST transformation\footnote{Without this field, the nil-potency would only be manifest on-shell \cite{Henneaux:1992ig}.}
\be
\delta_{\brst}^2=0.\label{nilpot}
\ee
\noindent This establishes a closed algebra
\bea
\left\{Q_{\brst},Q_{\brst}\right\}&=&0\nonumber\\
\left[iQ_G,Q_{\brst}\right]&=&Q_{\brst}\nonumber\\
\left[iQ_G,Q_G\right]&=&0\nonumber
\eea
\noindent of the residual local gauge symmetry.

A well-defined nilpotent charge directly splits the state space into three disjoint parts \cite{Peskin:ev,Henneaux:1992ig,Kugo:gm}. The states which are not annihilated by the BRST-transformation form a subspace $Q_1$, carrying BRST-charge. By acting on these states, daughter states in a subspace $Q_2$ are generated which are annihilated by the BRST-charge. The last possibility are states which are also annihilated by the BRST-charge but are not generated from parent states. These form a subspace $Q_0$. Physical states must be gauge invariant and are therefore annihilated by $Q_\brst$ \cite{Weinberg:1996kr}. In addition, any states in $Q_2$ do not contribute to matrix elements. Therefore the physical subspace is
\be
H_{phys}=\overline{\mathrm{Ker} Q_\brst/\mathrm{Im} {Q_\brst}}=\overline{Q_0}.\nonumber
\ee
It is this subspace in which the perturbatively physical transverse gauge bosons exist, while forward polarized gluons and anti-ghosts belong to $Q_1$ and backward polarized gluons and ghosts belong to $Q_2$. This can be seen directly using the Nakanishi-Lautrup formulation of the gauge-fixed Lagrangian \cite{Peskin:ev}. Due to the relation of $Q_2$ and $Q_1$, the unphysical degrees of freedom are connected by BRST transformations and are thus metric partners. They are said to be confined by the quartet mechanism \cite{Kugo:gm}. Hence in perturbation theory the physical subspace $Q_0$ contains only transverse gluons, and perturbatively unphysical degrees of freedom are confined\footnote{In principle it is possible to have states in $Q_0$ with non-vanishing ghost number, which would render the theory ill-defined \cite{Kugo:gm}. This seems not to be the case for Yang-Mills theories.}. One of the confinement mechanisms proposed, the Kugo-Ojima scenario discussed in subsection \ref{sskugoojima}, requires also transverse gauge bosons to belong to either $Q_2$ or $Q_1$ and thus provides confinement.

\subsection{Slavnov-Taylor Identities}\label{sssti}

It is possible to directly construct identities relating different Green's functions by calculating the BRST transform of an operator expression, using that the vacuum belongs to $Q_0$. These are the Slavnov-Taylor identities (STI) \cite{Taylor:ff,Slavnov:fg}. These identities are a result of gauge invariance. A failure in fulfilling them therefore indicates a violation of gauge invariance. This property makes them an important technical tool to check the consistency of calculations and they will be used in this way extensively in chapters \ref{cdse} to \ref{cft}.

In this work two STIs are of particular importance. The first is obtained when forming the expectation value of the BRST transform of $\bar c^a\pdm A_\mu^a$ and yields \cite{Bohm:yx} after usage of the equation of motions
\be
p_\nu D_{\mu\nu}^{ab}(p)=-i\xi_g p_\mu\delta^{ab}\label{stigluon}
\ee
\noindent with the gluon propagator $D_{\mu\nu}$. By virtue of Lorentz invariance, the Landau gauge limit hence requires $D_{\mu\nu}$ to be of the form
\bea
D_{\mu\nu}(p)=P_{\mu\nu}(p)\frac{Z(p^2)}{p^2}\\\label{vacgluonprop}
P_{\mu\nu}(p)=\delta_{\mu\nu}-\frac{p_\mu p_\nu}{p^2},\label{transverseproj}
\eea
\noindent i.e. to be transverse. $Z$ is the associated scalar dressing function. This simple Lorentz structure is one of the advantages of Landau gauge. The other prominent feature of \pref{stigluon} is its independence of higher $n$-point Green's functions. In general, the identity for an $n$-point Green's function depends on $n'$-point Green's functions with $n'>n$. Hence an infinite set of coupled equations arises. Therefore, their use is limited in the case of non-perturbative calculations\footnote{In perturbative calculations such contributions can be neglected as they are of higher order in the expansion parameter.}, as there is not yet any possibility to assess a-priori the contribution of the $n'$-point Green's functions. In special kinematic regions, however, those unknown contributions may drop out, providing relations between $n$-point functions only.

A further result which can be obtained from \pref{stigluon} in connection with the equation of motion of the ghost is that the ghost-gluon vertex is undressed for a vanishing incoming ghost momentum in Landau gauge \cite{Taylor:ff,Marciano:su}
\be
\lim _{p\to 0} \Gamma^{c\bar cA\indexsep abc}_\nu(p,q,-p-q) =  i g_d f^{abc} q_\nu.\label{taylor}
\ee
\noindent It is thus not divergent and its renormalization constant $\widetilde{Z}_1$ can and will be set to 1 here. This is maybe the most important property of Landau gauge. It makes it much more tractable than other gauges \cite{Alkofer:2003jr}. The ghost-gluon vertex will be further investigated in subsection \ref{sszwanzigergribov} and used in chapters \ref{cdse} to \ref{cft}.

The second identity is concerned with this ghost-gluon vertex. It can be obtained from the BRST-transform of $c^a\bar c^b\bar c^c$ \cite{vonSmekal:1997is,Watson:phd} and reads
\be
ik_\mu\Gamma^{\bar c cA}_\mu(p,-q) G(q^2)+iq_\mu\Gamma^{\bar c cA}(p,-k)_\mu G(k^2)=p^2\widetilde{Z}_1\frac{G(k^2)G(q^2)}{G(p^2)}+\eta\label{stigghv}.
\ee
\noindent $\eta$ summarizes contributions from $n'$-point Green's functions with $n'>3$. Color indices have been removed by assuming a tree-level\footnote{This is exact in perturbation theory.} color structure. This assumption is discussed in more detail in chapter \ref{cdse}. $G$ is the ghost dressing function, defined via its propagator as
\be
D_G^{ab}(p^2)=-\frac{G(p^2)}{p^2}\delta^{ab}.\label{ghostprop}
\ee
\noindent The identity \pref{stigghv} is consistent with the bareness of the ghost-gluon vertex \pref{taylor} if the contributions from the higher Green's functions vanish in this limit.

\section{Confinement}\label{sconfinement}

As one of the main observables for this work is the absence or presence of confinement it is necessary to detect it. Two fundamentally different approaches to this question will be discussed. The first is concerned with the mere statement of confinement. This will be investigated in subsection \ref{sscriterions}. The other approach consists of criteria deduced from possible confinement mechanisms. Three of them will be discussed in sections \ref{sskugoojima} to \ref{ssgribovstingl}. The number of proposed mechanisms for confinement is large. Only those which generate criteria that can be tested using the objects obtained in this work will be discussed here. A more general overview can be found in \cite{Alkofer:2000wg}.

In any case, cluster decomposition \cite{Weinberg:mt,Haag:1992hx} must be violated for colored objects to allow for a long range force. This necessarily implies the existence of a massless excitation in the complete state space. On the other hand since all experimental results show validity of cluster decomposition, the physical mass spectrum of QCD must have a mass gap for colorless objects. Otherwise it would be possible to scatter a colorless object into far apart colored objects, the so-called ``behind-the-moon'' problem. However, the existence of a massless excitation alone does not suffice for confinement of color, as it is necessary to show that it is not part of the physical spectrum.

\subsection{Criteria for Confinement}\label{sscriterions}

There are two criteria which will be used here. One implies the other. The basic object is in both cases the spectral density $\rho$ \cite{Itzykson:rh} of a given particle. For a particle to exist as a physical final state having a K\"allen-Lehmann representation, it has to have a positive semi-definite spectral function\footnote{In a theory with indefinite metric, as Yang-Mills theory, for unstable particles such that the width exceeds the mass the spectral function is also not necessarily positive semi-definite. They do not occur in final states, since they decay when letting $t\to\infty$.}. This is known as the Osterwalder-Schrader axiom of reflection positivity \cite{Haag:1992hx,Osterwalder:dx}. A violation of positivity can be tested using the Schwinger function to be discussed in chapter \ref{cderived}.

A stronger condition is violation of the Oehme-Zimmermann super-convergence relation \cite{Oehme:bj}. The spectral representation of a particle of mass $m_0$ is \cite{Itzykson:rh}
\be
D(p)=\frac{Z}{p^2+m_0^2}+\int_{m^2}^{\infty}d\kappa^2\frac{\rho(\kappa^2)}{p^2+\kappa^2},\label{oehmerep}
\ee
\noindent with $m>m_0$ being the threshold mass for contributions above the single-particle pole, the multi-particle threshold. $Z$ is the positive overlap between an in-state and the field described by the propagator $D$. If the propagator vanishes at 0, 
\be
\lim_{p\to 0}p^2D(p)=0,\label{oehme}
\ee
\noindent then the spectral function must be at least partly negative, thus implying the first condition above. Secondly, a vanishing propagator at $p=0$ implies the absence of a K\"allen-Lehmann representation. The particle can thus be not a physical particle anymore: The particle is confined. This gives the second condition for confinement.

For massless particles the interpretation of \pref{oehme} is more direct. As $p^2=0$ is just the statement of the particle being on-shell, it implies the vanishing of the on-shell propagator: The particle does not propagate and is thus confined.

Note here a subtle difference. It is possible to think of confined particles as either confined due to dynamic effects or due to being unphysical. For example magnetic confinement of neutrons in a magnetic field is such a case: The particles are physical but they are bound in a way disallowing separation. The confinement e.g.\/ of ghosts differs from this. These are unphysical particles and are thus confined in a different sense. Note that in an unbroken non-abelian gauge theory, charged currents are gauge-dependent and can thus not be observed directly. Observing a quark thus corresponds e.g.\/ to an observation of a colorless object with non-integer electric charge.

One of the major questions concerning confinement is which of both possibilities applies to colored objects. The Kugo-Ojima criterion described in the next subsection favors the latter option, as colored objects will turn out to be automatically BRST-charged. The more common attitude is the first option.

\subsection{Kugo-Ojima Scenario}\label{sskugoojima}

The Kugo-Ojima confinement scenario \cite{Kugo:gm} puts forward the idea that all colored objects form BRST-quartets and therefore do not belong to the physical state space. The metric partners of transverse gluons would be ghost-gluon bound states.

Thus, the confinement mechanism is essentially the same as in the case of ghosts in perturbation theory. This scenario is based on a rigorous derivation and requires three preconditions. One of them is an unbroken BRST charge also in the non-perturbative regime. Whether this is the case is not known\footnote{It is even unknown how to define a non-perturbative BRST charge. Due to the non-trivial topology of the gauge group this necessarily has to be done patch-wise in gauge space.}. The second is the failure of the cluster decomposition theorem and hence the existence of a massless excitation. This is also unknown and can not yet be attacked in the approach used here. The third ingredient is an unbroken global color charge. In Landau gauge, this condition can be put into the form \cite{Kugo:1995km}
\be
\lim_{p^2 \to 0} p^2 D_G(p^2)\to\infty,\label{kugo}
\ee
\noindent where $D_G$ is the propagator of the Faddeev-Popov ghost \pref{ghostprop}. This scenario also necessarily implies that \pref{oehme} holds for the gluon. Both of these conditions \pref{oehme} and \pref{kugo} will be checked.

Note that this scenario does not imply that colored objects do not have an asymptotic field. However, as they are BRST charged a configuration with arbitrarily many colored objects belongs to one equivalence class, and only the colorless objects contribute to matrix elements.

\subsection{Zwanziger-Gribov Scenario}\label{sszwanzigergribov}

The central idea of the Zwanziger-Gribov scenario \cite{Zwanziger:2003cf,Gribov:1977wm,Zwanziger:2001kw,Zwanziger:2002ia} is that zero-modes at the common boundary of the first Gribov horizon and the fundamental region dominate the infrared properties and thus generate confinement. Both regions necessarily have a common boundary. This boundary is convex, compact, and includes the origin \cite{Zwanziger:2003cf}.

As gauge space is infinite-dimensional, all the volume will be concentrated at the boundary, and the system is dominated by it. Since the Faddeev-Popov-determinant \pref{fpdet} vanishes there, the corresponding excitations have to be long-range. Using stochastic quantization and performing a Landau gauge limit under certain assumptions, it can be shown that \pref{kugo} follows. In addition, it follows that the infrared limit is dominated by the ghost-term of \pref{fppart} alone \cite{Zwanziger:2001kw}. As this term can be written as a BRST-transform, Yang-Mills theory in this limit is a topological field theory of Schwarz type with no propagating modes \cite{Birmingham:1991ty}. Thus colored objects do not appear, implying confinement. Using these results, \pref{oehme} for the gluon is also obtained, thus leading to the same criteria as the Kugo-Ojima scenario. This analogy, if it is more than mere coincidence, is not yet understood.

It is also intuitively clear that a strongly divergent ghost propagator at zero momentum can mediate confinement. After Fourier-transformation such an infrared divergence relates to long-ranged spatial correlations. These are stronger than the ones induced by a Coulomb force since the divergence in momentum space is stronger than that of a massless particle.

A bare ghost-gluon vertex in the infrared is sufficient to generate this behavior. Such a vertex would also be consistent with the perturbative renormalization group \cite{Zwanziger:2003cf}. This Zwanziger-hypothesis has been checked numerically in context with this work \cite{schleifenbaum:diploma} and turns out to be well fulfilled.

Recently, investigations in Coulomb gauge indicate, that a similar connection between the dynamics on the Gribov horizon and confinement of gluons holds there as well \cite{Greensite:2004bz}.

\subsection{Gribov-Stingl Scenario}\label{ssgribovstingl}

The last scenario investigates a manifestation of confinement different from the two previous ones. The Gribov-Stingl scenario \cite{Gribov:1977wm,Habel:1990tw,Habel:1989aq} puts forward the idea that confined particles have one or more pairs of complex conjugate poles, and thus cannot be physical states. In general, such a pole structure leads to violation of causality \cite{Peskin:ev}. However, for a special structure of the propagators it can be shown that it is possible to reconcile such a pole structure with causality on the level of the $S$-matrix \cite{Habel:1990tw}. The essential argument is that due to the absence of real poles, application of the LSZ-reduction formula \cite{Peskin:ev} leads to vanishing matrix elements for states with colored objects \cite{Habel:1990tw}. Thus colored objects can only exist as short-lived quantum excitations. In addition, colorless objects only have real poles but no continuum, if they cannot decay into colorless objects \cite{Habel:1990tw}.

\section{Interplay with Quarks}\label{ssfermions}

As quarks will be neglected throughout this work, a short comment to which extent the results may be affected by the presence of quarks is in order. Calculations in the vacuum show that the presence of quarks does not qualitatively alter the gauge propagators, as long as there are less than 5 light flavors \cite{Fischer:2003rp}. This is not expected to change at finite temperature, as quarks acquire an effective mass increasing with temperature and thus behave more perturbatively \cite{Appelquist:vg}. Especially they should not contribute to the infinite temperature limit of the gluon propagator. Thus the results presented will probably not be changed qualitatively by quarks; this is only a conjecture, though. Especially in the vicinity of the phase transition quarks are relevant. Lattice calculations indicate a possible change of the order of the phase transition when including quarks \cite{Karsch:2003jg}, but this issue is still under debate.

The impact on quarks by the results found here may be more significant but also more intricate to understand. Lattice calculations show a drastic change in the properties of quarks above the phase transition. The free energy of two static quarks changes substantially and may change from a linear rise with distance to a logarithmic or even flat shape \cite{Kaczmarek:2004gv}. However, as the mechanism of quark confinement is not understood yet \cite{Alkofer:2003jk}, assessing the consequences of this change is not simple. The most prominent change is the restoration of chiral symmetry \cite{Karsch:2003jg}, which surprisingly occurs at the same temperature as the changes in the gauge sector. 

One possible chain of arguments to understand this coincidence is based on evidence from lattice calculations that the underlying degrees of freedom confining quarks and gluons are topological objects \cite{Engelhardt:2003wm}. These carry topological charge, and can thus be brought into connection with zero modes of the Dirac-operator by the Atiyah-Singer index theorem \cite{Atiyah:1968mp}. The density of such modes again is related to the chiral condensate by the Banks-Casher formula \cite{Banks:1979yr}. If this chain of arguments is correct, it gives a connection between confinement and chiral symmetry breaking. As will turn out in this work, at least part of the gluons are confined even at high temperatures. It is therefore still unclear what the relation above the critical temperature is and how the quark properties are affected.


%% file: temperature.tex
\chapter{Derivation of the Dyson-Schwinger Equations}\label{cdse}

Knowledge of all of the Green's functions would grant complete knowledge of a theory \cite{Haag:1992hx}. The inverse 2-point Green's functions are the propagators. These are the main objects of interest here, as they carry information concerning confinement and other non-perturbative properties. The equations determining these Green's functions are the Dyson-Schwinger equations \cite{Dyson:1949ha} (DSEs), which can be obtained using the functional equations of motion. This will be described in section \ref{sdsevac}. Since the aim of this work is to investigate the equilibrium high temperature phase, temperature will be introduced into the DSEs in section \ref{sdseft}. As there are an infinite number of coupled DSEs, it is generally not possible to solve these simultaneously. To obtain approximate solutions, truncations are necessary and these are discussed in section \ref{strunc}. The connections to perturbation theory and renormalization are investigated in sections \ref{sperturb} and \ref{srenormalization}.

This work is based on a calculational scheme which has been applied successfully in the vacuum to both Yang-Mills theory and full QCD, as will be described in section \ref{ssolvac}.

\section{Vacuum Formulation}\label{sdsevac}

\begin{figure}
\epsfig{file=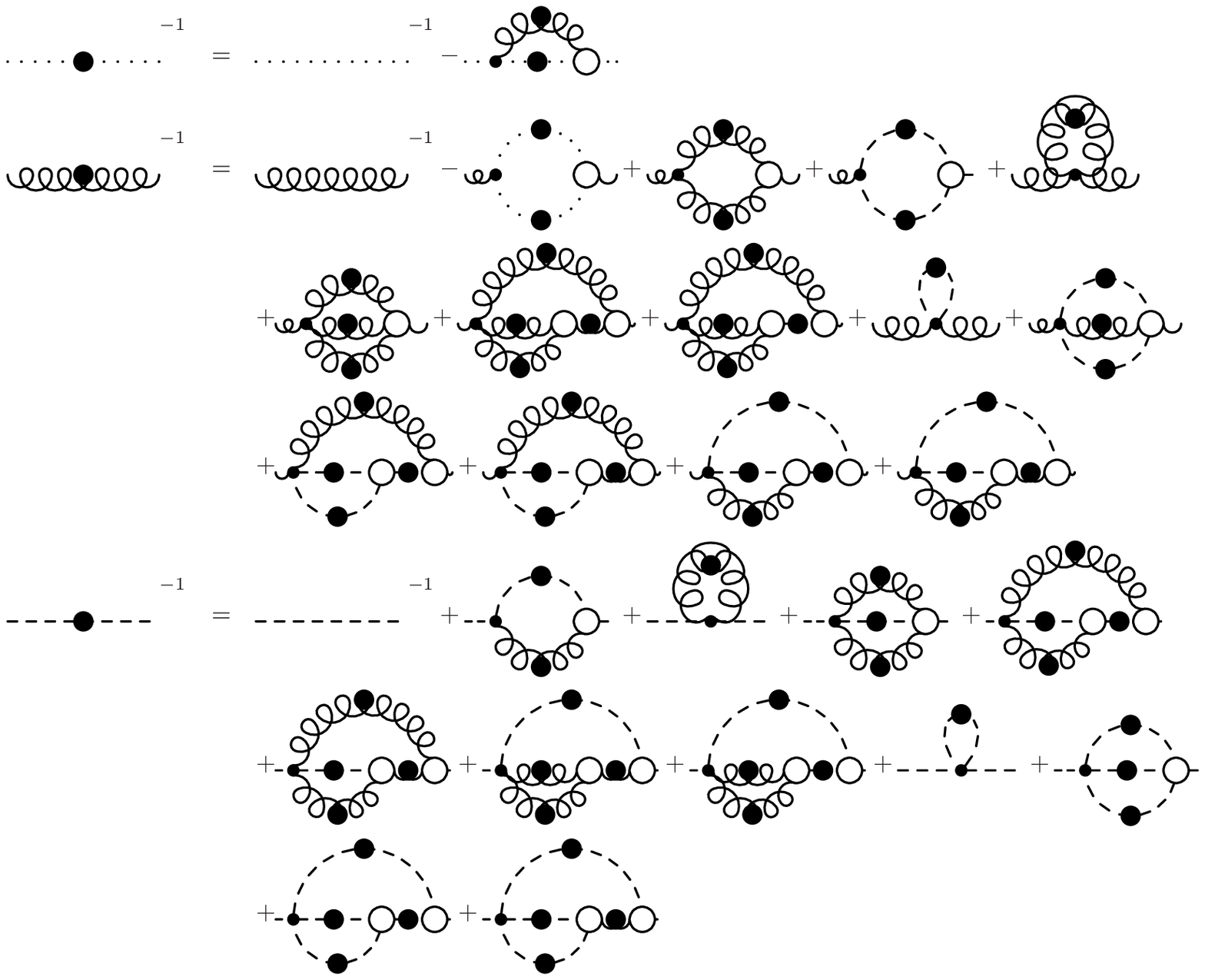,width=0.9\linewidth}
\caption{The Dyson-Schwinger equations for the 2-point functions described by the Lagrangian \pref{l3d}. Dotted lines are ghosts, wiggly lines are gluons and dashed lines are Higgs. Lines with a dot are full propagators, vertices with a small black dot are bare vertices and white circles are full 3- and 4-point functions.}
\label{figfullsys}
\end{figure}

The most straightforward way to derive the DSEs for a generic field $\phi^a$ is by using the fact that the integral of a total derivative vanishes \cite{Alkofer:2000wg}
\bea
0&=&\int{\cal D}\phi^a\frac{\delta}{\delta\phi^a(y)}e^{-S+\int d^dx\phi^a(x)j^a(x)}\nonumber\\
S&=&\int d^dx{\cal L}\nonumber.
\eea
\noindent $S$ is the action, $j^a$ is the source of $\phi^a$ and the integral is over full field space. Performing the derivative and pulling the resultant factor out of the integral by replacing $\phi^a$ with $\delta/\delta j^a$, the prescription to calculate the full one-point Green's function is obtained as
\be
\left(\left(-\frac{\delta S}{\delta \phi^a(x)}\Big|_{\phi^a(x)=\frac{\delta}{\delta j^a(x)}}+j^a(x)\right)Z[j^a]\right)_{j^a=0}=0.\label{dse}
\ee
\noindent Further derivatives with respect to the fields generate the Green's functions of arbitrarily high order. They form an infinite set of coupled non-linear integral equations. For Yang-Mills theory and full QCD, these are known for the propagators, see e.g. \cite{Alkofer:2000wg}. For the Lagrangian \pref{l3d} these are derived for all 2-point Green's functions in appendix \ref{adse}. A graphical representation of these is given in figure \ref{figfullsys}.

In the case of Yang-Mills theory or QCD, it is not necessary to integrate over all field-space for Green's functions. As the Faddeev-Popov determinant \pref{fpdet} appears explicitly in \pref{lpart}, it is sufficient to integrate only over the first Gribov horizon, since the determinant vanishes on its boundary \cite{Zwanziger:2003cf}. Hence the DSEs have the same form whether the integration is over all of gauge-space or only over the first Gribov horizon. Therefore it is necessary to restrict the space of solutions to those from inside the first Gribov horizon. This is discussed in section \ref{strunc}.

\section{Dyson-Schwinger Equations at Finite Temperature}\label{sdseft}

\begin{figure}
\epsfig{file=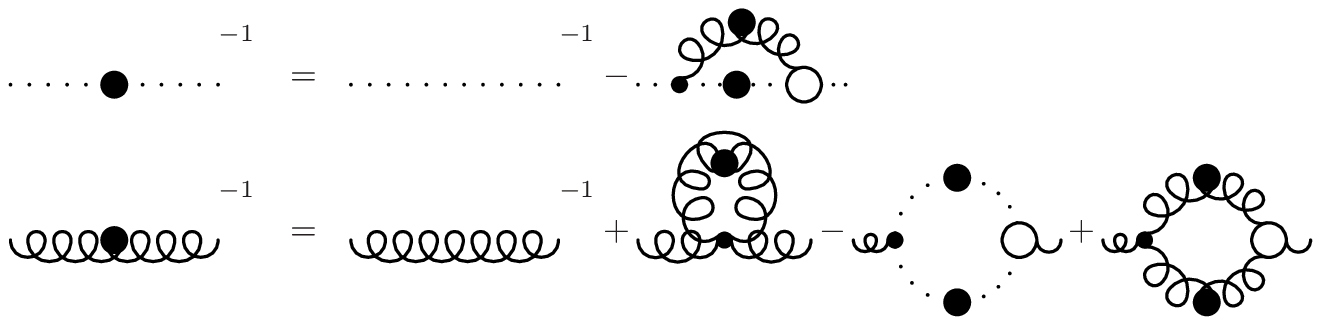,width=\linewidth}
\caption{The truncated Dyson-Schwinger equations for the 2-point functions of Yang-Mills theory at $T=0$. Dotted lines are ghosts and wiggly lines are gluons.  Lines with a dot are full propagators, vertices with a small black dot are bare vertices and white circles are full vertices.}
\label{figt0sys}
\end{figure}

The starting point of the analysis are the DSEs of Yang-Mills theory described by \pref{lym} with the quark fields set to zero. Besides neglecting all equations for $n$-point Green's functions with $n>2$, all genuine full two-loop graphs in figure \ref{figfullsys} are neglected, too. The consequences of this and further truncations will be discussed in detail in section \ref{strunc}. The remaining system is then represented in figure \ref{figt0sys} and given by
\bea
D^{ab-1}_G(p)&=&-\delta^{ab}p^2\nonumber\\
&+&\int\frac{d^dq}{(2\pi)^d}\Gamma_\mu^{\tl\indexsep c\bar cA\indexsep dae}(-q,p,q-p)D^{ef}_{\mu\nu}(p-q)D^{dg}_G(q)\Gamma^{c\bar cA\indexsep bgf}_\nu(-p,q,p-q)\label{vacgheq}\\
D^{ab-1}_{\mu\nu}(p)&=&\delta^{ab}(\delta_{\mu\nu}p^2-p_\mu p_\nu)+T^{GG}_{\mu\nu}\nonumber\\
&-&\int\frac{d^dq}{(2\pi)^d}\Gamma_\mu^{\tl\indexsep c\bar cA\indexsep dca}(-p-q,q,p) D_G^{cf}(q) D_G^{de}(p+q) \Gamma_\nu^{c\bar c A\indexsep feb}(-q,p+q,-p)\nonumber\\
&+&\frac{1}{2}\int\frac{d^dq}{(2\pi)^d}\Gamma^{\tl\indexsep A^3\indexsep acd}_{\mu\sigma\chi}(p,q-p,-q)D^{cf}_{\sigma\omega}(q)D^{de}_{\chi\lambda}(p-q)\Gamma^{A^3\indexsep bfe}_{\nu\omega\lambda}(-p,q,p-q).\nonumber\\\label{vacgleq}
\eea
\noindent Here $\Gamma^{c\bar cA}$ is the full ghost-gluon vertex and $\Gamma^{A^3}$ is the full three-gluon vertex. The respective tree-level quantities are denoted by a superscript `$\tl$' and given in \pref{tlcca} and \pref{tlggg}. $T^{GG}_{\mu\nu}$ is the tadpole term. The full vertices will be discussed further in section \ref{strunc}. Note that these are in general Minkowski-space equations, but equations \pref{vacgheq} and \pref{vacgleq} have already been rotated to Euclidean space.

To obtain the equilibrium Green's functions at a temperature $T$, the Matsubara or imaginary time formalism is implemented \cite{Kapusta:tk,Das:gg}. This amounts to a compactification of time and entails a genuine Euclidean formulation. All objects depend on the three-momenta $\vec p$ and the Fourier-component $p_0$ separately. Both ghosts and gluons have to obey periodic boundary conditions \cite{Bernard:1974bq}, thus 
\be
p_0=2\pi n T,\quad n\epsilon\mathbb{Z}.
\ee
\noindent Hence in both cases a soft mode exists, defined by Euclidean $p^2\ll T$, i.e.\/ those with $p_0=0$ in contrast to the hard modes with $p_0\neq 0$. Note that although Lorentz invariance is no longer manifest in this formulation, it is not lost \cite{Weldon:aq}. Furthermore, as the gluon is a vector-particle, its propagator exhibits two independent tensor-structures and two independent dressing functions\footnote{In general, three dressing functions exist, but only two are independent due to the STI \pref{stigluon}.} \cite{Kapusta:tk} instead of one as in \pref{vacgluonprop},
\bea
D_{\mu\nu}(p)&=&P_{T\mu\nu}(p)\frac{Z(p_0^2,\vec p^2)}{p^2}+P_{L\mu\nu}(p)\frac{H(p_0^2,\vec p^2)}{p^2}\label{gluonprop}\\
P_{T\mu \nu }(p)&=&\delta _{\mu \nu }-\frac{p_{\mu }p_{\nu }}{\vec p^{2}}+\delta _{\mu 0}\frac{p_{0}p_{\nu }}{\vec p^{2}}+\delta _{0\nu }\frac{p_{\mu }p_{0}}{\vec p^{2}}-\delta _{\mu 0}\delta _{0\nu }\left(1+\frac{p_{0}^{2}}{\vec p^{2}}\right)\label{tproj}\\
P_{L\mu \nu }(p)&=&P_{\mu \nu }(p)-P_{T\mu \nu }(p)\label{lproj}.
\eea
\noindent The projectors $P_{T\mu\nu}$ and $P_{L\mu\nu}$ are transverse and longitudinal with respect to the heat bath or alternatively with respect to the three spatial dimensions. Both are four-dimensionally transverse, so \pref{gluonprop} still satisfies the STI \pref{stigluon}. This is necessary, as STIs are still valid at finite temperature \cite{Das:gg}. For $T=0$, $Z=H$ and \pref{vacgluonprop} is recovered. Note that \pref{tproj} projects solely on the space-space components and the $p_0=0$-component of \pref{lproj} solely on the 00-component of the propagator. Hence the soft mode of $Z$ is purely chromomagnetic, while the one of $H$ is purely chromoelectric. For the ghost, being a scalar, one dressing function is sufficient at finite temperature.

\begin{figure}
\epsfig{file=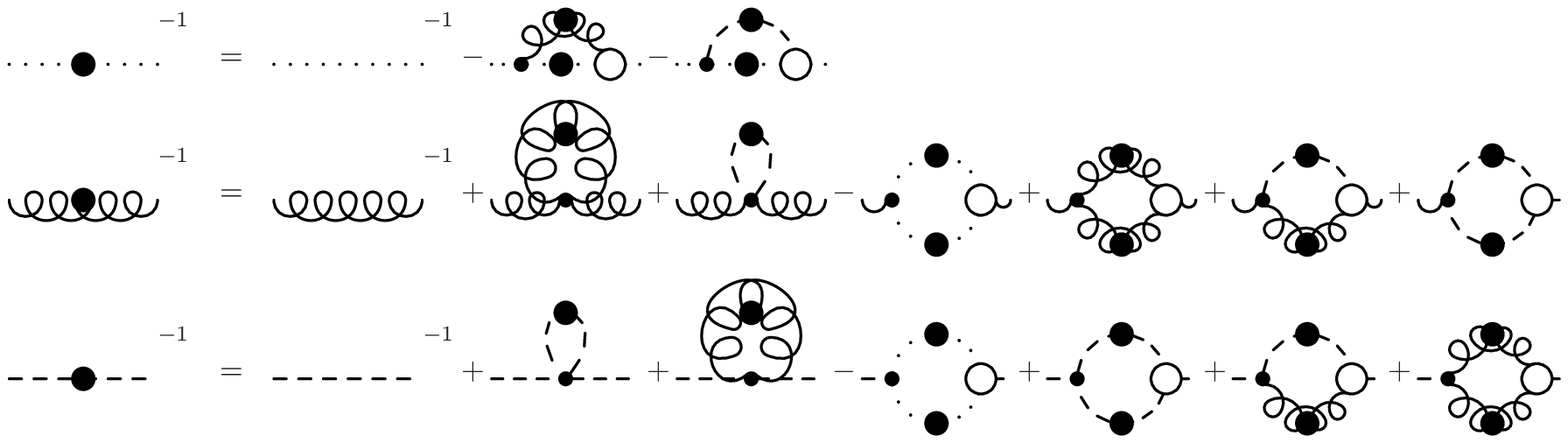,width=\linewidth}
\caption{The truncated Dyson-Schwinger equations for the 2-point functions of Yang-Mills theory at $T\neq0$. Dotted lines are ghosts, wiggly lines are 3d-transverse gluons and dashed lines are 3d-longitudinal gluons.  Lines with a dot are full propagators, vertices with a small black dot are bare vertices and white circles are full vertices.}
\label{figftsys}
\end{figure}

In general, all full Green's functions obtain a much more complicated tensor structure at finite temperature, e.g.\/ the ghost-gluon vertex obtains four instead of two tensor structures. This generates a significant amount of technical problems. As the main assumption of the truncation scheme presented in section \ref{strunc} is the negligibility of dressings other than those of propagators, the irrelevance of such effects is assumed as well. Hence the tree-level tensor structure is used for all full vertices. The last issue to be addressed before writing down the equations is the one determining the gluon. It is a matrix equation, but only two components are independent. The most direct way to obtain two scalar equations for the two dressing functions is to contract the gluon equation once with \pref{tproj} and once with \pref{lproj}. The more convenient way is to contract with
\bea
P_{T\mu\nu}^\zeta=\zeta P_{T\mu\nu}+(\zeta-1)A_{T\mu\nu}\label{gtproj}\\
P_{L\mu\nu}^\xi=\xi P_{L\mu\nu}+(\xi-1)A_{L\mu\nu}.\label{glproj}
\eea
\noindent Varying the parameters $\zeta$ and $\xi$ allows to investigate the amount of gauge invariance violation, as discussed in section \ref{strunc}. The tensors \pref{gtproj} and \pref{glproj} reduce to \pref{tproj} and \pref{lproj} at $\zeta=\xi=1$. The selection of $A_{T/L\mu\nu}$ is governed by practical aspects and will be different for the high temperature case and the finite temperature case. Thus writing down the final equations will have to await chapters \ref{c3d} and \ref{cft}, respectively. Graphically the equations for the three scalar functions $G$, $Z$ and $H$ at finite temperature are given in figure \ref{figftsys}.

\section{Truncations and Constraints}\label{strunc}

As already indicated, the DSEs to be solved in the following chapters are truncated. Although in the following several arguments will be made to support the truncations, there is no method (yet) known which permits an a priori controlled truncation. In addition, no possibility is yet known to conserve local internal symmetries at least on the level of the truncation. Even for global symmetries this leads to enormous technical complications, see e.g.\/ \cite{vanHees:2001ik}. Furthermore any such truncation necessarily violates unitarity. Also no scheme is yet known which at the same time conserves energy and agrees with perturbation theory in the far ultraviolet. Hence it is currently only possible to assume that all these effects do not contribute significantly. This can only be done by comparing a posteriori either to experiments or different methods. As the objects treated here are gauge-variant quantities, only other calculational methods are available for comparison. Only lattice calculations have  been performed in the region of interest up to now and only for SU$(N_c)$ gauge theories. It will turn out that the results agree remarkably well, considering the drastic assumptions made. By systematically assessing the error due to gauge invariance violation, it is found that the effects of the truncation are of a quantitative nature only. Hence the truncation scheme to be described seems to be applicable.

The most extreme truncation is the neglect of the equations of all $n$-point Green's functions with $n>2$. This truncation is well justified in the ultraviolet due to asymptotic freedom. In the infrared, assuming the correctness of the Zwanziger-Gribov scenario, this is also justified and the results support this assumption self-consistently. At mid-momenta, significant deviations are to be expected. Indeed this is the region where the largest deviation from lattice results will be found. The further truncation is to neglect all genuine full 2-loop contributions within the equations for the 2-point Green's functions. Largely the same arguments apply here: The 2-loop contributions are sub-dominant in the ultraviolet and most probably sub-leading in the infrared. The latter was checked and found to be correct for tree-level instead of full vertices \cite{Watson:phd}. In addition it was found that only considerable fine-tuning of the vertices allowed the 2-loop contributions to become as leading as the 1-loop contributions in the infrared \cite{Bloch:2003yu}.

The next assumption is that the color structure is the same as in perturbation theory. All investigations concerning this point have supported this assumption \cite{Boucaud:1998xi}. This also automatically entails the vanishing of all 1-point Green's functions due to the antisymmetry of the color structure of the vertices.

The last ingredient of the truncation is the construction of the remaining full vertices. The ghost-gluon-vertex will be kept at its tree-level form \pref{tlcca}, motivated by the Zwanziger-Gribov scenario. This is exact for vanishing incoming ghost momenta as discussed in subsection \ref{sssti}. A bare vertex is also supported by numerical studies in three and four dimensions \cite{schleifenbaum:diploma} and lattice calculations in four dimensions \cite{Cucchieri:2004sq}. Furthermore, at least in the vacuum, the qualitative nature of the infrared solution is independent of the detailed structure of the ghost-gluon vertex to a large extent \cite{Lerche:2002ep}. It has also been shown that, under weak assumptions, the qualitative infrared solution for the ghost is independent of the truncation \cite{Watson:2001yv}.

The various three-gluon vertices for 3d-longitudinal and 3d-transverse gluons are not fixed yet. These will be constructed by requiring minimal gauge invariance violation. This will be addressed in chapters \ref{c3d} and \ref{cft}, completing the truncation scheme.

Concerning the artifacts of the truncation, it is not possible to solve the problem of unitarity violation. Unitarity will always be violated as long as the full infinite system is not solved\footnote{In perturbation theory, these violations are of higher order in the expansion parameter and can thus be neglected. The same applies to the results here in the realm of applicability of perturbation theory, as the same results are obtained in this domain.}. The problem of energy conservation will be addressed in section \ref{std}.

The last point to be discussed is the violation of gauge invariance. There are two aspects to be treated. The first is the problem of Gribov copies. Due to the arguments given in section \ref{squant}, this can be resolved by requiring
\be
G(p^2)\ge 0,\quad Z(p^2)\ge 0,\quad H(p^2)\ge 0,\label{gribov}
\ee
\noindent since this guarantees to stay within the first Gribov horizon. Note that by condition \pref{gribov}, the ghost propagator is negative definite and cannot have a positive semidefinite spectral function.

The second problem is much harder to address. Even provided the Gribov problem is solved, gauge invariance is violated, as the STIs are no longer fulfilled. Indeed it is not even possible to test whether the STIs are fulfilled, as in general the STIs for an $n$-point Green's function contain contributions from $n'$-point Green's functions with $n'>n$. These are negligible in perturbation theory because they are of higher order in the expansion parameter, but this is not true in general in non-perturbative calculations. In principle it would be possible to test the STIs when truncating them to the same level as the Green's functions, but it turns out that this leads to inconsistencies, see e.g.\/ \cite{vonSmekal:1997is}. Nonetheless trying to construct vertices which as best as possible fulfill the truncated STIs, it is found that the results are only weakly affected compared to tree-level vertices \cite{vonSmekal:1997is}. This gives confidence that these violations are small.

There are two consequences of these gauge invariance violations. The first is the appearance of spurious divergences in cases where the degree of divergence is lowered by gauge invariance compared to naive power counting. This is the case for the gluon self-energy. These spurious divergences have to be removed. This will be done using the tadpole terms, which are not left as free parts of the equations, but are chosen to compensate such spurious divergences and other artifacts of the truncation, thus mimicking their role in perturbation theory.

The second consequence is the violation due to finite contributions of the STIs. In the remainder of the work, the main equation to test the amount of gauge invariance violation will be the STI for the gluon propagator, \pref{stigluon}. It is for this reason the projectors $\pref{gtproj}$ and $\pref{glproj}$ are used instead of $\pref{tproj}$ and $\pref{lproj}$. If \pref{stigluon} was exactly fulfilled, the results obtained would be independent of $\zeta$ and $\xi$. If the results do only weakly depend on these parameters, gauge invariance violations are most likely small \cite{Fischer:2002hn,Alkofer:2002ne,Fischer:2003zc}. The variational range for $\zeta$ and $\xi$ in the case of a violation cannot be extremely large, since otherwise the projection will be primarily on the gauge-violating longitudinal part and will not give rise to further useful information.

\section{Perturbation Theory}\label{sperturb}

The DSE approach followed here aims at the full 2-point Green's functions. Hence it is necessary that perturbative results are embedded in the final results. For sufficiently large momenta the Green's functions must reduce to their perturbative counterparts, due to the asymptotic freedom of Yang-Mills theories. As the DSEs are truncated at one-loop level, this reduction can only be correct to leading order (LO) in $g_d^2/p^{4-d}$, where $p$ is the momentum scale. Subleading contributions will necessarily deviate from perturbation theory. This also guarantees that the violation of gauge invariance will be not worse than in LO perturbation theory, and thus at least the same level of gauge invariance is achieved.

The high temperature limit in chapter \ref{c3d} will give an explicit example of this. In case of the finite temperature corrections in chapter \ref{cft}, the results are restricted to momenta of the order of $T$, see section \ref{ssmallpapprox}. Therefore, perturbation theory will only be reproduced if these momenta are already in the perturbative regime.

In the case of 4d vacuum calculations, agreement with LO resummed perturbation theory turns out to be a significant task and requires modifications of the three-gluon vertex \cite{vonSmekal:1997is,Fischer:2002hn}. Comparison to perturbation theory is at the current level of truncation the only possibility to compare to experiment, and thus an important constraint.

\section{Renormalization}\label{srenormalization}

As in perturbation theory, the usual ultraviolet divergences of Yang-Mills theory are encountered and must be regularized and renormalized \cite{Bohm:yx}. A wide variety of possibilities to deal with the divergences at the perturbative level exist, especially with dimensional regularization for gauge theories. However, most of these concepts, including the latter, are not applicable to non-perturbative calculations \cite{Collins:xc,Zinn-Justin:mi}.

Due to the lack of a symmetry conserving regularization scheme for DSEs which is of technically acceptable complexity, an alternative route is chosen. It is always possible to use gauge non-invariant regularization prescriptions if appropriate compensating counter-terms are chosen \cite{Collins:xc}. This approach will be employed in chapter \ref{cft}. There the effects of a non-gauge-invariant regularization will be absorbed into the tadpoles.

The remaining divergencies are then regularized and renormalized  using counter-terms. In general a counter-term for e.g.\/ renormalizing the wave function is introduced in a Lagrangian by performing the replacement
\be
{\cal L}=(\pdm\phi)^2\to{\cal L}_R=(\pdm\phi)^2+\delta Z(\pdm\phi^2)=:Z(\pdm\phi)^2.\nonumber
\ee
\noindent Here $\phi$ is an arbitrary field and the counter-term $\delta Z$ has been chosen such as to cancel the divergences \cite{Collins:xc}. This guarantees multiplicative renormalizability even in the non-perturbative regime by explicitly constructing the wave-function renormalization constant $Z$. At finite temperature no new divergences arise compared to the vacuum divergence structure of the phase the theory is in \cite{Das:gg}. Nevertheless, the finite parts of the counter-terms may depend on temperature.

When treating finite temperature effects in chapter \ref{cft} a few subtleties concerning renormalization in the employed truncation scheme arise. These are treated in section \ref{scttrunc}.

\section{Solutions in the Vacuum}\label{ssolvac}

Solutions of the DSEs in the vacuum \cite{Alkofer:2000wg,vonSmekal:1997is,Fischer:2002hn} have already convincingly demonstrated the non-triviality of the infrared regime and also showed good agreement with lattice calculations. This supports the applicability of the truncation scheme. In this section these results will be described briefly to put this work in the appropriate context and demonstrate that the method is sufficiently stable for an extension to finite temperature.

\subsection{Yang-Mills Theory}

\begin{figure}
\epsfig{file=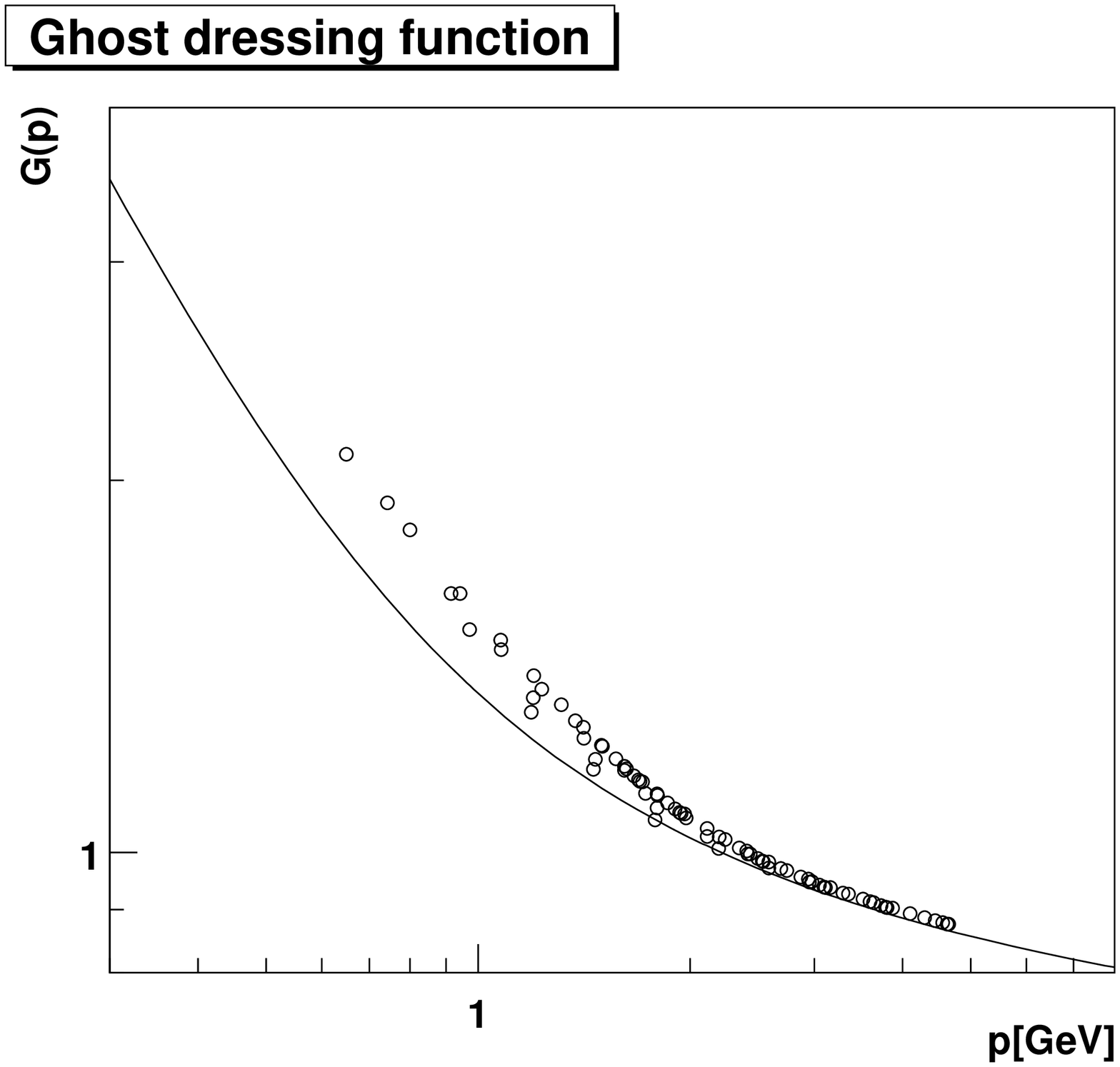,width=0.5\linewidth}\epsfig{file=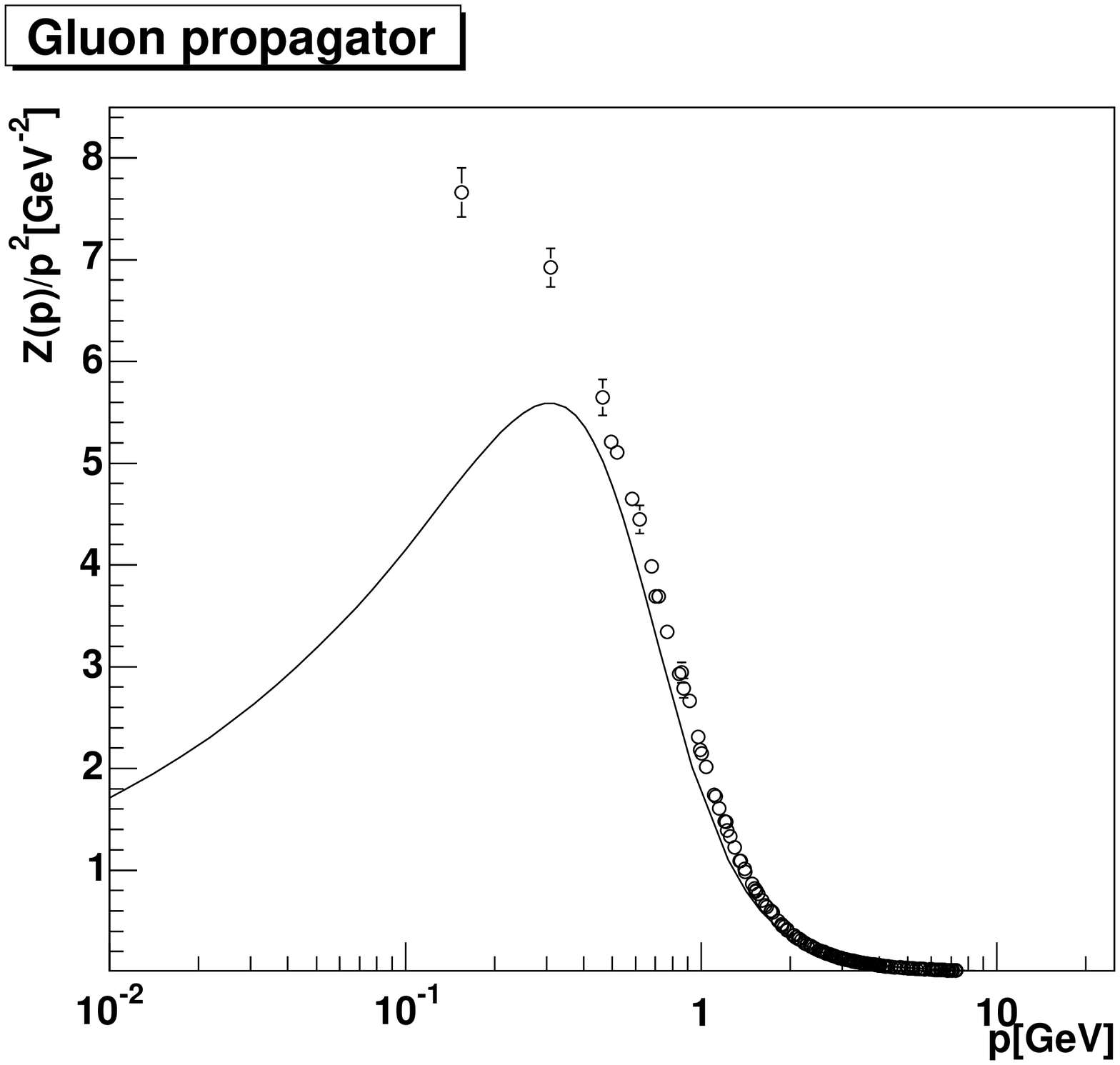,width=0.5\linewidth}
\caption{The left panel shows the ghost dressing function and the right panel the gluon propagator in vacuum Yang-Mills theory \cite{Fischer:2002hn}. The results are compared to lattice data of the ghost dressing function \cite{Langfeld:2002dd} and the gluon propagator \cite{Bowman:2004jm}. The indicated errors are statistical only.}
\label{figymvac}
\end{figure}

For the pure Yang-Mills sector, results have been obtained with increasing precision over time\footnote{The earliest attempt neglected the ghost contribution \cite{Mandelstam:1979xd}. This ``Mandelstam approximation'' leads to results contradicting recent lattice results and is therefore dismissed today.}. The DSE results satisfy \pref{oehme} and \pref{kugo} and thus exhibit manifest gluon confinement, in accordance with the Kugo-Ojima and Zwanziger-Gribov scenario \cite{Alkofer:2000wg,vonSmekal:1997is,Fischer:2002hn}. Also, the analytical structure has been understood to some extent \cite{Alkofer:2003jk}. The results \cite{Fischer:2002hn} are shown in figure \ref{figymvac} compared to lattice results \cite{Langfeld:2002dd,Bowman:2004jm}.

The lattice points farthest in the infrared suffer from finite volume effects and are expected to bend down for larger lattice volumes. In 3d-calculations, where significantly larger lattices can be used, this is indeed the case \cite{Cucchieri:2003di} and will be seen  when comparing to lattice results in section \ref{slattice3d}. Recently, the results have also been confirmed by exact renormalization group methods \cite{Gies:2002af}. These results give confidence that the method can be applied to the finite temperature case as well.

A further result of the vacuum studies is that the quantity 
\be
\alpha(\mu)=\alpha(s) G(s,\mu)^2 Z(s,\mu),\label{runningcoupling}
\ee
\noindent where $\mu$ is the renormalization scale and $s$ the subtraction point, is a renormalization group invariant, and agrees with the running coupling in the perturbative regime. Although it is under debate what the non-perturbative definition of a coupling constant is, if any, it is possible to investigate this quantity. It does not exhibit a Landau pole and has an infrared fixed point of $\alpha(0)N_c=8.915$ \cite{Fischer:2002hn}. This result has also been confirmed by exact renormalization group calculations \cite{Gies:2002af}.

\subsection{Full QCD}

\begin{figure}
\epsfig{file=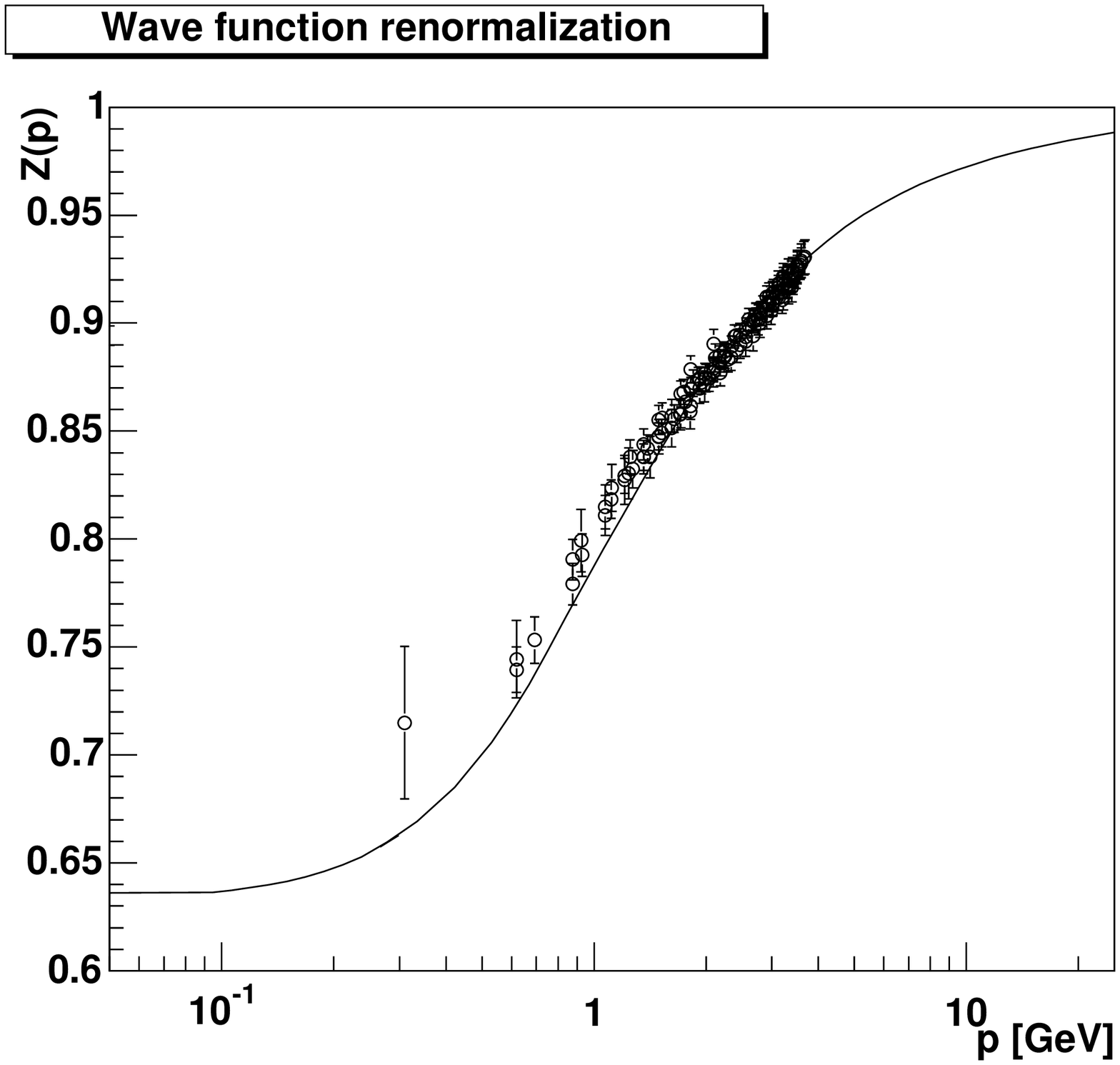,width=0.5\linewidth}\epsfig{file=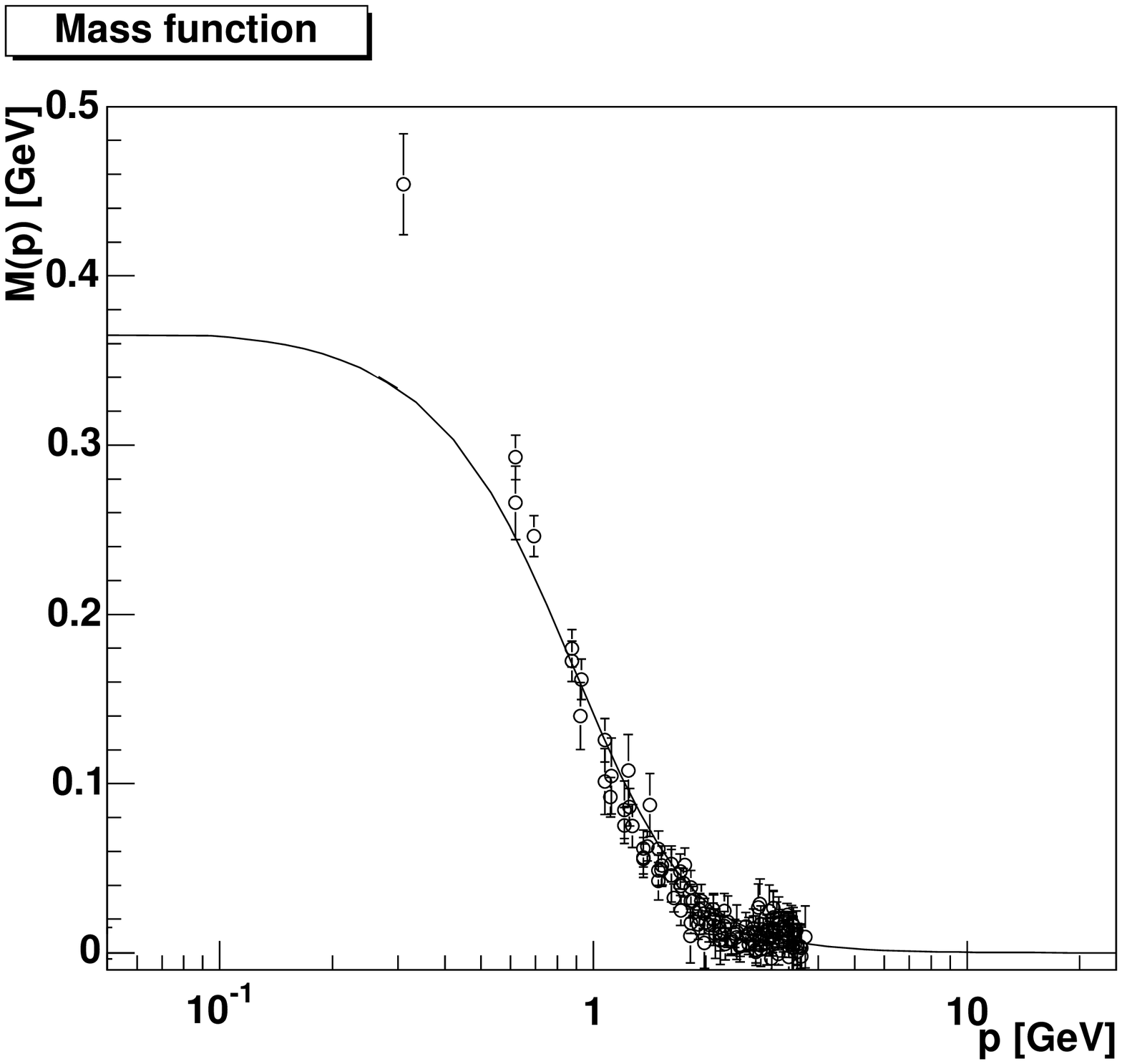,width=0.5\linewidth}
\caption{The quark propagator in the chiral limit \cite{Fischer:2003rp} compared to lattice results using the asqtad action \cite{Bowman:2002kn}. The left panel shows the quark wave-function renormalization compared to lattice results. The right panel shows the quark mass function compared to chirally extrapolated lattice results. The indicated errors are statistical only.}
\label{figquarks}
\end{figure}

DSEs have also been applied to full QCD in two different ways. A more phenomenological ansatz has employed model gluon propagators. This has led to extensive and successful investigations of hadron phenomenology. For a review see e.g.\/ \cite{Alkofer:2000wg,Roberts:2000aa}. As the approach followed here is more bottom-up, this aspect will not be discussed further. The other approach couples the Yang-Mills sector described above to the quarks and solves the corresponding DSEs in a similar truncation scheme \cite{Fischer:2003rp}. The general structure of the Euclidean quark propagator is
\be
S(p)=\frac{Z(p)}{-i\gamma_\mu p_\mu+M(p)}
\ee
\noindent with the wave-function renormalization $Z$ and the mass function $M$. For less than 4 light or massless quarks (as it is realized in nature), the Yang-Mills solutions remain essentially the same, besides the according changes in the ultraviolet anomalous dimensions. Thus even the effect of massless quarks is small. The two independent dressing functions $Z$ and $M$ are shown in figure \ref{figquarks}. Chiral symmetry breaking is manifest. However, the amount of symmetry breaking is sensitive to the specific quark-gluon vertex, as the quark results are in general. In the present case a modified Curtis-Pennington vertex has been employed \cite{Fischer:2003rp}. Also, the problem of quark confinement is not yet understood in this ansatz, as the information obtainable from the quark propagator alone are not conclusive \cite{Alkofer:2003jk}.


%% file: 3dlimit.tex
\chapter{Infinite-Temperature Limit}\label{c3d}

The infinite-temperature limit of Yang-Mills theory is presented in this chapter\footnote{Most of the results presented are published in \cite{Maas:2004se}.}. Although the limit itself is purely academic, it is valuable not only due to technical simplifications. Firstly, properties surviving in this limit will also be present at lower temperatures. Especially persisting non-trivial effects are genuine features of the high-temperature phase. Secondly, it is found in the finite-temperature calculations of chapter \ref{cft} as well as in lattice calculations \cite{Cucchieri:2001tw} that the propagators are already close to their asymptotic values for quite small temperatures, as low as a few times $T_c$.

The chapter starts with the derivation of the infinite-temperature limit in section \ref{s4dto3d}. The emerging theory is a 3d-Yang-Mills theory with an additional adjoint Higgs field. The infrared and ultraviolet properties of this theory will be inspected in section \ref{sanalytic}. It is then studied numerically in three truncation schemes. The first is the ghost-loop-only scheme, to test the Zwanziger-Gribov scenario, in section \ref{sghostloop3d}. The next schemes address the pure Yang-Mills theory in section \ref{syangmills} and finally the full theory in section \ref{sfull3d}. The numerical method deployed is described in appendix \ref{anum}.

A comparison to lattice results will be made in section \ref{slattice3d}. The 3d-theory is not only relevant to the high temperature behavior of Yang-Mills theory. As an example, a recently obtained relation between Landau gauge in 3d and Coulomb gauge in 4d \cite{Zwanziger:2003de} will be addressed in section \ref{scoulomb}.

As this chapter deals nearly exclusively with a 3d-theory, the notation $p\equiv|\vec p|$ is used, if not noted otherwise.

\enlargethispage*{1cm}

\section{From 4d to 3d}\label{s4dto3d}

To obtain the infinite-temperature limit, temperature is introduced into the vacuum equations \pref{vacgheq} and \pref{vacgleq} as detailed in the previous chapter. To obtain scalar equations for the dressing functions $Z$ and $H$, the additional tensor structures in \pref{gtproj} and \pref{glproj} are chosen most conveniently in 4d as
\bea
A_{T\mu\nu}&=&\left(\delta _{\mu \nu }-\left(1+\frac{p_{0}^{2}}{\vec p^{2}}\right)\delta _{\mu 0}\delta _{0\nu }\right)\label{d3tgenp}\\
A_{L\mu\nu}&=&\left(1+\frac{p_{0}^{2}}{\vec p^{2}}\right)\delta _{\mu 0}\delta _{0\nu}.\label{d3lgenp}
\eea
\noindent In \cite{Maas:2002if} it was demonstrated that, in zeroth order, the infinite-temperature limit can be found by neglecting any contributions from Matsubara frequencies different from zero. This yields an effective 3d-Yang-Mills theory with an additional adjoint Higgs. The Higgs field is the $A_0$ field of the 4d-theory, and therefore the number of degrees of freedom is conserved in this process.

By considering the structure of the projectors \pref{gtproj} and \pref{glproj} in the case of $p_0=0$, which is the zeroth Matsubara frequency, it is also possible to find the connection of the 3d and 4d degrees of freedom. At $p_0=0$, $P_{L\mu\nu}^\xi$ becomes $\delta_{00}$, independent of $\xi$. It projects out the time-time component of the propagator, which belongs to the $A_0$ field. Therefore the 3d-longitudinal part of the 4d gluon propagator corresponds to the Higgs propagator. On the other hand, $P_{T\mu\nu}^\zeta$ becomes
\be
P_{ij}^\zeta=\delta_{\mu\nu}-\zeta\frac{p_i p_j}{p^2}\label{bpproj}
\ee
\noindent with zero time-time and time-space components. At $\zeta=3$ this is the Brown-Pennington projector of the 3d-theory \cite{Brown:1988bm}. \pref{bpproj} projects onto the 3d-subspace and thus establishes the connection between the 3d-transverse gluon and the gluon of the 3d-theory. Therefore the dressing functions $H$ and $Z$ of the gluon propagator \pref{gluonprop} describe the Higgs and the 3d gluon, respectively.

This amounts to integrating out the hard modes at tree-level. In general, this is not sufficient \cite{Kajantie:1995dw}. Higher order effects of the hard modes can potentially still influence the interactions of the soft modes. Therefore the general prescription to obtain the effective 3d-theory is to write down the most general Lagrangian allowed and match the parameters by comparing to the 4d-theory \cite{Kajantie:1995dw}. This can be done e.g.\/ by lattice calculations \cite{Cucchieri:2001tw} and perturbation theory \cite{Kajantie:1995dw}. In the present case a tree-level mass for the Higgs and a 4-Higgs coupling are additionally present. This also modifies the Dyson-Schwinger equations, and they therefore have to be rederived. As an explicit example of this process the generation of the tree-level mass will be demonstrated in section \ref{sftir}.

Hence the Lagrangian \pref{l3d} governs the 3d-theory and describes a Yang-Mills field coupled to an adjoint scalar field \cite{Alkofer:2000wg,Kajantie:1995dw}. All occurring constants are effective constants, which arise by integrating out the hard modes. The effective constants can only be derived by calculating the full theory. Therefore the values obtained from lattice and perturbative calculations will be used here, as listed in \cite{Cucchieri:2001tw}. In most results these constants are irrelevant and will drop out, except for the objects discussed in chapter \ref{cderived}.

The only exception is the 4-Higgs coupling constant $h$, which can be uniquely determined already in the 3d-theory. In principle, the Higgs self-energy may contain linear divergences, if this theory stood on its own. However, the Higgs field is only a component of the 4d gluon field, and thus should not contain a linear divergence due to the STIs of the 4d-theory. Implementation of this requirement, in leading-order perturbation theory, fixes $h$ as
\be
h=-2g_3^2\frac{C_A}{C_A+2},\label{hvalue}
\ee 
\noindent which is detailed in appendix \ref{appUV}. $C_A$ is the second Casimir of the gauge group, see appendix \ref{aconventions}. Note that by \pref{hvalue} exact t'Hooft scaling \cite{'tHooft:1973jz} is also maintained, which would otherwise be broken by the Higgs-tadpoles.

Hence the 3d-theory is finite and therefore all renormalization constants can be set to 1, i.e.\/ no counter-terms are necessary. The divergences have not disappeared, though. By sending the temperature to infinity while maintaining for the renormalization scale $\mu\gg T$, renormalization takes place at $p\to\infty$ and can therefore be neglected at finite momenta. This intuitive argument is shown to be correct in chapter \ref{cft}, where the limit is taken explicitly.

\begin{figure}
\epsfig{file=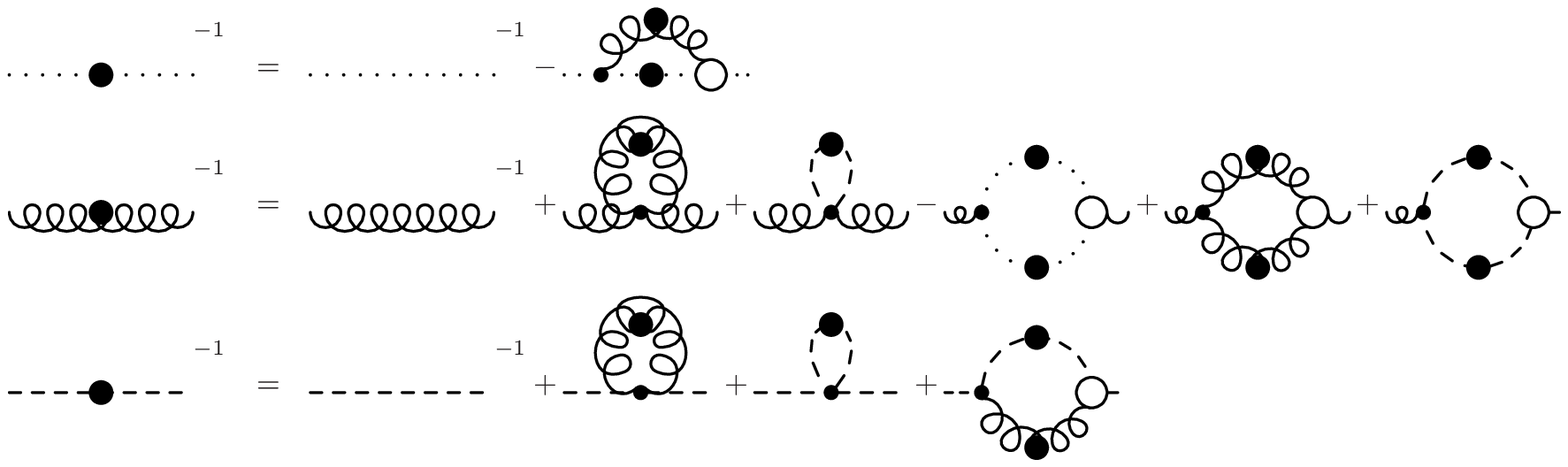,width=\linewidth}
\caption{The system under study in the infinite-temperature limit. Wiggly lines represent 3d-transverse gluons, dashed lines Higgs and dotted lines ghosts. Lines with a dot are full propagators. Black dots are bare vertices and white circles are full vertices, which have to be reconstructed in the truncation discussed.}
\label{figthreedsys}
\end{figure}

The DSEs for the Yang-Mills sector are already known, see e.g.\/ \cite{Alkofer:2000wg,Roberts:2000aa}. Adding the Higgs, the ghost equation will not be modified compared to pure Yang-Mills theory, since no tree-level Higgs-ghost-coupling is present. The remaining alterations of the equations due to the Higgs are derived in appendix \ref{adse}. They lead to the DSEs\enlargethispage*{1cm}
\bea
D^{ab}_G(p)^{-1}&=&-\delta^{ab}p^2\nonumber\\
&+&\int\frac{d^dq}{(2\pi)^d}\Gamma_\mu^{\tl\indexsep c\bar cA\indexsep dae}(-q,p,q-p)D^{ef}_{\mu\nu}(p-q)D^{dg}_G(q)\Gamma^{c\bar cA\indexsep bgf}_\nu(-p,q,p-q)\label{ghostd3}\\
D^{ab}_{H}(p)^{-1}&=&\delta^{ab}(p^2+m_h^2)+T^{HG\indexsep ab}+T^{HH\indexsep ab}\nonumber\\
&+&\int\frac{d^dq}{(2\pi)^d}\Gamma^{\tl\indexsep A\phi^2\indexsep eac}_\nu(-p-q,p,q)D_{\nu\mu}^{cg}(p+q)D^{fc}(q)\Gamma_\mu^{A\phi^2\indexsep gbf}(p+q,-p,-q)\nonumber\\\label{higgsd3}\\
D^{ab}_{\mu\nu}(p)^{-1}&=&\delta^{ab}(\delta_{\mu\nu}p^2-p_\mu p_\nu)+T_{\mu\nu}^{GG\indexsep ab}+T^{GH\indexsep ab}_{\mu\nu}\nonumber\\
&-&\int\frac{d^dq}{(2\pi)^d}\Gamma_\mu^{\tl\indexsep c\bar cA\indexsep dca}(-p-q,q,p) D_G^{cf}(q) D_G^{de}(p+q)\Gamma_\nu^{c\bar c A\indexsep feb}(-q,p+q,-p)\nonumber\\
&+&\frac{1}{2}\int\frac{d^dq}{(2\pi)^d}\Gamma^{\tl\indexsep A^3\indexsep acd}_{\mu\sigma\chi}(p,q-p,-q)D^{cf}_{\sigma\omega}(q)D^{de}_{\chi\lambda}(p-q)\Gamma^{A^3\indexsep bfe}_{\nu\omega\lambda}(-p,q,p-q)\nonumber\\
&+&\frac{1}{2}\int\frac{d^dq}{(2\pi)^d}\Gamma_\mu^{\tl\indexsep A\phi^2\indexsep acd}(p,q-p,-q)D^{de}(q)D^{cf}(p-q)\Gamma^{A\phi^2\indexsep bef}_\nu(-p,q,p-q),\nonumber\\\label{gluond3}
\eea
\noindent where $T^{ij}$ are the tadpole contributions. The first index gives the equation where the tadpole contributes, $G$ for gluon and $H$ for Higgs, and the second index the type of tadpole appearing. Tree-level quantities are again denoted by a superscript `$\tl$', and can be found in equations \prefr{tlcca}{tl4h} in appendix \ref{adse}. In these equations already the genuine two-loop contributions in the gluon and Higgs equations have been neglected. The ans\"atze for the 3-gluon and Higgs-gluon vertices are motivated by technical considerations and are thus deferred to sections \ref{syangmills} and \ref{sfull3d}. The graphical representation of this set of truncated equations is shown in figure \ref{figthreedsys}.

The Higgs propagator $D_H$ is linked to the dressing function $H$ by
\be
D_H(p^2)=\frac{H(p^2)}{p^2}\label{higgsdress}.
\ee
\noindent Replacing the propagators in \prefr{ghostd3}{gluond3} by their respective dressing functions, equations for the latter are obtained. These are divided by $p^2$ to make them dimensionless. To obtain a scalar equation for the gluon dressing function, equation \pref{gluond3} is contracted with \pref{bpproj} and divided by 2. This results in\enlargethispage*{1cm}
\bea
\frac{1}{G(p)}&=&1+\frac{g_3^2C_A}{(2\pi)^2}\int d\theta dq A_T(p,q)G(q)Z(p-q)\label{fulleqG3d}\\
\frac{1}{H(p)}&=&1+\frac{m_h^2}{p^2}+T^{HG}+T^{HH}+\nonumber\\
&&+\frac{g_3^2C_A}{(2\pi)^2}\int d\theta dq\Big(N_1(p,q)H(q)Z(p+q)+N_2(p,q)H(p+q)Z(q)\Big)\label{fulleqH3d}\\
\frac{1}{Z(p)}&=&1+T^{GH}+T^{GG}+\frac{g_3^2C_A}{(2\pi)^2}\int d\theta dq\Big(R(p,q)G(q)G(p+q)\nonumber\\
&&+M_L(p,q)H(q)H(p+q)+M_T(p,q)Z(q)Z(p+q)\Big).\label{fulleqZ3d}
\eea
\noindent Here $\theta$ is the angle between $\vec p$ and $\vec q$. $A_T$, $N_1$, $N_2$, $R$, $M_L$ and $M_T$ are the integral kernels for the employed truncation and are given in appendix \ref{s3dkernels}. The tadpoles in the gluon equation are now also contracted. Although $N_1$ and $N_2$ can be rearranged in one kernel by an integral transformation, they are left separated for a better comparison to the finite temperature case. $C_A$ is the second Casimir of the adjoint representation of the gauge group. It only appears in the combination $g_3^2C_A$, thus any change of the gauge group can be cast into a change of $g_3^2$. Especially 't Hooft-scaling is therefore manifest.

\section{Asymptotic Analysis}\label{sanalytic}

In this section, equations \prefr{fulleqG3d}{fulleqZ3d} will be solved analytically for asymptotically small and large momenta. In all these calculations, a massive Higgs will be assumed, as this is the case for Yang-Mills theories. However, a massless Higgs offers a rich infrared phase structure. Since this is somewhat out of the main line of interest it is deferred to appendix \ref{asmasslesshiggs}.

\subsection{Ultraviolet Analysis}\label{ssuvanalysis}

Since asymptotic freedom is expected to hold also in the 3d case \cite{Feynman:1981ss}, the dressing functions and vertices should reduce to their one-loop counterpart and ultimatively to their tree-level value for sufficiently large momenta.

As $g_3^2$ has dimension of mass and does not enter in the loop integrals in \prefr{fulleqG3d}{fulleqZ3d}, by dimensional analysis the 1-loop contributions must be proportional to $g_3^2/p$. All loop-integrals are therefore sub-leading in the ultraviolet compared to the tree-level contribution. In the case of the Higgs, this is only true if also $p\gg m_h$. This has been verified by explicit calculations, see appendix \ref{appUV}. The 3d-theory gives therefore a very vivid example of asymptotic freedom. Thus all dressing functions $D$ acquire their tree-level value in the ultraviolet,
\be
\lim_{p\to\infty}D(p)=1.\label{uvtree}
\ee
\noindent Stated otherwise, to leading order for $p\gg g_3^2,m_h$
\be
D(p)\approx 1+\frac{cg_3^2C_A}{p}.\label{uvnlo}
\ee
\noindent The constant $c$ turns out to be positive for all dressing functions. Thus, all dressing functions approach the tree-level behavior from above.

\subsection{Infrared Analysis}\label{ssiranalysis}

To obtain analytical solutions in the infrared, the first step is to note that all integral kernels contain a term $(p\pm q)^2$ or $q^2$ in the denominator. Therefore the integrands are strongly peaked for $p\approx q$. To obtain analytic infrared solutions it is hence permissible to replace the dressing functions by their asymptotic infrared form. Since the infrared limit is a critical limit of the theory, the ansatz of power-laws is justified. The ans\"atze for the dressing functions are
\bea
G(p)=A_g p^{-2g}\label{iransatz1}\\
Z(p)=A_z p^{-2t}\\
H(p)=A_h p^{-2l}.\label{iransatz}
\eea
\noindent In principle, three possibilities for the infrared behavior can be distinguished: Dominance of tree-level-, tadpole- and loop-terms.

Tree-level dominance is motivated by the naive idea of a free gas of gluons in the asymptotic temperature limit. A first tempting ansatz is $g=t=0$, and $l=-1$ due to the explicit mass term for the Higgs. This corresponds to a purely perturbative or Coulomb phase. This is already impossible in the ghost equation, since in this case the ghost self-energy would diverge as $1/p$ in the infrared and hence superseding the tree-level term. The only possibility would be a ghost-gluon vertex suppressed at least like $p$ in the infrared, which would be unexpected for tree-level propagators. Similar arguments apply to the gluon equation. Hence no Coulomb phase exists in 3d.

If the gluon would behave as a massive particle in the infrared then the ghost equation would allow for a tree-level ghost. This is expected in a Higgs-phase, which would generate a magnetic screening mass. Such a term could only be generated by a tadpole diagram, since otherwise any of the possible vertices would have to diverge at least as strong as $1/p$ in the infrared (from the ghost-loop) or like $1/p^5$ (from the gluon- or Higgs-loop). This is not very probable with massive or tree-level particles. This should then also have to be true for arbitrary projections of the gluon equation, especially for $\zeta=3$. Since the tadpoles, even without truncation, drop out identically, a contradiction arises, and this solution is excluded.

Attempting to save this option by using the ghost-loop to generate a mass also fails, since the required ghost exponent would lead to an non-renormalizable ultraviolet divergence of the integral. The only way to generate such a Higgs phase\footnote{Besides spontaneous breaking of the gauge symmetry on the level of the Lagrangian by giving the Higgs a vacuum expectation value.} would be by appropriate fine-tuning of the ghost-gluon vertex and the ghost. This could of course happen, but is unlikely. Besides, it is hard to see how the vacuum corresponding to such propagators avoids the violation of Elitzur's theorem \cite{Elitzur:im}. Hence the tadpole must at least be compensated in the gluon equation and $t\le -1$, but integral convergence will require $t<-1$, see appendix \ref{air}.

Thus, the remaining option is loop dominance. Motivated by the reasoning of Zwanziger \cite{Zwanziger:2003cf}, ghost dominance is assumed, and the solutions indeed satisfy it. Hence neglecting all contributions without ghost lines, it is then only necessary to specify the ghost-gluon vertex. As discussed in section \ref{strunc} a bare ghost-gluon vertex is chosen. As it is already required that
\be
t<-1,\label{tassump}
\ee
\noindent confinement is present due to the Oehme-Zimmermann super-convergence relation \pref{oehme}.

It is then possible to calculate the infrared limit of the DSEs \prefr{fulleqG3d}{fulleqZ3d}. This is done in appendix \ref{air} and leads to
\bea
\frac{y^g}{A_g}&=&y^{-(4-d)/2-g-t}I_{GT}(g,t)A_gA_z\label{ghostir}\\
\frac{y^t}{A_z}&=&1+y^{-(4-d)/2-2g}I_{GG}(g,\zeta)A_g^2\label{gluonir}\\
\frac{y^l}{A_h}&=&1+\frac{m_h^2+\delta m_3^2}{y}\label{higgsir},
\eea
\noindent where $y=p^2$ and $d$ is the dimension. A subtraction in \pref{ghostir} has been performed and only the finite part has been retained. The expressions for $I_{GT}$ and $I_{GG}$ stemming from the ghost-self energy and the ghost-loop, are calculated in appendix \ref{air}. In the Higgs equation, a finite renormalization of the mass has been allowed for. The mass renormalization will be discussed in subsection \ref{sfull3d} and is given in equation \pref{higgstadpole}. Equation \pref{higgsir} is then solved immediately by setting $l=-1$ and 
\be
A_h=\frac{1}{m_h^2+\delta m_3^2},\label{ah}
\ee 
\noindent since 1 can be neglected for $p\ll m_h$. This already indicates that a qualitative change occurs in the high temperature limit, as in the vacuum $t=l$. This also immediately shows that the Higgs particle decouples, at least in the infrared, from the Yang-Mills sector. This agrees with corresponding findings on the lattice \cite{Cucchieri:2001tw}.

By dimensional consistency in the ghost equation \pref{ghostir}, a relation between $g$ and $t$ follows directly \cite{Zwanziger:2001kw} as
\be
g=-\frac{1}{2}\left(t+\frac{4-d}{2}\right)=_{d\to 3}-\frac{1}{2}\left(t+\frac{1}{2}\right)\label{gt}.
\ee
\noindent The result at asymptotic temperature is hence different from the relation \cite{vonSmekal:1997is}
\be
g=-2t\label{gt4d}
\ee
\noindent which is found at zero temperature. The additional power of $p$ introduced by this change compensates the dimension of the effective coupling constant in 3d or the temperature in general finite-temperature 4d calculations.

Due to \pref{tassump} it then follows directly, that $g$ is less than 0, and thus any solution will automatically satisfy \pref{kugo} and therefore the Kugo-Ojima and the Zwanziger-Gribov conditions are both fulfilled. Also the tree-level contribution in the gluon equation in \pref{gluonir} can now be neglected, since the ghost-loop diverges in the infrared. Hence dominance of the gauge-fixing term as predicted by the Zwanziger-Gribov scenario is confirmed. 

\begin{figure}[t]
\epsfig{file=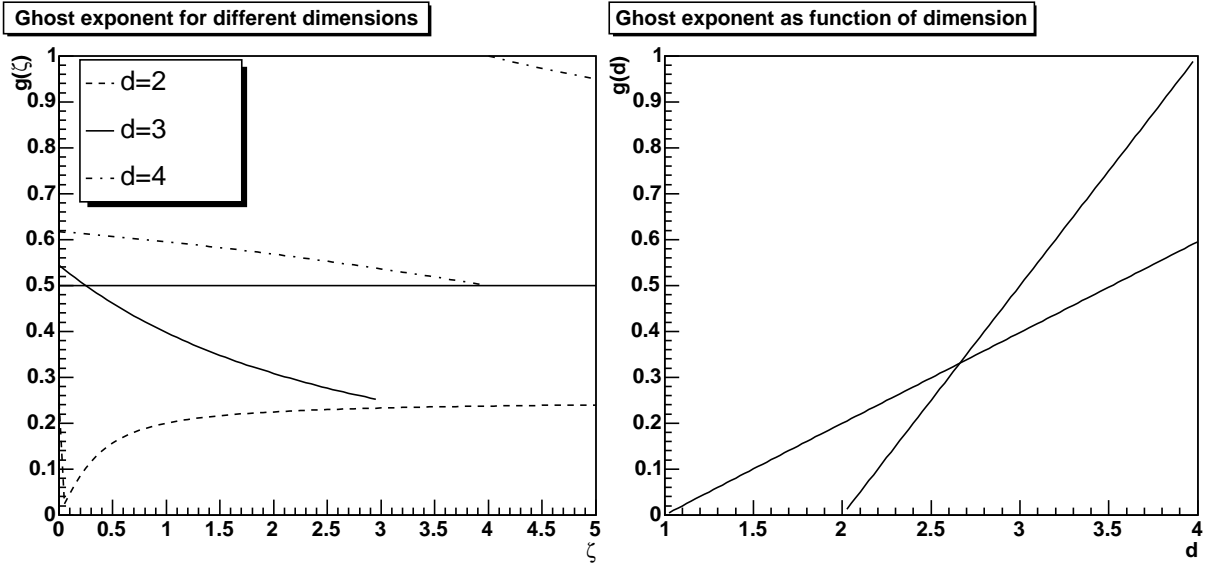,width=\linewidth}
\caption{The left panel shows the two solution branches of equation \pref{irconsistanyd} as a function of $\zeta$ within the region allowed by \pref{gribov} and integral convergence, see equation \pref{dglimits} and \cite{Zwanziger:2001kw}. This excludes the second branch in $d=2$. Dashed is 2d, solid is 3d and dashed-dotted is 4d. Note that at $d=4$ there seems to be only one solution branch. It has been argued that the second branch is $g=1$ \cite{Zwanziger:2001kw}. The right panel shows the solution of \pref{irconsistanyd} as a function of $d$ for $\zeta=1$.}\label{figexp}
\end{figure}

Dividing equations \pref{gluonir} and \pref{ghostir} and eliminating $t$ using \pref{gt}, a conditional equation for $g$ is obtained as
\be
1=-\frac{2^{2g-d}\sqrt{\pi}(\frac{d}{2}+g)(2+d(\zeta-2)-4g(\zeta-1)-\zeta)\csc\left(\frac{\pi(d-4g)}{2}\right)\sin(\pi g)\Gamma(\frac{d}{2}+g)}{\left((d-1)^2g\Gamma\left(\frac{1+d-2g}{2}\right)\Gamma(2g)\right)}.\label{irconsistanyd}
\ee
\noindent This equation has solutions in any dimension, see figure \ref{figexp}. For $d=3$ it simplifies to
\be
1=\frac{32g(1-g)(1-\cot^2(g\pi))}{(1+2g)(3+2g)(2+2g(\zeta-1)-\zeta)}.\label{irconsist}
\ee
\noindent This equation has two solution branches. One is $\zeta$-independent and yields
\be
(g,t)=(\frac{1}{2},-\frac{3}{2})\label{exponents}
\ee
\noindent while the other one varies with $\zeta$. In the special case of $\zeta=1$, the other branch yields
\be
(g,t)\approx(0.3976,-1.2952)\label{sexponents}
\ee
\noindent thus reproducing the results of a previous analysis \cite{Zwanziger:2001kw} which only regarded $\zeta=1$. The second solution is a function varying significantly with $\zeta$. It is therefore necessary to fix the allowed range of $g$. By virtue of the Gribov condition \pref{gribov} and \pref{tassump} as well as convergence of the integrals in the infrared, the allowed range for $g$ in 3d is
\be
\frac{1}{4}<g\le\frac{3}{4}.\label{gallowed}
\ee
\noindent It is further restricted by the requirement that a Fourier transform of the ghost propagator should exist, at least in the sense of a distribution, requiring $g\le 1/2$. This restricts the range of allowed $\zeta$-values for the varying branch to
\be
1/4\le \zeta<3.\label{zetaallowed}
\ee
\noindent At the lower boundary both solutions merge into one. Note that the position space ghost propagator for $g=1/2$ is a half-sided distribution, since its Fourier-transformed exists only in the sense of a limiting procedure. This is analogous to older expectations for the gluon propagator in 4d \cite{Brown:1988bm}. 

Comparing the solution \pref{exponents} with the 4d-case \cite{vonSmekal:1997is} where
\be
\frac{1}{2}<g\le 1,\label{4dexponents}
\ee
\noindent the ghost exponent is only very weakly different from the 4d case, and the difference is even less pronounced for the gluon exponent.

In equations \pref{ghostir} and \pref{gluonir} the  coefficients $A_g$ and $A_z$ only appear in the product $A_g^2A_z$. Therefore only this combination is determined by the infrared analysis and one of both coefficients has to be determined during the numerical solution procedure. This is a non-trivial problem, see appendix \ref{anum}. Expressing $A_z$ by $A_g$ yields
\be
\frac{1}{A_z}=\frac{A_g^2C_Ag_3^2}{(4\pi)^\frac{3}{2}}\frac{2^{4(g-1)}(2+2g(\zeta-1)-\zeta)\Gamma(2-2g)\sin^2(\pi g)}{\cos(2\pi g)(g-1)g^2\Gamma(\frac{3}{2}-2g)}\label{ag}.
\ee
\noindent This result is used to check whether a correct solution is found during the numerical calculations. It also emphasizes that there is only one free parameter of the infrared solution. This one degree of freedom is necessary when using the full solution to perform the exact cancellation of the tree-level term in the ghost equation, which is the single crucial point in the derivation.

With these results and appropriate regularization, it is possible to calculate the other contributions in all equations in the infrared limit as well. No contradiction is found. It would therefore be safe to proceed and solve the equations at all momenta numerically.

However, both solutions can be realized at all momenta. Therefore the question arises which is the correct infrared solution or are both physical, characterizing different phases? In principle, the solution having the lower thermodynamic potential should be the correct one. This is investigated in chapter \ref{cderived} and prefers the varying branch at $\zeta=1$. Likewise the comparison with lattice calculations in subsection \ref{slattice3d} prefer the varying branch at $\zeta=1$. Furthermore, under the assumption of continuity in $d$, the solution in 3d and 4d should lie on the same solution branch. From the right panel of figure \ref{figexp}, it can be inferred that again the varying solution lies on the same branch as the solution found in 4d. Therefore is is probable that the varying branch is the physical solution at $\zeta=1$. However, it is possible, that the presence of one of the solutions is only an artifact of the truncation. This cannot be ruled out in the current approximation scheme.

\section{Ghost-Loop-Only Truncation}\label{sghostloop3d}

The ghost-loop-only truncation retains only tree-level expressions and loops with explicit ghost-lines. It thus implements the hypotheses of Zwanziger of dominance of the gauge-fixing contribution at all momenta. Calculations in 4d with this truncation \cite{Atkinson:1998zc} give qualitatively the same results as with the gluon-loop included \cite{Fischer:2002hn}. 

\begin{figure}
\epsfig{file=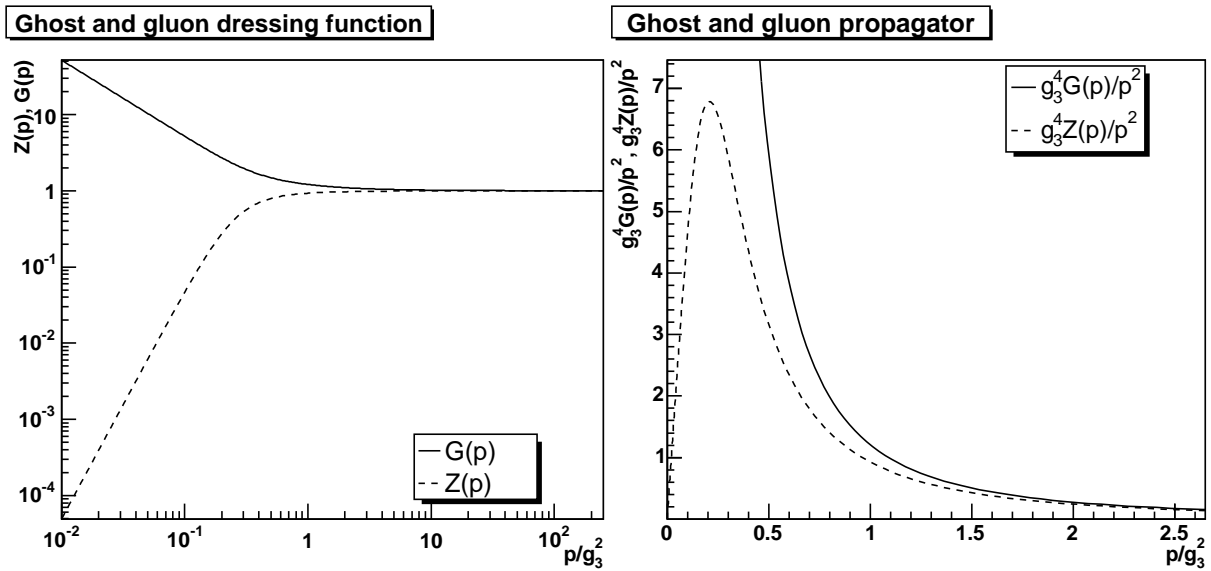,width=\linewidth}
\caption{The solutions of the DSEs in the ghost-loop-only truncation scheme at $\zeta=3$. The left panel shows the dressing functions and the right panel the propagators. The dashed line refers to $Z$ and the solid line to $G$. All quantities have been made dimensionless by dividing out appropriate powers of $g_3^2$.}\label{figgonly3}
\end{figure}

The result of the calculations for both the ghost and the gluon at $\zeta=3$ are shown in figure \ref{figgonly3}. In this case only the $g=1/2$ infrared solution exists.  The gluon propagator, for kinematical reasons, exhibits a significant maximum, with a center around $p/g_3^2=0.25$. The Higgs is in this case trivially tree-level, and therefore not shown. The infrared coefficients for all presented truncations are typically of order $A_gg_3^{-2g}={\cal O}(10^{-1})$ and $A_zg_3^{-2t}={\cal O}(10)$.

Note that the 3d-theory is finite, and therefore no dimensional transmutation occurs. Hence the coupling constant is the only dimensionful quantity left. Measuring then all dimensionful quantities in units of $g_3^2$, the theory is without scale and should be independent of the value of the coupling constant. In this truncation scheme, this is the case, which can been seen by rescaling the integration momenta in \prefr{fulleqG3d}{fulleqZ3d} appropriately. Thus, the plots shown in figure \ref{figgonly3} are the results for any positive value of the coupling constant. Even for very small values strong non-perturbative effects for momenta smaller than $g_3^2$ appear clearly. Hence this theory is never purely perturbative.

For values of $\zeta$ different from 3, spurious divergences are present, and it is therefore necessary to include and adjust the tadpole term $T^{GG}$ such that these terms are canceled. To identify the correct contribution, the integration kernel $R$ is split as
\be
R(\zeta)=R_0+(\zeta-3)R_3+R_D(\zeta),\label{rsplit}
\ee
\noindent where $R_0$ and $R_3$ are independent of $\zeta$ and do not give rise to divergences, and $R_D$ contains the divergent part for $\zeta\neq 3$. In perturbative calculations, the last term would be exactly canceled by the tadpole. However, it is not possible just to remove $R_D$ for two reasons. First, $R_D$ contributes dominantly to the infrared, and the subtraction should not alter the infrared. Secondly, $R_D$ is necessary to compensate an infrared divergence in the integration (not in the external momenta) of $R_3$ at $\zeta\neq3$. Setting
\be
T^{GG}=-\frac{g_3^2C_A}{(2\pi)^2}\int dqd\theta R_D(p,q)\left(2G(|\vec p|+|\vec q|)-1\right)\label{naivesub}
\ee
\noindent would be sufficient. This quantity contains no finite part for $\zeta=3$ and contributes only a small finite part compared to $R_0$ and $R_3$ otherwise.

However, even for a small deviation from $\zeta=3$, no further solution can be found. This is not unexpected, since the ghost-loop should only dominate in the infrared and is only there restricted to be purely transverse. It is not transverse in the ultraviolet, as perturbative calculations show. However, as the corresponding compensating parts proportional to $\delta_{\mu\nu}$ of the gluon-loop are neglected in this truncation, a failure was likely. Therefore using \pref{naivesub} does not permit to go beyond $\zeta=3$ in the ghost-loop-only truncation. This behavior is not altered when using the second solution branch of the infrared. It is necessary to use a different subtraction scheme which removes these terms at finite momenta. This is possible by using
\bea
&R(p,q)G(q)G(p+q)\to\nonumber\\
&R_0(p,q)G(q)G(p+q)+((\zeta-3)R_3(p,q)+R_D(p,q))A_g^2q^{-2g}(p+q)^{-2g}\nonumber\\
&+(\zeta-3)R_3(p,q)\label{rsubtract}
\eea
\noindent or equivalently setting
\bea
T^{GG}&=&-\frac{g_3^2C_A}{(2\pi)^2}\int dqd\theta\Big(R_D(p,q)(G(q)G(p+q)-A_g^2q^{-2g}(p+q)^{-2g})\nonumber\\
&&+(\zeta-3)R_3(p,q)\big(G(q)G(p+q)-A_g^2q^{-2g}(p+q)^{-2g}-1\big)\Big)\label{ggtad1}.
\eea
\noindent The last term in \pref{rsubtract} ensures the correct ultraviolet behavior to reproduce the 1-loop result.

Note that by using \pref{rsubtract} instead of \pref{naivesub}, also the solution at $\zeta=3$ is changed, since the second term in \pref{rsubtract} misses the correct mid-momenta behavior. This change is however negligible, and the results would hardly be distinguishable from the plots shown in figure \ref{figgonly3}. This is the expected trade-off of the truncation, a deficiency at mid-momenta.

\begin{figure}
\epsfig{file=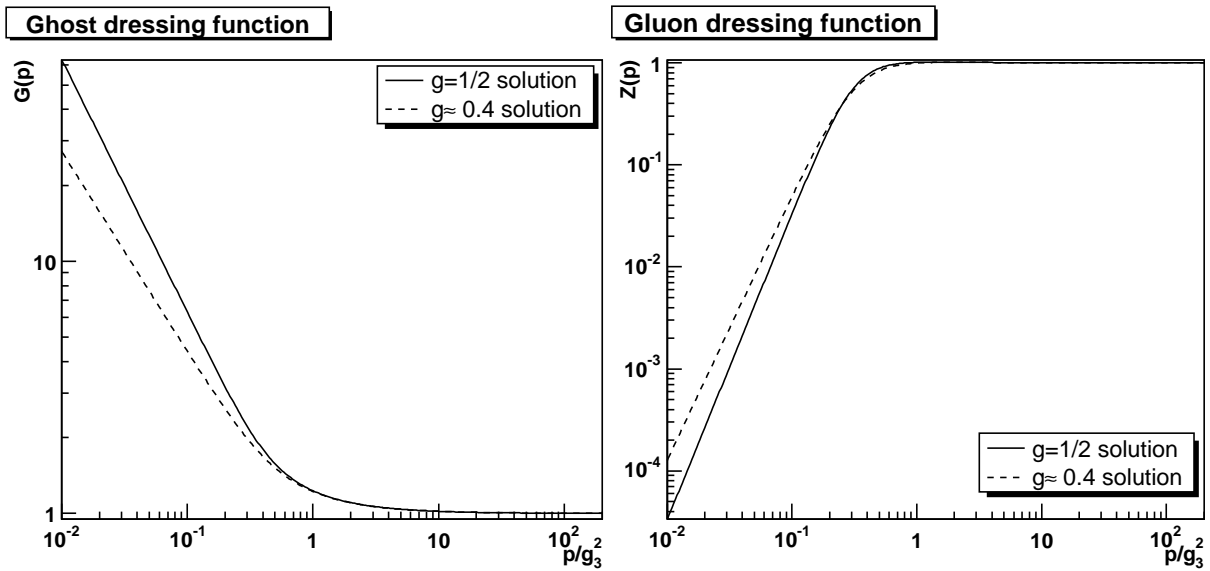,width=\linewidth}
\caption{The ghost and gluon dressing function at $\zeta=1$. The left panel shows the ghost dressing function, the right panel shows the gluon dressing function. The solid line denotes the solution for $g=1/2$ and the dashed line gives the other solution branch with $g \approx 0.4$.}
\label{gonlyzeta1}
\end{figure}

Both solutions at $\zeta=1$ are shown in figure \ref{gonlyzeta1}. The main features visible are the following: The ghost dressing function is nearly featureless. It approaches its tree-level behavior in the ultraviolet from above. The gluon dressing function has a (non-visible) peak, and it approaches its tree-level behavior also from above.

\begin{figure}[ht]
\epsfig{file=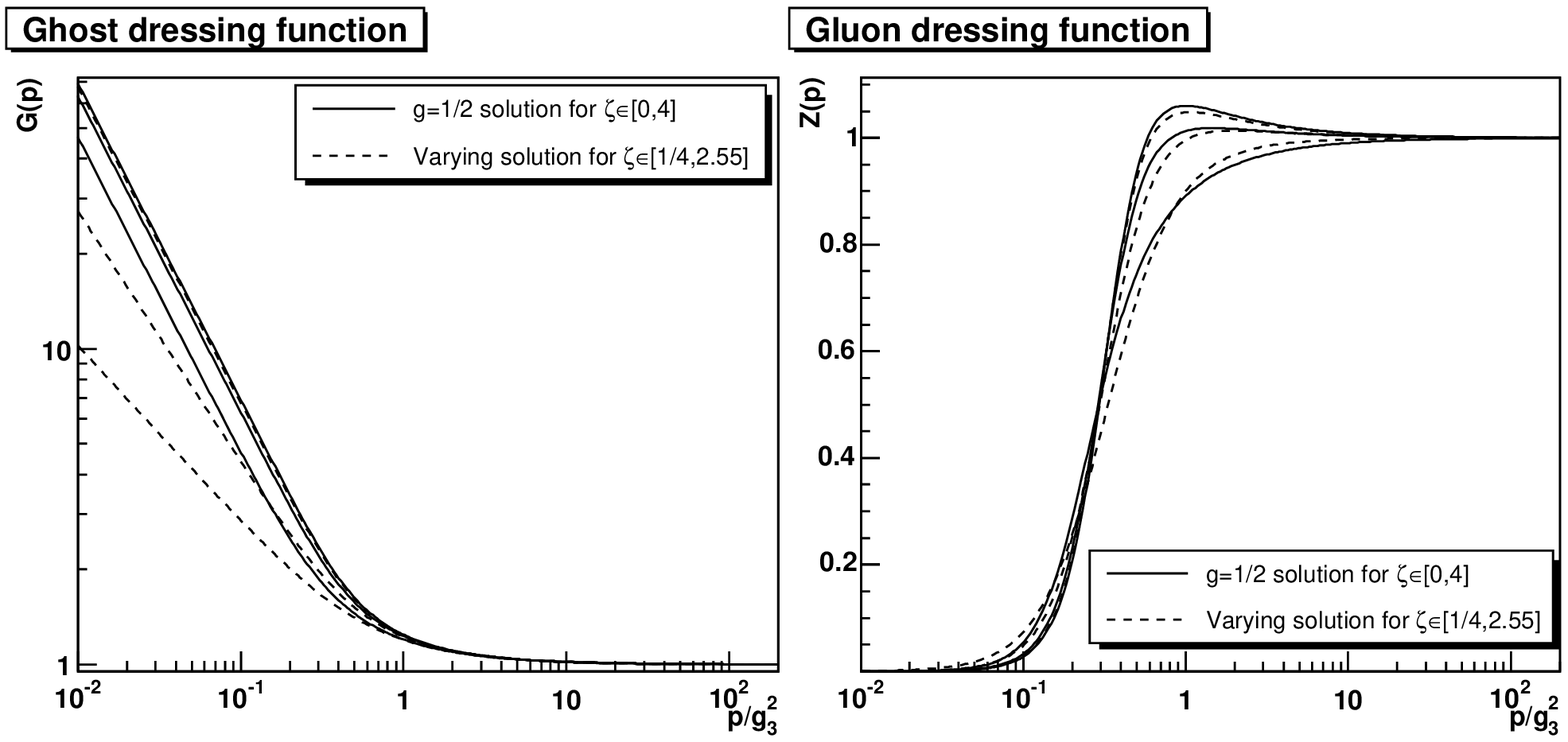,width=\linewidth}
\caption{The ghost and gluon dressing function in ghost-loop-only approximation for different values of $\zeta$.  The left panel shows the ghost dressing function, the right panel the gluon dressing function. The solid lines represent the solutions with $g=1/2$, and the dashed lines for the varying branch. At the peak in the gluon, the middle lines correspond to solutions at $\zeta=1$. The upper and lower line at mid-momenta give the solutions at $\zeta=0$ and $4$ for the $g=1/2$-branch and at $\zeta=1/4$ and $2.55$ for the other branch. Note the linear scale for $Z$ in the right panel.}
\label{figgonlyzeta}
\end{figure}

It is now possible to study different values of $\zeta$. This gives a systematic error for the solution, since it corresponds to varying the amount of gauge symmetry violation. This is shown in figure \ref{figgonlyzeta}. The variation for the $g=1/2$ branch is obtained by varying $\zeta$ between 0 and 4. It is possible to obtain solutions for other $\zeta$, but at some point the projection would be on the gauge violating longitudinal parts only and would hence give no information anymore on the transverse part. For the other branch $\zeta$ is varied from 1/4 up to 2.55, the latter corresponding to $g=0.2708$ and $t=-1.042$. In principle, it should be possible to extend the variation over the complete range \pref{zetaallowed}. However, it was not possible numerically to go to larger $\zeta$, since the available numerical precision of 16 significant digits was not sufficient. Once higher precision is available the remaining solutions should be obtainable. The reason is that as $t$ comes close to $-1$, the dressing function more and more resembles a massive solution. This corresponds to a jump from 0 to a finite quantity in the corresponding propagator. Hence, the zero is only reached far in the infrared, and it is numerically necessary to have at the same time very large momenta from the integration and very small ones from the infrared, a combination limited by precision. This is also the case when including the gluon- or Higgs-loop.

Looking now in detail at the $\zeta$-dependence, it can be observed that there are only quantitative, but no qualitative changes. The vanishing of the peak in the gluon dressing function for $\zeta>2$ is a purely perturbative phenomenon, since the contribution of the ghost-loop changes sign at $\zeta=2$ in the ultraviolet. The changes in the infrared are partly due to changes of the coefficients $A_g$ and $A_z$. The variable branch shows a much larger variation as expected. The deviation is comparatively small when going to smaller $\zeta$. It increases significantly when increasing $\zeta$, since the gluon propagator becomes more and more like the one of a massive particle. The propagator always vanishes at zero, though.

\section{Yang-Mills Theory}\label{syangmills}

The next step is to restrict the calculation to the Yang-Mills subsector without Higgs. This is then a theory defined on its own. Starting again at $\zeta=3$ no solution can be found. Even trying a significant diversity of subtraction schemes to do calculations at other values of $\zeta$ does not help, nor does employing the other solution branch.

Closer inspection reveals the source for these problems. The ghost- and gluon-loop contribute with opposite sign. This is correct in the ultraviolet and irrelevant in the infrared, where the gluon-loop is subleading. However, at momenta of the order of $g_3^2$, the gluon loop is able to dominate the ghost-loop and the tree-level term, and thus leads to a violation of the Gribov condition \pref{gribov}. Therefore, within this truncation scheme, the tree-level 3-gluon vertex is not an acceptable truncation. The reason is that either the bare three-gluon vertex overestimates the true vertex at mid-momenta, the bare ghost-gluon vertex underestimates the true one at mid-momenta, or the contributions of the two-loop graphs are important. A finite contribution from the tadpole would not help, since the problem persists even at $\zeta=3$ where its contribution vanishes identically\footnote{Albeit a scaleless integral also in 3d, it is not necessarily a purely divergent quantity. In 4d, it is possible to absorb any finite contributions in the renormalization constant. Lacking a renormalization process, this is not possible in 3d.}. This is clearly an artifact of the truncation. Any of these options deserves further attention, and as a first step, currently the ghost-gluon vertex is under closer investigation \cite{schleifenbaum:diploma}.

For the purpose of this work, the specific reason for this mid-momenta deficiency is of minor importance. For further calculations it is only necessary to improve the vertex such that it provides a sufficient suppression at mid-momenta. This situation is similar to the case of 4d, where without a change of either vertex, the correct ultraviolet behavior is not reproduced \cite{Fischer:2003rp}. Since the only relevant effect is at mid-momenta, such a vertex construction has to fulfill two criteria: It has to be tree-level in the ultraviolet and not more strongly divergent than $1/p^{t+1}$ on either leg in the infrared.

The simplest solution is to multiply the bare 3-gluon vertex with an additional function delivering this suppression. This leads to the ansatz
\bea
&\Gamma^{A^3\indexsep abc}_{\beta\sigma\mu}(-q,q+p,-p)\to\nonumber\\
&\Gamma_{\beta\sigma\mu}^{\tl\indexsep A^3\indexsep abc}(-q,q+p,-p)(A(q,G,Z)A(q+p,G,Z)A(p,G,Z))^\delta\nonumber\\
&A(q,G,Z)=\frac{1}{Z(q)G(q)^{2+\frac{1}{2g}}}.\label{g3vertex}
\eea
\noindent This ansatz is motivated by the vertex used in 4d \cite{Fischer:2002hn}, with the anomalous dimensions set to 0, as appropriate for 3d. Then an appropriate structure has been selected so that $A$ is constant in the infrared and tree-level in the ultraviolet. The additional parameter $\delta$ smoothly interpolates between a tree-level vertex and  a large suppression by selecting it from $[0,\infty)$. This ansatz conserves Bose symmetry, as opposed to the case of 4d, where this was not possible without spoiling the ultraviolet properties.

Alternative concepts have been studied. The simplest possibility to remedy the situation, albeit at the expense of an incorrect ultraviolet behavior, is to multiply the vertex just with a constant, thus introducing a vertex renormalization $Z_1$. Qualitatively the same and quantitatively similar solutions have been found for $Z_1\lessapprox 0.6$. The Bose non-symmetric ansatz $A(q,G,Z)A(q+p,G,Z)$, similar to the 4d-case, has also been made. Although it delivers a smaller suppression at the same value of $\delta$ as \pref{g3vertex}, the solution is possible for sufficiently large values of $\delta$. It does not differ qualitatively from the results obtained using \pref{g3vertex}. Even the quantitative changes are small. The last option investigated was modeling the two-loop terms by adding a term which is Gaussian in momentum to \pref{fulleqZ3d}. As long as its strength around $g_3^2$ was sufficient, the system was solvable, again with only minor modifications. Thus it is likely that the exact solution does not change the results qualitatively. The approach taken with \pref{g3vertex} is supported by lattice calculations of the dressing functions, see section \ref{slattice3d}.

\begin{figure}[t]
\epsfig{file=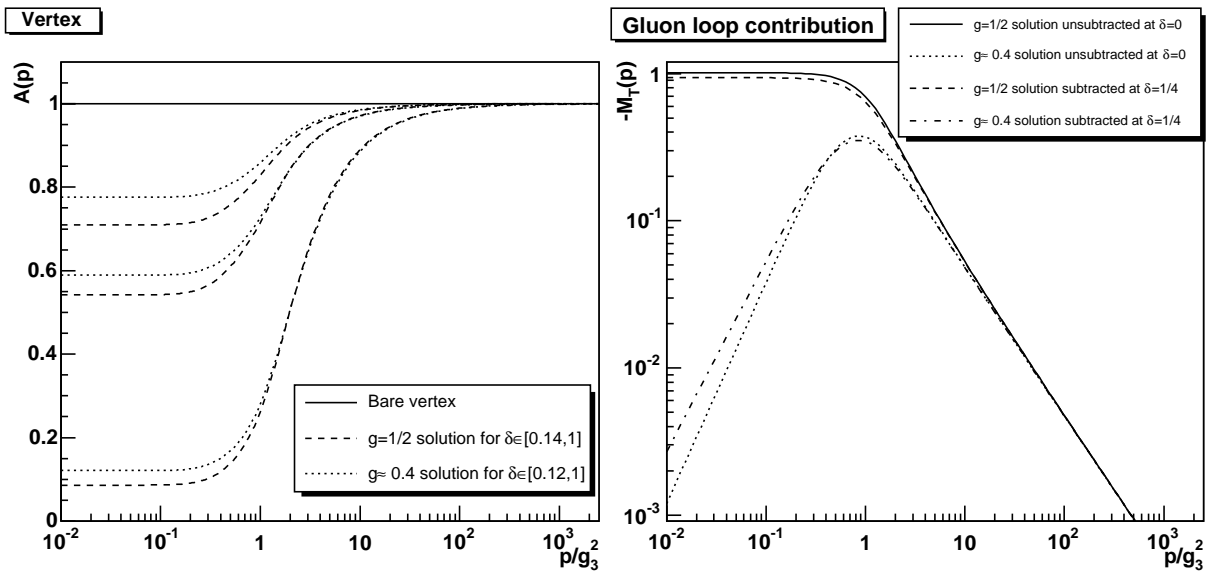,width=\linewidth}
\caption{The vertex and the gluon loop with and without modifications. The left panel shows the function $A$ from \pref{g3vertex}. The solid line is $\delta=0$, the tree-level vertex. The dashed line is the $g=1/2$ solution and the dotted line the $g\approx 0.4$ solution. From bottom to top these are $\delta=1$, $\delta=1/4$ and in the last case $\delta=0.14$ and $\delta=0.12$, respectively. The right panel shows at $\zeta=3$ the full gluon loop compared to the gluon loop suppressed by the vertex \pref{g3vertex} and subtracted by the gluon part of the tadpole \pref{ggtad2}, again for both solutions with the same type of line. The larger values are always without suppression.}\label{figvertex}
\end{figure}

An interesting possibility to assess the effect of different vertices is to change $\delta$. While $\delta=1$ corresponds to a significant suppression, the smallest values with still a stable solution are 0.131 for the $g=1/2$ solution and 0.114 for the other one. As a convenient value which still provides very stable numerics and at the same time not too large suppression, $\delta=1/4$ is arbitrarily chosen in the following. The structure of the vertex and its impact on the gluon loop are studied for different values of $\delta$ in figure \ref{figvertex}.

\begin{figure}[ht]
\epsfig{file=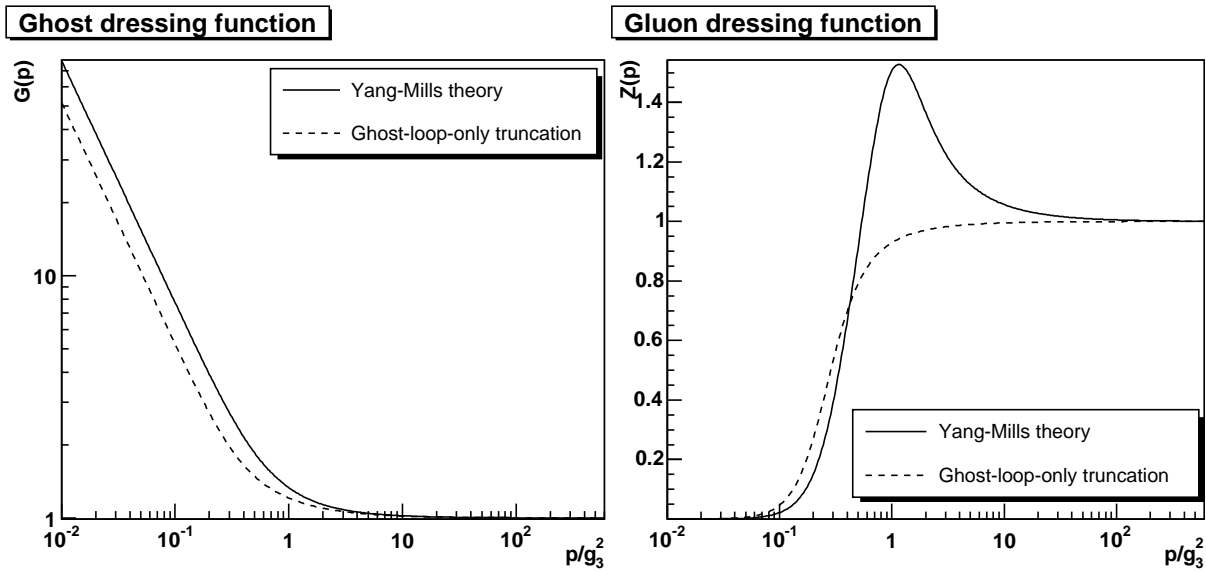,width=\linewidth}
\caption{Solution of the Yang-Mills sector at $\zeta=3$. The left and right panel shows the ghost and gluon dressing function, respectively. The dashed curve gives the ghost-loop-only solution for comparison. The solid curve gives the full Yang-Mills solution at $\delta=1/4$. Note the linear scale for $Z$ in the right panel.}\label{figym}
\end{figure}

It is now possible to solve the equations at $\zeta=3$. The results are shown in comparison to the ghost-loop-only results in figure \ref{figym}. The only qualitative difference compared to the ghost-loop-only truncation is that a peak in the gluon dressing function is now present for all $\zeta$ due to the now correctly reproduced LO perturbation theory for $p\gg g_3^2$.

After this first view at the full Yang-Mills solution, it is again interesting to study its dependence on $\zeta$. In principle, it would be sufficient again to split off the divergent part $M_{TD}$ of $M_T$ and subtract it in the same way as in \pref{naivesub}, by setting
\bea
T^{GG}&=&-\frac{g_3^2C_A}{(2\pi)^2}\int dqd\theta\Big(R_D(p,q)\left(2G(|\vec p|+|\vec q|)-1\right)\nonumber\\
&+&(A(q,G,Z)A(q+p,G,Z)A(p,G,Z))^\delta M_{TD}(p,q)\left(2Z(|\vec p|+|\vec q|)-1\right)\Big)\label{naivesub2}.
\eea
\noindent The naive expectation is that after the inclusion of the gluon loop the spurious terms would be eliminated. This does not occur, since by introducing an ad-hoc suppression, these cancellations are affected. Hence the same problem arises again, and it is necessary to correct both loops individually. This can be done by 
\bea
T^{GG}&=&-\frac{g_3^2C_A}{(2\pi)^2}\int dqd\theta\Big(R_D(p,q)(G(q)G(p+q)-A_g^2q^{-2g}(p+q)^{-2g})\nonumber\\
&+&(\zeta-3)R_3(k,q)\big(G(q)G(p+q)-A_g^2q^{-2g}(p+q)^{-2g}-1\big)\nonumber\\
&+&(A(q,G,Z)A(q+p,G,Z)A(p,G,Z))^\delta M_{TD}(p,q)Z(q)Z(p+q)\Big)\label{ggtad2},
\eea
\noindent i.e. by completely removing the term $M_{TD}$ and keeping the same subtraction as before for the ghost-loop. Removing $M_{TD}$ is not harmful to the infrared, since the gluon-loop is subdominant there. It is also irrelevant in the ultraviolet, where it contributes only a pure divergence and would therefore anyhow be canceled by the tadpole. Thus, only the mid-momenta behavior is altered.

\begin{figure}[ht]
\epsfig{file=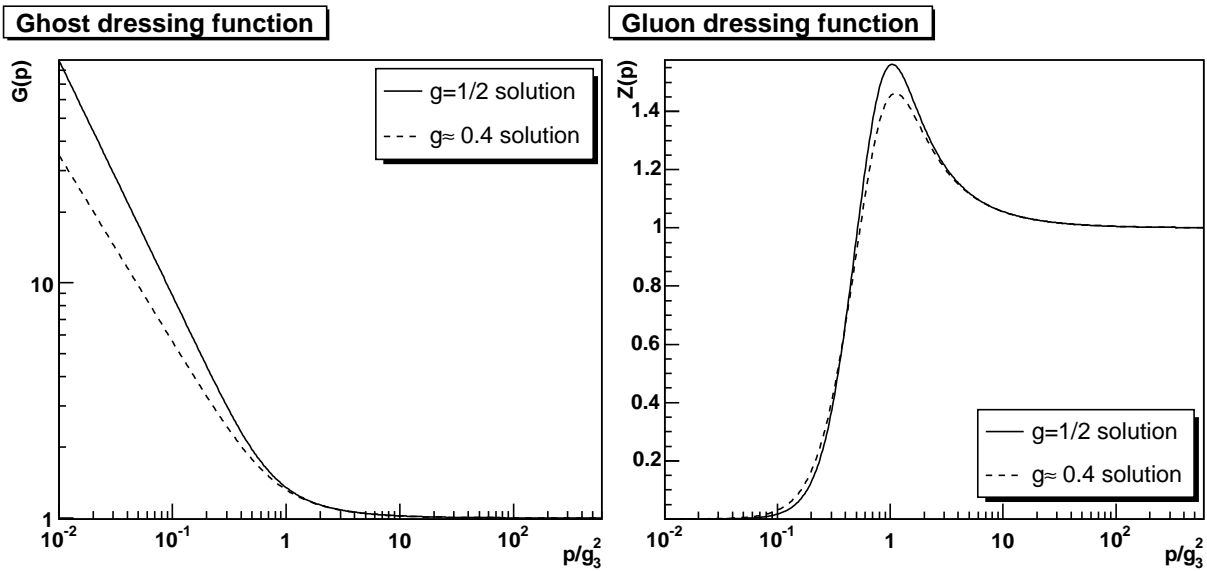,width=\linewidth}
\caption{Solution of the Yang-Mills sector at $\zeta=1$. The left panel shows the ghost dressing function and the right panel the gluon dressing function. The solid curve denotes the $g=1/2$ solution, while the dashed curve displays the $g\approx 0.4$ solution. Both are at $\delta=1/4$.}\label{figymzeta1}
\end{figure}

It is now possible to solve for any $\zeta$. Both infrared solutions are found to connect to full solutions and a comparison at $\zeta=1$ and $\delta=1/4$ is shown in figure \ref{figymzeta1}. The difference is again only quantitative in nature, and even this difference is quite small.

\begin{figure}[ht]
\epsfig{file=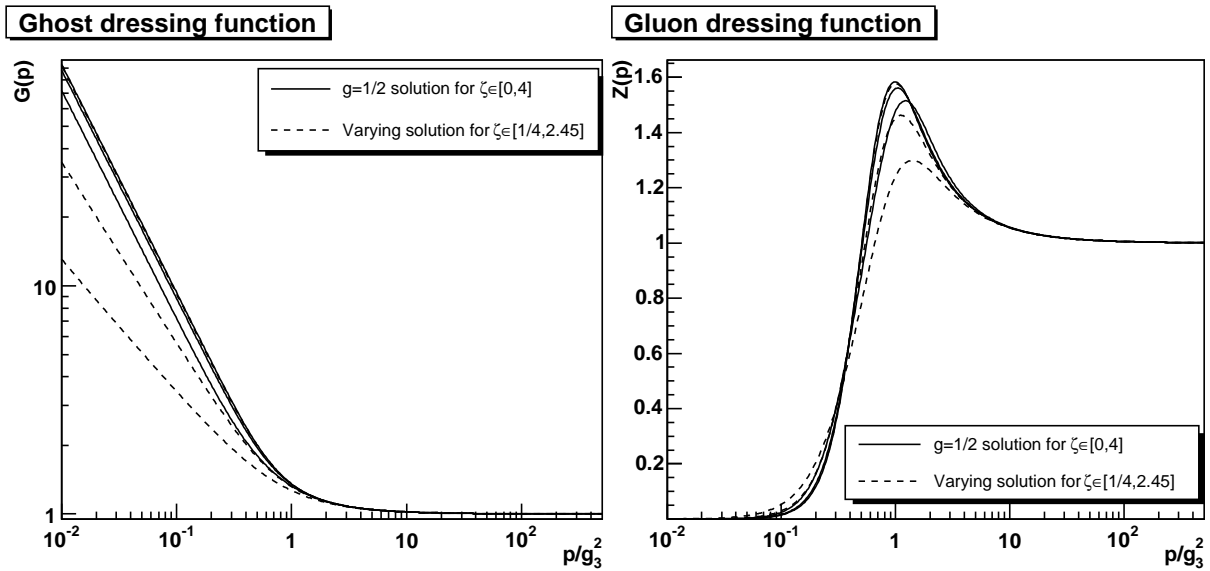,width=\linewidth}
\caption{The ghost and gluon dressing functions of Yang-Mills theory for different values of $\zeta$. The left panel shows the ghost dressing function, and the right panel the gluon dressing function. The solid line gives the solution for $g=1/2$ and the dashed line is for the other solution branch. At the peak in the gluon dressing function, the middle lines represent the solution at $\zeta=1$. The upper and lower lines at mid-momenta give the solutions at $\zeta=0$ and $\zeta=4$ for the $g=1/2$-branch and at $\zeta=1/4$ and $\zeta=2.45$ for the other solution branch.}
\label{figymzeta}
\end{figure}

\begin{figure}[ht]
\epsfig{file=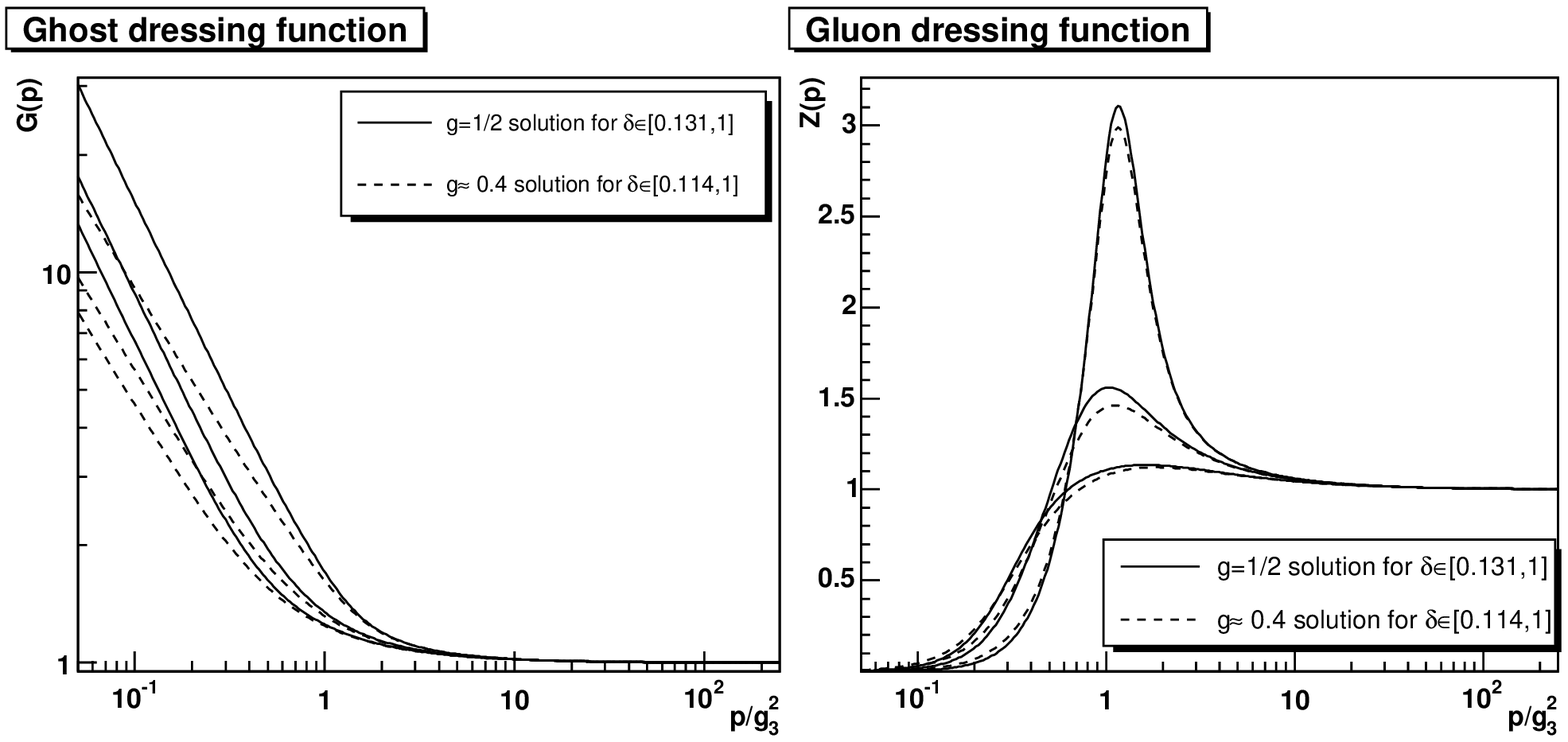,width=\linewidth}
\caption{The ghost and gluon dressing functions of Yang-Mills theory for different values of $\delta$ at $\zeta=1$. The left panel shows the ghost dressing function, and the right panel the gluon dressing function. The solid line gives the solution for $g=1/2$ and the dashed line for $g\approx 0.4$. At the peak in the gluon dressing function, the middle lines represent the solution at $\delta=1/4$. The upper and lower lines at mid-momenta give the solutions at $\delta=0.131$ and $\delta=1$ for the $g=1/2$-branch and at $\delta=0.114$ and $\delta=1$ for the other solution branch.}
\label{figymdelta}
\end{figure}

It remains to assess the dependence on the parameters $\zeta$ and $\delta$ as a systematic error estimate. While the dependence on $\zeta$ summarizes the impact of violation of gauge symmetry, the dependence on $\delta$ exhibits the uncertainty in the correct mid-momenta behavior of the vertices. Both problems are linked, and do not cause independent errors. However, both generate different changes when varied. Figure \ref{figymzeta} and \ref{figymdelta} display the dependence on $\zeta$ and $\delta$, respectively.

The uncertainty in the structure of the vertices is much more significant at mid-momenta than the violation of the STI \pref{stigluon}. On the other hand, the uncertainty in the vertex only amounts to a shift in the infrared but not to a change of the infrared exponents as in the case of the violation of the STI.

The height and the position of the peak depends weakly on the vertex construction. The height of the peak can be traced to the amount of suppression the vertex gives. When the suppression is weak enough, the peak tends to become a divergence with sign change. Therefore, this truncation scheme can give no definite answer to the size of the peak. It turns out that the peak only starts to get very large when $\delta$ gets close to the critical value. Before that, it is stable at height around 1.5 and below. Hence, it is to be expected that the correct suppression would also not deliver a significantly larger peak. This is confirmed by lattice results, as will be shown in subsection \ref{slattice3d}.

Comparing these results with those of the 4d-theory \cite{vonSmekal:1997is}, little difference is found. On qualitative grounds, both are the same, besides renormalization. They only differ in their quantitative nature. Thus the 3d Yang-Mills theory is a strongly interacting, confining theory, confirming the expectation \cite{Feynman:1981ss}.

\section{Full 3d-Limit}\label{sfull3d}

Finally, by adding the Higgs, the full system of equations \prefr{fulleqG3d}{fulleqZ3d} is implemented. Although the contribution of the Higgs-loop in \pref{fulleqG3d} is positive, it is not sufficient to allow for a solution with a bare 3-gluon vertex. Hence the vertex ansatz and also the tadpole construction will be kept the same as in the Yang-Mills case.

In addition, as no further input is available, the tree-level gluon-Higgs vertex \pref{tlgh} is employed. A closer inspection of the consequences of this choice is made in subsection \ref{sschwinger}.

The ratio $m_h/g_3^2$ remains to be fixed. In the pure 3d-theory, $m_h$ is a parameter and can therefore not be determined a priori. It stems from integrating out the hard modes when making the 3d-approximation \cite{Kajantie:1995dw}. By taking these modes into account in chapter \ref{cft}, it will be possible to investigate it more closely rather than making an ansatz. For the investigation in this chapter, the origin (and also the exact value) of the Higgs mass is not of direct importance, and hence will be fixed to $m_h/g_3^2=0.8808$, a value extracted by lattice calculations and perturbative fitting \cite{Cucchieri:2001tw}. 

By adding the Higgs, three more tadpole contributions arise, as can be seen from figure \ref{figftsys} and in appendix \ref{adse}. Since the Higgs has a tree-level mass, two of the tadpoles can have a non-vanishing finite part already in leading-order perturbation theory, see appendix \ref{appUV}. Hence, a more thorough discussion is needed here.

The first additional tadpole appears in the gluon equation. On the one hand, all diverging contributions must cancel again due to the STI, hence the structure of divergencies must be the same for all graphs. On the other hand, when projecting with the Brown-Pennington projector \pref{bpproj}, the tadpole identically drops out. Therefore all finite contributions of the tadpole must also be contained in the other loops as well. For a vanishing tree-level mass, the tadpole is canceled by divergent loop contributions. It is hence a good assumption that any finite part of the tadpole will also be canceled by the loop contributions. Therefore the approximation concerning the Higgs-tadpole will be
\be
T^{GH}=-\frac{g_3^2C_A}{(2\pi)^2}\int dqd\theta\Big(M_{LD}(k,q)H(q)H(k+q)\Big)\label{ghtad}
\ee
\noindent where $M_{LD}$ is the divergent part of $M_L$. If this were not justified, at $\zeta=3$ a trace of the finite contribution should be visible. Any such term would behave like $1/p^2$, a clear signature. The gluon- and Higgs-loop both are at most constant in the infrared and can thus not induce such a contribution. Only the ghost loop with its strongly divergent behavior could contain it. However, it seems unlikely that the finite part of the tadpole should be completely moved to the ghost loop, hence the assumption described above will be made.

\begin{figure}[t]
\epsfig{file=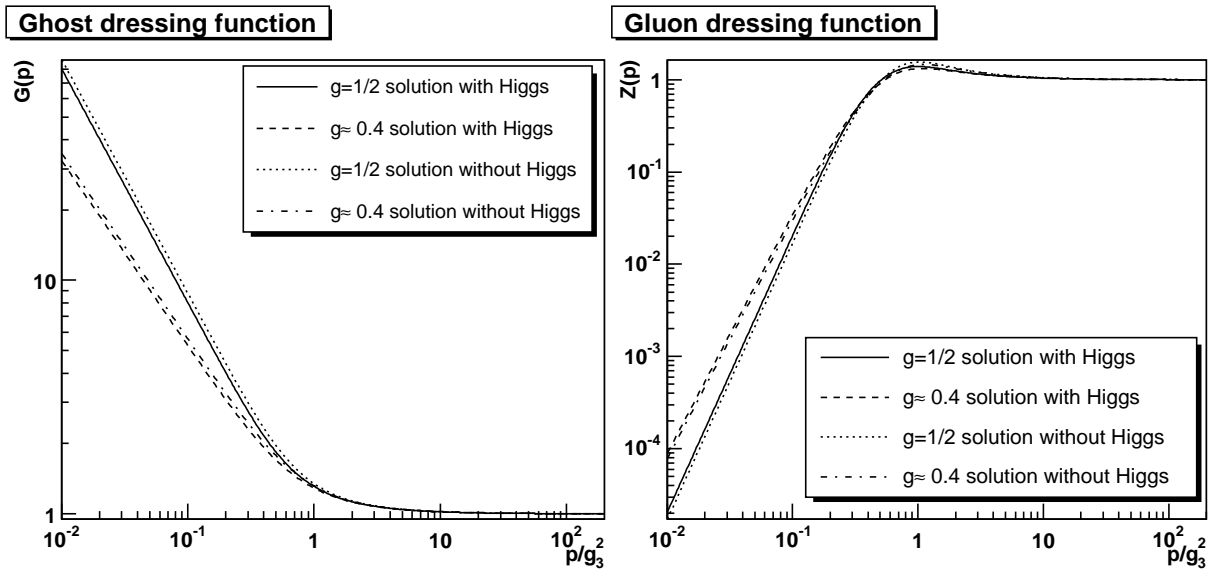,width=\linewidth}
\caption{The solution of the Yang-Mills system compared to the full system at $\zeta=1$. The solid line denotes the full solution at $g=1/2$ while the dotted line is the corresponding Yang-Mills solution. The dashed line gives the full solution for $g\approx 0.4$ while the dashed-dotted line is the corresponding Yang-Mills solution.}
\label{figymf}
\end{figure}

The Higgs equation with two tadpoles is more problematic, as the argument using the projection is not available. As the Higgs is a component of a 4d gluon, its self-energy has to be finite. Thus the divergent parts of the tadpoles have to cancel. This fixes the tree-level coupling $h$ at leading-order perturbation theory, as given by \pref{hvalue}. The only remaining question is that of a possible finite contribution. Such a contribution would only change the mass of the Higgs, and this occurs already at leading-order perturbation theory. However, the subtraction of two divergent quantities is not simple, and due to the error induced by the truncation somewhat arbitrary. Hence the mass-shift is essentially unknown. Nonetheless, the Higgs dressing function behaves similar to perturbation theory. Therefore assuming the finite part of the tadpole to be the same as in leading-order perturbation theory seems to be justified. Thus, the tadpoles are fixed to be
\be
T^{HG}+T^{HH}=\frac{g_3^2C_A}{p^2}\frac{m_h}{4\pi}=\frac{\delta m_3^2}{p^2},\label{higgstadpole}
\ee
\noindent defining the finite mass renormalization $\delta m_3$ of the bare Higgs mass. This amounts to a change of the mass of about 10\%. 

The impact of the Higgs on the Yang-Mills sector at $\zeta=1$ is shown in figure \ref{figymf}. The infrared behavior is only weakly affected and the situation is similar in the ultraviolet. The largest impact on the gluon dressing function is on the peak height and the impact on the peak position is significantly weaker. Altogether, the effect of the additional Higgs field is only quantitative, as expected from lattice calculations  \cite{Cucchieri:2001tw}.

\begin{figure}[t]
\epsfig{file=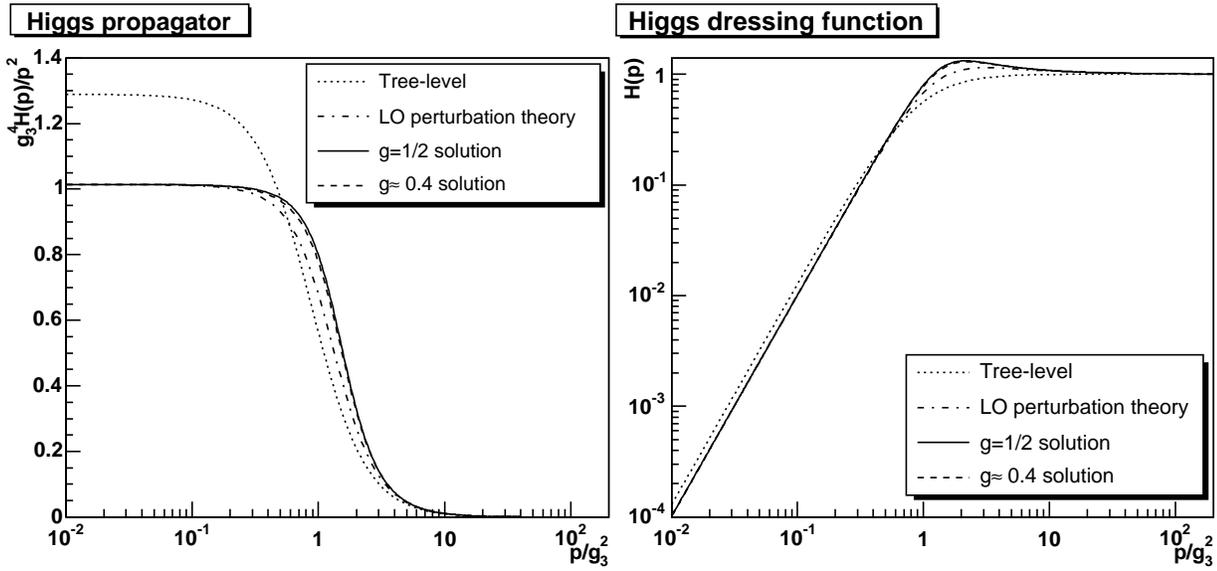,width=\linewidth}
\caption{The left  panel shows the Higgs propagator and the right panel its dressing function, both at $\zeta=1$. The solid line gives the solution for $g=1/2$, the dashed line for $g\approx 0.4$, the dashed-dotted line denotes the leading-order perturbative result and the dotted line the tree-level behavior.}
\label{fighiggsf}
\end{figure}

The Higgs propagator and dressing function are shown in figure \ref{fighiggsf}. Both show a behavior very similar to the leading-order perturbative result, and the difference to a tree-level propagator is mainly due to the mass-shift by the tadpole.

\begin{figure}[t]
\epsfig{file=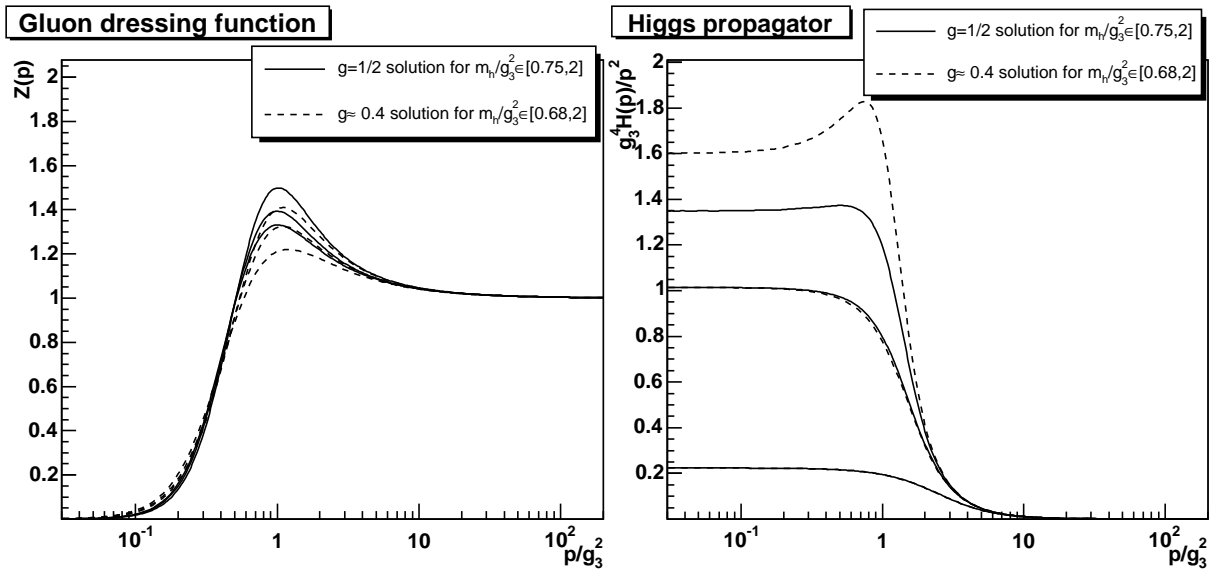,width=\linewidth}
\caption{The left panel shows the gluon dressing function and the right panel the Higgs propagator as a function of $m_h$. The gluon peak increases with decreasing mass, while the Higgs propagator decreases with increasing mass. Solid is the $g=1/2$ and dashed the $g\approx 0.4$ solution. The masses are $2g_3^2$, $0.88g_3^2$ for both solutions, and $0.75g_3^2$ and $0.68g_3^2$, respectively.}
\label{fighiggsmass}
\end{figure}

Although the mass is fixed in the current setting, the question of the dependence on the mass is still interesting. Making the tree-level mass larger, the solution approaches smoothly the Yang-Mills solution, with which it merges as $m_h\to\infty$. On the other hand, the mass cannot be made arbitrarily small. When the mass is reduced below $m_h/g_3^2\approx 0.7$, no solution can be found without changing $\delta$. The reason is that the Higgs self-energy is negative, and only while it is sufficiently small compared to the renormalized tree-level contribution, it is possible to maintain the Gribov condition \pref{gribov} at mid-momenta. This additional constraint only stems from the 4d origin of the theory and the fact that the Higgs is part of a 4d gluon, and is not inherent to the 3d-theory. In the 3d-theory, a negative dressing function of the Higgs would be allowed, and smaller masses could be reached. The height of the gluon peak directly corresponds to the size of the Higgs-self-energy and hence to the lowest masses achievable without violating \pref{gribov}. Thus, increasing $\delta$ compensates to some extent a lowering of the  mass, but this has only been pursued up to $\delta=1$ or a lowest mass of $m_h/g_3^2\approx 0.6$. The dependence of the Higgs propagator on the mass is depicted in figure \ref{fighiggsmass}. An effect at small Higgs masses had been anticipated from the effect of a massless Higgs on the infrared solutions, see appendix \ref{asmasslesshiggs}.

\begin{figure}[t]
\epsfig{file=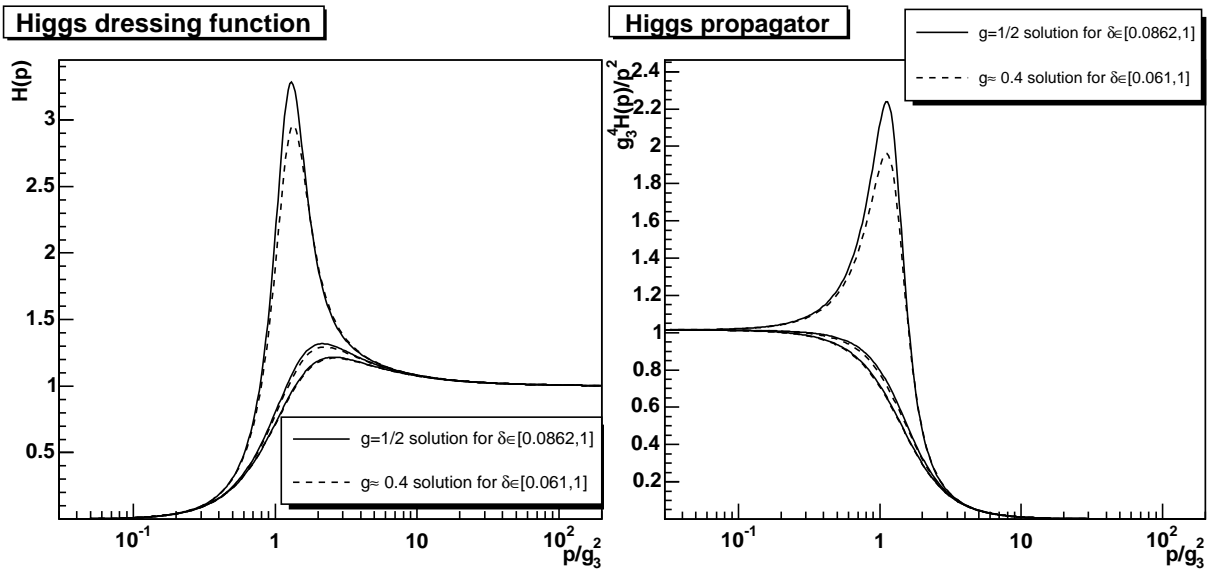,width=\linewidth}
\caption{The Higgs dressing function and propagator as a function of $\delta$ at $\zeta=1$. The solid line is the solution for $g=1/2$, the dashed line for $g\approx 0.4$. The higher the peak, the larger $\delta$. The lowest peak corresponds to $\delta=1$, the middle one to $\delta=1/4$ and the largest one to $\delta=0.0862$ and $\delta=0.061$, respectively.}
\label{fighiggsdelta}
\end{figure}

The dependence of the Higgs on $\zeta$ is very small, on the order of a few percent at mid-momenta. The $\zeta$-dependence of the Yang-Mills sector is also only affected to this extent by the presence of the Higgs. The dependence on $\delta$, which influences the size of the self-energy, is more pronounced. The latter is shown in figure \ref{fighiggsdelta}. Similar to the gluon dressing function, in the Higgs dressing function the peak increases with decreasing $\delta$, at some point also generating a peak in the propagator. This is due to the direct link of the Higgs self-energy to the peak volume of the gluon propagator. As the latter increases, the Higgs-self-energy increases. Since it is negative, it starts to compensate the tree-level term at mid-momenta, where the effect of the peak in the gluon dressing function is most pronounced and thus increases the peak in the Higgs dressing function. At some point a further effect sets in, when the system tries to compensate the increasing negative contribution from the gluon loop by an increasing positive contribution in the Higgs-loop. This is most easily achieved by an enhancement of the Higgs dressing function at mid-momenta, thus also increasing the peak in the Higgs dressing function. This process slows the growth and for sufficiently small values of $\delta$ even decreases the peak in the gluon dressing function. Therefore it becomes likely that a violation of the Gribov condition \pref{gribov} will occur due to the Higgs rather than due to the gluon. This behavior is much more pronounced in the $g=1/2$ solution. 

\begin{figure}[t]
\epsfig{file=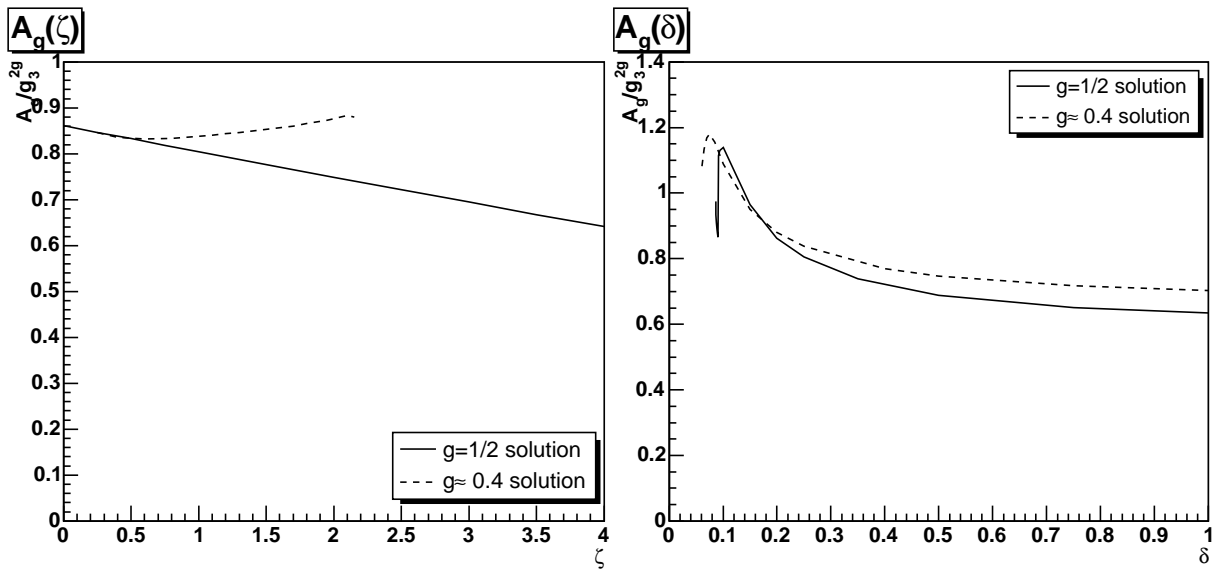,width=\linewidth}
\caption{The dependence of the infrared coefficient $A_g$ on $\zeta$ in the left panel and on $\delta$ in the right panel. Solid lines are the $g=1/2$ solution and dashed lines the other solution branch.}
\label{figagzd}
\end{figure}

Apart from the dependence on $\zeta$ of the exponents in the case of the second branch, also the dependence of the infrared coefficients on $\zeta$ and $\delta$ is of interest. $A_h$ is fixed by the renormalized mass and $A_z$ depends uniquely on $A_g$, hence only the dependence of $A_g$ is relevant. Its dependence is shown in figure \ref{figagzd}. The dependence on $\zeta$ is rather weak. Both solution branches evolve smoothly out of one solution at $\zeta=1/4$. The dependence on $\delta$ is much more pronounced, where the trend to become singular is visible in the strong increase at low values of $\delta$. The bending-over is the aforementioned effect of the peak growth switching from the gluon to the Higgs. 

\section{Comparison to Lattice Results}\label{slattice3d}

\begin{figure}
\epsfig{file=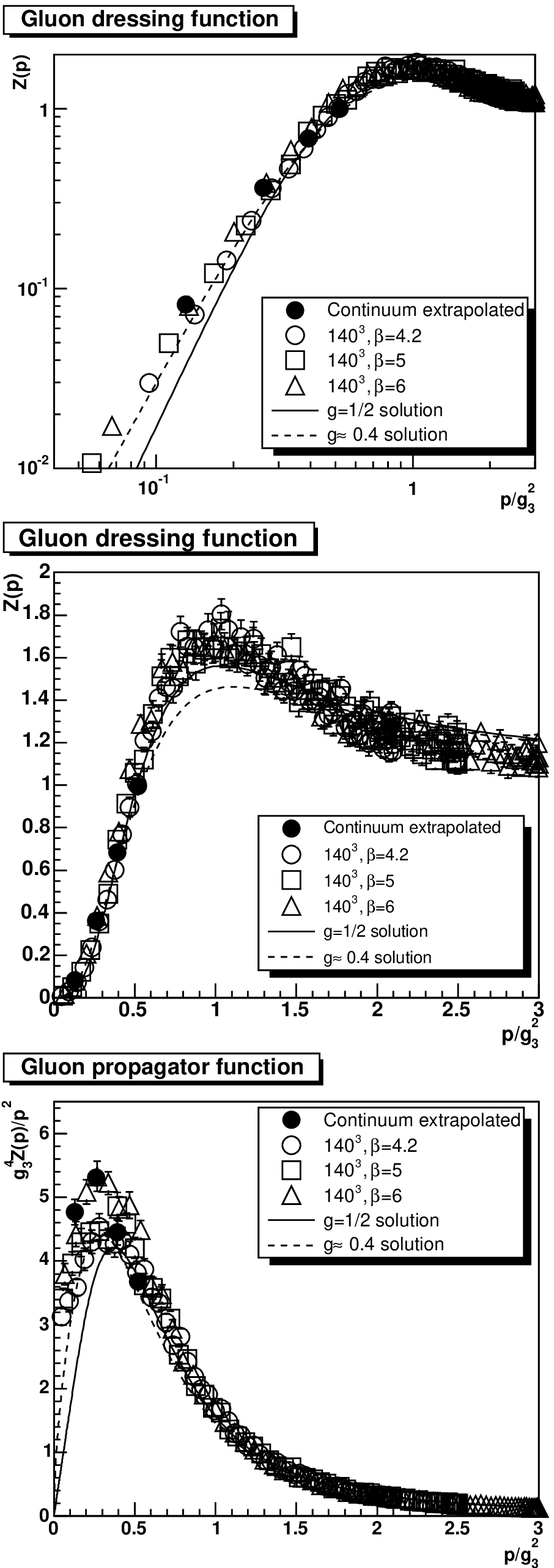,width=0.5\linewidth,height=1.2\linewidth}\epsfig{file=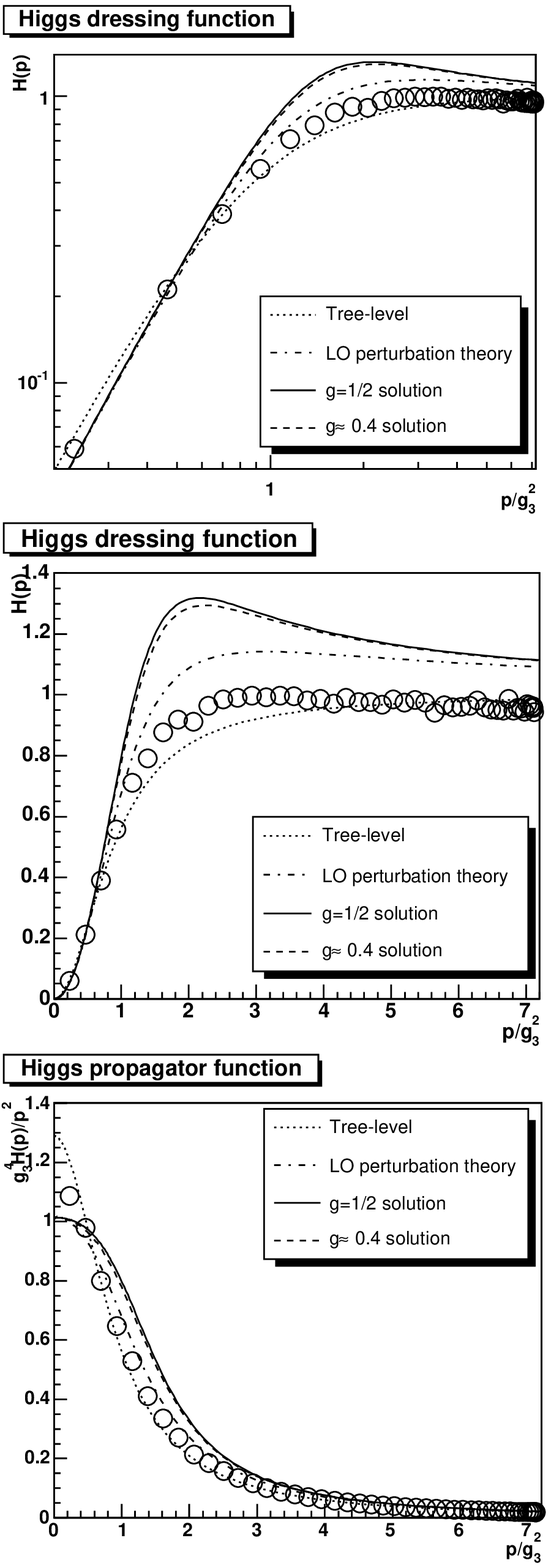,width=0.5\linewidth,height=1.2\linewidth}
\caption{The left panels show the gluon propagator and dressing function in Yang-Mills theory from lattice calculations and from section \ref{syangmills}. The continuum-extrapolated values are from \cite{Cucchieri:2001tw}, the others from \cite{Cucchieri:2003di}. The right panel shows the Higgs propagator and dressing function in the full theory from lattice calculations \cite{Cucchieri:2001tw} and from section \ref{sfull3d}. The errors indicated are statistical. }
\label{figlat}
\end{figure}

Currently, only lattice calculations for the gluon propagator \cite{Cucchieri:2003di,Cucchieri:2001tw} and the Higgs propagator \cite{Cucchieri:2001tw} are available, but not for the ghost propagator. Furthermore, lattice calculations do not yet see a significant difference in the gluon propagator with or without the Higgs \cite{Cucchieri:2001tw}. Hence the lattice calculations for the gluon propagator are done in pure 3d Yang-Mills theory. Therefore, these results have to be compared to the results from section \ref{syangmills} rather than to the full calculations in section \ref{sfull3d}. The comparison is made in figure \ref{figlat}.

Comparing the gluon propagator to the lattice results, the assumed suppression using $\delta=1/4$ seems to fit the data quite well. The dressing function is better fitted by the $g=1/2$ solution while the propagator and especially the infrared favors $g\approx 0.4$. At momenta $p\approx g_3^2$, the truncation scheme is not trustworthy and the better agreement of the $g=1/2$ solution with the lattice gluon dressing function is not conclusive. Although the lattice data for the propagator are quite scattered, they lie pretty close in case of the dressing function. The propagator is hence more sensitive than the dressing function. A better fit could be achieved by varying $\delta$. This is not the aim of this work, but may be of interest for phenomenological studies.

The lattice result for the Higgs propagator differ more from the DSE result than for the gluon propagator. Especially, the lattice Higgs propagator is much more similar to the tree-level result than to the leading-order perturbative result. The reason for the discrepancy between the current calculation and the lattice data may be twofold. The value at zero momentum indicates that \pref{higgstadpole} overestimates the mass renormalization and the mass found is too large by roughly 5\%. More severe is that the bare vertex used here seems to significantly overestimate the gluon-Higgs interaction compared to the fully dressed vertex of the lattice result. The self-energy must be significantly suppressed even for large values of $p$ to allow for such a deviation from the leading-order perturbative result, and must be nearly negligible. In principle, such an effect could be modeled by a construction similar to \pref{g3vertex} for the Higgs-gluon vertex. Again, this is not the aim of this work. However, this finding entails that strong subleading or even genuinely non-perturbative effects are also present in the Higgs sector. Even for quite large momenta of the order of $10g_3^2$, where the gluon propagator has already become perturbative, this is not the case for the Higgs propagator, which will only become perturbative at much larger momenta of the order of $100g_3^2$. Indeed, the overestimation of the mass already shifts the result towards the tree-level propagator, and a smaller mass would enhance the discrepancy. This is indeed quite an interesting effect, and it will be investigated in some more detail in section \ref{sschwinger}.

\section{Coulomb Gauge Instantaneous Potential}\label{scoulomb}

Recently, a connection between Coulomb-gauge in 4d and Landau-gauge in 3d has been established \cite{Zwanziger:2003de,Feuchter:2004gb}. It was shown \cite{Zwanziger:2003de} that under certain approximations, the static quark-anti-quark potential, which is linked to the 00-component of the Coulomb gauge gluon propagator, is given in four dimensions by
\be
V_{\mathrm{coul}}(p)=\frac{G(p)^2}{p^2}+V_{\mathrm{con}}(p)\label{vcoul}.
\ee
\noindent $V_{\mathrm{con}}$ designates contributions which stem from connected parts of expectation values of the Faddeev-Popov operator. Neglecting this part, the potential is solely given by the function $G$. It can be shown to be identical to the ghost dressing function of the 3d Landau gauge Yang-Mills theory \cite{Zwanziger:2003de}. While the solution $g\approx 0.4$ generates a potential which behaves as $\sim1/p^{3.6}$ and thus a little less than linear, the second branch generates a solution proportional to $1/p^4$ for small momenta, thus generating the behavior expected for a linear confining potential.

In principle it is expected that after Fourier-transformation \pref{vcoul} would give the leading contribution of the static potential. However, it is not possible in the given approximation scheme to obtain the Fourier transform. The Fourier integral is quadratically divergent in the ultraviolet. The reason is that $G$ goes to 1 in the ultraviolet, thus together with the measure it generates a contribution proportional to $p$. Using a prescription as will be used in section \ref{std}, it may still be possible to extract some information, but since not a constant but a function needs to be fitted, this would turn out to be significantly more complicated. It is improbable that the numerical accuracy would be sufficient to extract relevant information, as it is already hard to do so in the case of a constant quantity like the thermodynamic potential treated in section \ref{std}. This is due to the approximations involved when obtaining the infrared limit \pref{vcoul}.


%% file: ft.tex
\chapter{Finite-Temperature Effects}\label{cft}

Lowering the temperature down from infinity also the hard modes must be included. The corresponding DSEs are derived in section \ref{sftdse}. The precise implementation of the limiting process needs some deeper analysis of the running coupling, which will be conducted in section \ref{scc}. The infrared properties of the dressing functions will be investigated in section \ref{sftir}. The truncation will be discussed in section \ref{sfttrunc}. In general, an infinite number of Matsubara frequencies contributes at finite temperatures. At sufficiently high temperatures, most of them can be neglected, however. This corresponds to a small-momentum approximation, to be defined in section \ref{ssmallpapprox}. Also the renormalization process will be discussed there. Results on the ghost-loop-only truncation and the full theory will finally be shown in sections \ref{sftgonly} and \ref{sftfull}, respectively. The chapter will be closed in section \ref{sdselowt} by presenting results for low temperatures, which have been obtained in the context of this work. There, also a comparison to the high-temperature case will be drawn.

For convenience, the 3d-longitudinal and 3d-transverse gluons will be referred to as longitudinal and transverse gluons. In this chapter is again $p^2=p_0^2+\vec p^2\equiv p_0^2+p_3^2$. External momenta are referred to as $p$ while internal loop momenta and Matsubara summation momenta are referred to by $q$.

\section{Finite-Temperature Dyson-Schwinger Equations}\label{sftdse}

As described in section \ref{sdseft}, the equations are obtained from the vacuum equations \pref{vacgheq} and \pref{vacgleq} by application of the Matsubara formalism \cite{Kapusta:tk}. To obtain scalar equations for the (infinite) set of dressing functions $Z$ and $H$ of the gluon propagator \pref{gluonprop}, the gluon equation is contracted with the generalized projectors \pref{gtproj} and \pref{glproj}. Due to the kinematical singularity present, the choices \pref{d3tgenp} and \pref{d3lgenp} are extremely cumbersome. Therefore
\bea
A_{T\mu\nu}&=&\delta _{\mu \nu }-\delta _{\mu 0}\delta _{0\nu}\label{ftgenp1}\\
A_{L\mu\nu}&=&\delta_{\mu 0}\frac{p_0 p_\nu}{p^2}+\delta_{0\nu}\frac{p_\mu p_0}{p^2}\label{ftgenp2}
\eea
\noindent are used instead. This is a specialization of the most general projector
\be
P_{T/L\mu\nu}^\eta=\eta P_{T/L\mu\nu}+\left(1-\eta\right)\left(t_x\delta_{\mu\nu}+f\delta_{\mu 0}\delta_{0\nu}+f_t\delta_{\mu 0}\frac{p_0 p_\nu}{p^2}+c_t\delta_{0\nu}\frac{p_\mu p_0}{p^2}+t_z\frac{p_\mu p_\nu}{p^2}\right),\label{proj}
\ee
\noindent where $\eta$ can be either $\zeta$ or $\xi$. The choices of the constants $t_x$, $f$, $f_t$, $c_t$, and $t_z$ in \pref{ftgenp1} and \pref{ftgenp2} were made to obtain a well-defined 3d-limit. They are not unique, indicating a larger range of possibilities than just one parameter $\eta$ in general. Nevertheless, as the choices fulfill all requirements they are sufficient.

Employing the projectors \pref{gtproj} and \pref{glproj} yields
\bea
\frac{1}{G(p)}&=&\widetilde{Z}_3+\frac{g_4^2TC_A}{(2\pi)^2}\sum_{n=-\infty}^{\infty}\int d\theta dq \Big(A_T(p,q)G(q)Z(p-q)\nonumber\\
&&+A_L(p,q)G(q)H(p-q)\Big)\label{fulleqGft}\\
\frac{\xi}{H(p)}&=&\xi Z_{3L}+T^{HG}+T^{HH}+\frac{g_4^2C_A}{(2\pi)^2}\sum_{n=-\infty}^{\infty}\int d\theta dq\Big(P(p,q)G(q)G(p+q)\nonumber\\
&&+N_L(p,q)Z(q)Z(p+q)+N_1(p,q)H(q)Z(p+q)+N_2(p,q)H(p+q)Z(q)\nonumber\\
&&+N_T(p,q)H(q)H(p+q)\Big)\label{fulleqHft}\\
\frac{1}{Z(p)}&=&Z_{3T}+T^{GH}+T^{GG}+\frac{g_4^2C_A}{(2\pi)^2}\sum_{n=-\infty}^{\infty}\int d\theta dq\Big(R(p,q)G(q)G(p+q)\nonumber\\
&&+M_L(p,q)H(q)H(p+q)+M_1(p,q)H(q)Z(p+q)+M_2(p,q)H(p+q)Z(q)\nonumber\\
&&+M_T(p,q)Z(q)Z(p+q)\Big)+\frac{p_0^2(\zeta-1)}{2p^2}\left(Z_{3L}-\frac{1}{H(p)}\right).\label{fulleqZft}
\eea
\noindent The $\zeta$ and $\xi$ dependence is acquired by using \pref{ftgenp1} and \pref{ftgenp2}. At $\zeta=\xi=1$ the original form of the equations is recovered. For the correct solutions, the appearance of these terms would be compensated by corresponding structures in the loop contributions, and the result would be independent of $\zeta$ and $\xi$. For $p_0=0$, equation \pref{fulleqHft} is only superficially dependent on $\xi$. As all integral kernels are proportional to $\xi$, the dependence can be divided out for $\xi\neq 0$. Then, only the implicit dependence due to the interactions with the hard modes remains. The latter effect vanishes in the 3d-limit and the Higgs-equation \pref{higgsd3} is again independent of the projection and thus unique. The finite-temperature 4d-theory is no longer finite, but renormalizable. Thus explicit wave-function renormalization constants $\widetilde{Z}_3$, $Z_{3L}$, and $Z_{3T}$ have been introduced and will be discussed in section \ref{scttrunc} and \ref{srenormtrunc}.

The summation is over all Matsubara frequencies $q_0=2\pi T n$. $\widetilde{Z}_1=1$ has been employed \cite{Taylor:ff}. $Z_1$ cannot be calculated in the present approach, but it is finite and set to 1 \cite{Fischer:2002hn,Fischer:2003zc}. The kernels $A_T$, $A_L$, $R$, $M_T$, $M_1$, $M_2$, $M_L$, $P$, $N_T$, $N_1$, $N_2$, and $N_L$ are listed in appendix \ref{sftkernels}. The tadpoles $T^{ij}$ are again used to cancel spurious divergences in much the same way as in the previous section. However, also in the soft equations, they can possess finite parts at finite temperature. In case of the longitudinal equation, these can be absorbed in the mass renormalization discussed below. In the transverse equation, these are at $\zeta=3$ completely contained in the loop terms, and their continuation away from $\zeta=3$ is thus arbitrary. Also these contributions scale at best only as $1/p^2$ and are thus irrelevant to the infrared. Therefore, they are dropped, especially as no simple prescription as \pref{finite} below can remove the related spurious divergences

Note the considerably higher symmetry between the equations for $H$ and $Z$ than in equations \prefr{fulleqG3d}{fulleqZ3d}. Close inspection of the equations reveals further that the dressing functions can only depend on $\left|p_0\right|$ and $\left|\vec p\right|$. The corresponding symmetry, under $p_0\to -p_0$, is used to reduce the number of equations significantly.

\section{Temperature Dependence of the Coupling}\label{scc}

The equations \prefr{fulleqGft}{fulleqZft} depend on the coupling constant $g_4$ as the only parameter. In turn $g_4$ depends on the renormalization scale $\mu$, which can be chosen arbitrarily at any fixed temperature $T$. It directly enters into the definition of the infinite-temperature limit, as $g_3$ depends on $g_4$. In the simplest case, $g_3^2\sim g_4^2(\mu)T$. Comparing equations \prefr{fulleqG3d}{fulleqZ3d} and \prefr{fulleqGft}{fulleqZft}, the constant of proportionality has to be 1 in this truncation scheme to obtain a smooth infinite-temperature limit. Hence it only remains to choose the temperature dependence of $\mu$. There are two options to consider.

If $\mu$ is fixed, the 3d-coupling grows without limit as $T$ grows. This does not necessarily pose a problem, as the infinite-temperature propagators are independent of $g_3$ as long as expressed as a function of $p/g_3^2$. Under such circumstances, all momenta below $T$ are effectively infrared, and the non-perturbative regime would extend to all momenta. 

Alternatively, it is possible to use the limit prescription $g_4^2(\mu(T))T=c_\infty$, where $c_\infty$ is an arbitrary constant, effectively performing a renormalization group transformation when changing the temperature. This fixes $g_3^2$ to be proportional to $\Lambda_{QCD}$, the dynamical scale of QCD \cite{Bohm:yx}, while the 4d-coupling vanishes like $1/T$. This defines a t'Hooft-like scaling in $T$. Note that $g_4\to 0$ for $T\to\infty$ corresponds to the conventional arguments of a vanishing coupling in the high-temperature phase \cite{Blaizot:2001nr}. However here, as in the large-$N$ limit, t'Hooft-scaling is performed, generating a well-defined theory. Hence this possibility defines a smooth 3d-limit with a finite 3d-coupling constant. Here $c_\infty=1$ is chosen. Nonetheless, the case of fixed $\mu$ is also of interest, as it permits to compare the results for different temperatures directly.

These possibilities all correspond to the 4d-situation, where it is possible to shift the onset of non-perturbative physics by renormalization group transformations. Under all circumstances, for any finite temperature, there is always a non-perturbative regime at sufficiently small momenta.

Furthermore, it would be useful to give explicit units for the temperature scale. In the vacuum the scale is fixed via a comparison of the running coupling to perturbation theory \cite{Fischer:2002hn}. This is not possible here because the Matsubara sum is truncated and the coupling cannot be calculated for momenta of the order of $2\pi NT$, where $N$ is the number of Matsubara frequencies included\footnote{Always, only the frequencies with $p_0\ge 0$ are counted.}

The simplest procedure is to compare $g_3$ to lattice calculations. There the effective 3d-coupling is found to be $g_3^2(2T_c)/2T_c=2.83$ \cite{Cucchieri:2001tw}. The phase transition temperature is $T_c=269\pm 1$ MeV \cite{Karsch:2003jg}. Albeit this will not be exactly $g_4^2T_c$ due to the truncation, a first approximation is to require the same temperature scale at $2T_c$. Using the fixed $g_4^2T=c_\infty$ prescription, then $1=c_\infty/(2T_c2.83)$. Since here $c_\infty=1$ in internal units, internal units have to be multiplied by 1.5 GeV to yield physical units. This temperature scale will be used in sections \ref{sftgonly} and \ref{sftfull}. The ratios of temperatures are independent of this prescription.

\section{Infrared Properties}\label{sftir}

Regarding the infrared properties, asymptotic freedom turns out to be advantageous also for the infrared. Just at the phase transition, the $n=1$ Matsubara frequency has already an effective `mass' $p_0=2\pi T_c\approx 1.7$ GeV. By the Appelquist-Carazzone theorem \cite{Appelquist:tg} the hard modes are suppressed by powers of $|\vec p|/p_0$ in the infrared. Thus hard modes do not contribute significantly. Studying each equation in detail, it turns out that this quite general statement is implemented very differently in each equation.

In case of the hard modes, there is no pure soft contribution due to momentum conservation at the vertices. Thus they decouple and a self-consistent solution is that all hard mode dressing functions are constant in the infrared.

In the case of the soft ghost equation \pref{fulleqGft}, all contributions become also constant in the infrared. Thus they can be canceled by the (renormalized) tree-level value. Only the subleading behavior remains, as in the case of the 3d- and 4d-theory. As the subleading contribution of the purely soft term dominates the hard terms, the same behavior as in the infinite-temperature limit emerges.

In case of the transverse equation \pref{fulleqZft}, the hard mode contributions give rise at best to mass-like $1/p^2$ terms. These terms are subleading compared to the soft ghost-loop. Thus the infrared is dominated by the latter and the same infrared solution as in the infinite-temperature limit is found.

The longitudinal equation \pref{fulleqHft} is finally quite different from its equivalent in the 3d- or 4d-case. In the prior case, it was dominated by its tree-level mass, while in the latter it is identical to the transverse one. In the present case, the absence of pure soft interactions with ghosts, which could generate divergent contributions as in the transverse equation, leads to dominance of the hard modes\footnote{The pure soft interactions represented by the kernels $N_1$ and $N_2$ only generate a constant term in the infrared. Without a tree-level mass, the soft tadpoles generate a contribution which can be removed by renormalization.}. These provide mass-like contributions in the infrared. Inspecting e.g. $P$ at $\xi=1$
\be
P(0,q_0,\vec q,\vec p)=\frac{\left|\vec q\right|^2q_0^2\sin\theta}{\left|\vec p\right|^2q^2(q_0^2+(\vec p+\vec q)^2)},\label{dynmass}
\ee
\noindent and using the fact that the hard dressing functions are constant in the infrared, directly leads to a mass-like behavior due to the explicit $1/|\vec p|^2$ factor and the finiteness of the remaining expression. Thus the 3d-mass of the Higgs is generated spontaneously by the interaction with the hard modes.

Hence, the infrared behavior of all soft mode dressing functions does not depend on the temperature, expect for changes in the infrared coefficients.

Note that these observations imply that by using a finite number of Matsubara frequencies it will not be possible to obtain the 4d-vacuum solutions. Any finite number of such modes will not be able to generate a divergence stronger than mass-like in the longitudinal equation to compensate the missing soft interactions with ghosts. Hence, the 4d-behavior can be established only by infinite summation. In this case the generated mass has to diverge even after renormalization at $p=0$ to obtain the vacuum solution, providing over-screening instead of screening. The possibility that the ghost-gluon-vertex changes due to temperature to couple soft ghost modes to soft longitudinal gluons at low temperature and not at high temperature is unlikely, since a bare ghost-gluon vertex is already sufficient at low temperatures and in the vacuum \cite{vonSmekal:1997is,Gruter:2004kd}.

\section{Truncation and Spurious Divergences}\label{sfttrunc}

Also for large but still finite $T$  the 3d-limit is to be made explicit. Thus all the problems encountered in the 3d-theory persist, including the necessity of a modified soft 3-gluon vertex. Hence all modifications of the pure soft terms will be left as in the case of chapter \ref{c3d}, except for the tadpoles in the Higgs equation \pref{fulleqH3d}. Due to the spontaneously generated mass, it will be necessary to alter this behavior. There are also additional spurious divergences due to the hard mode contributions. As they occur at mid-momenta, when viewed from the 4d-perspective using $p^2=p_0^2+\vec p^2$, it is expected that due to the truncation ansatz, they are much harder to compensate. This is indeed the case.

The spurious divergences in contributions of hard modes due to the integration are dealt with in exactly the same way as in 3d by adjusting the tadpole terms. Each integration kernel $K$ is split as
\be
K=K_0+K_D\label{isplit}
\ee
\noindent $K_0$ is finite and $K_D$ divergent upon integration. Each kernel $K_D$ will be compensated by a corresponding tadpole as for $M_{TD}$ and $M_{LD}$ in the 3d-theory in equations \pref{ggtad2} and \pref{ghtad}, respectively.

However, a new kind of spurious divergences appear when performing the Matsubara sum. In principle this is no problem when using a finite number of Matsubara frequencies, since then all expressions are finite. However, when including more and more Matsubara frequencies, it is found that their contribution in the gluon equations \pref{fulleqHft} and \pref{fulleqZft} at $|\vec p|<\max(q_0)$ scales as $q_0$, thus behaving as a quadratic divergence. This is an artifact of using a finite number of Matsubara frequencies. It does not vanish as long as the number of frequencies is finite, no matter how large the number. This behavior is spurious and must be removed.

The ghost equation does not contain spurious divergences\footnote{The problem is also not present in the mixed contributions $M_1$, $M_2$, $N_1$ and $N_2$ in the gluon equations. Hence these do not need to be subtracted.} and can thus serve as a starting point to trace the origin of those in the gluon equations. Analyzing the $\left|\vec p\right|=0$ limit for $p_0\neq 0$, it is directly found that if the equation looks like
\be
\frac{1}{G(p_0\neq 0,0)}=\frac{1}{A_g(p_0)}=\widetilde{Z}_3+g_4^2 T I,\label{ghosteq}
\ee
\noindent then the integral $I$ must scale like $1/p_0$ because of dimensional consistency. Here $\widetilde{Z}_3$ is the wave-function renormalization of the ghost. For any finite number of Matsubara frequencies, $\widetilde{Z}_3$ is finite. Hence the infrared behavior of the hard ghost dressing function is $1/(\widetilde{Z}_3+aT/p_0)$ with a suitable constant $a$. For sufficiently large $p_0$ this can be expanded to yield $1/\widetilde{Z}_3-aT/(p_0\widetilde{Z}_3^2)$ in leading order.

The effect is most directly seen in the $p_0=0$ longitudinal equation in the ghost-loop. For dimensional reasons, in the infrared the terms in the Matsubara sum must scale as $Tq_0/|\vec p|^2 A_g^2(q_0)$. Hence the divergence is directly visible due to the leading constant term in $A_g(q_0)$. The situation is analogous, but not as directly visible in the remaining equations.

This spurious divergence can be removed when removing the leading $1/\widetilde{Z}_3$ term in the ghost dressing function, i.e.\/ by the replacement
\bea
&G(q,q_0)G(p+q,q_0+p_0)\nonumber\\
&\to\left(G(q,q_0)-\frac{1}{\widetilde{Z}_3}\right)\left(G(q+p,q_0+p_0)-\frac{1}{\widetilde{Z}_3}\right),\label{finite}
\eea
\noindent and correspondingly for the gluon loops.

Including a finite number $N$ of Matsubara frequencies only is equivalent to studying a 3d-theory with 1 massless particle and $2N$ massive particles for each 4d particle species. Therefore the ultraviolet asymptotic analysis of subsection \ref{ssuvanalysis} applies, and $I$ in \pref{ghosteq} will again fall off polynomially in the ultraviolet. Hence \pref{finite} amounts to subtracting the asymptotic constant to which $G$ evolves and thus corrects for $N$ being finite. $1/\widetilde{Z}_3$ must become 0 when including an infinite number of Matsubara frequencies, as this is the correct perturbative 4d behavior for $p\gg T$. Hence the spurious divergences are in this case only an artifact of the finite number of Matsubara frequencies.

In the the longitudinal equation the Brown-Pennington value $\xi=0$ is pathologic. It removes all terms in the soft equation and so is not applicable on its own. Thus the replacement \pref{finite} has to be applied to the complete loop contributions $P$, $N_T$ and $N_L$. In the transverse equations all spurious divergences including the one of the Matsubara sum are again removed at $\zeta=3$. Therefore the replacement \pref{finite} will only be necessary in those contributions in $R$, $M_T$, and $M_L$ which are proportional to $(\zeta-3)$. Hence the integral kernels are split differently than \pref{isplit} as
\be
K(\zeta)=K_0+(\zeta-3)K_3+K_D(\zeta),\nonumber
\ee
\noindent similar to \pref{rsplit}. Here $K_0$ and $K_3$ are finite and independent of $\zeta$, and $K_D$ contains all divergences upon integration. The corresponding subtraction is then performed by the replacement
\be
KD(q)D(p+q)\to K_0D(q)D(p+q)+(\zeta-3)K_3\left(D(q)-\frac{1}{Z_3}\right)\left(D(p+q)-\frac{1}{Z_3}\right)\nonumber\\
\ee
\noindent where $D$ is a generic dressing function and $Z_3$ its wave-function renormalization. 

In the soft transverse equation it is additionally necessary to remove tadpole-like structures at $\zeta\neq 3$ in $K_3$. They contribute a logarithmically divergent mass-term, which is irrelevant in the infrared and in the ultraviolet. It cannot be renormalized, as a counter-term would be required, which is forbidden by gauge symmetry. Thus these contributions are absorbed by the tadpoles as well. Then, in the transverse equation for the soft-soft interactions, the subtractions are as in chapter \ref{c3d}. The soft-hard contributions, $R$, $M_T$, and $M_D$ are subtracted as
\bea
&KD(q)D(p+q)\to K_0D(q)D(p+q)\nonumber\\
&+(\zeta-3)\left(K_3-\frac{1}{p^2}\left(\lim_{p\to 0}p^2K_3\right)\right)\left(D(q)-\frac{1}{Z_3}\right)\left(D(p+q)-\frac{1}{Z_3}\right)\label{finite2}.
\eea

\section{Small-Momentum Approximation and Renormalization}\label{ssmallpapprox}

Performing the subtractions \pref{finite} and \pref{finite2}, the resulting equations are still logarithmically divergent for $N\to\infty$. This also includes the ghost equations. The integrals themselves are convergent quantities, but the Matsubara sum is not, as the terms scale like $1/q_0$. Thus, for an infinite number of Matsubara frequencies, the sum diverges. This is the 4d logarithmic divergence and thus the usual one of Yang-Mills theory \cite{Bohm:yx}.

In the case of a finite number of Matsubara frequencies, the sum is finite. Nonetheless, there are a few aspects to be dealt with in the following subsections.

\subsection{Small-Momentum Approximation}

There are three regions of momentum to be distinguished:

At $p\le 2\pi T$, all Matsubara sums behave essentially the same due to the Appelquist-Carrazone decoupling theorem. It is valid as the hard modes behave essentially tree-level-like. As the contributions of the hard modes behave as $1/q_0$, the final result will depend on the number of Matsubara frequencies included, and by adding an arbitrary number, any quantitative (but not qualitative) result can be generated. Thus, it is necessary to renormalize in order to be independent of the cutoff. The cutoff is in this case imposed by the number of Matsubara frequencies included. The renormalization procedure will be implemented in the next subsections. By this approach, the results can be made quite reliable in this regime.

At $2\pi T\le p\le 2\pi T (N-1)$, where $N$ is the number of Matsubara frequencies included, more and more Matsubara terms depart from their $1/q_0$ behavior to a $1/p$ behavior. As the external momentum $p$ becomes large compared to the effective mass $q_0$ of a hard mode, the mode becomes dynamical and behaves like a massive 3d-particle. The results are still quite reliable in this region as also in the case $N\to\infty$, for any finite momentum $p$ only a finite number of Matsubara frequencies are dynamical. It becomes less and less reliable when approaching the upper limit $2\pi TN$.

At $p\ge 2\pi T (N-1)$, the situation changes drastically. Opposite to the case $N=\infty$, all Matsubara modes are dynamical and their contributions will scale as $1/p$. Thus the number of Matsubara modes will now enter linearly instead of logarithmically. This is an artifact of cutting off the Matsubara sum. In this region the results are not reliable. However, as the sum is still finite and suppressed by $1/p$, the contribution is subleading with respect to the tree-level term and thus the system of equations can still be closed consistently\footnote{Otherwise the corresponding finite 3d-theory would be ill-defined, which is not the case.}.

By renormalization, these artifacts can be reduced, if not completely removed at sufficiently small momenta. In that sense the finite Matsubara sum approximation is indeed a small-(3-)momentum approximation.

\subsection{Renormalization of a Truncated Matsubara Sum}\label{scttrunc}

There are a few subtleties involved concerning the renormalization of a truncated Matsubara sum. These are discussed here.

Firstly, in the vacuum case \cite{Alkofer:2000wg,Fischer:2002hn}, renormalization was performed using a momentum subtraction scheme (MOM), i.e. by subtracting from the DSEs themselves at a fixed subtraction point $s$. This leads to a generic form of
\be
\frac{1}{D(p)}-\frac{1}{D(s)}=I(p)-I(s),\label{subrenormalization}
\ee
\noindent where $D$ is the corresponding dressing function and $I$ is the self-energy contribution. As the highest divergences encountered are logarithmic \cite{Bohm:yx}, \pref{subrenormalization} is finite and can be solved if a proper value of $D(s)$ is known. This can be achieved by choosing $s$ inside the perturbative region. This approach fails if the large momentum asymptotic value of $D$ is a constant different from 0, since \pref{subrenormalization} is ambiguous with respect to such a constant. In the case of a truncated Matsubara sum, the ultraviolet behavior is that of a massive 3d-theory. Thus the self-energy contributions vanish as $1/p$ in the ultraviolet, see subsection \ref{ssuvanalysis}. Therefore the dressing functions $D$ are dominated by the tree-level term
\be
\lim_{\left|\vec p\right|\to\infty}D(p_0,\left|\vec p\right|)\to\frac{1}{Z_3},
\ee
\noindent with $Z_3$ their wave-function renormalization. For a finite number of Matsubara frequencies $Z_3$ is finite and thus $D$ goes to a constant in the ultraviolet and \pref{subrenormalization} is not applicable. Therefore the renormalization is performed by explicit counter-terms. This is discussed in the next subsection.

A second point is the mass and mass renormalization necessary for the soft longitudinal mode. The soft mode with frequency $p_0=0$ of the $A_0$ component of the gauge field transforms homogeneously instead of inhomogeneously under gauge transformations \pref{gtrans1}. Therefore gauge symmetry permits to add a term
\be
m^2A_0^2(0,\vec p)\label{a0mass}
\ee
\noindent to the Lagrangian. In the vacuum, however, such a term is forbidden by manifest Lorentz invariance. At finite temperature in the Matsubara formalism, this is no longer the case, and such a term could in principle be present. This term replaces the $p_0^2A_0^2$ term of the hard modes, which stems from the $A_0\pd_0^2 A_0$ term in the Lagrangian \pref{lym} for $p_0\neq 0$.

Concerning the counter-terms, the wave-function renormalization is performed by adding the counter-term
\be
\delta Z_3(A_\mu^a\pdm\pdn A_\mu^a-A_\mu^a\pd^2 A_\mu^a)\nonumber
\ee
\noindent to the Lagrangian. In the case $p_0=0$, the first term is not present for the longitudinal mode $A_0$. Its place can be taken by a counter-term for the mass term \pref{a0mass}. This implies a relation between the wave-function counter-term $\delta Z_3$ and the mass counter-term $\delta m^2$, which cannot be exploited here due to the truncation of the Matsubara sum. Therefore an independent renormalization of the wave function and the mass of the soft longitudinal mode is necessary and will be performed in the next subsection.

A last point concerns the implications for the counter-terms due to the truncation of the Matsubara sum. As the divergence structure must be the same as in the vacuum, the counter-terms must be the same for all frequencies. This is no longer the case when the Matsubara sum is truncated, as can be seen directly by counting. In the sum for the soft mode, $p_0=0$, contributions from $2N-1$ Matsubara modes are present for each loop. For the hard mode with $p_0=2\pi T(N-1)$, only $N$ contributions are present, as $\left|p_0+q_0\right|\le 2\pi (N-1)T$. Thus different numbers of modes contribute and the counter-terms cannot be the same. This is always the case as long as $N<\infty$. To surpass the problem in a constructive manner, each mode will be renormalized independently.

Therefore, a counter-term Lagrangian is added, given by
\bea
{\cal L}&=&\delta m^2 A_0^a(0)^2+\sum_{q_0} \Big(\delta Z_{3T}(q_0) A^a_\mu(q_0) \Delta_{T\mu\nu}(q_0) A^a_\nu(q_0)\nonumber\\
&+&\delta Z_{3L}(q_0) A^a_\mu(q_0) \Delta_{L\mu\nu}(q_0) A^a_\nu(q_0)+\delta\widetilde{Z}_3(q_0) \bar c^a(q_0)\pd^2 c_a(q_0)\Big)\label{ctl}
\eea
\noindent where $A^a_\mu(q_0)$ are the modes of the gluon field and $\bar c(q_0)$ and $c(q_0)$ are the modes of the ghost and anti-ghost field, respectively. $\Delta_{T/L\mu\nu}$ are the appropriate tensor structures of derivatives. Consequently, the corresponding wave-function renormalization constants in the subtractions \pref{finite} and \pref{finite2} have to be replaced by their frequency-dependent versions.

This shortcoming is a consequence of the small momentum approximation yielding the incorrect ultraviolet properties: The treated system is for large momenta equivalent to a 3d-theory of $2N-1$ particles per 4d particle species. In such a theory all fields can be renormalized independently. This problem has to be investigated further to obtain reliable results on the large momentum behavior of the dressing functions.

\subsection{Implementation of Renormalization}\label{srenormtrunc}

In all equations but the one for $H(0,\vec p)$, using the counter-terms of \pref{ctl} amounts to replacing the tree-level term $1$ by $1+\delta Z_3$. This way explicitly multiplicative renormalization is obtained. In the equation for $H(0,\vec p)$, the tree-level term is replaced\footnote{Note that when using the projector \pref{ftgenp2}, also $\delta m^2$ is multiplied by $\xi$.} by $1+\delta Z_3L+\delta m^2/p^2$. Thus, it also generates a mass-renormalization. By the arguments following equation \pref{dynmass} just a renormalization is indeed necessary, and the mass is not generated purely by this procedure.

The last ingredient is the renormalization prescription. To investigate both cases discussed in section \ref{scc}, either a fixed subtraction point $s=s_0$ or to establish a well-defined 3d-limit $s=T$ is employed. Thus, the following prescription is applied to all dressing functions $D$, except $H(0,\vec p)$:
\be
D(s)=1.\nonumber
\ee
\noindent This also ensures $G(s)^2Z(s)=1$ as is required in the 4d-theory \cite{vonSmekal:1997is}. The soft ghost mode is not renormalized at 0 as in the vacuum calculation \cite{Fischer:2002hn}.

In case of $H(0,\vec p)$, two prescriptions are necessary. The first is\footnote{For numerical reasons, it is actually performed at $2\delta_i$, where $\delta_i$ is the numerical IR-cutoff of the integration. This is only a marginal difference.}
\be
\lim_{|\vec p|\to 0}|\vec p|^{-2}H(0,\vec p)=\frac{1}{m_{3d}^2}=\frac{1}{r^2g_4^4T^2+g_4^2TC_A\frac{rg_4^2T}{4\pi}}\nonumber
\ee
\noindent where $m_{3d}$ is the tadpole-improved mass of the 3d-theory. It depends on $r=m_h/g_3^2$, which is again taken to be the same value as in the previous chapter. For fixed $g_3$, this mass is independent of the temperature while for fixed $\mu$ it scales with temperature as expected. This again indicates the necessity of control over the infinite-temperature limit of the theory in terms of the effective coupling constant. In addition,
\be
\frac{1}{H(0,s)}=1+\frac{m_{3d}}{s^2}\nonumber
\ee
\noindent is required to fix the mass renormalization.

It should be noted that, by renormalizing at $T$, it is guaranteed that the correct 3d-limit is obtained. This prescription requires that at $T\to\infty$ all dressing functions approach 1 at infinity. This yields the 3d-results\footnote{This is only correct, if $g_4^2T\ll T$. This is immanent, if $g_4^2 T$ is fixed. If $\mu$ is fixed, this is only possible if $\mu$ is taken sufficiently large for a fixed temperature. Still deviations from 1 will be found for $|\vec p|\gg\mu,T$.} as in chapter \ref{c3d}. It would be easily possible to readjust the requirements on $H(0,\vec p)$ to comply with the lattice data in the infrared and ultraviolet. Here the exact correspondence to the 3d results in chapter \ref{c3d} is preferred, as this will make the 3d-limit explicit.

Therefore the explicit implementation of the renormalization prescription for the DSE of a dressing function $D$ with self-energy contributions $I$
\be
\frac{1}{D(p)}=1+I(p)\nonumber
\ee
\noindent is then
\bea
\frac{1}{D(p)}=1+\delta Z_3+I(p)\nonumber\\
\delta Z_3=-I(s)\nonumber
\eea
\noindent and in the case of $H(0,\vec p)$
\bea
\frac{1}{H(0,\vec p)}=1+\delta Z_{3L}+\frac{\delta m^2}{p^2}+I(p)\nonumber\\
\delta m^2=m_r^2-\lim_{p\to 0}p^2I(p)\nonumber\\
\delta Z_{3L}=-I(s)+\frac{lim_{p\to 0}p^2I(p)}{s^2},\nonumber
\eea
\noindent where $m_r=m_{3d}$ is the renormalized mass.

\section{Ghost-Loop-Only Truncation}\label{sftgonly}

\begin{figure}
\epsfig{file=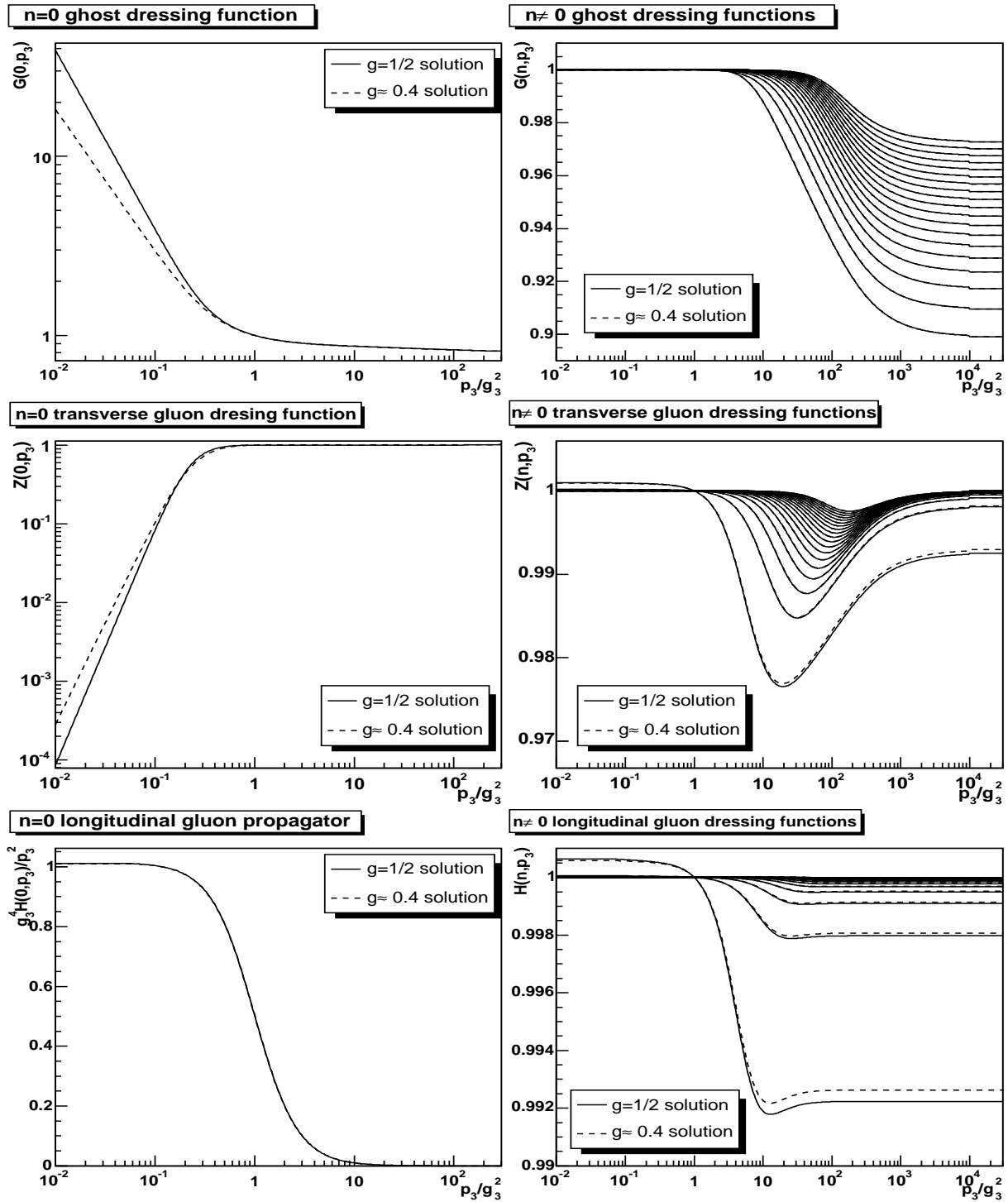,height=1.2\linewidth,width=\linewidth}
\caption{The left panels show for the soft modes from top to bottom the dressing functions of the ghost and transverse gluon and the propagator of the longitudinal gluon. The right panels show the hard mode dressing functions. The less the deviation from 1, the harder the mode is. The solid lines are the $g=1/2$ solution and the dashed lines the $g\approx 0.4$ solution. Both are  at $T=1.5$ GeV and $N=20$. Note the different momentum scale in the left and the right panels.}
\label{figftgonlystd}
\end{figure}

The first step is again the calculation of the ghost-loop-only solutions. Performing the calculations turns out to be numerically more complicated than in 3d due to the recursive definition of the elimination of the spurious divergences \pref{finite} and \pref{finite2}. This is discussed in more detail in appendix \ref{anum}. Nonetheless, both solutions can be found. They are shown in figure \ref{figftgonlystd} with the largest number of Matsubara frequencies currently available\footnote{The number of frequencies included is only a matter of CPU-time.} at $\zeta=\xi=1$. In general, all hard mode dressing functions are similar, but behave more and more tree-level like with increasing frequency $p_0=2\pi nT$. The different asymptotic values of the dressing functions for each mode are a direct manifestation of the artifacts of the truncation of the Matsubara sum, as discussed in section \ref{scttrunc}.

The small deviation from tree-level for the hard longitudinal modes is most likely due to the severe subtraction \pref{finite}. Within the current ansatz however, this is the best achievable. Note that due to $p^2=p_0^2+\vec p^2$, the infrared of the hard modes is actually mid-momenta with respect to 4d. Thus, these problems do not come unexpected, as the truncation is tailored to the infrared and thus to the soft modes. In addition, the transverse gluon wave-function renormalization is smaller than 1 for $n=0$. Thus the dressing function would diverge for large momenta in the limit $N\to\infty$. This is also an artifact of the truncation, as the ghost loops alone have the wrong sign at large momenta. This will be remedied by adding the gluon loops in the next section, and can be ignored for now.

\begin{figure}
\epsfig{file=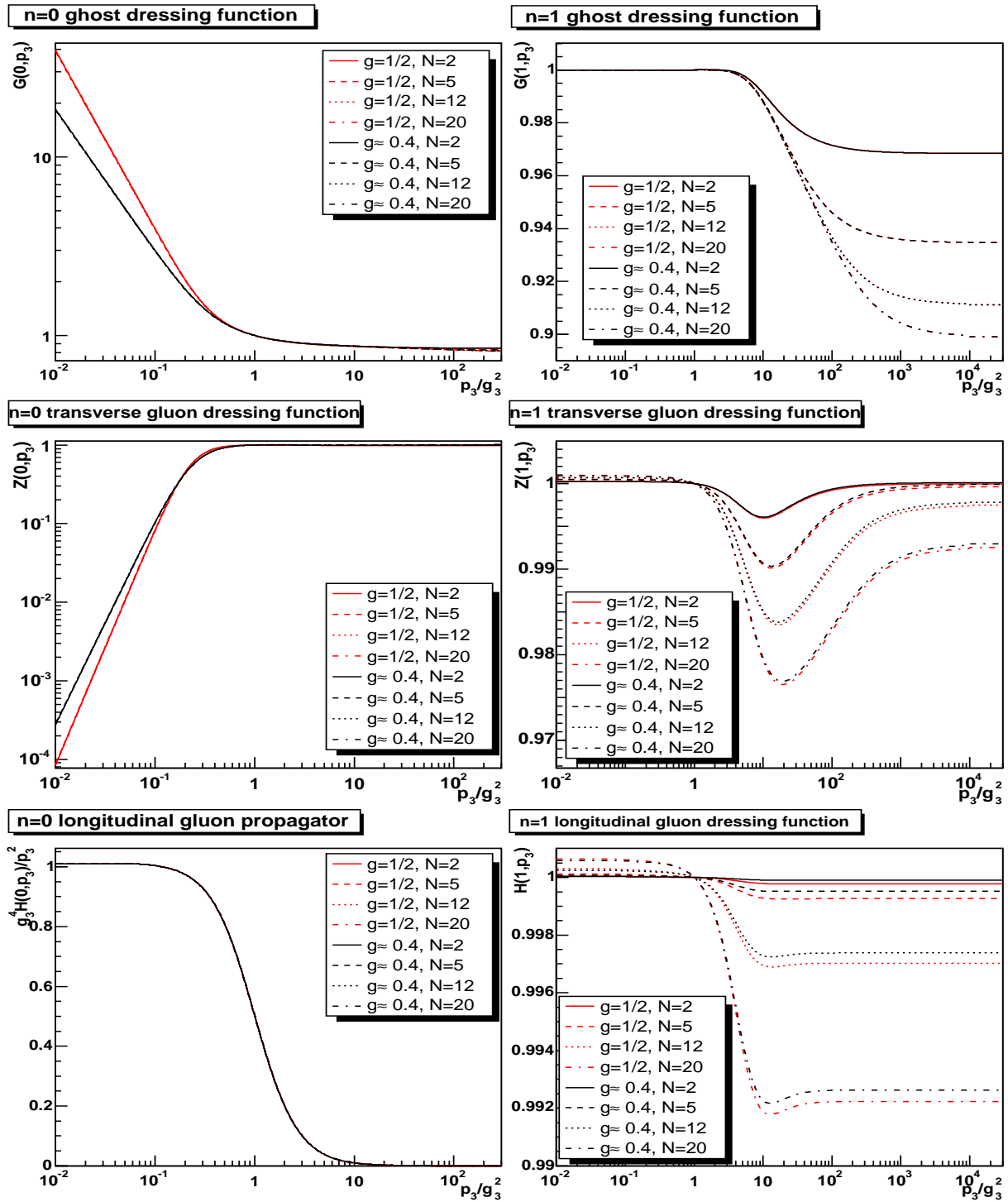,height=1.2\linewidth,width=\linewidth}
\caption{The dependence of the solutions on $N$ at $T=1.5$ GeV. The left panels show for the soft modes from top to bottom the dressing functions of the ghost and transverse gluon and the propagator of the longitudinal gluon. The right panels show the dressing functions for the $n=1$ hard mode. The \piclinecol lines show the solution $g=1/2$ and the black lines $g\approx 0.4$. Solid is $N=2$, dashed is $N=5$, dotted is $N=12$ and dashed-dotted is $N=20$.}
\label{figftgonlymstd}
\end{figure}

The dependence on the number of Matsubara frequencies is shown in figure \ref{figftgonlymstd}. As expected, in the infrared region, the soft mode dressing functions are not affected by the number of Matsubara frequencies. The onset of deviation of the hard mode dressing functions calculated with $N$ frequencies from the solutions with $M>N$ frequencies increases in 3-momenta with $N$. This is a manifestation of the small momentum approximation.

\begin{figure}
\epsfig{file=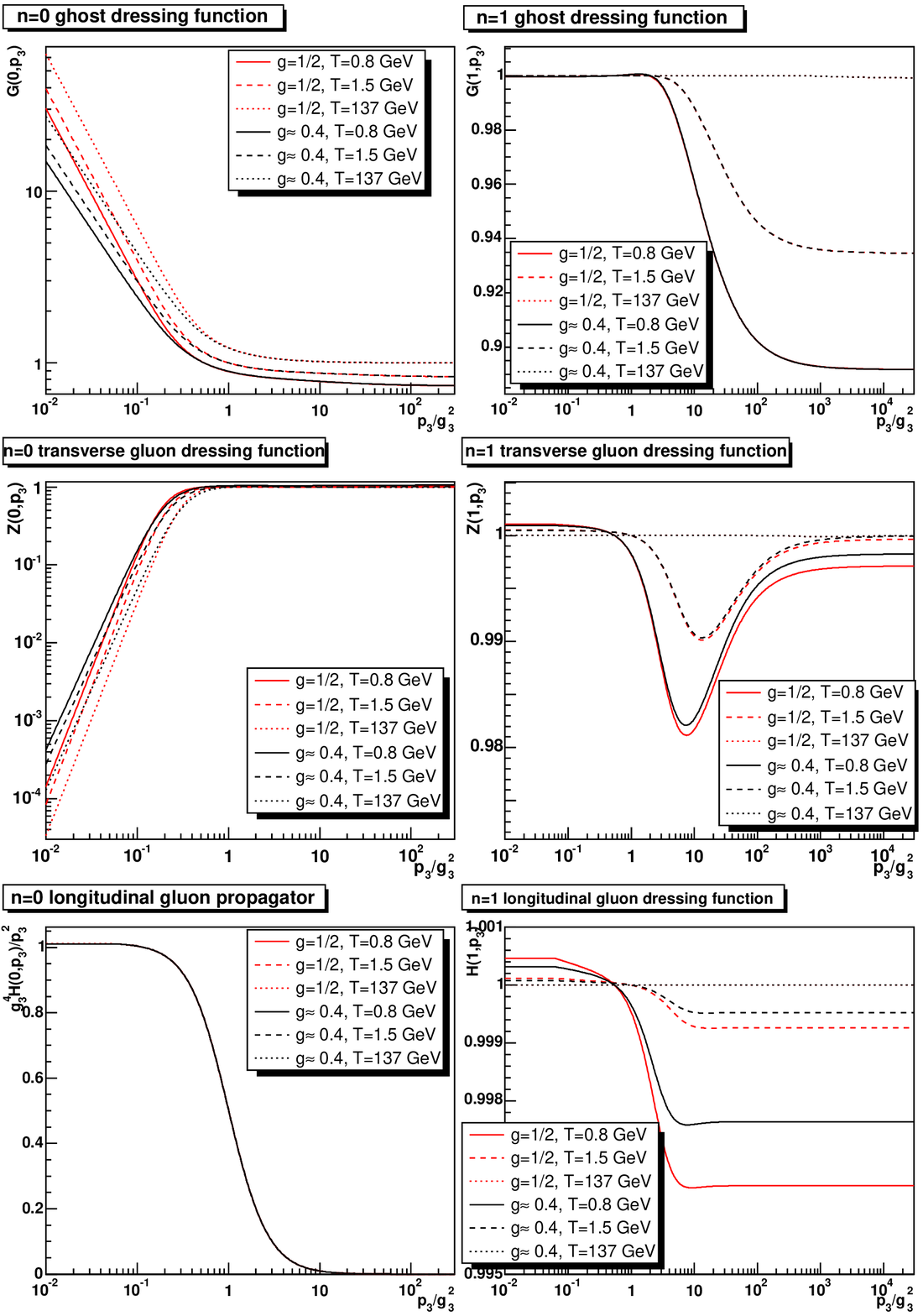,height=1.2\linewidth,width=\linewidth}
\caption{The dependence of the solutions on $T$ at fixed $g_4^2T$ and subtraction at $T$ at $N=5$. The left panels show for the soft modes from top to bottom the dressing functions of the ghost and transverse gluon and the propagator of the longitudinal gluon. The right panels show the dressing functions for the $n=1$ hard mode. The \piclinecol and black lines are the $g=1/2$ and the $g\approx 0.4$ solution, respectively. Solid is $T=0.8$ GeV, dashed is $T=1.5$ GeV, and dotted is $T=137$ GeV.}
\label{figftgonlygstd}
\end{figure}

Naturally, the dependence on the temperature is the next subject. Using the t'Hooft-like scaling, the results shown in figure \ref{figftgonlygstd} are obtained. While the soft mode dressing functions are relatively unaffected by temperature, significant changes are found for the hard mode dressing functions. These become increasingly tree-level with increasing temperature and thus the hard modes cease to interact. At low temperature they exhibit significantly more dynamics.

\begin{figure}
\epsfig{file=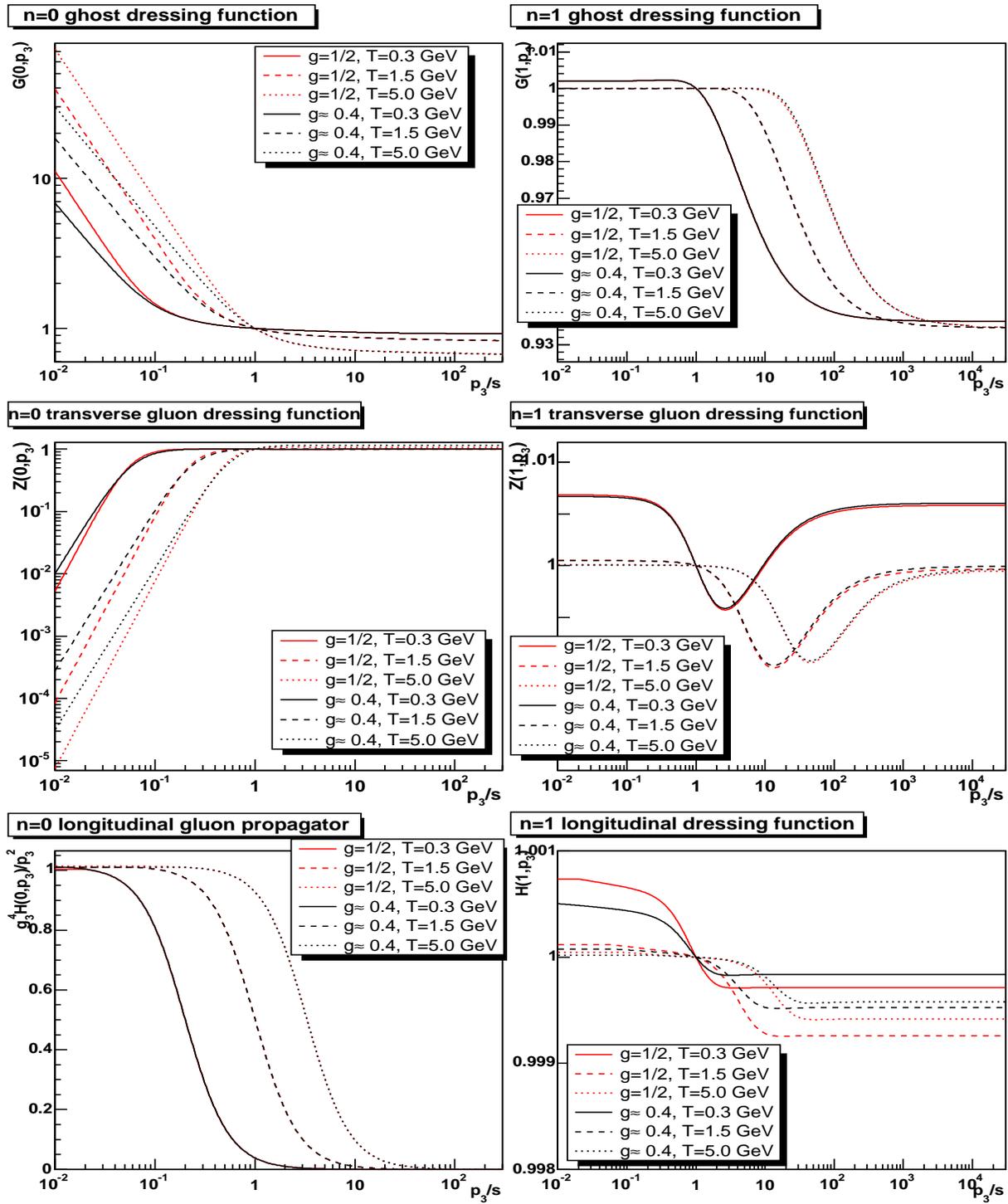,height=1.2\linewidth,width=\linewidth}
\caption{The dependence of the solutions on $T$ at fixed $\mu$ and $s$ at $N=5$. The left panels show for the soft modes from top to bottom the dressing functions of the ghost and transverse gluon and the propagator of the longitudinal gluon. The right panels show the dressing functions for the $n=1$ hard mode. The \piclinecol and black lines are the $g=1/2$ and the $g\approx 0.4$ solution, respectively. Solid is $T=0.3$ GeV, dashed is $T=1.5$ GeV, and dotted is $T=5.0$ GeV.}
\label{figftgonlytstd}
\end{figure}

The case for fixed $\mu$ is shown in figure \ref{figftgonlytstd}. The soft longitudinal gluon propagator is scaled by the interaction strength, since the mass of the soft longitudinal gluon is a dynamical effect. The results represent, how the system reacts on a change of $s/T$, which are the only dimensionful quantities involved as long as $g_4$ is fixed. The changes are governed by the renormalization prescription, and are qualitatively different from the previous case. As expected, the higher the temperature, the further out in momentum the non-perturbative effects in the soft mode dressing functions extend, only limited by the requirement to obey the renormalization prescription. At the same time the dynamics of the hard modes is pushed to higher momenta, although the dressing functions do not decrease. Indeed, this scenario describes rather a change of the relevant scale for the onset of non-perturbative effects than the effects of temperature.

At fixed $\mu$, it was possible\footnote{In the case of fixed $g_4^2T$, it is likely not yet found due to limitation in CPU time.} to follow the solution down to $T=91$ MeV. Lattice calculations find a phase transition temperature of $(269\pm 1)$ MeV \cite{Karsch:2003jg}. Thus super-cooling is possible, provided the temperature estimate of section \ref{scc} is correct within a factor of 2 and the phase transition temperature in the approximation scheme used here is not significantly lower than in lattice calculations.

\begin{figure}
\epsfig{file=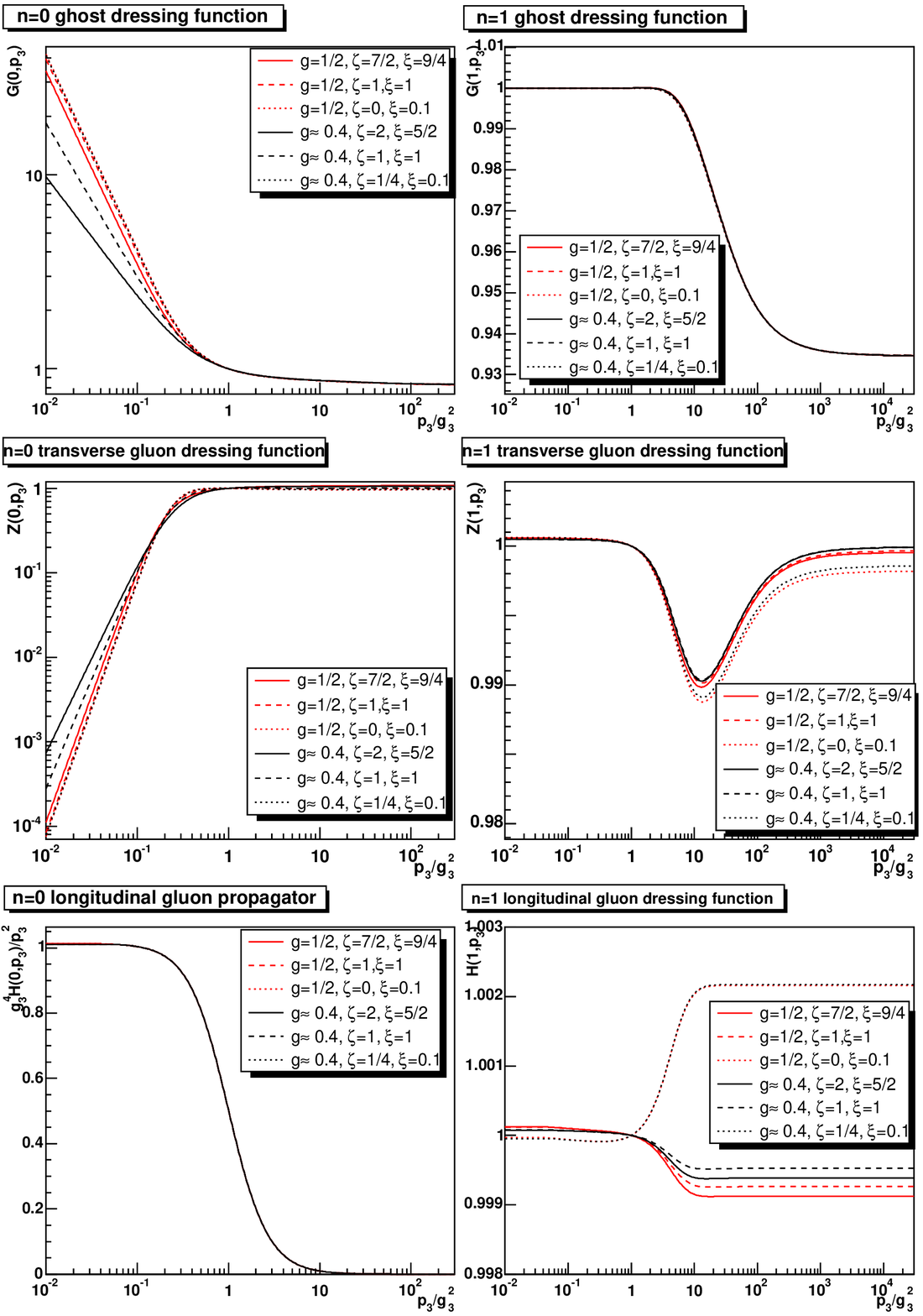,height=1.2\linewidth,width=\linewidth}
\caption{The dependence of the solutions on $\zeta$ and $\xi$ at $N=5$. The left panels show for the soft modes from top to bottom the dressing functions of the ghost and transverse gluon and the propagator of the longitudinal gluon. The right panels show the dressing functions for the $n=1$ hard mode. The \piclinecol and black lines are the $g=1/2$ and the $g\approx 0.4$ solution, respectively. Solid is $(g,\zeta,\xi)=(1/2,7/2,9/4)$ and $(0.4,2,5/2)$, dashed is $(1/2,1,1)$ and $(0.4,1,1)$, dotted is $(1/2,0,0.1)$ and $(0.4,1/4,0.1)$.}
\label{figftgonlyzx}
\end{figure}

The remaining issue is the dependence of the solutions on the projection and thus on $\zeta$ and $\xi$. It is shown in figure \ref{figftgonlyzx}. The largest effect is seen in the infrared region of the soft solutions of the alternating branch when varying $\zeta$, similar to the infinite-temperature limit. The hard modes are significantly less affected, with the largest impact on the hard longitudinal dressing functions when varying $\xi$. For the smallest $\xi$ values, the self-energy of the hard longitudinal mode seems to switch sign, generating a wave-function renormalization smaller than 1. This effect has not been observed in the full system and is thus likely an artifact of the ghost-loop-only truncation. Thus, the dependence on the projection is of similar extent as in the infinite-temperature limit.

\section{Full Theory}\label{sftfull}

\begin{figure}
\epsfig{file=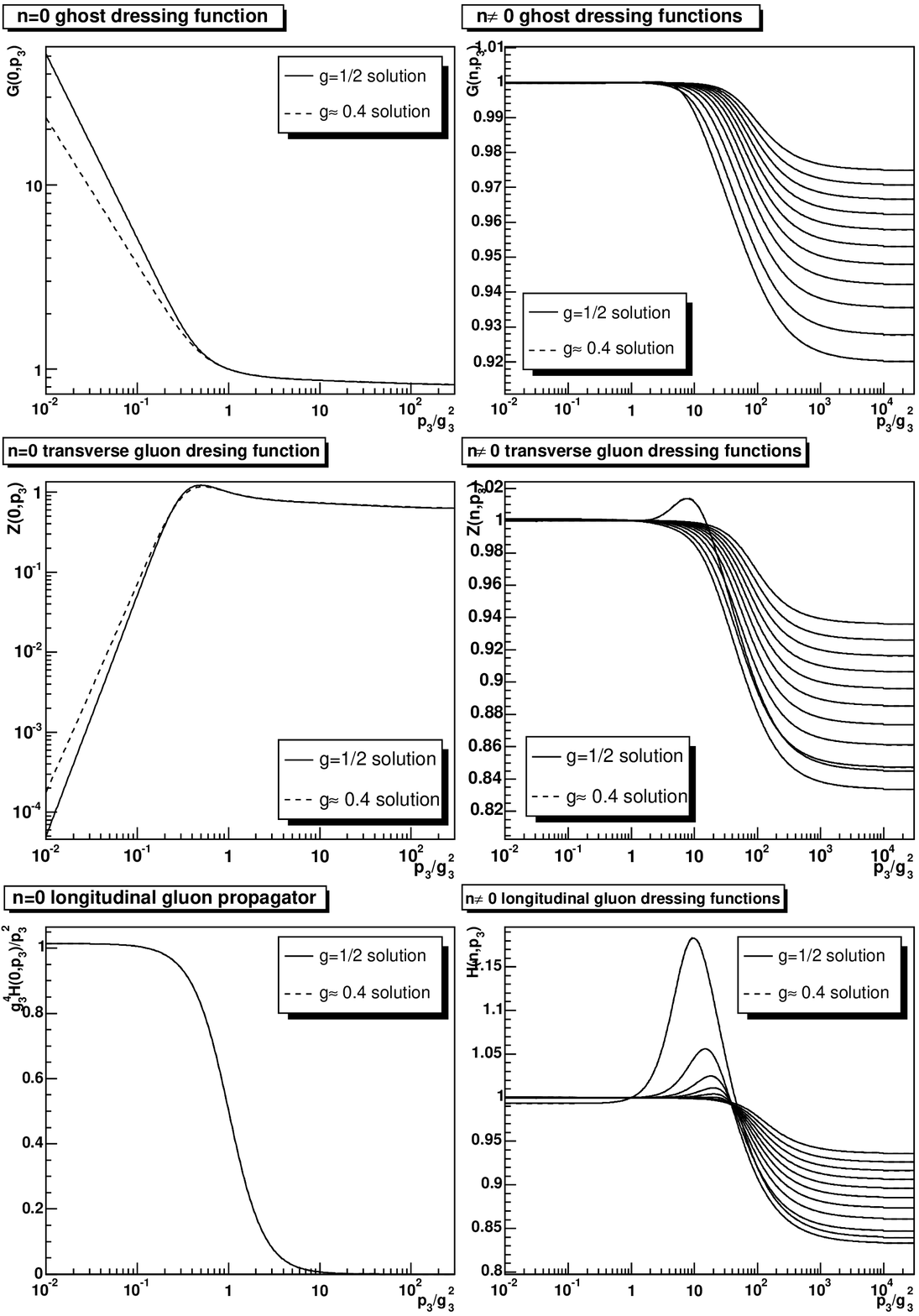,height=1.2\linewidth,width=\linewidth}
\caption{The left panels show for the soft modes from top to bottom the dressing functions of the ghost and transverse gluon and the propagator of the longitudinal gluon. The right panels show the hard mode dressing functions. The less the deviation from 1, the harder the mode is. The solid lines are the $g=1/2$ solution and the dashed lines the $g\approx 0.4$ solution. Both are  at $T=1.5$ GeV and $N=12$. Note the different momentum scale in the left and the right panels.}
\label{figftf}
\end{figure}

Finally, the full system of equations \prefr{fulleqGft}{fulleqZft} is implemented. For the largest available number of Matsubara frequencies, the result is shown in figure \ref{figftf}. There are several observations. First of all, the soft modes are again nearly unaffected in the infrared by the presence of the hard modes. The latter show a significant modification, compared to tree-level, although still only of the order of 30\%. Also, all wave-function renormalization constants are now larger than one, as is required. The hard mode dressing functions exhibit some structure. There are maxima in all dressing functions. This is most pronounced in the case of the dressing functions of the longitudinal gluon. These structures do not translate into a corresponding structure in the propagators, which are monotonically decreasing from a constant of order $1/p_0^2$ in the infrared to 0 in the ultraviolet. It is also nearly irrelevant for the hard mode dressing functions to which soft infrared solution they are coupled to.

\begin{figure}
\epsfig{file=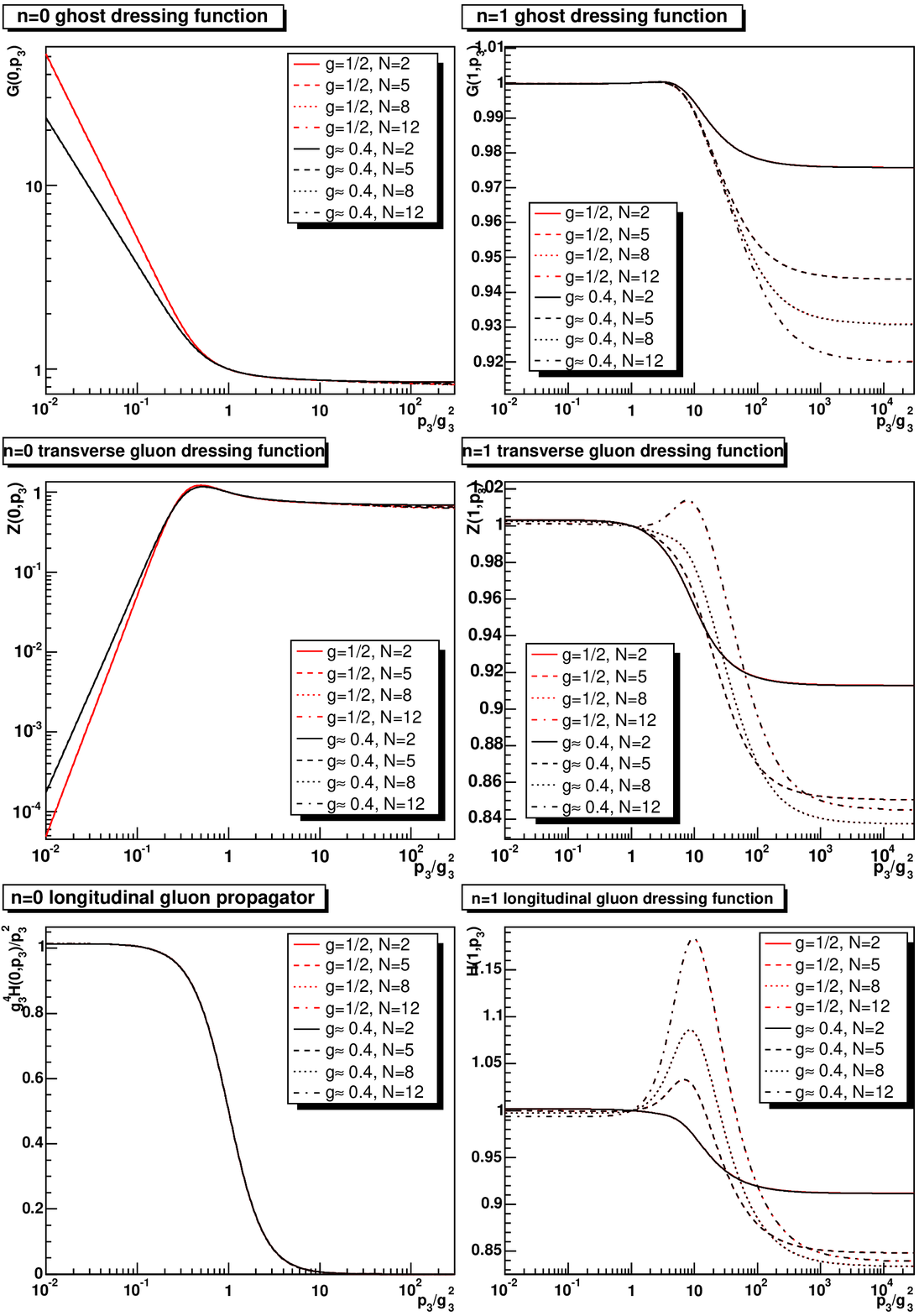,height=1.2\linewidth,width=\linewidth}
\caption{The dependence of the solutions on $N$ at $T=1.5$ GeV. The left panels show for the soft modes from top to bottom the dressing functions of the ghost and transverse gluon and the propagator of the longitudinal gluon. The right panels show the dressing functions for the $n=1$ hard mode. The \piclinecol lines show the solution $g=1/2$ and the black lines $g\approx 0.4$. Solid is $N=2$, dashed is $N=5$, dotted is $N=8$ and dashed-dotted is $N=12$.}
\label{figftfm}
\end{figure}

Figure \ref{figftfm} displays the dependence of the full solutions on the number of Matsubara frequencies. While the soft mode dressing functions are nearly unaffected, apart from the value of the renormalization constants, the effect on the hard mode dressing functions is significant. The ghost is quite insensitive, except for its wave-function renormalization. This is not the case for the gluons. The peaks at mid-momentum are sensitive to the number of Matsubara frequencies included. The effect is largest for the longitudinal gluon dressing functions. From the available number of Matsubara frequencies, it is hard to estimate whether the peak grows to a finite value when $N\to\infty$ or not. It cannot be excluded that the hard longitudinal gluon dressing functions violate the Gribov condition \pref{gribov} once sufficiently many Matsubara frequencies are included. This would be very similar to the case of the transverse gluon dressing function in the infinite-temperature limit and therefore would necessitate a similar vertex construction for the longitudinal-gluon-transverse-gluon vertex as for the soft transverse gluon vertex. If such a vertex would be necessary, the result would be a finite peak, very similar to the present situation. Thus, the results would be even quantitatively quite similar.

The disappearance of the peaks in the gluon dressing functions at $N=2$ can be directly related to the vanishing of hard-mode couplings due to the restriction $(p_0+q_0)/(2\pi T)<2$. Only hard-soft-mode couplings contribute. Thus, the peak is generated due to pure hard-hard interactions alone.

Again here and in the following only the $n=1$ modes are presented. The $n>1$ modes essentially follow their behavior, albeit much closer at tree-level.

\begin{figure}
\epsfig{file=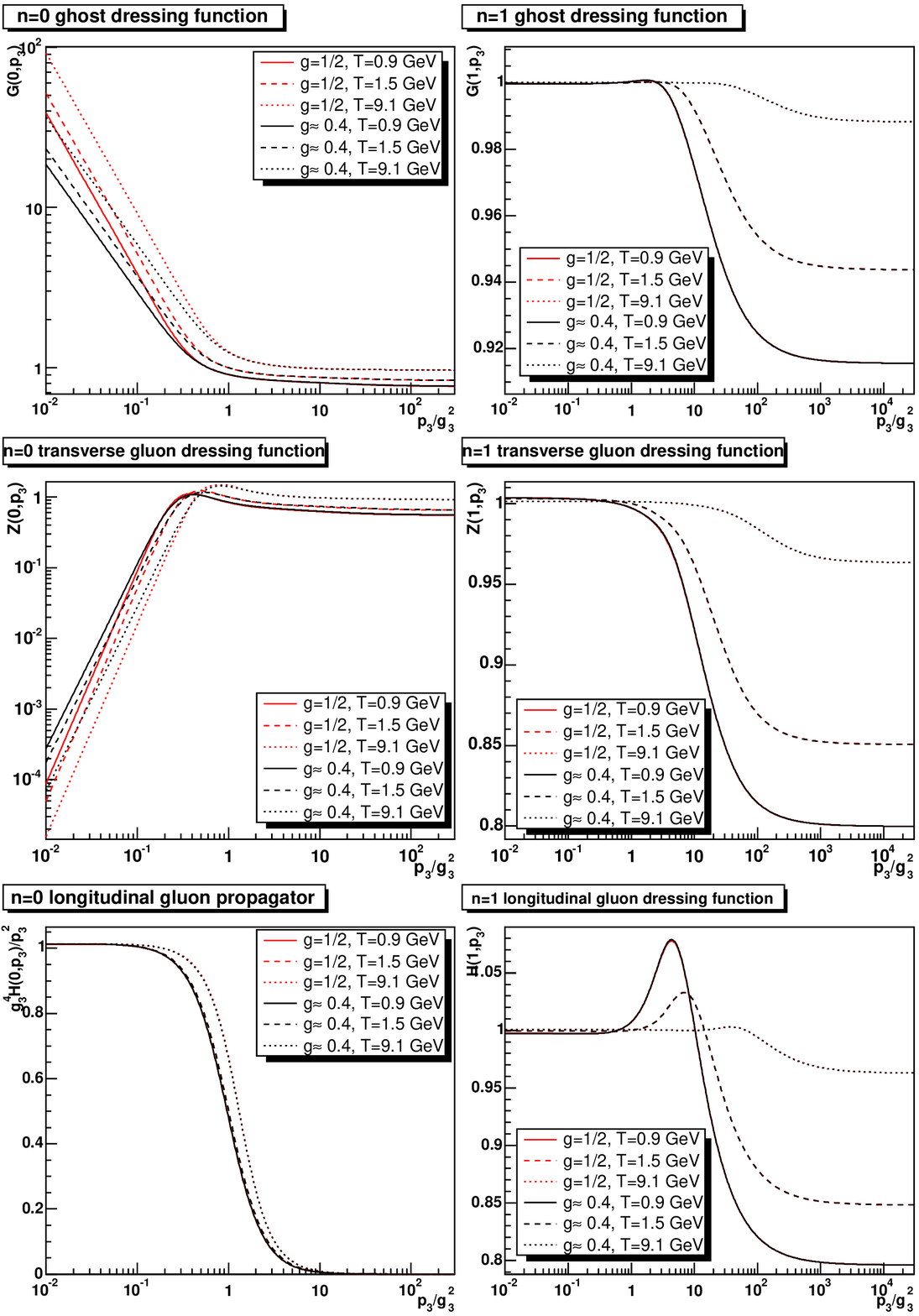,height=1.2\linewidth,width=\linewidth}
\caption{The dependence of the solutions on $T$ at fixed $g_4^2T$ and subtraction at $T$ at $N=5$. The left panel shows the soft modes, where apart from the longitudinal gluon the dressing functions are shown. In the latter case the propagator is shown. The right panels show the dressing functions for the $n=1$ hard mode. The \piclinecol and black lines are the $g=1/2$ and the $g\approx 0.4$ solution, respectively. Solid is $T=0.9$ GeV, dashed is $T=1.5$ GeV, and dotted is $T=9.1$ GeV.}
\label{figftfg}
\end{figure}

The dependence on temperature when using t'Hooft-like scaling is shown in figure \ref{figftfg}. Here, a significant effect also on the soft modes can be seen. While a direct influence on the infrared coefficients for the ghost and transverse gluon dressing functions is visible, for the dressing function of the longitudinal gluon only a slight change at mid-momenta occurs. Therefore, the dressing function of the longitudinal gluon is dominated by its renormalized mass also when changing the temperature. As expected, the hard mode dressing functions become more and more tree-level like when increasing the temperature. The peaks in the hard mode dressing functions shift to higher momenta, owing to the renormalization condition, and become smaller due to the increase of the effective mass $p_0$ of the hard modes. At low temperature, the opposite effect is observed. In general, the hard mode dressing functions become more dynamical as their effective mass decreases, albeit quite slowly. The insensitivity on the infrared solution of the soft sector is not changed when reducing the temperature, and it may need a significantly lower temperature to induce a change.

\begin{figure}
\epsfig{file=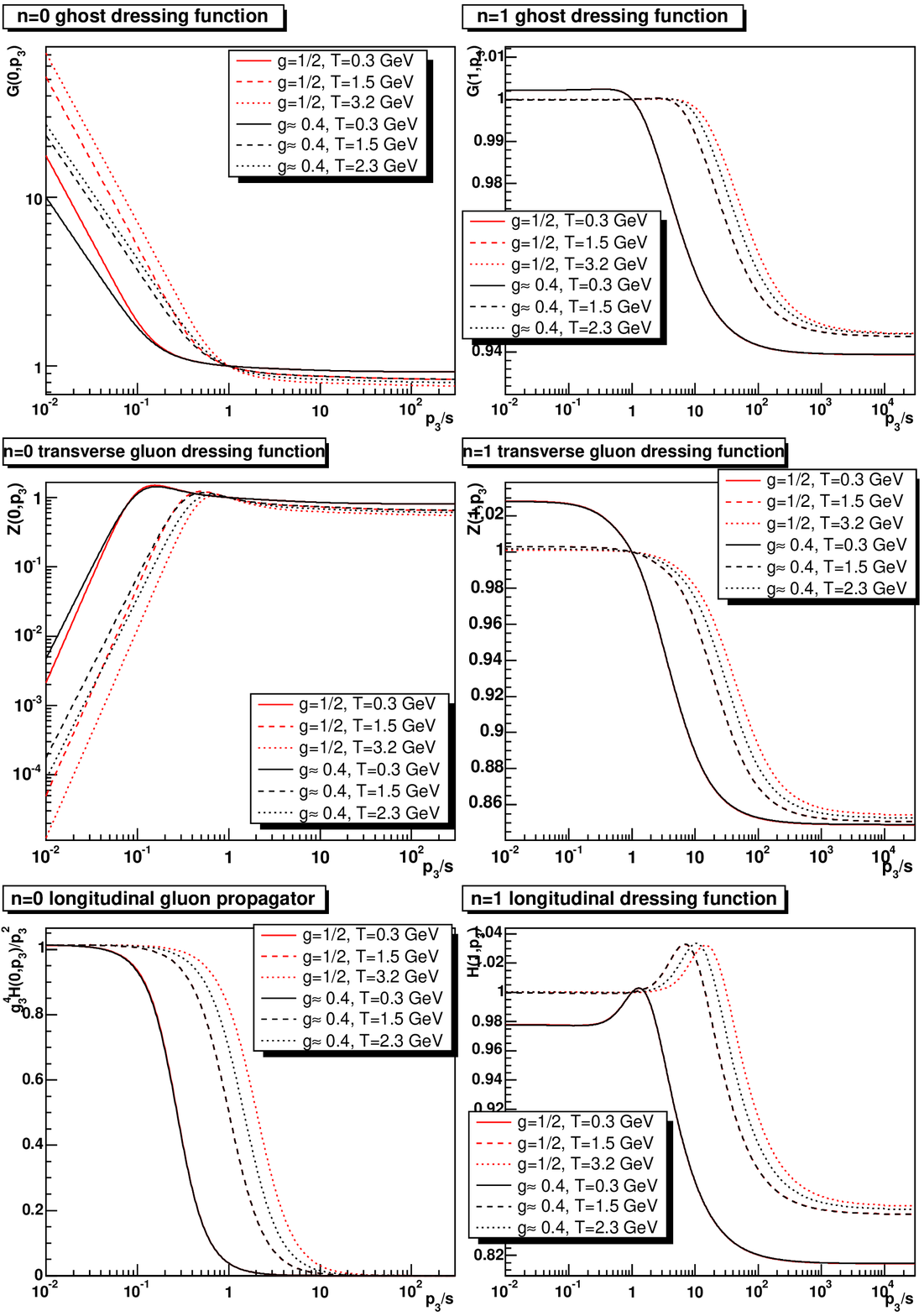,height=1.2\linewidth,width=\linewidth}
\caption{The dependence of the solutions on $T$ at fixed $\mu$ and $s$ at $N=5$. The left panel shows the soft modes, where apart from the longitudinal gluon the dressing functions are shown. In the latter case the propagator is shown. The right panels show the dressing functions for the $n=1$ hard mode. The \piclinecol and black lines are the $g=1/2$ and the $g\approx 0.4$ solution, respectively. Solid is $T=0.3$ GeV, dashed is $T=1.5$ GeV, and dotted is $T=3.2$ GeV for $g=1/2$ and $T=2.3$ GeV for $g\approx 0.4$.}
\label{figftft}
\end{figure}

Switching to a fixed renormalization scale and subtraction point, the solutions shown in figure \ref{figftft} are found. The effect is quite the same as in the ghost-loop-only truncation. Changing the temperature at fixed 4d-coupling, and thus increasing the 3d-coupling with temperature, shifts the onset of non-perturbative effects to larger momenta. Obeying the renormalization at fixed momentum then requires the changes in the dressing functions seen in the figure. In this case, it was also again possible to super-cool down to at least $T=152$ MeV.

\begin{figure}
\epsfig{file=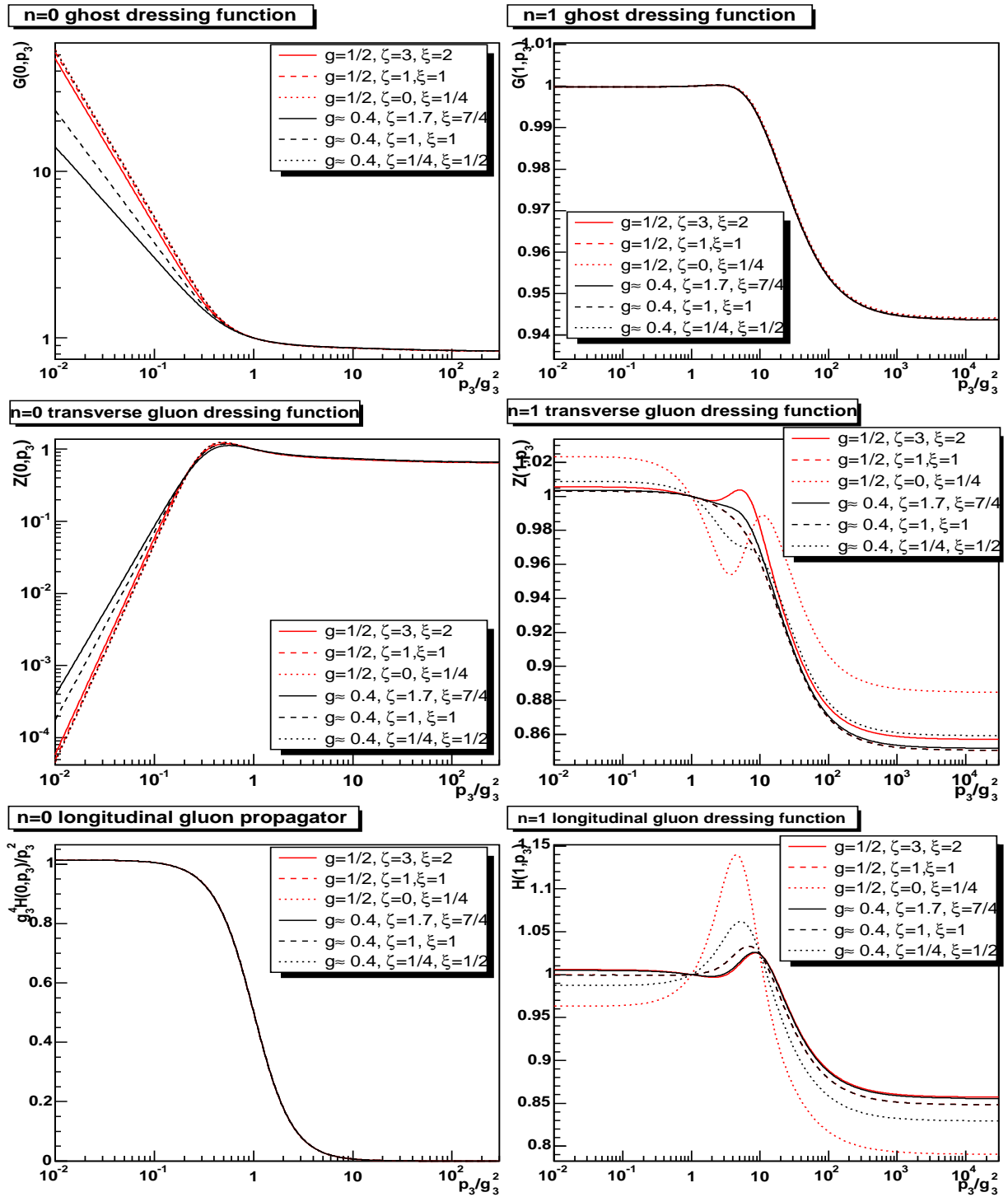,height=1.2\linewidth,width=\linewidth}
\caption{The dependence of the solutions on $\zeta$ and $\xi$ at $N=5$. The left panel shows the soft modes, where apart from the longitudinal gluon the dressing functions are shown. In the latter case the propagator is shown. The right panels show the dressing functions for the $n=1$ hard mode. The \piclinecol and black lines are the $g=1/2$ and the $g\approx 0.4$ solution, respectively. Solid is $(g,\zeta,\xi)=(1/2,3,2)$ and $(0.4,1.7,7/4)$, dashed is $(1/2,1,1)$ and $(0.4,1,1)$, dotted is $(1/2,0,1/4)$ and $(0.4,1/4,1/2)$.}
\label{figftzx}
\end{figure}

The dependence on $\zeta$ and $\xi$ is shown in figure \ref{figftzx}. Varying $\zeta$ leads to similar variations for the soft mode dressing functions as in the 3d-case. The largest effect is seen in the infrared. As the hard mode dressing functions are insensitive to the infrared behavior of the soft modes due to their effective mass, there are only weak variations of them with $\zeta$. Only at small $\zeta$ an additional structure appears in the transverse gluon dressing function. This is likely due to the cross-term in equation \pref{fulleqZft}. The effect of varying $\zeta$ on the longitudinal sector is negligible, as in the 3d-case. Correspondingly, the effect of varying $\xi$ is only significant for the longitudinal sector. The sensitivity is much less pronounced then in the case of varying $\zeta$. Especially the soft longitudinal dressing function is nearly unaffected by variation of $\xi$. The reason is that the soft equation does not depend explicitly on $\xi$, as it can be divided out of the equation for $\xi\neq 0$. Therefore, any $\xi$ dependence only enters indirectly by the weak dependence of the hard modes on $\xi$. 

The dependence on $\delta$ is qualitatively not different from the high temperature case for the soft modes. With decreasing temperature, the lowest attainable value of $\delta$ increases and is around $\delta\approx 0.11$ for the $g=1/2$ solution and $\delta\approx 0.15$ for the $g\approx 0.4$ solution at $T=1.5$ GeV. Surprisingly, the roles of the $g=1/2$ and $g\approx 0.4$ solutions are interchanged: At low temperatures, it is possible to reach lower values of $\delta$ in case of the half-integer solution. The hard mode dressing functions are nearly not affected by the change in $\delta$. Thus, there is no significant change concerning the dependence on $\delta$ at finite temperatures.

At this point, a comparison can be made to a different approach to obtain the high-temperature gauge propagators. The usual continuum method is the semi-perturbative hard thermal loop (HTL) approach \cite{Blaizot:2001nr}. It is based on resumming the hard mode contributions in self-energy diagrams, hence its name. In the transverse infrared sector it is plagued by severe problems, due to its perturbative nature. The final result is a transverse gluon propagator with a particle like pole at $p=0$, thus $t=0$. This is in sharp contrast to the results found here and, as discussed in chapter \ref{c3d}, such a behavior is not likely. Also the lattice results tend to support the results found here. Concerning the soft longitudinal mode and the hard modes, HTLs and the ansatz presented here find qualitatively similar results on the level of the propagators.

\section{Solutions at Small Temperature}\label{sdselowt}

\begin{figure}[ht]
\begin{center}\epsfig{file=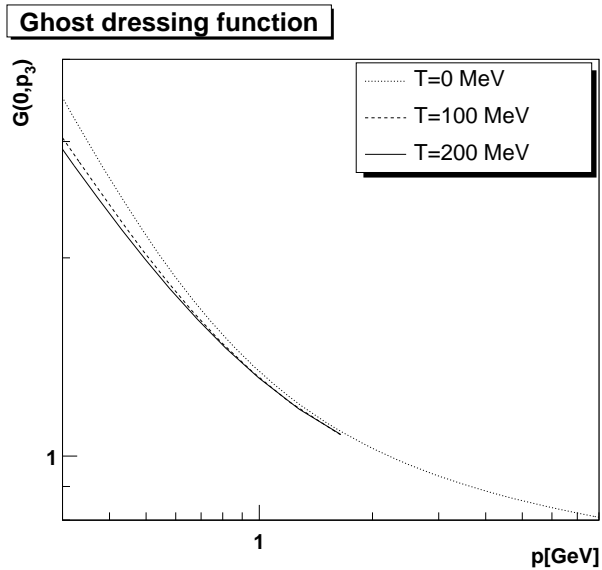,width=0.5\linewidth}\end{center}
\caption{The dressing function of the soft ghost mode at zero \cite{Fischer:2002hn} and finite temperature \cite{Gruter:2004kd}.}
\label{figglowt}
\end{figure}

As the temperature region of and above the phase transition is probed by experiments, the aim of the calculations presented here is to eventually describe the phase transition. This is not yet possible. Currently, the phase transition is approached from two sides. The first approach is from above, and is presented in this work. The second approach is from below and treated especially in \cite{Gruter:phd} and also to some extent in the context of this work \cite{Gruter:2004kd}. The main aim in these calculations is to investigate whether any qualitative changes are observed at small temperatures as compared to the vacuum solutions. Sufficient for this aim is to employ not a full continuum method but solving the DSEs approximately on a discretized space-time with a toroidal topology. This method had already been applied successfully to the vacuum \cite{Fischer:2002hn,Fischer:2003zc}.

\begin{figure}[t]
\epsfig{file=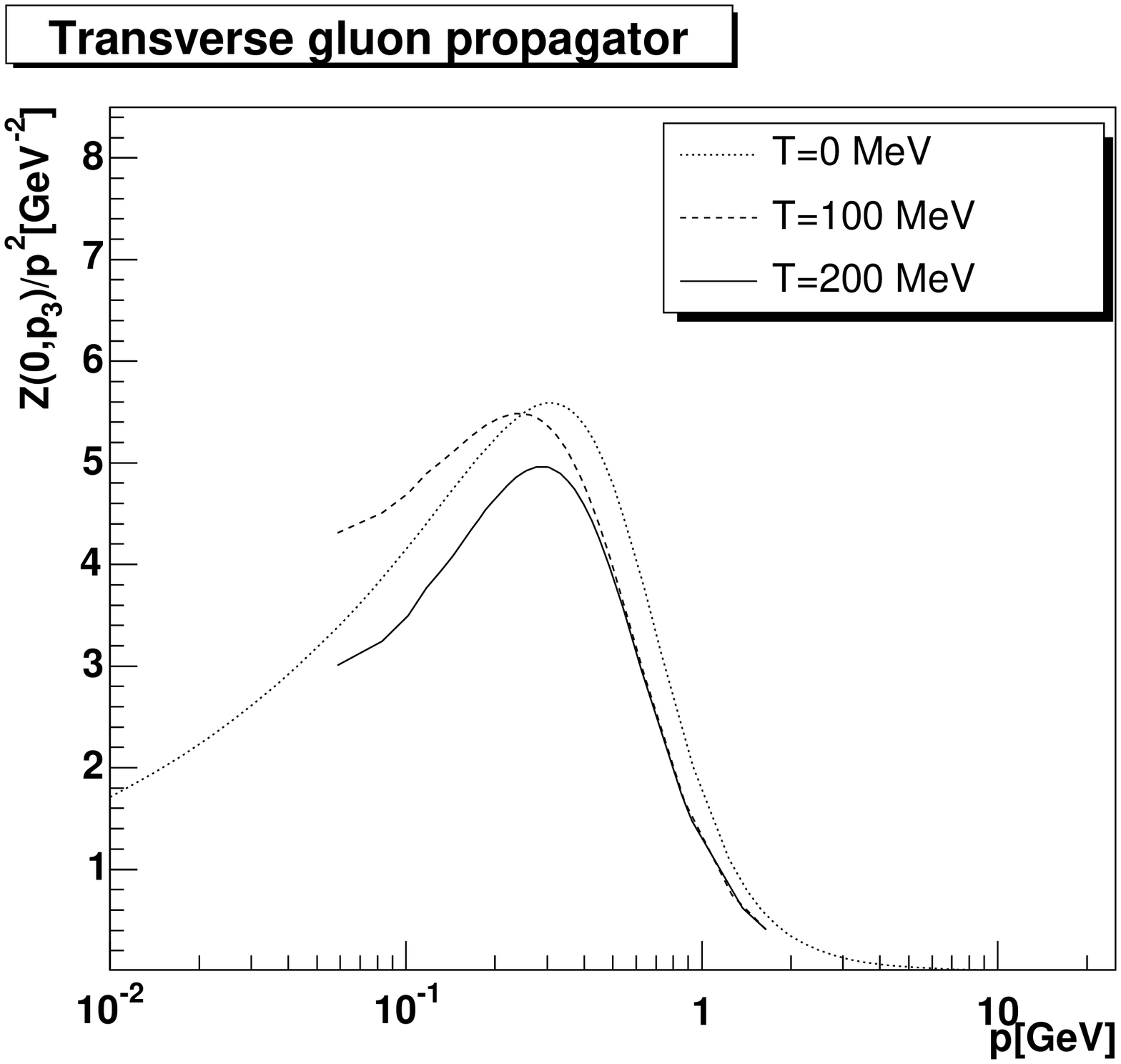,width=0.5\linewidth}\epsfig{file=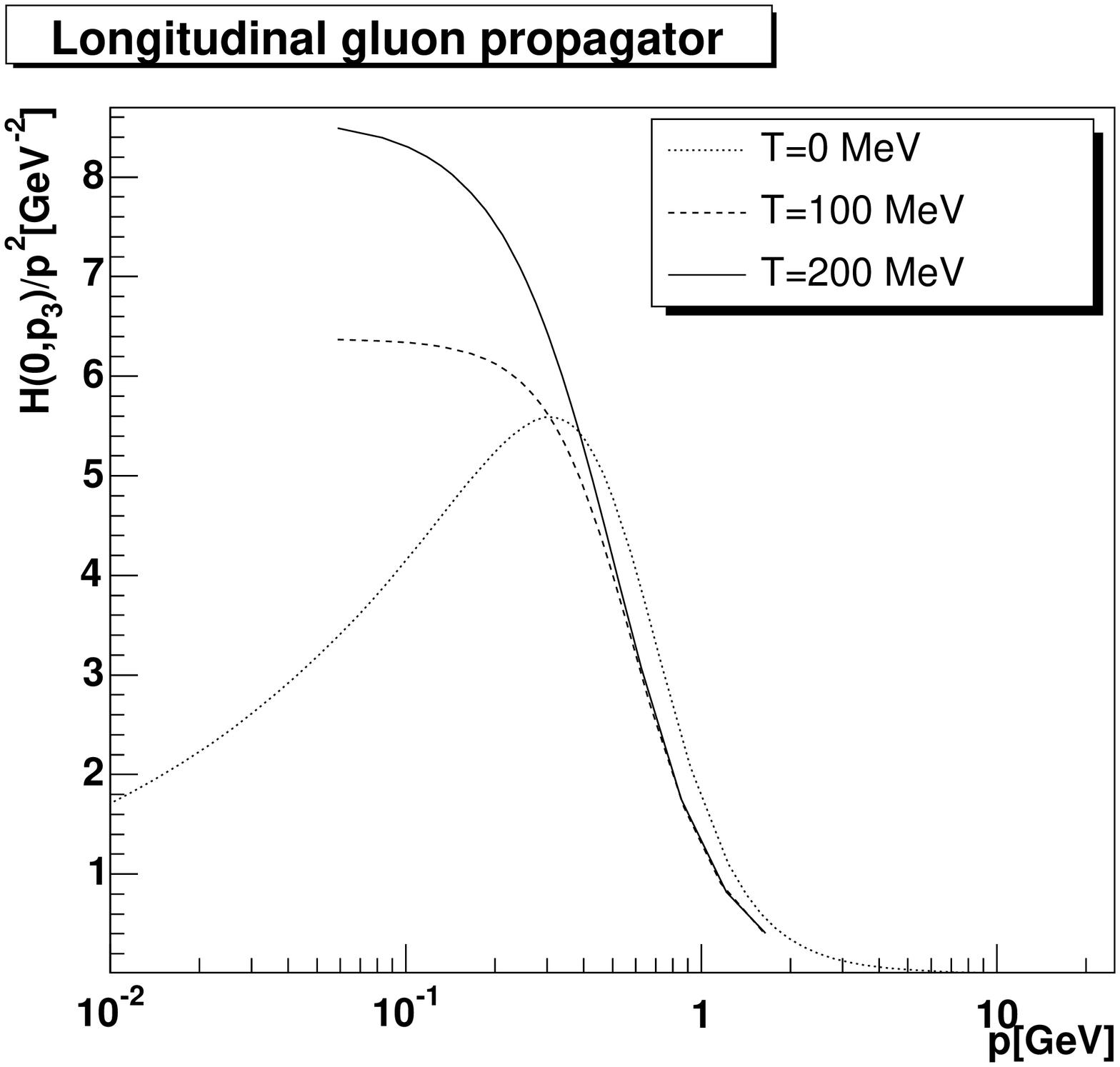,width=0.5\linewidth}
\caption{The propagator of the soft gluon mode at zero \cite{Fischer:2002hn} and finite temperature \cite{Gruter:2004kd}. The left panel shows the 3d-transverse part $Z$ of \pref{gluonprop} and the right panel the 3d-longitudinal part $H$.}
\label{figzhlowt}
\end{figure}

The results for the soft mode dressing functions are shown in figure \ref{figglowt} for the ghost propagator and in figure \ref{figzhlowt} for the two independent dressing functions of the gluon propagator \pref{gluonprop}, compared to the zero temperature solution. While the 3d-transverse part of the propagator becomes steeper in the infrared, the 3d-longitudinal part becomes more shallow. Thus, it experiences significant finite volume effects, as in the case of lattice calculations. Finite volume scaling shows that it is still confined, as in the case of the vacuum \cite{Gruter:2004kd}. Indeed it cannot be distinguished yet whether the exponents or merely the coefficients of the infrared solution change.

Interestingly, the 4d-transverse projection of \pref{gluonprop}, yields that the quantity $2Z+H$, is extremely temperature-independent \cite{Gruter:2004kd}. As this combination is the only one entering the thermodynamic potential to be discussed in section \ref{std}, and $G$ in itself is rather temperature independent, the corresponding thermodynamic potential is itself nearly independent of temperature and compatible with 0. This is what would be naively expected from a confined gluon system, as bound states are not included in the description here.

In addition, it is possible to trace the solution up to temperatures of 800 MeV \cite{Gruter:2004kd}, significantly above the expected phase transition temperature of $(269\pm 1)$ MeV according to lattice results \cite{Karsch:2003jg}. Therefore it is possible to super-heat the system.

If the temperature scale used in \cite{Gruter:2004kd} and here are compatible within a factor of 4, then within a substantial temperature range, two qualitatively different solutions exist. One solution has an over-screened and one a screened soft longitudinal gluon. Together with the observation of super-heating and super-cooling when comparing to lattice results, this implies a first order phase transition.

Compared to the high-temperature case presented before, the sector of ghosts and transverse gluons is not changed qualitatively. Quantitatively, especially the infrared exponents changed, though. The main difference is in the longitudinal sector, where a change from over-screening to screening occurred. Comparing the hard modes in both cases, no qualitative change is found in any sector. Therefore, the main difference of the two phases is only in the longitudinal/chromoelectric sector. Therefore the nature of the phase transition will likely be chromoelectric. This implication for the phase transition and the compatibility of these observations with lattice calculations will be discussed in section \ref{ssumprop}.


%% file: derived.tex
\chapter{Derived Quantities}\label{cderived}

The dressing functions obtained in chapters \ref{c3d} and \ref{cft} are only the first step from the basic Lagrangian towards observables. In this chapter, derivations towards such observables are performed. The screening masses and with them the analytic structure of the particles described will be investigated in section \ref{sschwinger}. A second natural object is the thermodynamic potential discussed in section \ref{std}. 

If not stated otherwise, throughout this chapter $C_A=N_c=3$, $\zeta=\xi=1$, $\delta=1/4$, and the lattice value for $m_h/g_3^2$ are used.

\section{Schwinger Functions and Analytic Structure}\label{sschwinger}

\subsection{Schwinger Functions}

Screening masses are most directly extracted from the analytic structure of the propagators. To obtain access to these analytic properties, the Schwinger function related to the dressing function $D$ is calculated. It is defined as \cite{Alkofer:2003jk}
\be
\Delta(t)=\frac{1}{\pi}\int_0^\infty dp_0\cos(tp_0)\frac{D(p_0)}{p_0^2},\label{schwinger}
\ee
\noindent i.e.\/ the Fourier transform of the propagator with respect to (Euclidean) time. Note that this definition is independent of the dimensionality of the underlying theory. Negative values for the Schwinger function can be traced to violations of positivity and therefore to absence of the particle represented by $D$ from the physical spectrum~\cite{Alkofer:2003jk}. 

The numerical implementation of \pref{schwinger} is non-trivial. To obtain sufficient precision, a FFT-algorithm using edge corrections and compensation for the infinite integration range of \pref{schwinger} together with at least 512 or more frequencies\footnote{Ten thousands to a million are much better.  The results presented here have been obtained using roughly $5\cdot10^5$ frequencies.} was employed \cite{Press:1997}. 

\begin{figure}
\epsfig{file=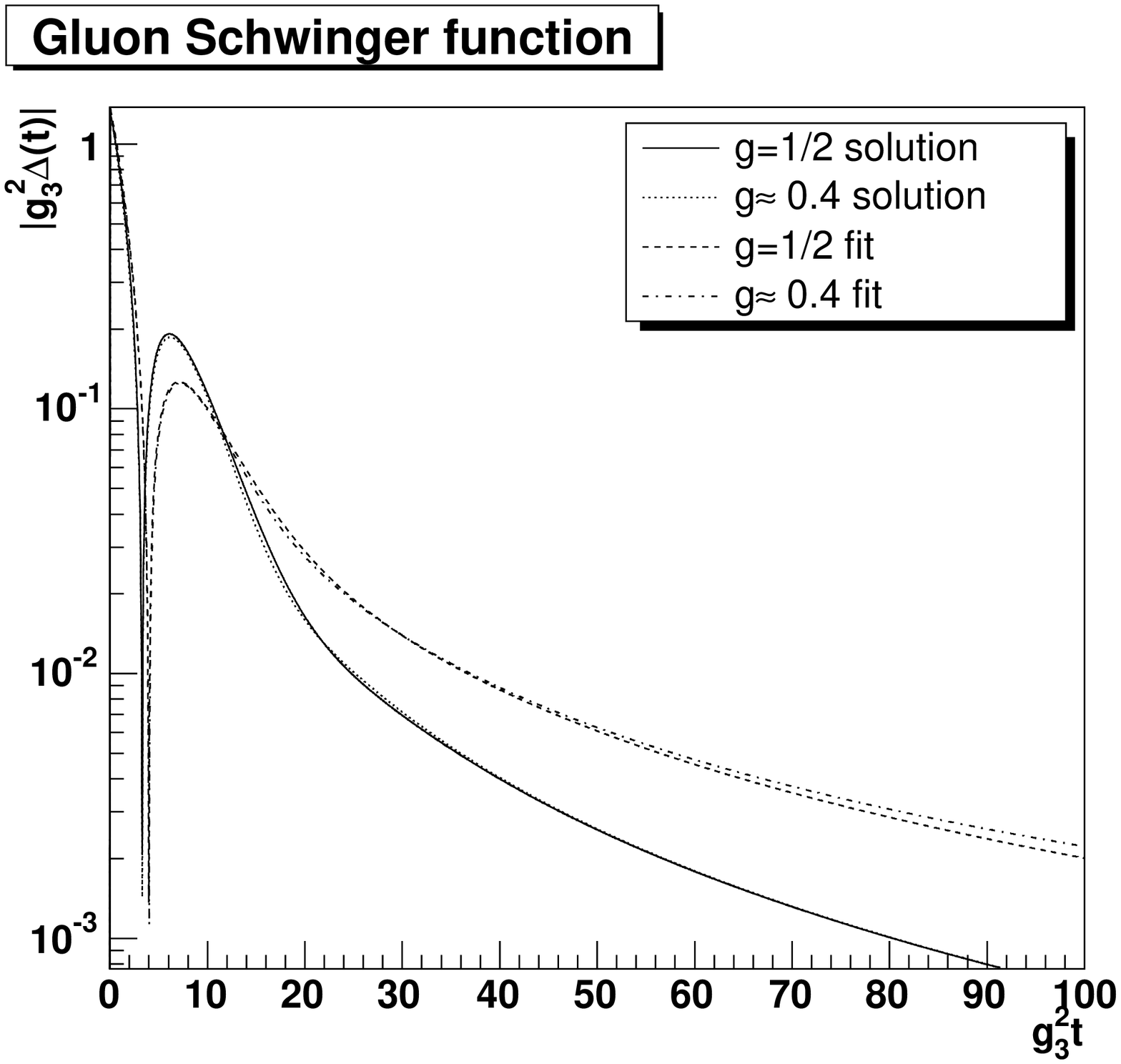,width=0.5\linewidth}\epsfig{file=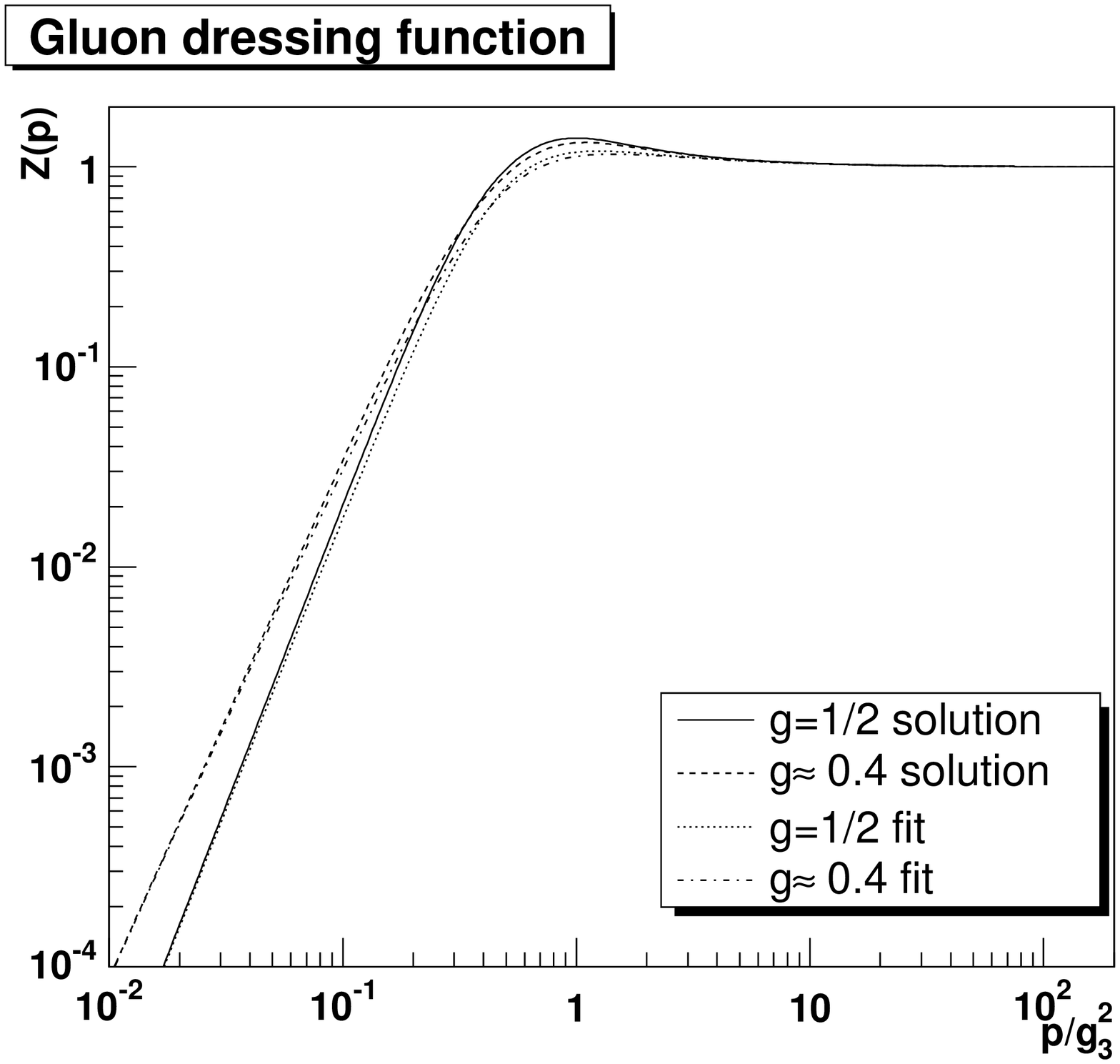,width=0.5\linewidth}
\caption{The left panel shows the Schwinger function of the gluon. The right panel shows the comparison of the fit-function \pref{gluonfit} to the full solutions
of the dressing function. The solid line gives the numerical result for $g=1/2$ while the dashed line is for $g\approx0.4$. The dotted and dash-dotted line denote their respective fits using the ansatz (\ref{gluonfit}).} 
\label{figzschwing}
\end{figure}

The time $t$ in \pref{schwinger} in the 3d theory of the infinite temperature limit is actually a space-direction in 4d. To such a direction also the arguments concerning positivity can be applied in Euclidean space-time. Therefore, the result for the gluon, presented in figure \ref{figzschwing}, clearly exhibits positivity violations. This is in accordance with the Oehme-Zimmermann super-convergence relation \pref{oehme} and thus has been expected. Furthermore, the position of the zeros can be interpreted as the confinement scale. For the $g=1/2$ solution, the zero occurs at $g_3^2t\approx 3.29$  and at $g_3^2t\approx 3.98$ for the other branch. This result is in agreement with recent lattice results~\cite{Cucchieri:2004mf}.

To be able to perform an analytic continuation of the gluon propagator into the complex $p^2$-plane, the Schwinger function is fitted.  This is performed by using the ansatz\footnote{In \cite{Alkofer:2003jk}, a parameterization with only a branch cut and no isolated pole provided a successful fit. Due to the different asymptotic behavior in 4 and 3 dimensions, a different ansatz is used here.}
\be
Z_f(p)=\frac{A_z p^{4g+1}}{1+f+A_z p^{4g+1}}\left(1+\frac{f}{1+fa_up}\right)=\frac{A_z p^{4g+1}(1+f+a_ufp)}{(1+a_ufp)(1+f+A_zp^{4g+1})}\label{gluonfit}
\ee
\noindent for the dressing function. $A_z$ is the infrared coefficient determined previously, and $a_u$ is the ultraviolet coefficient of leading-order resummed perturbation theory, as calculated in appendix \ref{appUV}. The fit parameters for both solutions are given in table \ref{tabglfit}. As demonstrated in figure \ref{figzschwing}, the Schwinger function is fitted very well. The gluon propagator and dressing function are also reasonably well described by the fit.

\begin{table}
\begin{center}
\begin{tabular}{|c|c|c|c|}
\hline
Solution & $A_zg_3^{-2t}$ & $a_ug_3^2$ & $f$ \cr
\hline
$g=1/2$ & 20.3 & $64/27$ & 1.32511 \cr
\hline
$g\approx 0.4$ & 13.4 & $64/27$ & 1.03148 \cr
\hline
\end{tabular}
\end{center}
\caption{The coefficients for the gluon fit (\ref{gluonfit}).}
\label{tabglfit}
\end{table}

\begin{figure}
\epsfig{file=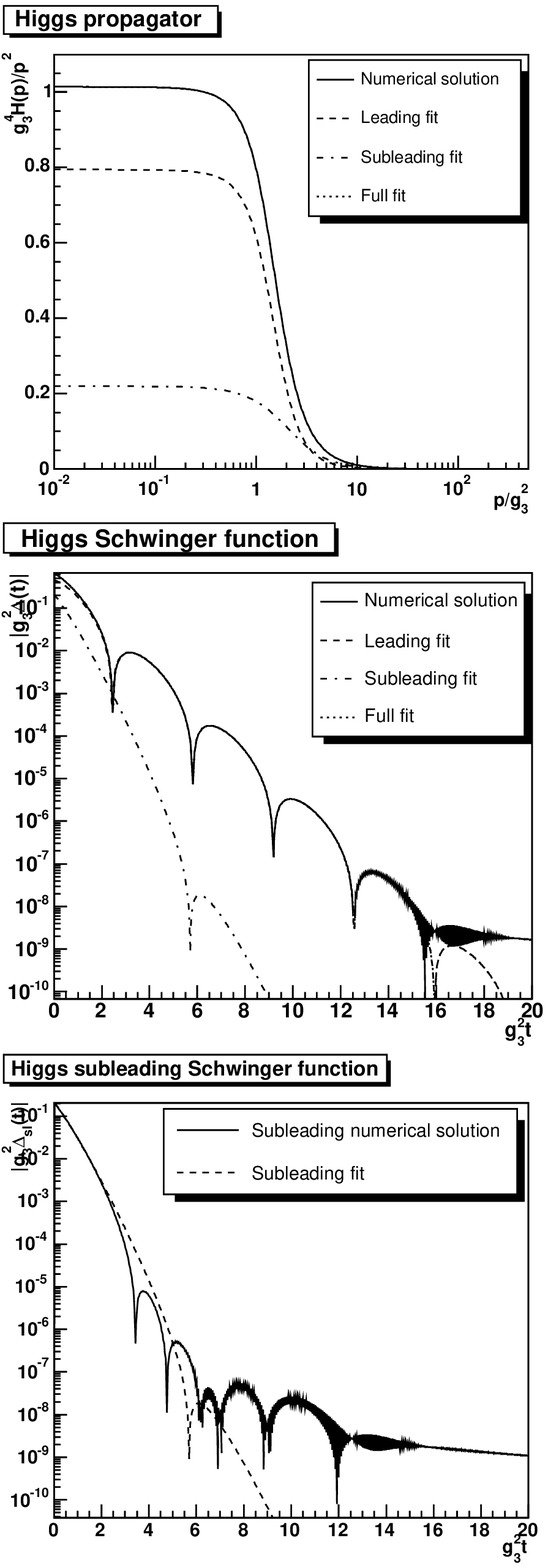,height=1.2\linewidth,width=0.5\linewidth}\epsfig{file=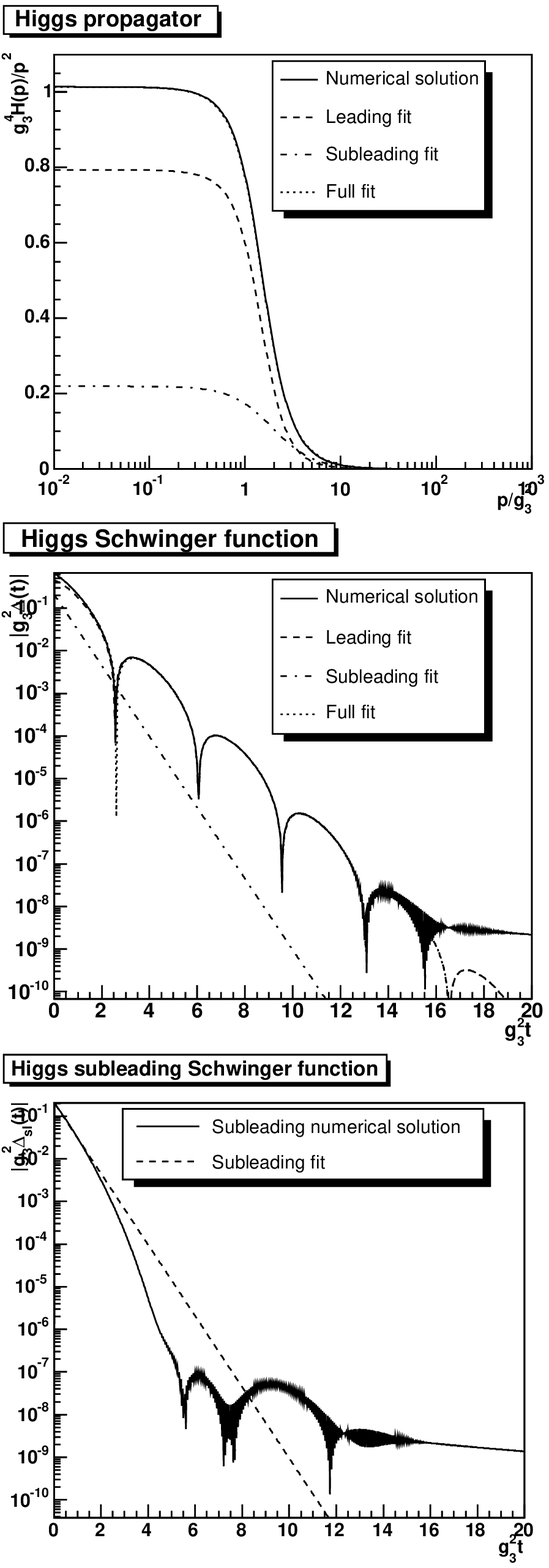,height=1.2\linewidth,width=0.5\linewidth}
\caption{The Higgs propagator is shown in the top panels and its Schwinger function in the middle panels compared to their fits. The bottom panels show the comparison of the numerical (solid) and the fitted (dashed) subleading Higgs contribution, see text. The left and right side shows the $g=1/2$ and $g\approx 0.4$ solutions, respectively. The solid lines represent the numerical solution, the dashed lines give the leading contribution and the dashed-dotted lines the first subleading contribution. The dotted lines underneath the solid lines give the sum of the leading and subleading contribution. }
\label{fighstdfit}
\end{figure}

\begin{figure}
\epsfig{file=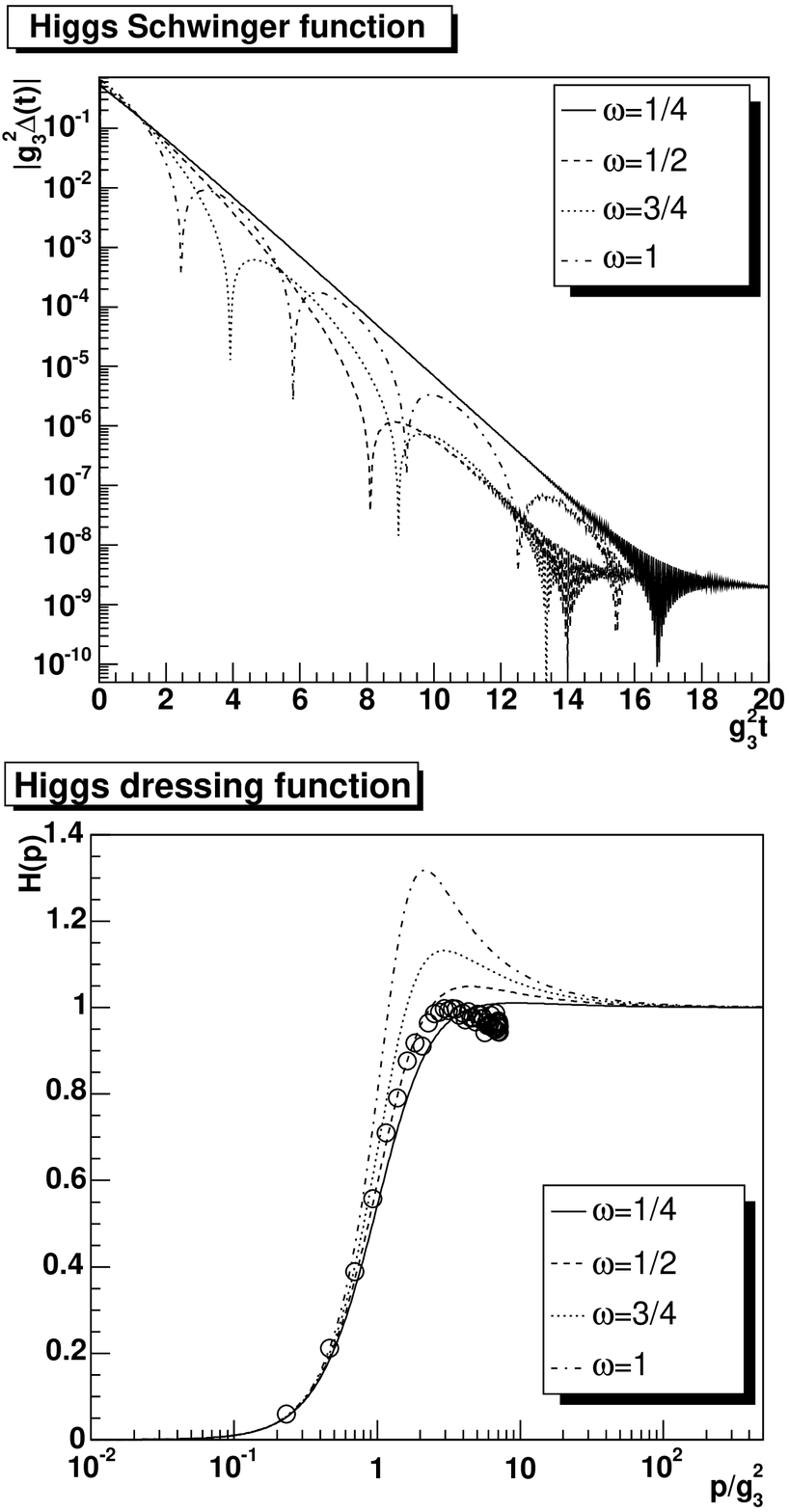,width=0.5\linewidth}\epsfig{file=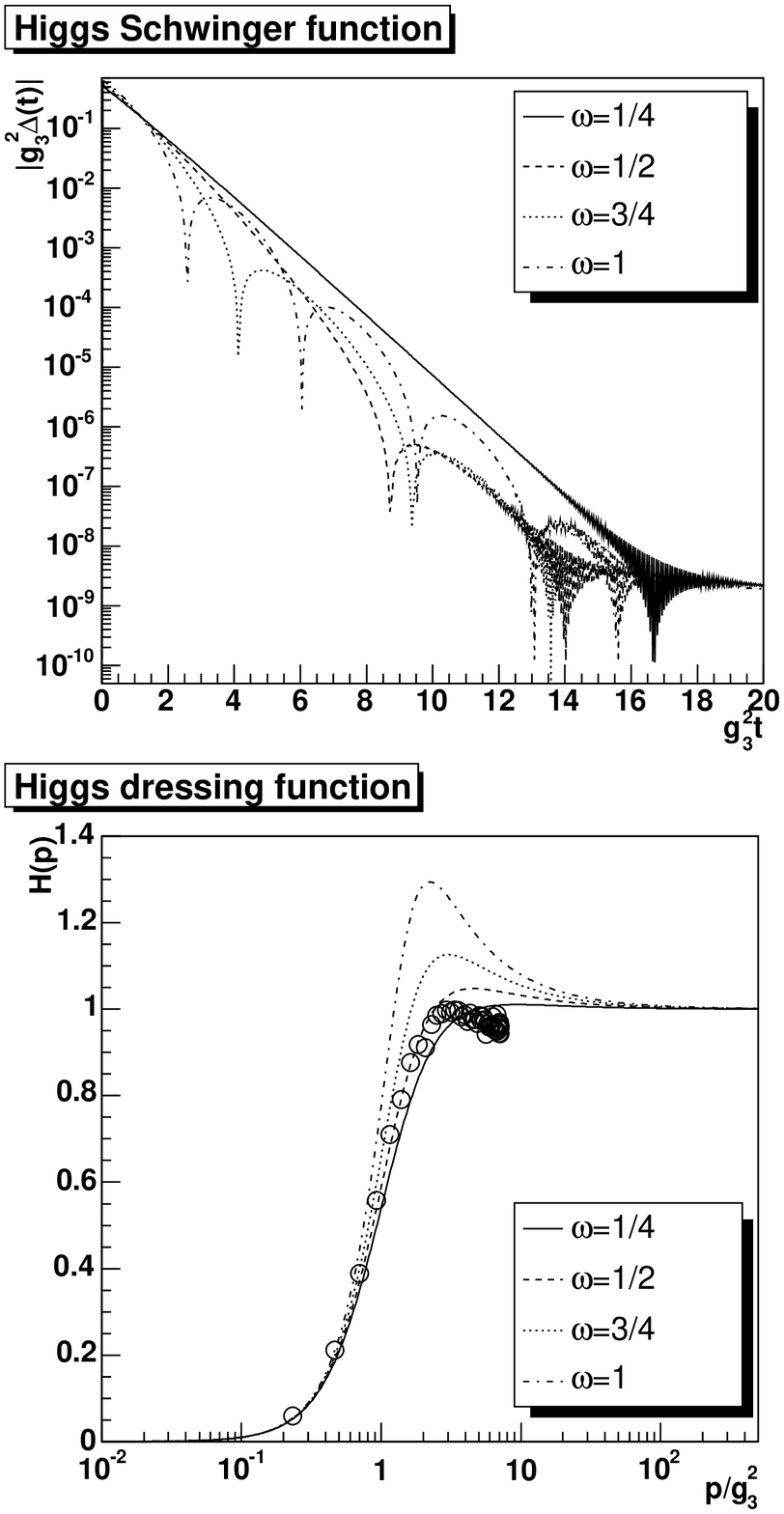,width=0.5\linewidth}
\caption{The bottom panel shows the Higgs propagator for various suppression factors $\omega$ of the Higgs-gluon vertex. The top panel shows the corresponding Schwinger functions. The left panels contain the $g=1/2$ solution while the right ones display the $g\approx 0.4$ solution. The lattice data are the same as on the right side of figure \ref{figlat}. The solid line is for $\omega=1/4$, the dashed line for $\omega=1/2$, the dotted line is for $\omega=3/4$ and the dashed-dotted line for $\omega=1$.}
\label{fighsalt}
\end{figure}

Similar to a meromorphic fit ansatz used in \cite{Alkofer:2003jk}, the Higgs propagator and its Schwinger function, the latter being analytically calculated from the former \cite{Gradstein:1981}, are described as
\bea
H_f(p)&=&\frac{e+f p^2}{p^4+2m^2\cos\left(2\phi\right)p^2+m^4},\label{higgsfit}\\
\Delta_f(t)&=&\frac{e}{2m^3\sin\left(2\phi\right)}e^{-tm\cos\left(\phi\right)}\left(\sin\left(\phi+tm\sin\left(\phi\right)\right)+\frac{fm^2}{e}\sin\left(\phi-tm\sin\left(\phi\right)\right)\right).
\nonumber
\eea
\noindent As demonstrated in figure \ref{fighstdfit}, these fits already describe the Schwinger function quite well, but miss around 20\% of the propagator at zero momentum. This indicates that further massive modes are present. Indeed, for the $g=1/2$ solution, a further term of the form (\ref{higgsfit}) has to be added. For the $g\approx 0.4$ solution, adding a term with one pole, 
\bea
H_f(p)=\frac{e}{p^2+m^2},\label{higgsaltfit}\\
\Delta_f(t)=\frac{e}{2m}e^{-mt},\label{higgssaltfit}
\eea
improves the fit also in this case. Both subleading fits are not very accurate for large $t$, and the results have to be taken with care. They indicate, nevertheless, the existence of subleading contributions due to the presence of further massive-particle-like contributions. The fit parameters of both solutions can be found in table \ref{tabhfit}.

\begin{table}
\begin{center}
\begin{tabular}{|c|c|c|c|c|}
\hline
Solution & $e$ & $fg_3^4$ & $\phi$ & $m/g_3^2$ \cr
\hline
$g=1/2$ & 4.0199 & 0.3545 & -0.67078 & 1.4998 \cr
\hline
subleading & 9.4493 & 0.6736 & -0.18387 & 2.561 \cr
\hline
$g\approx 0.4$ & 4.0697 & 0.37793 & -0.63975 & 1.5045 \cr
\hline
subleading & 0.81188 & & & 1.9223 \cr
\hline
\end{tabular}
\end{center}
\caption{The coefficients for the Higgs fit (\ref{higgsfit}) and the subleading one (\ref{higgsfit}) and (\ref{higgsaltfit}), respectively.} \label{tabhfit}
\end{table}

As can be seen from figure \ref{figlat}, the result for the Higgs propagator deviates significantly from the lattice results, the self-energy being significantly overestimated. While the gluon Schwinger function is reasonably independent of the truncation, this turns out not to be the case for the Higgs Schwinger function. In order to show this, the solutions for a bare Higgs-gluon vertex suppressed via a scaling factor $\omega$ are obtained\footnote{In the Landau gauge, due to the transversality of the gluon, this is equivalent to modifying the tensor structure of the vertex.}. The first result is that the gluon Schwinger function, in contrast to the Higgs one, is not susceptible to such a change. 

Figure \ref{fighsalt} displays corresponding results for different values of $\omega$. The position of the first zero tends to increase for decreasing $\omega$. At $\omega\approx 1/4$  any oscillation, at least within the available numerical precision, seems to be gone altogether. A fit to the Schwinger function using \pref{higgssaltfit} reveals again additional structure at very small $t$ which cannot be captured by such a simple fit. As the fit also misses some strength at zero momentum for the propagator, this again indicates the presence of further massive contributions in the propagator.

\begin{figure}
\epsfig{file=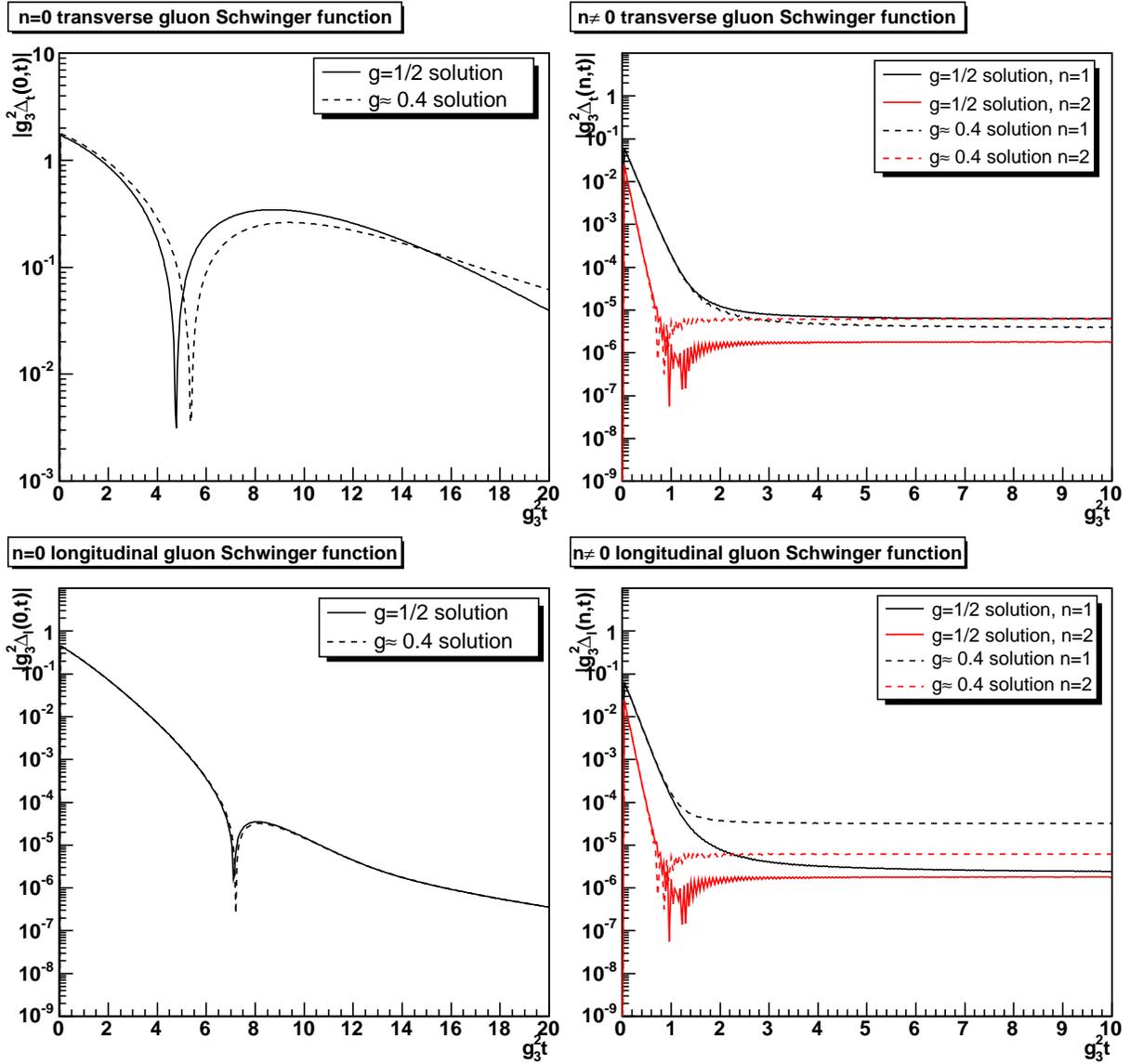,width=\linewidth}
\caption{The Schwinger functions at $T=1.5$ GeV and $N=5$. The left top panel shows the soft transverse gluon Schwinger function, the right top panel the hard transverse gluon Schwinger function for $n=1$ with a black line and for $n=2$ with a \piclinecol line. The bottom panels show the same for the longitudinal gluon Schwinger function. In all cases are solid lines the $g=1/2$ solution and dashed lines the $g\approx 0.4$ solution.}
\label{figftschwinger}
\end{figure}

When adding the hard modes, the qualitative picture does not change, as is visible in figure \ref{figftschwinger}. In this case the Schwinger function also depends on a `space'-variable with respect to 4d and $p_0=2\pi Tn$ acts as an effective mass. The gluon soft mode shows an oscillation as expected. The same is still true for the Higgs, thus this property of the truncation used here is still present. At this rather low temperature, it exhibits only one oscillation, and is thus qualitatively much more similar to the gluon. In both cases, the oscillation frequency has decreased due to the additional interactions with the hard modes. The interpretation of the oscillations in the longitudinal gluon Schwinger function are still inconclusive, though.

For the hard mode Schwinger functions no oscillations are seen. They cannot be excluded as they might probably not be resolvable numerically if the oscillation frequency is of the order of the mass. Up to the numerical accuracy, the hard modes present the behavior expected for a massive particle. This equally well applies to the higher modes, as is visible by the example of the next mode $n=2$.

Thus, the features found in the infinite temperature limit persist at finite temperatures. Note, however, that if the hard modes oscillate, this oscillation does not necessarily vanish when the $T\to\infty$ limit is taken. If the oscillation frequency only vanishes as fast as the interaction, then even for arbitrarily small interactions, oscillations remain. It will take more sophistication to settle the question of whether the hard modes show positivity violation or not.

\subsection{Analytic Properties in the Infinite-Temperature Limit}\label{analytic}

Although the high-temperature limit of the four-dimensional Minkowski theory is a genuinely Euclidian theory, it will be of interest for other applications to extract the analytic structure of the propagators investigated.

The gluon propagator exhibits similar behavior for both solutions, but there are also some significant differences. The denominator of the ansatz \pref{gluonfit} contains two factors, both of which could possibly give rise to a non-trivial analytic structure. The first part stems from the fit of the perturbative tail necessary to generate the maximum in the gluon dressing function. Since all fit parameters are positive, this factor does not give rise to a pole on the first Riemann sheet. However, it generates a pole on the second Riemann sheet, which will occur at $1/a_uf$, that is, at Euclidian momenta. This pole does not have a physical interpretation, and may well be an artifact of the fit, since \pref{gluonfit} is tailored to generate the correct leading-order perturbative behavior. Thus, it generates most likely a structure which has the Landau pole of perturbation theory on the second Riemann sheet. It is to be expected that this pole vanishes when using a more sophisticated fit.

The second factor generates a genuine isolated pole at $(-(1+f)/A_z)^{-1/t}$. This expression has only one value on the first Riemann sheet given by $(-0.0933+0.1615i)g_3^4$ for the $g=1/2$ and $(-0.1462+0.1248i)g_3^4$ for the other solution\footnote{This is of the order of $\Lambda_{QCD}$ when using the  't Hooft-like scaling of section \ref{scc}.}. In both cases, a pole close to the origin is generated, with an imaginary part larger than the real part in one case. For the first solution, the pole is found for an angle significantly above $\pi/4$ while in the second case somewhat below. In addition, the first solution generates two more poles on two more Riemann sheets, while the second solution, with a (most likely) irrational exponent, generates an infinite number of further poles on an infinite number of Riemann sheets. In both cases, the residue is complex. In addition, there is a cut along the complete negative real axis starting at zero. In this way it is similar to the results in four dimensions~\cite{Alkofer:2003jk}.

This analysis infers that the gluon propagator is violating positivity, and thus satisfies the requirements of the Kugo-Ojima and Zwanziger-Gribov confinement scenario.

\begin{table}
\begin{center}
\begin{tabular}{|c|c|c|}
\hline
Solution & Order & Pole$/g_3^4$ \cr
\hline
$g=1/2$ & Leading & $-0.5112\pm2.191i$ \cr
\hline
$g=1/2$ & Subleading & $-6.11972\pm2.35777i$ \cr
\hline
$g\approx 0.4$ & Leading & $-0.650078\pm2.16822i$ \cr
\hline
$g\approx 0.4$ & Subleading & $-3.69539$ \cr
\hline
\end{tabular}
\end{center}
\caption{Location of the poles of the Higgs propagator for $\omega=1$.}
\label{tabhiggspoles}
\end{table}

On the other hand, the Higgs propagator very likely does not have a branch cut but a number of simple poles whose locations are given in table \ref{tabhiggspoles}. The sensitivity to the Higgs-gluon vertex, however, necessitates further investigations before a firmer conclusion can be drawn.

\begin{figure}[ht]
\begin{center}\epsfig{file=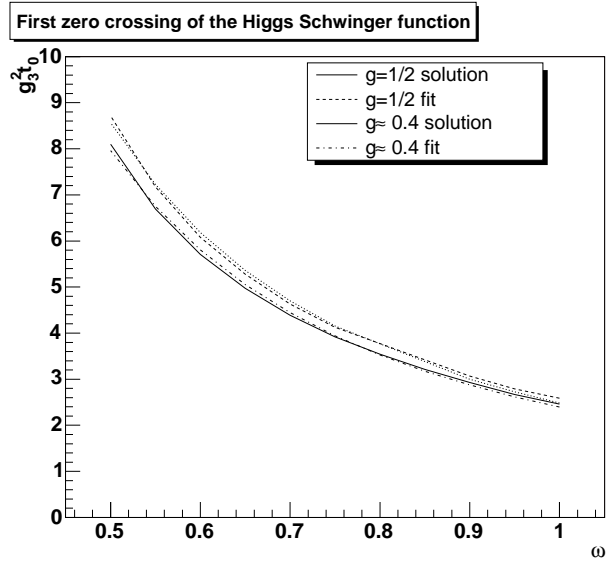,width=0.5\linewidth}\end{center}
\caption{The dependence of the first zero of the Higgs Schwinger function on the vertex suppression $\omega$. Solid is the $g=1/2$ solution and dotted its fit, dashed is the $g\approx 0.4$ solution and dashed-dotted its fit.}
\label{figcomega}
\end{figure}

The dependence of the first zero of the Higgs-Schwinger function on $\omega$ is shown in figure \ref{figcomega}. The result can be fitted using
\be
t_0=\frac{a}{\omega^b}\nonumber
\ee
\noindent where $t_0$ is the position of the first zero. The parameters are $(a,b)=(2.49,1.78)$ for the $g=1/2$ solution and $(2.40,1.73)$ for the $g\approx 0.4$ solution. Such a behavior would indicate a vanishing imaginary part and thus oscillations only when the Higgs ceases to interact. Thus, the oscillations would remain, albeit the poles move closer to the real axis. This might only be an indication of the possibility of the Higgs to decay, but the question remains into what, as gluons are confined. The existence of one or more complex conjugate poles on the same Riemann sheet can be interpreted in the framework of the Gribov-Stingl scenario \cite{Habel:1989aq} as a signal of confinement for the Higgs as well. However, as the Higgs is sensitive to the truncation, this is not conclusive, until more studies have been performed. 

\section{Thermodynamic Potential}\label{std}

\subsection{Soft Mode Contribution in the Infinite-Temperature Limit}

Knowledge of all Green's functions as functions of the temperature would permit to calculate the thermodynamic potential and therefore all thermodynamic quantities. Since not all of these are known, it is not possible to calculate an exact thermodynamic potential. Knowledge of the propagators is however sufficient to compute an approximation to the thermodynamic potential, using the Luttinger-Ward or Cornwall-Jackiw-Tomboulis (LW/CJT) effective action \cite{Luttinger:1960ua,Cornwall:1974vz}. 

This effective action can be extended to abelian and non-abelian gauge theories, see e.g.\/ \cite{Haeri:hi}. Its calculation is prohibitively complicated when using constructed vertices instead of bare or exact vertices. Therefore, only some qualitative features using the simplest approximation to the thermodynamic potential will be extracted here. It only depends on the propagators, and reads
\bea
\Omega&=&\frac 1 2 d(G) T\sum\int\frac{d^3q}{(2\pi)^3}\Big(-\ln\left(\frac{Z(q)}{Z_0(q)}\right)+\left(\frac{Z(q)}{Z_0(q)}-1\right)\nonumber\\
&&+\ln\left(\frac{G(q)}{G_0(q)}\right)-\left(\frac{G(q)}{G_0(q)}-1\right)-\frac{1}{2}\ln\left(\frac{H(q)}{H_0(q)}\right)+\frac{1}{2}\left(\frac{H(q)}{H_0(q)}-1\right)\Big)\label{cjt},
\eea
where $d(G)=\delta^{aa}$ is the dimension of the gauge group. $Z_0$, $G_0$, and $H_0$ are the tree-level dressing functions
\bea
Z_0(p)=G_0(p)=1\nonumber\\
H_0(p)=\frac{p^2}{p^2+m_h^2}.\nonumber
\eea
Working in the infinite-temperature limit, the thermodynamic potential divided by $T^4$ becomes interesting. This motivates a rescaling of the integration momenta to obtain
\be
\frac{\Omega}{T^4}=\frac{g_3^6}{T^3}a\label{eps}
\ee
where the dimensionless constant $a$ depends only on the dimensionless ratio $m_h/g_3^2$ and thus becomes independent of temperature in the limit $T\to\infty$. Herein, the explicit value of $g_3$ enters. Using the t'Hooft-like scaling presented in section \ref{scc}, $g_3^2\sim\Lambda_{QCD}$ results. Hence (\ref{eps}) scales as $1/T^3$ and does not contribute to the thermodynamic pressure significantly compared to the hard modes. The expected Stefan-Boltzmann behavior must then be obtained from the hard modes alone, requiring 
\be
\frac{\Omega}{T^4}=g^6_4(\mu)\left(a+\frac{a_h}{g_4^6(\mu)}\right) = \frac{g_3^6}{T^3}a+a_h\nonumber
\ee
where $a_h$ stems from the hard modes. However, the soft modes may still contribute significantly to thermodynamic properties, especially the trace anomaly $\theta=\epsilon-3p$, near the phase transition \cite{Zwanziger:2004np}.

The calculation of \pref{cjt} turns out to be plagued by spurious divergences, since the DSEs with the modified 3-gluon-vertex are no longer exact stationary
solutions of \pref{cjt}. By close examination of \pref{cjt}, and the fact that all dressing functions $D$ can be expanded in the infinite temperature limit for $p\gg g_3^2$ as
\be
D(p)=1+\sum_{n=1}^{\infty}a_n\left(\frac{g_3^2}{p}\right)^2,\nonumber
\ee
\noindent the divergence structure can be isolated. The tree-level value drops out in each term independently. Expanding the logarithms also in $g_3^2/q$ directly shows that the LO cancels between the linear and the logarithmic terms as well. This is no longer the case in next-to-leading order (NLO), and this will only be possible by requiring a correct solution of the Dyson-Schwinger equations to NLO in perturbation theory and using the appropriate 2-particle irreducible expression \cite{Cornwall:1974vz} for the thermodynamic potential. As the NLO term is proportional to $(g_3^2/q)^2$, it generates a linear divergence in $\Omega$ together with the measure. Although there is no a priori reason for this NLO contribution to be the same for both solutions, as this term is truncation dependent, still this is the case\footnote{Since the two solutions only differ in their respective infrared behavior, this may have only a minor influence on the perturbative tail.}. The NNLO will generate a logarithmic divergence and only from NNNLO all terms will be UV-finite. The general structure of the result is therefore
\be
\frac{\Omega}{g_3^6T}=a+b\log\frac{\Lambda}{c}+d\Lambda\label{fullfit}
\ee
\noindent where $\Lambda$ is the ultraviolet cutoff when calculating \pref{cjt}.

While the linear term is easily identifiable, this is not the case for the logarithmic one, especially as it is always possible to move contributions from $c$ to $a$ and back. This makes it in some sense arbitrary to calculate $a$. Assuming that the divergent contributions will be small at sufficiently small cutoff, but $a$ has already reached its final value, a first fit is performed using
\be
\frac{\Omega}{g_3^6T}=a+d\Lambda.\nonumber
\ee
\noindent In a second step, with $a$ held fixed, the full fit using \pref{fullfit} is performed. This gives the final values. For this approach to be valid, the logarithmic term must be much smaller than $a$. At least for calculating the differences of both solutions, this is the case. Also $c$ turns out to be always the same value for all fits, giving confidence in the fit procedure.

In the case of the ghost-loop-only truncation, the difference is independent of the cutoff and is $-0.001862(0)$. As all differences were performed by subtracting the $g=1/2$ solution from the second solution, this result would favor the $g=1/2$ solution.

In all other cases the prescribed subtraction was necessary. Note that, although $d$ should be 0 when calculating the differences, this is only true up to numerical errors. Hence also for differences it was necessary to extract $d$ and perform the corresponding fits\footnote{The required accuracy for these calculations is very high. An expansion in up to 500 Chebychef polynomials, see appendix \ref{anum}, was necessary.}. 48 fit points have been included from the $\Lambda$ interval [54763.70,4830021]. Within these values, the integrand and the integral did not show significant numerical fluctuations. The statistical error on $a$ is obtained by calculating the mean square error of the full fit. Besides the statistical error, systematic errors were investigated using variations of the 3-gluon vertex and the gluon-Higgs vertex by varying $\delta$ and $\omega$. The results are listed in table \ref{tres}. Note that the calculated difference and the fitted difference are always the same within statistical errors.

\begin{table}
\begin{center}
\begin{tabular}{|c|c|c|c|c|c|c|}
\hline
Truncation & Object & Sol. & $a$ & $b$ & $c/g_3^2$ & $dg_3^2$ \cr
\hline
Ghost-loop & $\Delta\Omega$ & & -0.001862(0) & & & \cr
\hline
Ghost-loop & $\Omega$ & 1 & -0.2095(186) & -0.258 & $189\cdot 10^3$ & -0.0025 \cr
\hline
Ghost-loop & $\Omega$ & 2 & -0.2076(607) & -0.258 & $189\cdot 10^3$ & -0.00250 \cr
\hline
Yang-Mills & $\Delta\Omega$ & & 0.03942(0) & 0.00265 & $189\cdot 10^3$ & $1.25\cdot 10^{-11}$ \cr
\hline
Yang-Mills & $\Omega$ & 1 & 1.735(135) & 1.92 & $189\cdot 10^3$ & 0.0175 \cr
\hline
Yang-Mills & $\Omega$ & 2 & 1.696(135) & 1.92 & $188\cdot 10^3$ & 0.0175 \cr
\hline
Full & $\Delta\Omega$ & & 0.05276(0) & 0.00280 & $189\cdot 10^3$ & $5.54\cdot 10^{-12}$ \cr
\hline
Full & $\Omega$ & 1 & 3.586(254) & 3.67 & $189\cdot 10^3$ & 0.0322 \cr
\hline
Full & $\Omega$ & 2 & 3.533(254) & 3.67 & $189\cdot 10^3$ & 0.322 \cr
\hline
YM contribution & $\Delta\Omega$ & & 0.02083(0) & 0.00162 & $189\cdot 10^3$ & $9.92\cdot 10^{-12}$ \cr
\hline
YM contribution & $\Omega$ & 1 & 0.8215(983) & 1.33 & $189\cdot 10^3$ & 0.0108 \cr
\hline
YM contribution & $\Omega$ & 2 & 0.8007(983) & 1.33 & $189\cdot 10^3$ & 0.0108 \cr
\hline
$\delta=1$ & $\Delta\Omega$ & & 0.08433(0) & 0.00910 & $189\cdot 10^3$ & $4.90\cdot 10^{-11}$ \cr
\hline
$\delta=1$ & $\Omega$ & 1 & 2.442(255) & 3.55 & $189\cdot 10^3$ & 0.0322 \cr
\hline
$\delta=1$ & $\Omega$ & 2 & 2.338(255) & 3.54 & $189\cdot 10^3$ & 0.0322 \cr
\hline
$\omega=1/4$ & $\Delta\Omega$ & & 0.03820(0) & 0.00258 & $189\cdot 10^3$ & $9.37\cdot 10^{-12}$ \cr
\hline
$\omega=1/4$ & $\Omega$ & 1 & 1.708(136) & 1.93 & $189\cdot 10^3$ & 0.0171 \cr
\hline
$\omega=1/4$ & $\Omega$ & 2 & 1.669(136) & 1.93 & $189\cdot 10^3$ & 0.0171 \cr
\hline
$\delta=1,\omega=1/4$ & $\Delta\Omega$ & & 0.08282(0) & 0.00999 & $189\cdot 10^3$ & $6.96 10^{-11}$ \cr
\hline
$\delta=1,\omega=1/4$ & $\Omega$ & 1 & 0.4518(4210) & 1.79 & $189\cdot 10^3$ & 0.0171 \cr
\hline
$\delta=1,\omega=1/4$ & $\Omega$ & 2 & 0.3689(1362) & 1.78 & $189\cdot 10^3$ & 0.0171 \cr
\hline
\end{tabular}
\end{center}
\caption{The thermodynamic potential $\Omega$ and the difference between both solutions $\Delta\Omega$ in the different truncation schemes. The fit was performed using \pref{fullfit}. The quoted errors are statistical only. Solution 1 is $g=1/2$ and solution 2 is $g\approx 0.4$. Note that $\ln(4830021/189000)\approx 3.25$. YM contribution is the Yang-Mills part of the full solution. The $\delta$-, $\omega$-, and $\delta\omega$-variations stem from the variation of the 3-gluon vertex, the gluon-Higgs vertex and the combined variations.}
\label{tres}
\end{table}

These results strongly indicate that the $g\approx 0.4$ solution is the thermodynamically preferred one, as this result is stable with respect to variation of the parameters. This agrees with the comparison to lattice results in section \ref{slattice3d}. However, as it still cannot be excluded that either of the solutions is a truncation artifact, this is not a final statement. Moreover, the value of $a$ seems to be ${\cal O}(1)$. However, the systematic study of the vertex influence indicates that the value can only be determined within a factor of 2 or worse. The thermodynamic potential seems to be generated mainly in the chromoelectric sector, but also this is only an indication.

\subsection{Hard Mode Contribution}

To obtain the contribution from the hard modes, it is at least problematic to use \pref{cjt}. The LW/CJT-action \pref{cjt} considers only the interaction part. It vanishes identically for a free system, i.e.\/ for a system containing only tree-level dressing functions. This free contribution has thus to be added explicitly. Therefore the hard mode contribution can essentially be calculated to be the same as that of a non-interacting system of free gluons. The LW/CJT-expression \pref{cjt} does not capture the tree-level contribution of the soft modes, and it thus can be added here. This yields a thermodynamic potential of a gas of massless gluons with small corrections due to the explicit soft-mode contributions and the residual interactions of the hard modes. It is for $N_c=3$ \cite{Karsch:2003jg} 
\be
\lim_{T\to\infty}\frac{\Omega}{T^4}\approx-\frac{16\pi^2}{90}.\label{sblimit}
\ee
\noindent Using the t'Hooft-scaling for the interaction strength, the corrections due to the soft contributions vanish like $1/T$. The only remaining contribution in the infinite temperature limit is the Stefan-Boltzmann-limit \pref{sblimit}, in agreement with results from lattice calculations \cite{Karsch:2003jg}. Therefore the thermodynamical properties of the high-temperature limit are governed by the hard modes, although they do not participate in the dynamical interactions.


%% file: summary.tex
\chapter{Conclusions, Summary, and Outlook}\label{csummary}

\section{Concluding Remarks}

Before discussing the implications of the results found in chapters \ref{c3d}, \ref{cft}, and \ref{cderived} on the structure of the Yang-Mills (and QCD) phase diagram, it is worthwhile to take a step back to gather and assess the findings. Especially the reliability of the results deserves special attention.

The primary objects investigated in chapters \ref{c3d} and \ref{cft} are the propagators of Yang-Mills theory in the high temperature phase. These can be grouped into three classes. One consists of the soft modes of the ghost and the 3d-transverse gluon, constituting the Yang-Mills sector of the limiting 3d-theory. The second is the soft 3d-longitudinal gluon, corresponding to the Higgs sector of the 3d-theory. From the point of view of the 4d-theory, these are the soft chromomagnetic and chromoelectric degrees of freedom, respectively. The last class contains all the hard modes.

The first class on its own constitutes a 3d Yang-Mills theory. Its properties are found to be very similar to the corresponding 4d-theory. An infrared divergent ghost propagator is accompanied by an infrared vanishing gluon propagator. Therefore the gluon is confined and the results fulfill the Kugo-Ojima and Zwanziger-Gribov scenarios. Also the Schwinger function and analytic structure of the gluon propagator are similar to the 4d case.

The systematic study of the errors due to the truncation find a very weak quantitative effect only. Especially the confining properties are extremely robust against variations in the truncation. Finally the comparison with lattice calculations on exceptionally large lattices shows a very good agreement. Thus it is very likely that these results on a 3d Yang-Mills theory are reliable.

The situation is more subtle concerning the Higgs in the 3d-theory. Its propagator shows the typical behavior of a massive particle and it is dominated by its mass. However, the comparison to lattice data indicates that to some extent subleading or probably non-perturbative effects are relevant. This is much more pronounced when regarding the Schwinger function. Here significant changes are found when modifying the Higgs-gluon vertex. Thus the qualitative behavior of the Higgs propagator seems to be under control and the presence of non-trivial effects is quite certain. The consequences of these effects on the other side are far less conclusive and require further investigation. Especially the question whether the Higgs is confined is of great interest. On the other hand, a quite firm statement is that it has only little effect on the Yang-Mills sector.

Combining both classes, a qualitatively stable infinite temperature limit is found. Only the chromoelectric sector exhibits screening while at least in the chromomagnetic sector confinement prevails. Therefore the infinite temperature limit is non-trivial and the high temperature phase strongly interacting. {\it Hence the simple picture described in the introduction in section \ref{stdqcd} is not applicable. This and the detailed analysis of the infinite temperature limit are the central results of this work.}

Less reliable is the exploratory study of finite temperature effects in chapter \ref{cft} and thus the properties of the hard modes. The success of the truncation scheme in the 3d-theory is based on its exactness in the ultraviolet and its presumed exactness in the infrared. Thus the consequences of the deficiencies at intermediate momenta are strongly constrained. This possibility is lost by cutting off the Matsubara sum. As a consequence the ultraviolet properties of the hard modes are incorrect and a rooting in perturbation theory is prevented. This amplifies the second problem. The hard modes do not reach into the 4d-infrared due to their effective mass. Therefore the advantages of the truncation scheme are lost to a large extent and the equations for the hard modes become unreliable.

Nonetheless, the results found are not irrelevant. At sufficiently large temperatures the truncation-dependent self-energies of the hard modes are suppressed by the large effective mass and they are dominated by the truncation-independent tree-level terms. Therefore the truncation artifacts vanish when the temperature goes to infinity, reproducing the reliable 3d-theory. The hard part is then to assess down to which temperature the results are still at least qualitatively reliable.

As it is found that the system is only very weakly dependent on temperature, qualitative conclusions can probably be drawn even at quite small temperatures. This is due to the fact that the infrared properties of the soft modes are nearly independent of the hard modes. In general the hard modes effectively decouple, because of their effective mass, which is large  compared to $\Lambda_{QCD}$ even at the phase transition. The only exception is the generation of the screening mass for the soft 3d-longitudinal gluon, which is found on quite general grounds and thus can be considered also a reliable conclusion. It is also supported by the systematic error estimations performed, which do not find qualitative effects but again only small quantitative ones.

Thus, although the quantitative results may be subject to change, the qualitative result that, {\it probably down to temperatures within the same order of magnitude as the phase transition, the high temperature phase consists of strongly interacting soft modes exhibiting confining properties and inert hard modes} is expected to be quite reliable. This is also supported by lattice results, which find that the infinite-temperature limit is effectively reached at about $2T_c$ \cite{Cucchieri:2001tw} and the qualitative features of the infinite temperature limit up to the highest temperatures, for which propagators are available, of $6T_c$ \cite{Nakamura:2003pu}. {\it This constitutes the second major result found in this work.}

Concerning as the last point the thermodynamic potential, no final conclusion can be drawn. The results indicate that the gross thermodynamic properties far away from the phase transition are completely dominated by the hard modes and that a Stefan-Boltzmann behavior is reached in the infinite temperature limit. Considering the crude assumptions made and the difficulties encountered, this result is indicative at best. Nonetheless, if it can be substantiated, it would allow to understand how a Stefan-Boltzmann-like behavior can emerge from a non-trivial theory. At the same time, the results also indicate that at low temperatures non-perturbative effects likely will play a role in the thermodynamics and will especially be relevant in the vicinity of the phase transition and thus to experiment.

\section{Implications for the Phase Structure of QCD}\label{ssumprop}

Gathering the results found in chapters \ref{c3d}, \ref{cft}, and \ref{cderived} and the solutions at zero and small temperatures presented in sections \ref{ssolvac} and \ref{sdselowt}, a coherent picture emerges. Still, this picture is overshadowed by the problems of truncation artifacts, especially at finite temperatures.

Combining these results has impact on the understanding of the phase transition. The main difference between the low-temperature and the high-temperature phase is not primarily one between a strongly interacting and confining system and one with only quasi-free particles. The chromoelectric gluons, whose infrared behavior change from over-screening to screening, come somewhat close to such a picture, and the hard modes certainly do. The latter become free due to the vacuum property of asymptotic freedom and the dependence of their energy on the temperature, thus in an expected although not entirely trivial manner. The chromomagnetic gluons remain over-screened in the infrared and are thus confined. At least in the transverse sector, the results provide strong evidence for the Zwanziger-Gribov and/or the Kugo-Ojima scenarios.

The order parameter for the phase transition is then necessarily only a chromoelectric one. This is in agreement with studies of Wilson loops in lattice calculations. There it is known that only the temporal (chromoelectric) Wilson lines show a behavior typical of deconfinement, while the spatial (chromomagnetic) ones do not~\cite{Bali:1993tz}. Note that the order parameters used to study the deconfinement transition on the lattice are typically chromoelectric ones like the Polyakov lines~\cite{Karsch:2003jg}. This leads to the conjecture that if the corresponding chromomagnetic Polyakov lines were to be used almost no change would be found.

Including the observation of super-heating in the low temperature regime \cite{Gruter:2004kd} and super-cooling in the high temperature regime, the scenario emerging is a chromoelectric phase transition of first order. This phase transition connects two different strongly interacting phases of Yang-Mills theory. Especially quantum fluctuations always dominate thermal fluctuations and at least part of the gluon spectrum is always confined.

In the vicinity of the phase transition, the non-perturbative effects found are likely also relevant to thermodynamic properties, especially to the pressure and the trace anomaly \cite{Zwanziger:2004np}. This underlines the importance of non-perturbative effects at least for the temperature range relevant to experiment.

It is still unknown which effects can be induced by quarks. In the infinite temperature limit, arguments exist \cite{Appelquist:vg} that they will not contribute due to their fermionic nature. This nature requires anti-periodic boundary conditions \cite{Kapusta:tk}, and therefore generates effective `masses' of $(2n+1)\pi T$ with $n$ integer. Thus all quarks become infinitely massive in the infinite temperature limit, irrespective of their intrinsic mass and thus decouple due to asymptotic freedom. Lattice calculations indicate \cite{Karsch:2003jg} that the quarks possibly are able to change the nature of the phase transition up to a cross-over. It is hard to imagine how such a partially confined phase would behave, and thus the topic is of high interest.

In addition, the observation of a drastic change in the inter-quark potential \cite{Kaczmarek:2004gv} and its connection to the residual confinement of gluons still needs to be understood. However, it cannot yet be firmly concluded that the inter-quark potential does not rise only logarithmically at high temperatures, as it would be natural to expect from a dimensionally reduced theory. On the other hand, quark confinement is still not understood at zero temperature, thus posing a more significant task.

\section{Summing Up and Looking Ahead}

Summarizing, in this work evidence is presented in agreement with lattice calculations that the finite temperature phase transition of Yang-Mills theory does not lead into a trivial phase. By investigating first the dimensionally reduced theory, thereby providing results on genuine 3d-theories of Yang-Mills and Yang-Mills-Higgs type, the non-triviality of soft interactions was found. Extending the range to finite temperatures, this was confirmed and the decoupling of the hard modes explicitly demonstrated. This establishes the main features of the high temperature phase of Yang-Mills theories in Landau gauge. Furthermore, using the Schwinger functions, the absence of at least part of the gluons from the physical spectrum was found, making manifest the existence of residual confinement. The results found comply with Zwanziger-Gribov or Kugo-Ojima type confinement mechanisms. As a final point, a first step towards thermodynamic observables was performed.

Taking a look ahead, two issues are of high interest. One from an experimental and one from a conceptual point of view.

In heavy-ion collisions, not only finite temperature is of interest, but especially for low-energy collisions results at finite density are of importance. Such experiments are performed at the SPS at CERN and have been performed at the AGS at Brookhaven and will also be conducted at the planned FAIR facility at GSI. Finite density is also relevant for the physics of compact stellar objects as well. For sufficiently large chemical potentials (densities), at least the quark sector will likely be weakly interacting, although non-perturbatively, see e.g.\/ \cite{Rischke:2003mt}. However, it has not yet been convincingly demonstrated how the quarks can generate gluon deconfinement. In addition, those chemical potentials probably only occur in nature during black hole formation. For chemical potentials relevant to experiments and possibly neutron stars, strong non-perturbative effects are expected. Thus a framework as the one used previously for full QCD \cite{Fischer:2003rp} seems to be necessary. First steps in this direction have been performed \cite{Epple:dipl} and are pursued further, they, however, are not yet as advanced as the finite temperature calculations presented here. More results are available from studies using model gluon propagators \cite{Roberts:2000aa} and semi-perturbative methods \cite{Rischke:2003mt}.

On the conceptual sides the problems encountered in chapter \ref{cft} to obtain the vacuum solution indicate that it will probably be necessary to choose other means to investigate the phase transition directly, e.g.\/ real-time methods \cite{Das:gg}. In addition, the inclusion of quarks is an urgent problem as their impact on thermodynamic properties is highly relevant for experiments, including RHIC at Brookhaven and ALICE, which is under construction  at the LHC at CERN. In addition, understanding of quark confinement in the vacuum might be supported by understanding the high temperature behavior of quarks and the phase transition. This would also lead to a deeper understanding of the relationship of chiral symmetry and confinement of fundamental quarks, for which evidence exists. Indeed, their phase transitions seem to be interlinked \cite{Karsch:2003jg}. This immediately leads to the problem of the connection to topological excitations, which lattice calculations indicate to play a dominant role in the physics of confinement and chiral symmetry breaking. If they are relevant, these excitations must also be present in the current approach, as their consequences are. Therefore, it would be desirable to disentangle their contributions and identify their role. It is at this point also completely unclear, how their dynamics change above the phase transition and their connection to the at least partial confinement of gluons.

To reach these aims, implementation of new methods and close cooperation with other approaches, especially lattice gauge theory, will be necessary. Nevertheless, over the last decade significant progress has been made in the understanding of low energy QCD, providing hope that it will eventually be understood.

%% file: ack.tex
\newpage

\thispagestyle{empty}

{\bf \Large Acknowledgments}\newline

Many people have contributed to this work and made it possible. I apologize to anyone whom I forgot.

First of all I would like to thank Professor Wambach for the opportunity to work on this subject and his interest and support of this work as well as enlightening discussions. Also, I would like to thank him for the possibility to visit several conferences and schools, which considerably broadened my perspective in physics.

I would like to thank Professor Reinhard Alkofer for the warm hospitality at T\"ubingen and his interest and engagement in this work and many fruitful and supportive discussions.

I thank Professor Grewe for being second examiner on this work and his interest in its content.

I would like to thank Burghard Gr\"uter and Christian S. Fischer for a pleasant collaboration and many discussions.

I thank the NHQ, TNP, and high-energy groups at Darmstadt University of Technology for an enjoyable working atmosphere. Especially, I would like to thank Dr.\ Michael Buballa, Dr.\ Bernd-Jochen Schaefer, Dominik Nickel, Wolfgang Schleifenbaum, and Dr.\ Thomas Roth for valuable discussions. I thank Mathias Wagner, Professor Robert Roth, Carsten Isselhorst and especially Dr.\ Thomas Roth for computer administration and support.

For enlightening discussions I would like to thank Professors Daniel Zwanziger, Craig D. Roberts, Pierre van Baal and Jean Zinn-Justin. Also I would like to thank Dr.\ Peter Petreczky, PhD Peter Watson, and Claus Feuchter for very helpful discussions.

I thank my parents for all their patience, advise, love and support.

I thank Dr.\ A.\ Rost and the team of station 12 and the oncological ambulance at the Klinikum Darmstadt for treating my fianc$\mathrm{\acute{e}}$e - as far as it seems - successfully for cancer. Without this success, this work would most likely have taken much longer, if it had been possible at all.

The most I thank my beloved fianc$\mathrm{\acute{e}}$e Renate Knobloch for all her love she gave me during the years and her support while writing this thesis.

This work was supported by the BMBF under grant numbers 06DA917 and 06DA116, by the European Graduate School Basel-T\"ubingen (DFG contract GRK683), and by the Helmholtz association (Virtual Theory Institute VH-VI-041).

$\quad$\\
This work is dedicated to the memory of my grandfather Clemens Margraf,\\ *10.12.1925, $^+$27.03.2003.

\newpage


%% file: conventions.tex
\chapter{Conventions}\label{aconventions}

Calculations are done in natural units,
\be
\hbar=c=k_{B}=1.\nonumber
\ee
\noindent The Minkowski-metric is
\be
\eta=\mathrm{diag}(1,-1,-1,-1).\nonumber
\ee
\noindent The Euclidean metric after factorization of -1 is
\be
\eta_e=\mathrm{diag}(1,1,1,1).\nonumber
\ee
\noindent Thus $p^2=p_0^2+\vec p^2$. Sometimes $|\vec p|=p_3$ is used. The individual components of $\vec p$ do never appear explicitly.

The second Casimir $C_A$ of a compact, semi-simple Lie-group $G$ in the adjoint representation with structure constants $f^{abc}$ is defined by
\be
f^{acd}f^{acd}=C_A.\nonumber
\ee
\noindent For SU$(N_c)$ groups, $C_A=N_c$. 

Fourier transformations are performed with all momenta incoming,
\bea
f(p)&=&\int d^dx f(x)e^{ixp}\nonumber\\
f(x)&=&\int\frac{d^dp}{(2\pi)^d} f(p)e^{-ixp}.\nonumber
\eea


%% file: dse.tex
\chapter{Derivation of the Dyson-Schwinger Equations}\label{adse}

Starting from \pref{dse}, it is possible to derive the DSEs for the Lagrangian \pref{l3d}. In this context also the Legendre transform of $W$ plays a role, where $W$ is defined by the generating functional \pref{lpart}. The Legendre transform can be obtained by
\be
W(j^a)=-\Gamma(\phi^a)+\int d^dxj^a(x)\phi^a(x)\label{effpot}
\ee
\noindent which entails
\bea
\phi^a=\frac{\delta W}{\delta j^a}\nonumber\\
j^a=\frac{\delta\Gamma}{\delta\phi^a}\nonumber.
\eea
\noindent In the case of Grassmann fields $u$ like ghosts and fermions, two independent sources are necessary. This modifies the above to
\bea
Z=\int{\cal D}u^a{\cal D}\bar u^a e^{-S[u^a,\bar u^a]+\int d^dx(\bar\eta^a(x) u^a(x)+\bar u^a(x)\eta^a(x))}\nonumber\\
u^a(x)=\frac{\delta W}{\delta\bar\eta^a(x)}\quad\quad\quad\quad\bar u^a(x)=-\frac{\delta W}{\delta\eta^a(x)}\nonumber\\
W(\eta^a,\bar\eta^a)=-\Gamma(u^a,\bar u^a)+\int d^dx\left(\bar\eta^a(x)u^a(x)+\bar u^a(x)\eta^a(x)\right)\nonumber\\
\eta^a(x)=\frac{\delta\Gamma}{\delta\bar u^a(x)}\quad\quad\quad\quad\bar\eta^a(x)=-\frac{\delta\Gamma}{\delta u^a(x)}\nonumber,
\eea
\noindent where all derivatives with respect to Grassmann variables act in the direction of ordinary derivatives.

The general procedure to obtain the corresponding DSEs is to calculate equation \pref{dse} and then take the derivative once more with respect to the field or with respect to the anti-field in case of anti-commuting fields. The additional source terms then yield the full propagators, while the right-hand-sides of the equations are found by the derivative of the action. 

Since in the course of the derivation, the source in equation \pref{dse} becomes the inverse full propagator, it makes sense to already rewrite \pref{dse} as
\be
j^a(x)Z=\frac{\delta S}{\delta\phi^a(x)}\Big|_{\phi^a(x)=\frac{\delta}{\delta j^a(x)}}Z\label{mdse}
\ee
\noindent at the sources set equal to 0 in the end. This entails also that all single derivatives of the effective action, being classical fields, vanish at the end.

The inverse propagators of the ghost, the gluon and the Higgs are defined as
\bea
\frac{\delta^2\Gamma}{\delta c^b(y)\delta\bar c^a(x)}&=D^{ab}_G(x-y)^{-1}\label{adighprop}\\
\frac{\delta^2\Gamma}{\delta A_\nu^b(y)\delta A_\mu^a(y)}&=D^{ab}_{\mu\nu}(x-y)^{-1}\\
\frac{\delta^2\Gamma}{\delta\phi^b(y)\delta\phi^a(x)}&=D_H^{ab}(x-y)^{-1},\label{adihprop}
\eea
\noindent The propagators are then given by
\bea
\frac{\delta^2 W}{\delta\eta^b(y)\delta\bar\eta^a(x)}&=D^{ab}_G(x-y)\label{adghprop}\\
\frac{\delta^2 W}{\delta j_\nu^b(y)j_\mu^a(x)}&=D^{ab}_{\mu\nu}(x-y)\\
\frac{\delta^2 W}{\delta k^b(y)\delta k^a(x)}&=D_H^{ab}(x-y)\label{adhprop}.
\eea
\noindent $\eta$ is the source of the ghost field, $\bar\eta$ of the anti-ghost field, $j$ of the gluon, and $k$ of the Higgs. \prefr{adghprop}{adhprop} are inverse to \prefr{adighprop}{adihprop} which can be proven, e.g. for the ghost propagator by
\be
\int d^dz\frac{\delta^2 W}{\delta\eta^c(z)\delta\bar\eta^a(x)}\frac{\delta^2\Gamma}{\delta c^b(y)\delta\bar c^c(z)}=\int d^dz\frac{\delta c^a(x)}{\delta\eta^c(z)}\frac{\delta\eta^c(z)}{\delta c^b(y)}=\frac{\delta c^a(x)}{\delta c^b(y)}=\delta^{ab}\delta(x-y).\nonumber
\ee
\noindent Note that all mixed propagators vanish when the sources are set to 0, e.g.
\be
\frac{\delta^2\Gamma}{\delta\bar c^g(w)\delta A_\nu^f(z)}\Big |_{j=\eta=\bar\eta=0}=0,\nonumber
\ee
due to conservation of ghost number and spin.

Higher $n$-point functions, the vertices, are defined accordingly as
\bea
\frac{\delta^3\Gamma}{\delta c^a(x)\delta\bar c^b(y)\delta A^c_\mu(z)}=\Gamma_\mu^{c\bar cA\indexsep abc}(x,y,z)\nonumber
\eea
\noindent and correspondingly for the other vertices.

There are a few useful identities in the course of the derivation. These are
\bea
\frac{\delta^2 W}{\delta j^e_\mu(x)\delta\bar\eta^d(x)}&=&-\int d^dzd^dw\frac{\delta^2 W}{\delta j_\nu^f(z)\delta j^e_\mu(x)}\frac{\delta^2\Gamma}{\delta\bar c^g(w)\delta A_\nu^f(z)}\frac{\delta^2 W}{\delta\bar\eta^g(w)\delta\eta^d(x)}\nonumber\\
\pdm^x\frac{\delta}{\delta\eta^c(x)}&=&\int d^dz \left(\pdm^x\frac{\delta\Gamma}{\delta\eta^c(x)\delta\bar c^e(z)}\right)\frac{\delta}{\delta\eta^e(z)}.\nonumber
\eea
Further, since
\bea
&&\frac{\delta^3 W}{\delta A_\nu^b(y)\delta j^c_\sigma(x)\delta j^e_\sigma(x)}=\frac{\delta}{\delta A_\nu^b(y)}\int d^dzd^dw \frac{\delta^2 W}{\delta j_\sigma^c(x)\delta j_\rho^f(z)}\frac{\delta^2\Gamma}{\delta A_\rho^f(z)\delta A_\omega^g(w)}\frac{\delta^2 W}{\delta j_\omega^g(w)\delta j_\sigma^e(x)}\nonumber\\
&=&2\frac{\delta^3 W}{\delta A_\nu^b(y)\delta j^c_\sigma(x)\delta^e_\sigma(x)}+\int d^dzd^dw \frac{\delta^2 W}{\delta j_\sigma^c(x)\delta j_\rho^f(z)}\frac{\delta^3\Gamma}{\delta A_\nu^b(y)\delta A_\rho^f(z)\delta A_\omega^g(w)}\frac{\delta^2 W}{\delta j_\omega^g(w)\delta j_\sigma^e(x)}\nonumber
\eea
\noindent it follows that
\be
\frac{\delta^3 W}{\delta A_\nu^b(y)\delta j^c_\sigma(x)\delta j^e_\sigma(x)}=-\int d^dzd^dw \frac{\delta^2 W}{\delta j_\sigma^c(x)\delta j_\rho^f(z)}\frac{\delta^3\Gamma}{\delta A_\nu^b(y)\delta A_\rho^f(z)\delta A_\omega^g(w)}\frac{\delta^2 W}{\delta j_\omega^g(w)\delta j_\sigma^e(x)},\label{w3id}
\ee
\noindent when the sources are set to 0. If not all sources and fields are of the same type, then it is necessary to sum over all possible intermediate fields $\omega^a$ and their sources $l^a$, where the index $a$ includes also the field type. Thus
\be
\frac{\delta^3 W}{\delta \omega^b(y)\delta l^c(x)\delta l^e(x)}=-\int d^dzd^dw \frac{\delta^2 W}{\delta l^c(x)\delta l^f(z)}\frac{\delta^3\Gamma}{\delta \omega^b(y)\delta \omega^f(z)\delta \omega^g(w)}\frac{\delta^2 W}{\delta l^g(w)\delta l^e(x)}.
\ee
\noindent It is also important to note that
\be
\pd_\sigma^x\frac{\delta^2 W}{\delta j^c_\sigma(x)\delta j^d_\mu(x)}\nonumber
\ee
\noindent is not zero, despite the translational invariance of the two-point functions. The derivative of the propagator at 0 does not necessarily vanish especially, if the propagator does only exist in the sense of a distribution.

As the effective potential defined by \pref{effpot} is bosonic, it can only depend on even numbers of ghost and anti-ghost fields. Therefore all vertices with only one ghost or anti-ghost leg vanish, when the sources are set to zero.

After a tedious calculation, the DSEs in position space are obtained. Adding suitable combinations of permuted diagrams, it is in all cases possible to make the appearance of one tree-level vertex explicit. Performing a Fourier-transformation, and keeping the redundant momentum at the vertices explicit, the DSEs in momentum space are for the ghost\footnote{Note that some diagrams which vanish for tree-level vertices had not been included in \cite{Maas:2004se}.}\enlargethispage*{1cm}
\bea
&D^{ab}_G(p)^{-1}=-\delta^{ab}p^2\nonumber\\
&+\int\frac{d^dq}{(2\pi)^d}\Gamma_\mu^{\tl\indexsep c\bar cA\indexsep dae}(-q,p,q-p)D^{ef}_{\mu\nu}(p-q)D^{dg}_G(q)\Gamma^{c\bar cA\indexsep bgf}_\nu(-p,q,p-q),\label{tlgheq}
\eea
\noindent the gluon
\bea
&D^{ab}_{\mu\nu}(p)^{-1}=\delta^{ab}(\delta_{\mu\nu}p^2-p_\mu p_\nu)\nonumber\\
&-\int\frac{d^dq}{(2\pi)^d}\Gamma_\mu^{\tl\indexsep c\bar cA\indexsep dca}(-p-q,q,p) D_G^{cf}(q) D_G^{de}(p+q) \Gamma_\nu^{c\bar c A\indexsep feb}(-q,p+q,-p)\nonumber\\
&+\frac{1}{2}\int\frac{d^dq}{(2\pi)^d}\Gamma^{\tl\indexsep A^3\indexsep acd}_{\mu\sigma\chi}(p,q-p,-q)D^{cf}_{\sigma\omega}(q)D^{de}_{\chi\lambda}(p-q)\Gamma^{A^3\indexsep bfe}_{\nu\omega\lambda}(-p,q,p-q)\nonumber\\
&+\frac{1}{2}\int\frac{d^dq}{(2\pi)^d}\Gamma_\mu^{\tl\indexsep A\phi^2\indexsep acd}(p,q-p,-q)D_H^{de}(q)D_H^(cf)(p-q)\Gamma^{A\phi^2\indexsep bef}_\nu(-p,q,p-q)\nonumber\\
&+\frac{1}{2}\int\frac{d^d q}{(2\pi)^d}\Gamma_{\mu\nu\sigma\rho}^{\tl\indexsep A^4\indexsep abcd}(p,-p,q,-q)D^{cd}_{\sigma\rho}(q)\nonumber\\
&+\frac{1}{6}\int\frac{d^dqd^dk}{(2\pi)^{2d}}\Gamma_{\mu\sigma\xi\chi}^{\tl\indexsep A^4\indexsep acde}(p,-q,-p+q-k,k)\cdot\nonumber\\
&\cdot D_{\chi\lambda}^{dh}(p-q-k)D_{\xi\rho}^{cf}(q)D_{\sigma\omega}^{eg}(k)\Gamma^{A^4\indexsep hbfg}_{\lambda\nu\rho\omega}(p-q-k,-p,q,k)\nonumber\\
&+\frac{1}{2}\int\frac{d^dqd^dk}{(2\pi)^{2d}}\Gamma^{\tl\indexsep A^4\indexsep acde}_{\mu\delta\gamma\sigma}(p,-q,q-k-p,k)D_{\gamma\lambda}^{dh}(k+p-q)\Gamma_{\nu\rho\omega}^{A^3\indexsep bfg}(-p,p+k,-k)\cdot\nonumber\\
&\cdot\Gamma_{\lambda\chi\xi}^{A^3\indexsep hij}(k+p-q,q,-k-p) D_{\rho\xi}^{fj}(p+k)D_{\sigma\omega}^{eg}(k)D_{\delta\chi}^{ci}(q)\nonumber\\
&+\frac{1}{2}\int\frac{d^dqd^dk}{(2\pi)^{2d}}\Gamma^{\tl\indexsep A^4\indexsep acde}_{\mu\delta\gamma\sigma}(p,-q,q-k-p,k)D_{\gamma\lambda}^{dh}(k+p-q)\Gamma_{\nu\omega}^{A^2\phi\indexsep bgf}(-p,-k,p+k)\cdot\nonumber\\
&\cdot\Gamma_{\lambda\chi}^{A^2\phi\indexsep hij}(k+p-q,q,-k-p) D_H^{fj}(p+k)D_{\sigma\omega}^{eg}(k)D_{\delta\chi}^{ci}(q)\nonumber\\
&+\frac{1}{2}\int\frac{d^dq}{(2\pi)^d}\Gamma^{\tl\indexsep A^2\phi^2\indexsep abdf}_{\mu\nu}(p,-p,q,-q)D_H^{df}(q)\nonumber\\
&-\frac{1}{2}\int\frac{d^dqd^dk}{(2\pi)^{2d}}\Gamma_{\mu\sigma}^{\tl\indexsep A^2\phi^2\indexsep aedf}(p,q-p-k,-q,k)\nonumber\\
&\cdot D_{\sigma\rho}^{eg}(p-q+k)D_H^{dh}(q)D_H^{fi}(k)\Gamma_{\rho\nu}^{A^2\phi^2\indexsep gbhi}(p-q+k,-p,q,-k)\nonumber\\
&+\int\frac{d^dqd^dk}{(2\pi)^{2d}}\Gamma_{\mu\sigma}^{\tl\indexsep A^2\phi^2\indexsep aedf}(p,q-p-k,-q,k)D_{\sigma\rho}^{eg}(p+k-q)D_H^{fi}(k)D_H^{hk}(k+p)\cdot\nonumber\\
&\cdot D_H^{dj}(q)\Gamma^{A\phi^2\indexsep gjk}_\rho(p+k-q,q,-k-p)\Gamma^{A\phi^2\indexsep bhi}_\nu(-p,k+p,-k)\nonumber\\
&+\int\frac{d^dqd^dk}{(2\pi)^{2d}}\Gamma_{\mu\sigma}^{\tl\indexsep A^2\phi^2\indexsep aedf}(p,q-p-k,-q,k)D_{\sigma\rho}^{eg}(p+k-q)D_H^{fi}(k)D_{\omega\lambda}^{hk}(k+p)\cdot\nonumber\\
&\cdot D_H^{dj}(q)\Gamma^{A^2\phi\indexsep gkj}_{\rho\lambda}(p+k-q,-k-p,q)\Gamma^{A^2\phi\indexsep bhi}_{\nu\omega}(-p,k+p,-k)\nonumber\\
&+\int\frac{d^dqd^dk}{(2\pi)^{2d}}\Gamma_{\mu\rho}^{\tl\indexsep A^2\phi^2\indexsep afed}(p,k,q-p-k,-q)D_H^{eg}(p+k-q)D_{\sigma\rho}^{fi}(k)D_H^{hk}(k+p)\cdot\nonumber\\
&\cdot D_H^{dj}(q)\Gamma^{\phi^3\indexsep gjk}(p+k-q,q,-k-p)\Gamma^{A^2\phi\indexsep bih}_{\nu\rho}(-p,-k,k+p)\nonumber\\
&+\int\frac{d^dqd^dk}{(2\pi)^{2d}}\Gamma_{\mu\rho}^{\tl\indexsep A^2\phi^2\indexsep afed}(p,k,q-p-k,-q)D_H^{eg}(p+k-q)D_{\sigma\rho}^{fi}(k)D_{\lambda\omega}^{hk}(k+p)\cdot\nonumber\\
&\cdot D_H^{dj}(q)\Gamma_\omega^{A\phi^2\indexsep kgj}(-k-p,p+k-q,q)\Gamma^{A^3\indexsep bih}_{\nu\rho\lambda}(-p,-k,k+p),\label{tlgleq}
\eea
\noindent and the Higgs
\bea
&D_H^{ab}(p)^{-1}=\delta^{ab}(p^2+m_h^2)+\nonumber
\eea
\bea
&+\int\frac{d^dq}{(2\pi)^d}\Gamma^{\tl\indexsep A\phi^2\indexsep eac}_\nu(-p-q,p,q)D_{\nu\mu}^{cg}(p+q)D_H^{fc}(q)\Gamma_\mu^{A^2\phi\indexsep gbf}(p+q,-p,-q)\nonumber\\
&+\frac{1}{2}\int\frac{d^dq}{(2\pi)^d}\Gamma_{\mu\nu}^{\tl\indexsep A^2\phi^2\indexsep cdab}(q,-q,p,-p)D_{\mu\nu}^{cd}(q)\nonumber\\
&-\frac{1}{2}\int\frac{d^dqd^dk}{(2\pi)^d}\Gamma_{\mu\sigma}^{\tl\indexsep A^2\phi^2\indexsep cdae}(-p-q+k,-k,p,q)D_{\mu\nu}^{cg}(p+q-k)D_{\rho\sigma}^{id}(k)\cdot\nonumber\\
&\cdot D_H^{eh}(q)\Gamma_{\rho\nu}^{A^2\phi^2\indexsep igbh}(-k,p+q-k,-p,q)\nonumber\\
&+\frac{1}{2}\int\frac{d^dqd^dk}{(2\pi)^{2d}}\Gamma_{\mu\sigma}^{\tl\indexsep A^2\phi^2\indexsep cdae}(-q,k,p,-p+q-k)D_H^{ej}(p-q+k)D_{\mu\nu}^{cg}(q)D_{\rho\sigma}^{id}(k)\cdot\nonumber\\
&\cdot\Gamma_\nu^{A\phi^2\indexsep gjk}(q,p-q+k,-p-k)D_H^{kh}(p+k)\Gamma_\rho^{A\phi^2\indexsep ibh}(-k,-p,p+k)\nonumber\\
&+\frac{1}{2}\int\frac{d^dqd^dk}{(2\pi)^{2d}}\Gamma_{\mu\sigma}^{\tl\indexsep A^2\phi^2\indexsep cdae}(-q,k,p,-p+q-k)D_H^{ej}(p-q+k)D_{\mu\nu}^{cg}(q)D_{\rho\sigma}^{id}(k)\cdot\nonumber\\
&\cdot\Gamma_{\nu\lambda}^{A^2\phi\indexsep gkj}(q,-p-k,p-q+k)D_{\lambda\omega}^{kh}(p+k)\Gamma_{\rho\omega}^{A^2\phi\indexsep ihb}(-k,p+k,-p)\nonumber\\
&+\frac{1}{2}\int\frac{d^dqd^dk}{(2\pi)^{2d}}\Gamma_{\mu\chi}^{\tl\indexsep A^2\phi^2\indexsep cdae}(-q,q+k,p,-p-k)D^{cg}_{\mu\nu}(q)D_H^{eh}(p+k)D_{\lambda\chi}^{kd}(q+k)\cdot\nonumber\\
&\cdot\Gamma_{\sigma\nu\lambda}^{A^3\indexsep jgk}(k,q,-q-k)D_{\rho\sigma}^{ij}(k)\Gamma_\rho^{A\phi^2\indexsep ibh}(-k,-p,p+k)\nonumber\\
&+\frac{1}{2}\int\frac{d^dqd^dk}{(2\pi)^{2d}}\Gamma_{\mu\chi}^{\tl\indexsep A^2\phi^2\indexsep cdae}(-q,q+k,p,-p-k)D^{cg}_{\mu\nu}(q)D_H^{eh}(p+k)D_{\lambda\chi}^{kd}(q+k)\cdot\nonumber\\
&\cdot\Gamma_{\nu\lambda}^{A^2\phi\indexsep gkj}(q,-q-k,k)D_H^{ij}(k)\Gamma^{\phi^3\indexsep ibh}(-k,-p,p+k)\nonumber\\
&+\frac{1}{2}\int\frac{d^dq}{(2\pi)^d}\Gamma^{\tl\indexsep \phi^4\indexsep abcd}(p,-p,q,-q)D_H^{cd}(q)-\nonumber\\
&-\frac{1}{6}\int\frac{d^dqd^dk}{(2\pi)^{2d}}\Gamma^{\tl\indexsep \phi^4\indexsep agch}(p,-p+q-k\indexsep k\indexsep -q)D_H^{gd}(p-q+k)D_H^{he}(q)\cdot\nonumber\\
&\cdot D_H^{fc}(k)\Gamma^{\phi^4\indexsep debf}(p-q+k,q,-p,-k)\nonumber\\
&+\frac{1}{3}\int\frac{d^dqd^dk}{(2\pi)^{2d}}\Gamma^{\tl\indexsep \phi^4\indexsep aicj}(p,-p-k+q,k,-q)D_H^{id}(p+k-q)D_H^{jg}(q)D_H^{fc}(k)\cdot\nonumber\\
&\cdot\Gamma^{\phi^3\indexsep gdh}(q,p+k-q,-k-p)D_H^{eh}(p+k)\Gamma^{\phi^3\indexsep ebf}(p+k,-p,-k)\nonumber\\
&+\frac{1}{3}\int\frac{d^dqd^dk}{(2\pi)^{2d}}\Gamma^{\tl\indexsep \phi^4\indexsep aicj}(p,-p-k+q,k,-q)D_H^{id}(p+k-q)D_H^{jg}(q)D_H^{fc}(k)\cdot\nonumber\\
&\cdot\Gamma_\mu^{A\phi^2\indexsep hgd}(-k-p,q,p+k-q)D_{\mu\nu}^{eh}(p+k)\Gamma^{A\phi^2\indexsep ebf}_\nu(p+k,-p,-k).\label{tlhiggseq}
\eea
\noindent Several of the appearing vertices do not have a tree-level counterpart. The tree-level vertices for the ghost-gluon-, 3-gluon-, 4-gluon-, 2-gluon-Higgs-, 2-gluon-2-Higgs- and 4-Higgs- interactions have been used. They are given by
\bea
\Gamma_\mu^{\tl\indexsep c\bar cA\indexsep abc}(p,q,k)&=&ig_df^{abc}q_\mu\label{tlcca}\\
\Gamma_{\mu\nu\rho}^{\tl\indexsep A^3\indexsep abc}(p,q,k)&=&-ig_df^{abc}((q-k)_\mu\delta_{\nu\rho}+(k-p)_\nu\delta_{\mu\rho}+(p-q)_\rho\delta_{\mu\nu})\label{tlggg}\\
\Gamma_{\mu\nu\sigma\rho}^{\tl\indexsep A^4\indexsep abcd}(p,q,k,l)&=&g_d^2(f^{eab}f^{ecd}(\delta_{\mu\sigma}\delta_{\nu\rho}-\delta_{\mu\rho}\delta_{\nu\sigma})+f^{gac}f^{gbd}(\delta_{\mu\nu}\delta_{\sigma\rho}-\delta_{\mu\rho}\delta_{\nu\sigma})\nonumber\\
&&+f^{gad}f^{gbc}(\delta_{\mu\nu}\delta_{\sigma\rho}-\delta_{\mu\sigma}\delta_{\nu\rho}))\label{tl4g}\\
\Gamma_\mu^{\tl\indexsep A\phi^2\indexsep abc}(p,q,k)&=&ig_df^{abc}(q-k)_\mu\label{tlgh}\\
\Gamma_{\mu\nu}^{\tl\indexsep A^2\phi^2\indexsep abcd}(p,q,k,l)&=&g_d^2\delta_{\mu\nu}(f^{eac}f^{ebd}+f^{ead}f^{ebc})\label{tl2g2h}\\
\Gamma^{\tl\indexsep \phi^4\indexsep abcd}(p,q,k,l)&=&2h(\delta_{ab}\delta_{cd}+\delta_{ac}\delta_{bd}+\delta_{ad}\delta_{bc})\label{tl4h},
\eea
where the momentum-conserving $\delta$-functions have been suppressed. The complete graphical representation of \prefr{tlgheq}{tlhiggseq} is shown in figure \ref{figfullsys}.

%% file: kernels.tex
\chapter{Kernels}

\section{3d Kernels}\label{s3dkernels}

The integral kernels in (\ref{fulleqG3d}-\ref{fulleqZ3d}) are obtained by using tree-level vertices and performing the corresponding contractions. The ghost kernel is given by
\be
A_t(k,q,\theta)=-\frac{q^2\sin^3(\theta)}{(k^2+q^2-2kq\cos\theta)^2}.\nonumber
\ee
The contributions in the Higgs equation are
\bea
N_1(k,q,\theta)=-\frac{2q^2\sin^3(\theta)}{(k^2+q^2+2kq\cos\theta)^2}\nonumber\\
N_2(k,q,\theta)=-\frac{2\sin^3(\theta)}{k^2+q^2+2kq\cos\theta}.\nonumber
\eea
The kernels in the gluon equation are finally
\bea
&R(k,q,\theta)=-\frac{((\zeta-1)kq\cos(\theta)-q^2+\zeta q^2\cos^2(\theta))\nonumber
\sin\theta}{2k^2(k^2+q^2+2kq\cos\theta)}\nonumber\\
&M_L(k,q,\theta)=\frac{((\zeta-1)(k^2+4kq\cos\theta)-4q^2+4q^2\zeta\cos^2(\theta))\sin\theta}
{4k^2(k^2+q^2+2kq\cos\theta)}\nonumber\\
&M_T(k,q,\theta)=\frac{\sin\theta}{4k^2(k^2+q^2+2kq\cos\theta)}\cdot\nonumber\\
&\cdot\Big((k^2+2q^2)((\zeta-9)k^2-4q^2)+8(\zeta-3)(k^2+q^2)kq\cos\theta+(8\zeta q^4+(\zeta+7)k^4\nonumber\\
&+4(5\zeta-1)k^2q^2)\cos^2(\theta)+4(4\zeta q^2+(\zeta+3)k^2)\cos^3(\theta)+4\zeta k^2q^2\cos^4(\theta)\Big).\nonumber
\eea
The modified gluon vertex is introduced by multiplying $M_T$ with \pref{g3vertex}.

\section{Finite Temperature Kernels}\label{sftkernels}

At finite temperature, the kernels are obtained from tree-level vertices. The following abbreviations have been used:
\bea
q&=&\left|\vec q\right|\nonumber\\
k&=&\left|\vec k\right|\nonumber\\
x&=&q^2+q_0^2\nonumber\\
y&=&k^2+k_0^2\nonumber\\
u_3&=&(\vec q-\vec k)^2\nonumber\\
z_3&=&(\vec q+\vec k)^2\nonumber\\
u_4&=&(k_0-q_0)^2+u_3\nonumber\\
z_4&=&(k_0+q_0)^2+z_3.\nonumber
\eea
\noindent The kernels in the ghost equations are
\bea
A_T(k_0,q_0,k,q,\theta)&=&-\frac{k^2q^4{\sin (\theta )}^3}{xy{u_3}{u_4}}\nonumber\\
A_L(k_0,q_0,k,q,\theta)&=&-\frac{q^2\sin\theta{\left( {q}^2{k_0} + {k}^2{q_0} - kq\cos\theta\left( {k}_0 + {q_0} \right)\right) }^2}{xy{u_3}{{u_4}}^2}.\nonumber
\eea
\noindent The kernels in the transverse gluon equations are
\bea
R(k_0,q_0,k,q,\theta)&=&-\frac{q^3\left( -q + k\left( -1 + \zeta  \right) \cos\theta+ q\zeta {\cos (\theta )}^2 \right) \sin (\theta )}{2xyz_4}\nonumber\\
M_T(k_0,q_0,k,q,\theta)&=&\frac{q^2}{16xy z_3 z_4}\Big( \Big( 8q^4\left( -4 + \zeta  \right)  + k^4\left( -29 + 5\zeta  \right)\nonumber\\
&&+ 2k^2q^2\left( -46 + 15\zeta  \right)  \Big) \sin (\theta ) +4kq\Big( 4q^2\left( -3 + 2\zeta  \right)\nonumber\\
&&+ k^2\left( -9 + 5\zeta  \right)  \Big) \sin (2\theta ) + \Big( 8q^4\zeta  + k^4\left( 7 + \zeta  \right)\nonumber\\
&&+ k^2q^2\left( -4 + 23\zeta  \right)  \Big) \sin (3\theta ) + 2kq\left( 4q^2\zeta  + k^2\left( 3 + \zeta  \right)  \right) \sin (4\theta )\nonumber\\
&& + k^2q^2\zeta \sin (5\theta ) \Big)\nonumber\\
M_1(k_0,q_0,k,q,\theta)&=&\frac{q^2\sin\theta}{8x^2y{z_3}{z_4}}\Big( -4q^4\left( -4 + \zeta  \right) {{k_0}}^2 - k^4\left( -9 + \zeta  \right) {{q_0}}^2 + k^2q^2\left( 4{k_0} - {q_0} \right)\cdot\nonumber\\
&&\cdot \left( 4{k_0} + \left( -8 + \zeta  \right) {q_0} \right)  - 2kq\cos\theta\Big( -16q^2{{k_0}}^2\nonumber\\
&& + 2\left( 8k^2 - q^2\left( -8 + \zeta  \right)  \right) {k_0}{q_0} + k^2\left( -13 + \zeta  \right) {{q_0}}^2 \Big)  + \cos (2\theta )\cdot\nonumber\\
&&\cdot \left( 4q^4\zeta {{k_0}}^2 - 4k^2q^2\left( \left( 7 + \zeta  \right) {k_0} - 2{q_0} \right) {q_0} + k^4\left( 7 + \zeta  \right) {{q_0}}^2 \right)  + \nonumber\\
&&kq{q_0}\left( kq\zeta \cos (4\theta )q_0 + 2\cos (3\theta )\left( -2q^2\zeta {k_0} + k^2\left( 3 + \zeta  \right) {q_0} \right)  \right)  \Big)\nonumber
\eea
\bea
M_2(k_0,q_0,k,q,\theta)&=&\frac{q^2}{16xy{z_3}{{z_4}}^2}\Big(-\left(k^2q^2\zeta\sin(5\theta)\left(k_0-q_0\right)^2 \right)  +4kq\sin (2\theta )\cdot\nonumber\\
&&\cdot\left( k_0 - q_0 \right)\left( 2q^2\left( -4 + \zeta  \right) {k_0} + k^2\left( \left( -1 + \zeta  \right) {k_0} - \left( -7 + \zeta  \right) {q_0} \right)  \right)\nonumber\\
&&-2kq\sin (4\theta )\left( k_0 - {q_0} \right)\left( 2q^2\zeta {k_0} + k^2\left( \left( -1 + \zeta  \right) {k_0} - \left( 1 + \zeta  \right) {q_0} \right)  \right)\nonumber\\
&&+\sin\theta\Big( 4q^4\left( -8 + 3\zeta  \right) {{k_0}}^2 +k^4\Big( -3{{k_0}}^2 + 3\zeta {\left( {k_0} - {q_0} \right) }^2 + 6{k_0}{q_0}\nonumber\\
&&- 23{{q_0}}^2 \Big)  +2k^2q^2\left( \left( -10 + 7\zeta  \right) {{k_0}}^2 + \left( 34 - 8\zeta  \right) {k_0}{q_0} + \left( -4 + \zeta  \right) {{q_0}}^2 \right)  \Big)\nonumber\\
&&-\sin (3\theta )\Big( 4q^4\zeta {{k_0}}^2 + k^2q^2\Big( \left( 4 + 3\zeta  \right) {{k_0}}^2 - 2\left( 10 + \zeta  \right) {k_0}{q_0}\nonumber\\
&&- \left( -8 + \zeta  \right) {{q_0}}^2 \Big)  +k^4\Big( \left( -1 + \zeta  \right) {{k_0}}^2 - 2\left( -1 + \zeta  \right) {k_0}{q_0}\nonumber\\
&&+ \left( 3 + \zeta  \right) {{q_0}}^2 \Big)\Big)\Big)\nonumber\\
M_L(k_0,q_0,k,q,\theta)&=&\frac{q^2\sin\theta}{4x^2y{z_3}{{z_4}}^2}\Big( {{k}}^6q^2\left( -1 + \zeta  \right)  - 4q^4{\left( x + {{k_0}}^2 + {k_0}{q_0} \right) }^2 + 4k^2q^2\zeta {\cos\theta}^4\cdot\nonumber\\
&&\cdot{\left( 2q^2 + {q_0}\left( -{k_0} + {q_0} \right)  \right) }^2 +2k^4\Big( 2q^2\left( -1 + \zeta  \right) {{k_0}}^2 + 3q^2\left( -1 + \zeta  \right) {k_0}{q_0}\nonumber\\
&& +x\left( q^2\left( -3 + \zeta  \right)  - 2{{q_0}}^2 \right)  \Big)  +{{k}}^2q^2\Big( x^2\left( -9 + \zeta  \right)  + 4\left( -1 + \zeta  \right) {{k_0}}^4\nonumber\\
&&+ 2x\left( -7 + 3\zeta  \right) {k_0}{q_0} +12\left( -1 + \zeta  \right) {{k_0}}^3{q_0} + {{k_0}}^2\Big( 4q^2\left( -3 + \zeta  \right)\nonumber\\
&&+ \left( -21 + 13\zeta  \right) {{q_0}}^2 \Big)  \Big)  +4kq{\cos\theta}^3\left( 2q^2 + {q_0}\left( -{k_0} + {q_0} \right)  \right)\cdot\nonumber\\
&&\cdot\Big( 2q^2\zeta \left( x + {{k_0}}^2 + {k_0}{q_0} \right)  +k^2\Big( q^2\left( -2 + 4\zeta  \right)  + {q_0}\Big( {k_0} - \zeta {k_0} + {q_0}\nonumber\\
&&+ \zeta {q_0} \Big)  \Big)  \Big)  +2kq\cos\theta\Big( 2q^2\left( x + {{k_0}}^2 + {k_0}{q_0} \right)\Big( q^2\left( -5 + \zeta  \right)\nonumber\\
&&+ 2\left( -1 + \zeta  \right) {{k_0}}^2 + \left( -1 + 3\zeta  \right) {k_0}{q_0} +\left( -3 + \zeta  \right) {{q_0}}^2 \Big)  + {{k}}^4\left( -1 + \zeta  \right)\cdot\nonumber\\
&&\cdot\left( 4q^2 + {q_0}\left( -{k_0} + {q_0} \right)  \right)  +{{k}}^2\Big( -2\left( -1 + \zeta  \right) {{k_0}}^3{q_0} + \left( -1 + \zeta  \right) {{k_0}}^2\cdot\nonumber\\
&&\cdot\left( 10q^2 - {{q_0}}^2 \right)  +x\left( 2q^2\left( -7 + 3\zeta  \right)  + \left( -5 + \zeta  \right) {{q_0}}^2 \right)  +{k_0}{q_0}\Big( q^2\cdot\nonumber\\
&&\cdot\left( -9 + 13\zeta  \right)  + 2\left( 1 + \zeta  \right) {{q_0}}^2 \Big)  \Big)  \Big)  +{\cos\theta}^2\Big( 4q^4\zeta {\left( x + {{k_0}}^2 + {k_0}{q_0} \right) }^2\nonumber\\
&&+4k^2q^2\Big( q^4\left( -8 + 6\zeta  \right)  - 3\left( -1 + \zeta  \right) {{k_0}}^3{q_0} + 8q^2\left( -1 + \zeta  \right) {{q_0}}^2 +\nonumber\\
&&\left( -1 + 2\zeta  \right) {{q_0}}^4 + {{k_0}}^2\left( q^2\left( -6 + 8\zeta  \right)  - \left( -2 + \zeta  \right) {{q_0}}^2 \right)  +2{k_0}{q_0}\Big( q^2\cdot\nonumber\\
&&\cdot\left( -1 + 4\zeta  \right)  + \left( 1 + \zeta  \right) {{q_0}}^2 \Big)  \Big)  +k^4\Big( 4q^4\left( -5 + 6\zeta  \right)  - 4q^2{q_0}\Big( 3\left( -1 + \zeta  \right)\cdot\nonumber\\
&&\cdot{k_0} + {q_0} - 3\zeta {q_0} \Big)  + {{q_0}}^2\Big( \left( -1 + \zeta  \right) {{k_0}}^2 - 2\left( -1 + \zeta  \right) {k_0}{q_0}\nonumber\\
&& + \left( 3 + \zeta  \right) {{q_0}}^2 \Big)  \Big)\Big)  \Big).\nonumber
\eea
\noindent The kernels in the longitudinal gluon equations are
\bea
P(k_0,q_0,k,q,\theta)&=&\frac{q^2\sin\theta}{xy^2{z_4}}\Big( -2\left(\xi-1 \right) k_0^2{q_0}\left( {k_0} + {q_0} \right)  + k^2{q_0}\left( {k_0} - \xi {k_0} + \xi {q_0} \right)  + qk_0\cos\theta\nonumber\\
&&\cdot\left( q\xi \cos \theta k_0 + k\left( {k_0} + 2{q_0} - \xi \left( {k_0} + 4{q_0} \right)  \right)  \right)  \Big)\nonumber\\
N_T(k_0,q_0,k,q,\theta)&=&\frac{-q^2\sin\theta}{2xy^2{z_3}{z_4}}\left( 3k^2 + 4q^2 + 8kq\cos\theta + k^2\cos (2\theta ) \right) \Big( k_0\left( {k_0} + 2{q_0} \right)\Big( k^2\nonumber\\
&&+ {k_0}\left( {k_0} + 2{q_0} \right)  \Big)-\xi\Big( {{k_0}}^2\left( -q + {k_0} + 2{q_0} \right) \left( q + {k_0} + 2{q_0} \right)  + k^2\Big( {{k_0}}^2\nonumber\\
&& + 2{k_0}{q_0} - 2{{q_0}}^2 \Big)  \Big)  +q{k_0}\Big( q\xi \cos (2\theta )k_0 +2k\cos\theta\Big( k_0 + 2{q_0}\nonumber\\
&&- \xi \left( {k_0} + 4{q_0} \right)  \Big)  \Big)  \Big)\nonumber\\
N_1(k_0,q_0,k,q,\theta)&=&\frac{-q^2\sin^3\theta}{x^2y^2{z_3}{z_4}}\Big( 2q^4\xi {{k_0}}^4 + k^4\left( 2x^2\xi  - 2x\left(\xi-1  \right) {k_0}{q_0} - \left(\xi-1  \right) {{k_0}}^2{{q_0}}^2 \right)\nonumber\\
&&+k^2{{k_0}}^2\Big( {q_0}\left( {k_0} + 2{q_0} \right) \left( 2q^2 + {q_0}\left( {k_0} + 2{q_0} \right)  \right)  + \xi \Big( 4q^4 + q^2{q_0}\Big( -2{k_0}\nonumber\\
&&+ {q_0} \Big)  - {{q_0}}^2{\left( {k_0} + 2{q_0} \right) }^2 \Big)  \Big)  +kq{k_0}{q_0}\Big( kq\xi \cos (2\theta )k_0{q_0} -2\cos\theta\nonumber\\
&&\cdot\left( 2q^2\xi {{k_0}}^2 + k^2\left( q^2\left(4\xi-2  \right)  + {q_0}\left( \left(\xi-1  \right) {k_0} - 2{q_0} + 4\xi {q_0} \right)  \right)  \right)  \Big)  \Big)\nonumber\\
N_2(k_0,q_0,k,q,\theta)&=&\frac{-q^2\sin^3\theta}{xy^2{z_3}{{z_4}}^2}\Big( k^2{k_0}\left( {k_0} + {q_0} \right)\Big( 2k^4 + {k_0}\left( {k_0}+ 2{q_0} \right) \Big( 2q^2 + \left( {k_0} + {q_0} \right) \Big( {k_0}\nonumber\\
&&+ 2{q_0} \Big)  \Big)  +k^2\left( 6q^2 + \left( 3{k_0} + {q_0} \right) \left( {k_0} + 2{q_0} \right)  \right)  \Big)  +\xi \Big( 2k^8 + 2q^4{{k_0}}^4\nonumber\\
&&+ 2k^6\left( 4q^2 + {{k_0}}^2 + {k_0}{q_0} + 2{{q_0}}^2 \right)  -k^2{{k_0}}^2\Big( -4q^4 + {{k_0}}^4 + 6{{k_0}}^3{q_0}\nonumber\\
&&- q^2{{q_0}}^2 + 4{{q_0}}^4 + 4{k_0}{q_0}\left( q^2 + 3{{q_0}}^2 \right)  +{{k_0}}^2\left( -3q^2 + 13{{q_0}}^2 \right)  \Big)  +k^4\nonumber\\
&&\cdot\left( 2x^2 - {{k_0}}^4 - 6{{k_0}}^3{q_0} + {{k_0}}^2\left( 6q^2 - 3{{q_0}}^2 \right)  + {k_0}\left( -6q^2{q_0} + 2{{q_0}}^3 \right)  \right)\Big)\nonumber\\
&&+ kq\Big( kq\cos (2\theta )\left( 4k^4\xi  + \xi{{k_0}}^2{\left( {k_0} - {q_0} \right) }^2 +4k^2{k_0}\left( {k_0} + {q_0} - 2\xi {q_0} \right)  \right)\nonumber\\
&&+2\cos\theta\Big( k^2{k_0}\left( {k_0} + {q_0} \right)\left( 4k^2 + 2q^2 + \left( 3{k_0} + {q_0} \right) \left( {k_0} + 2{q_0} \right)  \right)\nonumber\\
&&+\xi\Big( 4k^6 + 2q^2{{k_0}}^3\left( {k_0} - {q_0} \right)  + 2k^4\left( 2x + {{k_0}}^2 - {k_0}{q_0} \right)  -k^2{k_0}\Big( {{k_0}}^3\nonumber\\
&&+ 4x{q_0} + 10{{k_0}}^2{q_0} + {k_0}\left( -4q^2 + 9{{q_0}}^2 \right)  \Big)  \Big)  \Big)  \Big)  \Big)\nonumber
\eea
\bea
&&N_L(k_0,q_0,k,q,\theta)=\frac{-q^2\sin\theta}{x^2y^2{z_3}{{z_4}}^2}\nonumber\\
&&\cdot\Big(2k^2q^2\cos^4\theta\Big( 4k^4\xi {{q_0}}^2 + \xi {{k_0}}^2{\left( 2q^2 + {q_0}\left( -{k_0} + {q_0} \right)  \right) }^2 +4k^2{k_0}{q_0}\Big( q^2\left( 2 - 4\xi  \right)  + {q_0}\Big( {k_0} + {q_0}\nonumber\\
&&- 2\xi {q_0} \Big)  \Big)  \Big)  +2kq{\cos (\theta )}^3\Big( 4k^6\xi {{q_0}}^2+ 2q^2\xi {{k_0}}^2\Big( 2q^4 + 2q^2{{k_0}}^2 + q^2{k_0}{q_0} - {{k_0}}^3{q_0} + 3q^2{{q_0}}^2+\nonumber
\eea
\bea
&&+ {{q_0}}^4 \Big)  +2k^4{q_0}\left( \left( 2 + \xi  \right) {{k_0}}^2{q_0} + 2\xi {q_0}\left( 2q^2 + {{q_0}}^2 \right)  +{k_0}\left( q^2\left( 6 - 12\xi  \right)  - \left( -2 + \xi  \right) {{q_0}}^2 \right)  \right)\nonumber\\
&&+k^2{k_0}\Big( -\left( \left(\xi-3  \right) {{k_0}}^3{{q_0}}^2 \right)  -2{{k_0}}^2{q_0}\left( q^2\left( -3 + 9\xi  \right)  + 5\left(\xi-1  \right) {{q_0}}^2 \right)-2\left( -1 + 2\xi  \right) {q_0}\Big( 6q^4\nonumber\\
&&+ 6q^2{{q_0}}^2 + {{q_0}}^4 \Big)  +{k_0}\Big( 8q^4\xi  - 22q^2\left( -1 + \xi  \right) {{q_0}}^2- 9\left(\xi-1  \right) {{q_0}}^4 \Big)  \Big)  \Big)  +q^2\Big( k^6\left(3\xi-1  \right) {{k_0}}^2\nonumber\\
&&+k^4{k_0}\Big( \left( -1 + 5\xi  \right) {{k_0}}^3- 2x\left( -1 + 3\xi  \right) {q_0} + 2\left( -1 + 3\xi  \right) {{k_0}}^2{q_0} +2\left( 1 - 3\xi  \right) {k_0}{{q_0}}^2 \Big)  -\nonumber\\
&&\left(\xi-1  \right){{k_0}}^2{\left( 2x{q_0} + {{k_0}}^2{q_0} + {k_0}\left( q^2 + 3{{q_0}}^2 \right)  \right) }^2 +k^2\Big( 2\xi {{k_0}}^6 + 4\xi {{k_0}}^5{q_0}+ 2x^2\xi {{q_0}}^2 + \left( 3 - 5\xi  \right)\nonumber\\
&&\cdot{{k_0}}^4{{q_0}}^2 -2{{k_0}}^3{q_0}\left( q^2\left(4\xi-2  \right)  + \left(9\xi-5  \right) {{q_0}}^2 \right)+{{k_0}}^2\Big( -\left( q^4\left(\xi-1  \right)  \right)  - 2q^2\left(7\xi-5  \right) {{q_0}}^2\nonumber\\
&&+ \left( 9 - 11\xi  \right) {{q_0}}^4 \Big)+2{k_0}{q_0}\left( -\left( q^4\left( -1 + \xi  \right)  \right)  + 2q^2{{q_0}}^2 + \left( 1 + \xi  \right) {{q_0}}^4 \right)  \Big)  \Big)  -2kq\cos\theta\nonumber\\
&&\cdot\Big( k^6\left(3\xi-1  \right)\cdot{k_0}{q_0} +k^4\Big( \left( -1 + 5\xi  \right) {{k_0}}^3{q_0} - 2x\xi {{q_0}}^2 +{k_0}{q_0}\Big( q^2\left(7\xi-3  \right)  + 2\left(\xi-1  \right){{q_0}}^2 \Big)\nonumber\\
&&+{{k_0}}^2\left( q^2\left( 2 - 6\xi  \right)  + 4\left( -1 + 2\xi  \right) {{q_0}}^2 \right)  \Big)  -k_0\Big( -2q^2x^2\left( -1 + 2\xi  \right) {q_0}+2{{k_0}}^4{q_0}\Big( q^2\left( 1 + \xi  \right)\nonumber\\
&&- 3\left( -1 + \xi  \right) {{q_0}}^2 \Big)  +{{k_0}}^5\left( 2q^2\xi  - \left( -1 + \xi  \right) {{q_0}}^2 \right)  +{{k_0}}^3\Big( q^4\left( 1 + \xi  \right)  + q^2\left( 13 - 11\xi  \right) {{q_0}}^2\nonumber\\
&&- 13\left(\xi-1  \right) {{q_0}}^4 \Big)  -x{k_0}\Big( q^4\left(\xi-1  \right)+ q^2\left(17\xi-13  \right) {{q_0}}^2 + 4\left(\xi-1  \right) {{q_0}}^4 \Big)  -2{{k_0}}^2{q_0}\Big( 4q^4\nonumber\\
&&\cdot\left(\xi-1  \right)  + q^2\left(14\xi-13  \right){{q_0}}^2 + 6\left(\xi-1 \right) {{q_0}}^4 \Big)\Big)  + k^2\Big( 2\xi {{k_0}}^5{q_0} +{{k_0}}^3{q_0}\Big( q^2\left(\xi-3  \right)  + 2\nonumber\\
&&\cdot\left(6\xi-5  \right){{q_0}}^2 \Big)  +{{k_0}}^4\left( q^2\left( 1 - 7\xi  \right)  + \left( -3 + 7\xi  \right) {{q_0}}^2 \right)  -2\xi {{q_0}}^2\left( 2q^4 + 3q^2{{q_0}}^2 + {{q_0}}^4 \right)-{{k_0}}^2\nonumber\\
&&\cdot\left( q^4\left( 1 + \xi  \right)  - 4q^2\left( -4 + 5\xi  \right) {{q_0}}^2 + \left( 9 - 7\xi  \right) {{q_0}}^4 \right)  -2{k_0}{q_0}\Big( q^4\left( 3 - 5\xi  \right)+ q^2\left( 4 - 3\xi  \right) {{q_0}}^2\nonumber\\
&&+ \left( 1 + \xi  \right) {{q_0}}^4 \Big)  \Big)  \Big)+ {\cos^2\theta}\Big( 2k^8\xi {{q_0}}^2 + 2q^4\xi {{k_0}}^2\Big( x + {{k_0}}^2+ {k_0}{q_0} \Big)^2 +2k^6{q_0}\Big( \left( 1 + \xi  \right) {{k_0}}^2{q_0}\nonumber\\
&&+ 2\xi {q_0}\left( 2q^2 + {{q_0}}^2 \right)  +{k_0}\Big( q^2\left( 6 - 14\xi  \right)+ \left( 1 + \xi  \right) {{q_0}}^2 \Big)  \Big)  +k^4\Big( -\left( \left( -3 + \xi  \right) {{k_0}}^4{{q_0}}^2 \right)\nonumber\\
&&-2{{k_0}}^3{q_0}\Big( q^2\left(17\xi-5  \right)+ \left(3\xi-5  \right) {{q_0}}^2 \Big)  +2\xi {{q_0}}^2\left( 8q^4 + 8q^2{{q_0}}^2 + {{q_0}}^4 \right)  +{{k_0}}^2\Big( 4q^4\left(5\xi-1  \right)\nonumber\\
&&+ 2q^2\left( 19 - 23\xi  \right) {{q_0}}^2 - 3\left( -3 + \xi  \right) {{q_0}}^4 \Big) +2{k_0}{q_0}\Big( q^4\left( 12 - 22\xi  \right)  + 2q^2
\cdot\left( 5 - 4\xi  \right) {{q_0}}^2\nonumber\\
&&+ \left( 1 + \xi  \right) {{q_0}}^4 \Big)  \Big)  +k^2{k_0}\Big( -\left( \left( -1 + \xi  \right) {{k_0}}^5{{q_0}}^2 \right)  +2{{k_0}}^4{q_0}\Big( q^2\left( 1 - 5\xi  \right)  - 3\left( -1 + \xi  \right) {{q_0}}^2 \Big)\nonumber\\
&&-8q^2\left(2\xi-1  \right) {q_0}\left( 2q^4 + 3q^2{{q_0}}^2 + {{q_0}}^4 \right)+{{k_0}}^3\left( 16q^4\xi  + 2q^2\left( 11 - 15\xi  \right) {{q_0}}^2 - 13\left(\xi-1  \right) {{q_0}}^4 \right)\nonumber\\
&&-4{{k_0}}^2{q_0}\Big( q^4\Big( -5+ 4\xi  \Big)  + q^2\left( -14 + 15\xi  \right) {{q_0}}^2 + 3\left( -1 + \xi  \right) {{q_0}}^4 \Big)  + 4{k_0}\Big( q^6\left( 1 + \xi  \right)  - 2q^4\nonumber\\
&&\cdot\left( -7 + 8\xi  \right) {{q_0}}^2 + q^2\left( 11 - 13\xi  \right) {{q_0}}^4 -\left( -1 + \xi  \right) {{q_0}}^6 \Big)  \Big)  \Big)  \Big). \nonumber
\eea
The vertex dressing in the soft-soft $M_T$-kernel is performed as in the 3d-case.

%% file: ir.tex
\chapter{Infrared Expressions}\label{air}

\section{Massive Higgs}

As already argued in section \ref{ssiranalysis}, the only solution without further assumption is that of ghost dominance. This then requires the calculation of $I_{GT}$ and $I_{GG}$ in \pref{ghostir} and \pref{gluonir} only. The latter can be obtained straightforwardly when using the ansatz \pref{iransatz} and the general formula
\be
\int\frac{d^dq}{(2\pi)^d}q^{2\alpha} (q-p)^{2\beta}=\frac{1}{(4\pi)^{\frac{d}{2}}}\frac{\Gamma(-\alpha-\beta-\frac{d}{2})\Gamma(\frac{d}{2}+\alpha)\Gamma(\frac{d}{2}+\beta)}{\Gamma(d+\alpha+\beta)\Gamma(-\alpha)\Gamma(-\beta)}y^{2\left(\frac{d}{2}+\alpha+\beta\right)},\label{dimregrule}
\ee
\noindent valid for finite integrals. It yields
\be
I_{GG}=-\frac{g_3^2C_A\pi}{(4\pi)^{\frac{d}{2}}}\frac{2^{4g-2d}(d-4g)(2+d(\zeta-2)-4g(\zeta-1)-\zeta)\Gamma(d-2g)\Gamma(2g-\frac{d}{2})}{(d-1)g^2\Gamma(\frac{1+d-2g}{2})^2\Gamma(g)^2}.\nonumber
\ee
\noindent Note that the integral is convergent if and only if
\be
\frac{d-1}{2}\ge g>\frac{d-2}{4},\label{dglimits}
\ee
\noindent where the equality on the upper boundary requires the result to exist only in the sense of a distribution.

Performing the same calculation for the ghost self-energy is more complicated. Since it is, in general, a  divergent quantity, equation \pref{dimregrule} cannot be applied directly. It is necessary to regularize and then renormalize the expression. This can be done in a momentum subtraction scheme~\cite{Zwanziger:2001kw} or via dimensional regularization. Here, the latter was performed by applying the standard rules of dimensional regularization~\cite{Peskin:ev}. This immediately gives a finite result in odd dimensions, as dimensional regularization is in this case already a renormalization prescription. However, by doing so, a divergent quantity has been removed which is formally eliminated by setting $-\widetilde Z_3$ equal to this quantity. Regularization and renormalization in even dimensions can be performed using the MS-prescription \cite{Collins:xc}. This procedure yields the same result as the momentum subtraction scheme. The range allowed for $g$ in \pref{dglimits} permits $I_{GT}$ not only to have a divergence of logarithmic or linear order, but also quadratic or cubic divergences in even or uneven dimensions, respectively. Using a subtraction scheme, it would be necessary to include the next term in the Taylor expansion. On the other hand, dimensional renormalization directly yields
\be
I_{GT}=\frac{g_3^2C_A}{(4\pi)^{\frac{d}{2}}}\frac{2^{1-2g}(4^g(d-3)d+2^{1+2g}(1+g-dg))\Gamma(\frac{d}{2}-g)\Gamma(-g)\Gamma(2g)}{(2-d+2g)(d+2g)\Gamma(\frac{d}{2}-2g)\Gamma(g)\Gamma(\frac{d}{2}+g)}.\nonumber
\ee
\noindent Note that this expression becomes negative already for values allowed by \pref{dglimits}, e.g. for $g\ge3/4$ in three dimensions, thus reducing the allowed range and leading to the plots in figure \ref{figexp}. 

\section{Massless Higgs}\label{asmasslesshiggs}

If the Higgs is massless, then the Higgs-loop in the gluon equation \pref{fulleqZ3d} is no longer suppressed in the infrared and can in principle be as leading as the ghost-loop or even dominate. Thus three solutions can be found in the infrared. As the arguments from section \ref{ssiranalysis} are still valid, all must satisfy $t<-1$.

In the first case, the Higgs and the ghost exchange their roles. The ghost behaves then like a tree-level particle in the infrared with $g=0$, while the Higgs diverges in the infrared and drives the gluon. In the second case, $l=g$, and the gluon is driven by the combination of both loops. In the third case, the ghost and the gluon behave as in the massive case and $l=0$: The Higgs behaves as a massless tree-level particle in the infrared. In all cases gluons are confined. The first solution may correspond to a phase structure similar to a Higgs-Anderson phase, but different to the conventional `perturbative' Higgs phases. The physical interpretation of the second scenario is lacking and the properties of this phase would be quite peculiar. The last case has similar physics as the massive case. Note that in the first two cases gauge symmetry is not necessarily intact any more, and may be broken by the interaction with the Higgs field. This may restrict these phases to theories which do not emerge from a pure Yang-Mills theory, as this would be difficult to reconcile with Elitzur's theorem \cite{Elitzur:im}. Hence the Gribov condition \pref{gribov} will be dropped for the first two cases for now. As these cases offer a different structure than the massive case, they will be investigated a little more closely.

After employing a renormalization prescription in the Higgs equation as in the case of the ghost equation, the infrared system becomes
\bea
\frac{y^g}{A_g}&=&\widetilde{Z}_3(1-\delta_{gl})+y^{-\frac{1}{2}-g-t}A_gA_zI_{GT}(y)\nonumber\\
\frac{y^t}{A_t}&=&y^{-\frac{1}{2}-2g}A_g^2I_{GG}(y)+y^{-\frac{1}{2}-2l}A_h^2I_{GH}(y)\nonumber\\
\frac{y^l}{A_l}&=&y^{-\frac{1}{2}-g-l}A_hA_zI_{HT}(y).\nonumber
\eea
\noindent Keeping $d=3$ for simplicity, and noting that in both cases
\be
l=-\frac{1}{2}(t+\frac{1}{2})
\ee
\noindent and
\bea
I_{HT}&=&\frac{2^{2+4l}(8-11l-15l^2+14l^3)\Gamma(-2l)}{(l-3+32l^2+44l^3+16l^4)\Gamma(\frac{1}{2}-2l)}\nonumber\\
I_{GH}&=&-\frac{2^{-2+4l}(l-1+l\zeta)\Gamma(-2l)\sec(2l\pi)\sin^2(l\pi)}{(l-1)(4l-1)l\Gamma(\frac{1}{2}-2l)},\nonumber
\eea
\noindent the conditional equation for the first case is
\be
1=\frac{(1+l)(1+2l)(3+2l)(l-1+l\zeta)(\sec(2l\pi)-1)}{16(l-1)l(8+l(1+2l)(7l-11))}.\nonumber
\ee
\noindent This has no solution at $\zeta=1$ in $1/4< l\le 1/2$, but has e.g. the solution $l=0.362568$ at $\zeta=3$. This already indicates that in these situation the allowed projections differ from the normal case. Since for a broken gauge symmetry the STI \pref{stigluon} may not be valid anymore and the gluon may acquire longitudinal components, this equally well indicates that the current truncation is inadequate for this case.

In the second case, the condition for a solution becomes
\be
1=\frac{I_{GG}}{I_{GT}}+\frac{A_h^2}{A_g^2}\frac{I_{GH}}{I_{HT}}=\frac{A_g^2}{A_h^2}\frac{I_{GG}}{I_{HT}}+\frac{I_{GH}}{I_{HT}},\nonumber
\ee
\noindent yielding
\bea
1&=&\frac{I_{GH}I_{GT}}{I_{GT}I_{HT}-I_{GG}I_{HT}}\nonumber\\
&=&\frac{(3(\zeta-2)+2g(59-4g^2(\zeta-1)+\zeta-6g(10+\zeta))+64(g-1)g\csc^2(g\pi))}{8(1+g)(1+2g)(3+2g)(g-1+g\zeta)}\cdot\nonumber\\
&&\cdot(8+g(1+2g)(7g-11)).\nonumber
\eea
\noindent Only one infrared coefficient is left undetermined, thus it is likely that here the infrared behavior will be linked stronger to finite momentum solutions to allow for compensation both in the Higgs equation and the ghost equation in the infrared. Note that again a solution does not exist for $\zeta=1$ for $1/4< g=l\le 1/2$, but it exists for $\zeta=3$ with value $l=g=0.340408$. Thus also here the validity of the truncations made is debatable.

Nonetheless, these investigations show that additional massless degrees of freedom may change the infrared properties significantly. Especially in QCD it turns out that no solutions have been found yet with $N_f>4$ massless quarks \cite{Fischer:2003rp} in this approach, despite the possibility of chiral symmetry breaking.

%% file: uv.tex
\chapter{Perturbative Expressions}\label{appUV}

This appendix contains the perturbative calculations to leading order in the dimensionally reduced theory of chapter \ref{c3d}. Replacing all full quantities in the truncated DSEs (\ref{fulleqG3d}-\ref{fulleqZ3d}) by their tree-level values, i.e. setting all dressing functions equal to one and using the tree-level vertices (\ref{tlcca}-\ref{tl4h}) instead of the full vertices, the standard perturbation theory to one-loop order is obtained. It is then possible to calculate the leading-order perturbative dressing functions. For the ghost, this leads to
\be
G(p)^{-1}=1-\frac{g_3^2C_A}{16p}\label{apertgh}
\ee
to leading order, independent of the presence of a Higgs. Note that a Landau pole at $p={\cal O}(g_3^2)$ is present.

The calculation of the gluon self-energy is a little more complicated, since each single contribution is linearly divergent, although the sum is finite in three dimensions due to the STI \pref{stigluon}. Corresponding problems are most easily circumvented by contracting the gluon equation with the Brown-Pennington projector \pref{bpproj}. Performing the calculation in pure Yang-Mills theory yields
\be
Z(p)^{-1}=1-\frac{11g_3^2C_A}{64p}\label{apertgl}.
\ee
Including the Higgs yields
\be
Z(p)^{-1}=1-\frac{11g_3^2C_A}{64p}+\frac{g_3^2C_A}{16\pi p}\left(2\frac{m_h}{p}-\frac{p^2+4m_h^2}{p^2}\csc^{-1}\left(\sqrt{1+\frac{4m_h^2}{p^2}}\right)\right).\label{apertglwh}
\ee
Since non-perturbative effects already arise at order $g_3^4/p^2$ in the present truncation scheme, the only interesting part is the one for $p\gg g_3^2,m_h$, leading to
\be
Z(p)^{-1}=1-\frac{9g_3^2C_A}{64p}\label{apertgllead}.
\ee
Finally, the Higgs self-energy is
\be
H(p)^{-1}=1+\frac{m_h^2}{p^2}+\frac{g_3^2C_A}{p}\frac{m_h}{4\pi p}+\frac{g_3^2C_A}{p}\frac{1}{2\pi p} \left(\left(\frac{m_h^2+p^2}{p}-2p\right)\arcsin\left(\sqrt{\frac{p^2}{m_h^2+p^2}}\right)-m_h\right)\label{aperth},
\ee
\noindent where the third term is the tadpole contribution. The leading contribution is
\be
H(p)^{-1}=1-\frac{g_3^2C_A}{4p}.\label{aperthlead}
\ee
The coupling constant $h$ is determined by requiring that the sum of the tadpole kernels already generates a finite integral, i.e.\/ independent of the regularization scheme. This is obtained by requiring \pref{hvalue}.

Note that, in three dimensions, resummation produces effects only from order $g_3^4$ on. By dimensional arguments alone, only tree-level expressions in the loops can contribute at order $g_3^2$. Therefore \pref{apertgh}, \pref{apertgllead} and \pref{aperthlead} already constitute the resummed solutions. This is confirmed by the numerical calculations presented in section \ref{sfull3d}.

Note further that this also ensures that gauge symmetry is intact to one-loop order in the regime of applicability of leading-order resummed perturbation theory. The violation of gauge symmetry in this approach is not stronger than in ordinary leading-order perturbation theory.

%% file: num.tex
\chapter{Numerical Method}\label{anum}

In order to numerically solve the DSEs, the method described in \cite{Fischer:2002hn,Fischer:2003zc,Atkinson:1998zc,Hauck:1998fz} was extended and improved in detail. The dressing functions were split into three parts. For sufficiently small momenta of the order of $10^{-2}g_3^2$, they are replaced by their infrared behavior \prefr{iransatz1}{iransatz}. For sufficiently large momenta, of the order $10^3g_3^2$, they are replaced by their leading order perturbative form due to \pref{apertgh}, \pref{apertgllead} and \pref{aperthlead}. In the remaining part, the dressing functions were separated into an analytic factor, which interpolates between the analytic infrared and ultraviolet behavior, and a modification function. The logarithm of this modification function is then expanded in Chebychef polynomials. This factorization is not necessary for the $g=1/2$-solution, albeit it lowers the needed CPU time significantly. However, it was not possible to find the other solution branch prior to this improvement. The analytic interpolation functions are
\bea
G(q)&=&1+\frac{g_3^{4g}+q^{2g}}{q^{2g}}\frac{1}{g_3^2+q}\label{anghost}\\
Z(q)&=&\frac{q^{-2t}}{\frac{2}{g_3^{2t}}+q^{-2t}}\left(1+\frac{g_3^2}{g_3^2+q}\right)\label{angluon}\\
H(q)&=&\frac{q^{-2l}}{m_h^{-2l}+q^{-2l}}\label{anhiggs}.
\eea
The coefficients of the Chebychef polynomials are determined using a global rather than a local Newton method \cite{Kelley}. This made it possible to solve the equations without any further prior knowledge about the solution. For a solution, it is required that relations \pref{ag} and \pref{ah} are fulfilled at least up to $10^{-3}$ besides fulfilling the system of equations \prefr{fulleqG3d}{fulleqZ3d} to a much higher precision.

It turns out that simultaneous fits of the Chebychef coefficients and $A_g$, $A_z$, and $A_h$ are not sufficiently stable. Hence an iterated procedure has been used where first a global Newton method was employed and then new infrared coefficients were extracted from the requirement that the functions have to be continuous in the infrared. This was iterated until convergence of the infrared coefficients was achieved. A detailed account of the method will be given in \cite{Maas:ccp}.

To add the hard modes posed another problem. Due to the recursive nature of the subtraction prescription \pref{finite} of the spurious divergences, the Matsubara sum easily switches from logarithmic to quadratic divergence. The only way found up to now to solve this problem is to recalculate the renormalization constants in each Newton iteration step. By this in principle each Newton step again starts as a first step, but already with quite good starting values. This is sufficient to stabilize the algorithm and provides the solutions found in chapter \ref{cft}.
 

%% file: bib.tex

%% file: thesis.bbl
\begin{thebibliography}{}

\bibitem{Bohm:yx}
M.~Bohm, A.~Denner and H.~Joos,
``Gauge Theories Of The Strong And Electroweak Interaction,'' Stuttgart, Germany: Teubner (2001) 784 p.

\bibitem{Gravitation}
C.~W.~Misner, K.~S.~Thorne and J.~A~.Wheeler
``Gravitation,'' W. H. Freeman and Company, U.S.A. (1997) 1279 p.

\bibitem{Hicks:2004vd}
K.~Hicks,
arXiv:hep-ph/0408001 and references therein.

\bibitem{Hagiwara:2002fs}
K.~Hagiwara {\it et al.}  [Particle Data Group Collaboration],
Phys.\ Rev.\ D {\bf 66} (2002) 010001.

\bibitem{Montvay:1994cy}
I.~Montvay and G.~Munster,
``Quantum fields on a lattice,'' Cambridge, UK: Univ. Pr. (1994) 491 p. (Cambridge monographs on mathematical physics) and references therein.

\bibitem{Cabibbo:1975ig}
N.~Cabibbo and G.~Parisi,
Phys.\ Lett.\ B {\bf 59} (1975) 67.

\bibitem{Wong:1995jf}
C.~Y.~Wong,
``Introduction to high-energy heavy ion collisions,'' Singapore, Singapore: World Scientific (1994) 516 p.

\bibitem{Andronic:2004tx}
A.~Andronic and P.~Braun-Munzinger,
arXiv:hep-ph/0402291; 
and references therein.

\bibitem{Karsch:2003jg}
F.~Karsch and E.~Laermann,
arXiv:hep-lat/0305025 and references therein.

\bibitem{Fodor:2001pe}
Z.~Fodor and S.~D.~Katz,
JHEP {\bf 0203} (2002) 014
[arXiv:hep-lat/0106002];
Z.~Fodor and S.~D.~Katz,
JHEP {\bf 0404} (2004) 050
[arXiv:hep-lat/0402006];
P.~de Forcrand and O.~Philipsen,
Nucl.\ Phys.\ B {\bf 642} (2002) 290
[arXiv:hep-lat/0205016].

\bibitem{Braun-Munzinger:2001mh}
P.~Braun-Munzinger and J.~Stachel,
J.\ Phys.\ G {\bf 28} (2002) 1971
[arXiv:nucl-th/0112051].

\bibitem{Rajagopal:2000wf}
K.~Rajagopal and F.~Wilczek,
arXiv:hep-ph/0011333 and references therein.

\bibitem{Kapusta:tk}
J.~I.~Kapusta,
``Finite Temperature Field Theory'', Cambridge University Press (1989).

\bibitem{Alkofer:2000wg}
R.~Alkofer and L.~von Smekal,
Phys.\ Rept.\  {\bf 353}, 281 (2001)
[arXiv:hep-ph/0007355].

\bibitem{Yang:ek}
C.~N.~Yang and R.~L.~Mills,
Phys.\ Rev.\  {\bf 96} (1954) 191.

\bibitem{Rivers:hi}
R.~J.~Rivers,
``Path Integral Methods In Quantum Field Theory,''
Cambridge, UK: Univ. Pr. (1987) 339 p. (Cambridge monographs on mathematical physics).

\bibitem{Damgaard:1987rr}
P.~H.~Damgaard and H.~Huffel,
Phys.\ Rept.\  {\bf 152} (1987) 227;
G.~Parisi and Y.~s.~Wu,
Sci.\ Sin.\  {\bf 24} (1981) 483.

\bibitem{Zwanziger:1981kg}
D.~Zwanziger,
Nucl.\ Phys.\ B {\bf 192} (1981) 259;
L.~Baulieu and D.~Zwanziger,
Nucl.\ Phys.\ B {\bf 193} (1981) 163.

\bibitem{Zwanziger:2003cf}
D.~Zwanziger,
Phys.\ Rev.\ D {\bf 69} (2004) 016002
[arXiv:hep-ph/0303028].

\bibitem{Gribov:1977wm}
V.~N.~Gribov,
Nucl.\ Phys.\ B {\bf 139} (1978) 1.

\bibitem{Singer:dk}
I.~M.~Singer,
Commun.\ Math.\ Phys.\  {\bf 60} (1978) 7.

\bibitem{Hirschfeld:yq}
P.~Hirschfeld,
Nucl.\ Phys.\ B {\bf 157} (1979) 37;
R.~Friedberg, T.~D.~Lee, Y.~Pang and H.~C.~Ren,
Annals Phys.\  {\bf 246} (1996) 381.

\bibitem{vanBaal:1997gu}
P.~van Baal,
arXiv:hep-th/9711070;
M.~Semenov-Tyan-Shanskii and V.~Franke, Zap.\ Nauch.\ Sem.\ Leningrad. Otdeleniya Matematicheskogo Institutia im V.~A.~Stekolov, AN SSSR, Vol.\ 120 (1982), 159 (In English translation: New York, Plenum Press 1986);
G.~Dell'Antonio, D.~Zwanziger, Proceedings of the NATO Advanced Workshop on Probabilistic Methods in Quantum Field Theory and Quantum Gravity, Cargese. New York, Plenum Press (1986), 21.

\bibitem{Zwanziger:1993dh}
D.~Zwanziger,
Nucl.\ Phys.\ B {\bf 412} (1994) 657.

\bibitem{Cucchieri:1997ns}
A.~Cucchieri,
Nucl.\ Phys.\ B {\bf 521} (1998) 365;
T.~D.~Bakeev, E.~M.~Ilgenfritz, V.~K.~Mitrjushkin and M.~Mueller-Preussker,
Phys.\ Rev.\ D {\bf 69} (2004) 074507
[arXiv:hep-lat/0311041].

\bibitem{Faddeev:fc}
L.~D.~Faddeev and V.~N.~Popov,
Phys.\ Lett.\ B {\bf 25} (1967) 29.

\bibitem{Alkofer:2003jr}
R.~Alkofer, C.~S.~Fischer, H.~Reinhardt and L.~von Smekal,
Phys.\ Rev.\ D {\bf 68} (2003) 045003
[arXiv:hep-th/0304134].

\bibitem{Peskin:ev}
M.~E.~Peskin and D.~V.~Schroeder,
``An Introduction To Quantum Field Theory,''
Reading, USA: Addison-Wesley (1995) 842 p.

\bibitem{Becchi:1975nq}
C.~Becchi, A.~Rouet and R.~Stora,
Annals Phys.\  {\bf 98} (1976) 287;
I.~V.~Tyutin,
Lebedev Institute preprint (1975), unpublished;
M.~Z.~Iofa and I.~V.~Tyutin,
Teor.\ Mat.\ Fiz.\  {\bf 27} (1976) 38.

\bibitem{Henneaux:1992ig}
M.~Henneaux and C.~Teitelboim,
``Quantization of gauge systems,'' Princeton, USA: Univ. Pr. (1992) 520 p;
G.~Barnich, F.~Brandt and M.~Henneaux,
Phys.\ Rept.\  {\bf 338} (2000) 439
[arXiv:hep-th/0002245].

\bibitem{Kugo:gm}
T.~Kugo and I.~Ojima,
Prog.\ Theor.\ Phys.\ Suppl.\  {\bf 66} (1979) 1.

\bibitem{Weinberg:1996kr}
S.~Weinberg,
``The quantum theory of fields. Vol. 2: Modern applications,'' Cambridge, UK: Univ. Pr. (1996) 489 p.

\bibitem{Taylor:ff}
J.~C.~Taylor,
Nucl.\ Phys.\ B {\bf 33} (1971) 436.

\bibitem{Slavnov:fg}
A.~A.~Slavnov,
Theor.\ Math.\ Phys.\  {\bf 10} (1972) 99
[Teor.\ Mat.\ Fiz.\  {\bf 10} (1972) 153];
G.~'t Hooft,
Nucl.\ Phys.\ B {\bf 33} (1971) 173;
G.~'t Hooft and M.~J.~G.~Veltman,
Nucl.\ Phys.\ B {\bf 50} (1972) 318;
B.~W.~Lee and J.~Zinn-Justin,
Phys.\ Rev.\ D {\bf 5} (1972) 3121.

\bibitem{vonSmekal:1997is}
L.~von Smekal, R.~Alkofer and A.~Hauck,
Phys.\ Rev.\ Lett.\  {\bf 79} (1997) 3591
[arXiv:hep-ph/9705242];
L.~von Smekal, A.~Hauck and R.~Alkofer,
Annals Phys.\  {\bf 267} (1998) 1
[Erratum-ibid.\  {\bf 269} (1998) 182]
[arXiv:hep-ph/9707327].

\bibitem{Marciano:su}
W.~J.~Marciano and H.~Pagels,
Phys.\ Rept.\  {\bf 36} (1978) 137.

\bibitem{Watson:phd}
P. Watson, PhD thesis, University of Durham, UK, 2000

\bibitem{Weinberg:mt}
S.~Weinberg,
``The Quantum Theory Of Fields. Vol. 1: Foundations,'' Cambridge, UK: Univ. Pr. (1995) 609 p.

\bibitem{Haag:1992hx}
R.~Haag,
``Local quantum physics: Fields, particles, algebras,'' Berlin, Germany: Springer (1992) 356 p. (Texts and monographs in physics).

\bibitem{Itzykson:rh}
C.~Itzykson and J.~B.~Zuber,
``Quantum Field Theory,'' New York, Usa: Mcgraw-hill (1980) 705 P.(International Series In Pure and Applied Physics).

\bibitem{Osterwalder:dx}
K.~Osterwalder and R.~Schrader,
Commun.\ Math.\ Phys.\  {\bf 31} (1973) 83;
K.~Osterwalder and R.~Schrader,
Commun.\ Math.\ Phys.\  {\bf 42}, 281 (1975).

\bibitem{Oehme:bj}
R.~Oehme and W.~Zimmermann,
Phys.\ Rev.\ D {\bf 21} (1980) 1661;
R.~Oehme and W.~Zimmermann,
Phys.\ Rev.\ D {\bf 21} (1980) 471.

\bibitem{Kugo:1995km}
T.~Kugo,
arXiv:hep-th/9511033.

\bibitem{Zwanziger:2001kw}
D.~Zwanziger,
Phys.\ Rev.\ D {\bf 65} (2002) 094039
[arXiv:hep-th/0109224].

\bibitem{Zwanziger:2002ia}
D.~Zwanziger,
Phys.\ Rev.\ D {\bf 67} (2003) 105001
[arXiv:hep-th/0206053].

\bibitem{Birmingham:1991ty}
D.~Birmingham, M.~Blau, M.~Rakowski and G.~Thompson,
Phys.\ Rept.\  {\bf 209} (1991) 129.

\bibitem{schleifenbaum:diploma}
W. Schleifenbaum, diploma thesis, TU Darmstadt, 2004.\\
W.~Schleifenbaum, A.~Maas, J.~Wambach and R.~Alkofer,
arXiv:hep-ph/0411052.

\bibitem{Greensite:2004bz}
J.~Greensite, S.~Olejnik and D.~Zwanziger,
arXiv:hep-lat/0408023;
J.~Greensite, S.~Olejnik and D.~Zwanziger,
arXiv:hep-lat/0407032, and references therein.

\bibitem{Habel:1990tw}
U.~Habel, R.~Konning, H.~G.~Reusch, M.~Stingl and S.~Wigard,
Z.\ Phys.\ A {\bf 336} (1990) 435.

\bibitem{Habel:1989aq}
U.~Habel, R.~Konning, H.~G.~Reusch, M.~Stingl and S.~Wigard,
Z.\ Phys.\ A {\bf 336} (1990) 423;
M.~Stingl,
Z.\ Phys.\ A {\bf 353} (1996) 423
[arXiv:hep-th/9502157].

\bibitem{Fischer:2003rp}
C.~S.~Fischer and R.~Alkofer,
Phys.\ Rev.\ D {\bf 67} (2003) 094020
[arXiv:hep-ph/0301094].

\bibitem{Appelquist:vg}
T.~Appelquist and R.~D.~Pisarski,
Phys.\ Rev.\ D {\bf 23} (1981) 2305.

\bibitem{Kaczmarek:2004gv}
O.~Kaczmarek, F.~Karsch, F.~Zantow and P.~Petreczky,
arXiv:hep-lat/0406036.

\bibitem{Alkofer:2003jk}
R.~Alkofer, W.~Detmold, C.~S.~Fischer and P.~Maris,
arXiv:hep-ph/0309078;
R.~Alkofer, W.~Detmold, C.~S.~Fischer and P.~Maris,
Phys.\ Rev.\ D {\bf 70} (2004) 014014
[arXiv:hep-ph/0309077].

\bibitem{Engelhardt:2003wm}
M.~Engelhardt, M.~Quandt and H.~Reinhardt,
Nucl.\ Phys.\ B {\bf 685} (2004) 227
[arXiv:hep-lat/0311029];
A.~Di Giacomo,
Nucl.\ Phys.\ Proc.\ Suppl.\  {\bf 121} (2003) 320,
and references therein.

\bibitem{Atiyah:1968mp}
M.~F.~Atiyah and I.~M.~Singer,
Annals Math.\  {\bf 87} (1968) 484.

\bibitem{Banks:1979yr}
T.~Banks and A.~Casher,
Nucl.\ Phys.\ B {\bf 169} (1980) 103.

\bibitem{Dyson:1949ha}
F.~J.~Dyson,
Phys.\ Rev.\  {\bf 75} (1949) 1736;
J.~S.~Schwinger,
Proc.\ Nat.\ Acad.\ Sci.\  {\bf 37} (1951) 452;
Proc.\ Nat.\ Acad.\ Sci.\  {\bf 37} (1951) 455.

\bibitem{Das:gg}
A.~K.~Das,
``Finite Temperature Field Theory'', Singapore: World Scientific (1997).

\bibitem{Bernard:1974bq}
C.~W.~Bernard,
Phys.\ Rev.\ D {\bf 9} (1974) 3312.

\bibitem{Weldon:aq}
H.~A.~Weldon,
Phys.\ Rev.\ D {\bf 26} (1982) 1394.

\bibitem{vanHees:2001ik}
H.~van Hees and J.~Knoll,
Phys.\ Rev.\ D {\bf 65} (2002) 025010
[arXiv:hep-ph/0107200];
H.~van Hees and J.~Knoll,
Phys.\ Rev.\ D {\bf 65} (2002) 105005
[arXiv:hep-ph/0111193];
H.~van Hees and J.~Knoll,
Phys.\ Rev.\ D {\bf 66} (2002) 025028
[arXiv:hep-ph/0203008];
Z.~Aouissat,
Nucl.\ Phys.\ A {\bf 642} (1998) 210
[arXiv:hep-ph/9903316].

\bibitem{Bloch:2003yu}
J.~C.~R.~Bloch,
Few Body Syst.\  {\bf 33} (2003) 111
[arXiv:hep-ph/0303125].

\bibitem{Boucaud:1998xi}
P.~Boucaud, J.~P.~Leroy, J.~Micheli, O.~Pene and C.~Roiesnel,
JHEP {\bf 9812} (1998) 004
[arXiv:hep-ph/9810437];
P.~Boucaud, J.~P.~Leroy, J.~Micheli, O.~Pene and C.~Roiesnel,
JHEP {\bf 9810} (1998) 017
[arXiv:hep-ph/9810322];
H.~Nakajima and S.~Furui,
Nucl.\ Phys.\ Proc.\ Suppl.\  {\bf 129} (2004) 730
[arXiv:hep-lat/0309165];
S.~Furui and H.~Nakajima,
arXiv:hep-lat/0309166.

\bibitem{Cucchieri:2004sq}
A.~Cucchieri, T.~Mendes and A.~Mihara,
arXiv:hep-lat/0408034;
A.~Mihara, A.~Cucchieri and T.~Mendes,
arXiv:hep-lat/0408021.

\bibitem{Lerche:2002ep}
C.~Lerche and L.~von Smekal,
Phys.\ Rev.\ D {\bf 65} (2002) 125006
[arXiv:hep-ph/0202194].

\bibitem{Watson:2001yv}
P.~Watson and R.~Alkofer,
Phys.\ Rev.\ Lett.\  {\bf 86} (2001) 5239
[arXiv:hep-ph/0102332];
R.~Alkofer, L.~von Smekal and P.~Watson,
arXiv:hep-ph/0105142.

\bibitem{Fischer:2002hn}
C.~S.~Fischer  and R.~Alkofer,
Phys. Lett. {\bf B536}, 177 (2002)
[arXiv:hep-ph/0202202].

\bibitem{Alkofer:2002ne}
R.~Alkofer, C.~S.~Fischer and L.~von Smekal,
Acta Phys.\ Slov.\  {\bf 52}, 191 (2002) [arXiv:hep-ph/0205125];
C.~S.~Fischer, R.~Alkofer and H.~Reinhardt,
Phys.\ Rev.\ {\bf D65}, 094008 {2002} [arXiv:hep-ph/0202195];

\bibitem{Fischer:2003zc}
C.~S.~Fischer,
PhD thesis, U. of T\"ubingen, arXiv:hep-ph/0304233.

\bibitem{Collins:xc}
J.~C.~Collins,
``Renormalization. An Introduction To Renormalization, The Renormalization
Group, And The Operator Product Expansion,'' Cambridge, Uk: Univ. Pr. (1984) 380p.

\bibitem{Zinn-Justin:mi}
J.~Zinn-Justin,
``Quantum Field Theory And Critical Phenomena,'' Oxford, UK: Clarendon (1989) 914 p. (International series of monographs on physics, 77).

\bibitem{Elitzur:im}
S.~Elitzur,
Phys.\ Rev.\ D {\bf 12} (1975) 3978.

\bibitem{Mandelstam:1979xd}
S.~Mandelstam,
Phys.\ Rev.\ D {\bf 20} (1979) 3223.

\bibitem{Langfeld:2002dd}
K.~Langfeld, J.~C.~R.~Bloch, J.~Gattnar, H.~Reinhardt, A.~Cucchieri and T.~Mendes,
arXiv:hep-th/0209173.

\bibitem{Bowman:2004jm}
P.~O.~Bowman, U.~M.~Heller, D.~B.~Leinweber, M.~B.~Parappilly and A.~G.~Williams,
Phys.\ Rev.\ D {\bf 70} (2004) 034509
[arXiv:hep-lat/0402032].

\bibitem{Cucchieri:2003di}
A.~Cucchieri, T.~Mendes and A.~R.~Taurines,
Phys.\ Rev.\ D {\bf 67} (2003) 091502
[arXiv:hep-lat/0302022].

\bibitem{Gies:2002af}
H.~Gies,
Phys.\ Rev.\ D {\bf 66} (2002) 025006
[arXiv:hep-th/0202207];
J.~M.~Pawlowski, D.~F.~Litim, S.~Nedelko and L.~von Smekal,
arXiv:hep-th/0312324.

\bibitem{Roberts:2000aa}
C.~D.~Roberts and S.~M.~Schmidt,
Prog.\ Part.\ Nucl.\ Phys.\  {\bf 45} (2000) S1
[arXiv:nucl-th/0005064].

\bibitem{Bowman:2002kn}
P.~O.~Bowman, U.~M.~Heller, D.~B.~Leinweber and A.~G.~Williams,
Nucl.\ Phys.\ Proc.\ Suppl.\  {\bf 119} (2003) 323
[arXiv:hep-lat/0209129].

\bibitem{Maas:2004se}
A.~Maas, J.~Wambach, B.~Gr\"uter and R.~Alkofer,
Eur.\ Phys.\ J.\ {\bf C37}, No.3, (2004) 335
[arXiv:hep-ph/0408074].

\bibitem{Cucchieri:2001tw}
A.~Cucchieri, F.~Karsch and P.~Petreczky,
Phys.\ Rev.\ D {\bf 64} (2001) 036001
[arXiv:hep-lat/0103009].

\bibitem{Zwanziger:2003de}
D.~Zwanziger,
arXiv:hep-ph/0312254.

\bibitem{Maas:2002if}
A.~Maas, B.~Gr\"uter, R.~Alkofer and J.~Wambach,
arXiv:hep-ph/0210178.

\bibitem{Brown:1988bm}
N.~Brown and M.~R.~Pennington,
Phys.\ Rev.\ D {\bf 38} (1988) 2266.

\bibitem{Kajantie:1995dw}
K.~Kajantie, M.~Laine, K.~Rummukainen and M.~E.~Shaposhnikov,
Nucl.\ Phys.\ B {\bf 458} (1996) 90.

\bibitem{'tHooft:1973jz}
G.~'t Hooft,
Nucl.\ Phys.\ B {\bf 72} (1974) 461.

\bibitem{Feynman:1981ss}
R.~P.~Feynman,
Nucl.\ Phys.\ B {\bf 188} (1981) 479.

\bibitem{Atkinson:1998zc}
D.~Atkinson and J.~C.~R.~Bloch,
Mod.\ Phys.\ Lett.\ A {\bf 13} (1998) 1055
[arXiv:hep-ph/9802239].

\bibitem{Feuchter:2004gb}
C.~Feuchter and H.~Reinhardt,
arXiv:hep-th/0402106.

\bibitem{Blaizot:2001nr}
J.~P.~Blaizot and E.~Iancu,
Phys.\ Rept.\  {\bf 359} (2002) 355
[arXiv:hep-ph/0101103] and references therein.

\bibitem{Appelquist:tg}
T.~Appelquist and J.~Carazzone,
Phys.\ Rev.\ D {\bf 11} (1975) 2856.

\bibitem{Gruter:2004kd}
B.~Gr\"uter, R.~Alkofer, A.~Maas and J.~Wambach,
arXiv:hep-ph/0408282.

\bibitem{Gruter:phd}
B.~Gr\"uter, PhD thesis, 2005.

\bibitem{Press:1997}
W.~H.~Press, S.~A.~Teukolsky, W.~T.~Vetterling, B.~P.~Flannery,
``Numerical recipes in C'', Cambridge University Press, Cambridge, 1997.

\bibitem{Cucchieri:2004mf}
A.~Cucchieri, T.~Mendes and A.~R.~Taurines,
arXiv:hep-lat/0406020.

\bibitem{Gradstein:1981}
I.~S.~Gradstein, I.~M.~Ryshik,
``Tables of series, products and integrals'', Vol. 1 and 2, Verlag Harri Deutsch, Frankfurt/Main, 1981.

\bibitem{Luttinger:1960ua}
J.~M.~Luttinger and J.~C.~Ward,
Phys.\ Rev.\  {\bf 118} (1960) 1417;

\bibitem{Cornwall:1974vz}
J.~M.~Cornwall, R.~Jackiw and E.~Tomboulis,
Phys.\ Rev.\ D {\bf 10} (1974) 2428.

\bibitem{Haeri:hi}
B.~J.~Haeri,
Phys.\ Rev.\ D {\bf 48} (1993) 5930
[arXiv:hep-ph/9309224].
M.~E.~Carrington, G.~Kunstatter and H.~Zaraket,
arXiv: hep-ph/0309084;
M.~E.~Carrington,
arXiv:hep-ph/0401123;
A.~Arrizabalaga and J.~Smit,
Phys.\ Rev.\ D {\bf 66} (2002) 065014
[arXiv:hep-ph/0207044].

\bibitem{Zwanziger:2004np}
D.~Zwanziger,
arXiv:hep-ph/0407103.

\bibitem{Bali:1993tz}
G.~S.~Bali, J.~Fingberg, U.~M.~Heller, F.~Karsch and K.~Schilling,
Phys.\ Rev.\ Lett.\  {\bf 71} (1993) 3059
[arXiv:hep-lat/9306024].

\bibitem{Nakamura:2003pu}
A.~Nakamura, T.~Saito and S.~Sakai,
Phys.\ Rev.\ D {\bf 69} (2004) 014506
[arXiv:hep-lat/0311024].

\bibitem{Rischke:2003mt}
D.~H.~Rischke,
Prog.\ Part.\ Nucl.\ Phys.\  {\bf 52} (2004) 197
[arXiv:nucl-th/0305030];
P.~T.~Reuter, Q.~Wang and D.~H.~Rischke,
arXiv:nucl-th/0405079.

\bibitem{Epple:dipl}
D.~Epple, diploma thesis, University of T\"ubingen, 2003.

\bibitem{Hauck:1998fz}
A.~Hauck, L.~von Smekal and R.~Alkofer,
Comput.\ Phys.\ Commun.\  {\bf 112} (1998) 166
[arXiv:hep-ph/9804376];
J.~C.~R.~Bloch,
arXiv:hep-ph/0208074.

\bibitem{Kelley}
C.~Kelley, 
``Iterative methods for linear and non-linear equations'', Philadelphia: Siam (1995).

\bibitem{Maas:ccp}
A.~Maas, in preparation.

\end{thebibliography}
